\documentstyle[prd,aps]{revtex}
\begin{document}
\draft
\title{
Out of Equilibrium Relativistic Quantum Field Theory --- 
Perturbation Theory and Progress of Phase Transition} 
\author{A. Ni\'{e}gawa} 
\address{Department of Physics, Osaka City University, Sumiyoshi-ku, 
Osaka 558, JAPAN \\ 
E-Mail: niegawa@sci.osaka-cu.ac.jp} 
\maketitle 
\begin{center} 
Abstract 
\end{center} 
This paper describes perturbative framework, on the basis of 
closed-time-path formalism, for studying quasiuniform relativistic 
quantum field systems near equilibrium and nonequilibrium 
quasistationary systems. At the first part, starting from first 
principles, we construct perturbative schemes for relativistic 
complex-scalar-field theory. We clarify what assumption is involved 
in arriving at a standard perturbative framework and to what extent 
the $n \, (\geq 4)$-point initial correlation functions that are 
usually discarded in the standard framework can in fact be 
discarded. Two calculational schemes are introduced, the one is 
formulated on the basis of the initial particle distribution 
function and the one is formulated on the basis of the \lq\lq 
physical'' particle distribution function. Both schemes are 
equivalent and lead to a generalized relativistic kinetic or 
Boltzmann equation. At the second part, using the perturbative 
loop-expansion scheme for an $O (N)$ linear $\sigma$ model, we 
analyze how the chiral phase transition proceeds through disoriented 
chiral condensates. The system of coupled equations that governs the 
spacetime evolution of the condensate or order-parameter fields is 
derived. The region where the curvature of the \lq\lq potential'' is 
negative is dealt with by introducing the random-force fields. 
Application to simple situations is made. 
\newpage 
\setcounter{equation}{0} 
\setcounter{section}{0} 
\section{Introduction} 
\def\theequation{\mbox{\arabic{section}.\arabic{equation}}} 
Since mid-fifties, efforts have been made to incorporate quantum 
field theory with nonequilibrium statistical mechanics 
\cite{sch,chou,lan,ume}. Necessity of this incorporation originated 
from the field of solid-state physics. Since then, rapid progress of 
the studies of the early Universe and the quark-gluon plasma have 
further activated this field of research (cf., e.g., 
\cite{TFT,TFT1}). The quark-gluon plasma is a system of (anti)quarks 
and gluons, which is governed by quantum chromodynamics (QCD). 
According to QCD, there are two types of phase transitions; a chiral 
phase transition and a confined-deconfined phase transition. 
According to lattice analyses, critical points for these two 
transitions are very close. The quark-gluon plasma is the system 
that is in the deconfined and chiral-symmetric phases and is 
expected to be produced in relativistic heavy-ion collisions and to 
have existed in the early Universe. Such quark-gluon plasmas evolve 
into a confined (or hadronic) and chiral-symmetry breaking phases 
through the phase transitions. The study of this evolution on the 
basis of QCD is very involved and some tractable effective theories 
have been proposed. Among those is a linear $\sigma$ model, which 
was proposed long ago \cite{Gell} and is revived as a toy model that 
mimic the aspect of the chiral phase transition of 
QCD\footnote{Lattice simulations of finite-temperature QCD 
indicates that the chiral phase transition is of second order (see, 
e.g., \cite{uka}). The phase transition described by the linear 
$\sigma$ model is of second order.}. A numerous work was reported so 
far on the \lq\lq static'' properties of this model or its variants. 
(See, e.g., \cite{pis,chiku}. Earlier work is quoted therein.) 
Recently, analyses of dynamical evolution of the system from a 
symmetric phase to a broken symmetric phase have begun (see, e.g., 
\cite{tsunami}). Such an analysis is still involved and different 
approximations have been employed by various authors. 

The purpose of this paper is twofold. In the first part, we aim at 
laying perturbative framework for dealing with out-of-equilibrium 
relativistic quantum complex-scalar-field systems. In the second 
part, taking up an fermionless $O (N)$ linear $\sigma$ model, we lay 
down the framework, based on the perturbative loop expansion of the 
effective action, for analyzing how the phase transition proceeds 
through disoriented chiral condensates (see, e.g., \cite{bjo}). We 
use the standard framework of nonequilibrium statistical 
quantum-field theory that is formulated \cite{sch,chou,lan,hu} by 
employing the closed-time path, $- \infty \to + \infty \to - 
\infty$, in a complex-time plane, which is referred to as the 
closed-time-path (CTP) formalism. Specializing the density matrix, 
which characterizes the initial states of the systems, to the one 
for the Gibbs distribution, the framework reduces to the standard 
framework of equilibrium thermal quantum field theories 
\cite{lan,le-b}. Throughout this paper, we are interested in 
quasiuniform systems near equilibrium or nonequilibrium 
quasistationary systems, which we simply refer to as 
out-of-equilibrium systems. This means, in particular, that the 
phase transition dealt with in the second part proceeds slowly. 

The out-of equilibrium systems are characterized by two different 
spacetime scales: microscopic or quantum-field-theoretical and 
macroscopic or statistical. The first scale, the 
microscopic-correlation scale, characterizes the range of radiative 
correction to reactions taking place in the system while the second 
scale measures the range of interaction of particles. For a 
weak-coupling theory, in which we are interested in this paper, the 
former scale is much smaller than the latter scale. A well-known 
intuitive picture (cf., e.g., \cite{hu}) for dealing with such 
systems is to separate spacetime into many \lq\lq cells'' whose 
characteristic size, $L^\mu$ $(\mu = 0, ..., 3)$, is in between the 
microscopic and macroscopic scales. It is assumed that the 
correlation between different cells is negligible in the sense that 
microscopic or elementary reactions can be regarded as taking place 
in a single cell. On the other hand, in a single cell, relaxation 
phenomena are negligible. Thus, in dealing with a reaction, all the 
vertices in a Feynman diagram for the reaction are in a single cell. 
In more precise, propagators are Fourier transformed over a cell and 
carry an \lq\lq index'' that labels the cell. 

Above intuitive picture may be implemented as follows. Let $\Delta 
(x, y)$ be a generic propagator or two-point function. An inverse 
Fourier transformation with respect to the relative coordinates $x - 
y$ yields 
\begin{equation} 
\Delta (X; P) \equiv \int d^{\, 4} (x - y)  \, e^{i P \cdot (x - y)} 
\, \Delta (x, y) , 
\label{Fourier} 
\end{equation} 
where $X = (x + y) / 2$ is the midpoint, which is chosen as a label 
for the cells. Sometimes, $X$ is called the macroscopic spacetime 
coordinates. It is obvious from the above argument that 
(\ref{Fourier}) is meaningful for $|P^\mu| \gtrsim 1 / L^\mu$. The 
self-energy part $\Sigma (x, y)$ enjoys similar property. 

Development of the phase transition is described by an evolution 
equation for a set of condensate or order-parameter fields 
$\vec{\varphi} (x)$. As mentioned above, we assume that the phase 
transition proceeds slowly. By this we mean that the $x$ dependence 
of $\vec{\varphi} (x)$ is weak, or more precisely, $x$ of 
$\vec{\varphi} (x)$ may be treated as a macroscopic spacetime 
coordinates. It should be noted here that, as the system approaches 
the critical point of the phase transition, the microscopic 
correlation scale diverges. This means that the formalism applies to 
the systems away from the critical region. Recalling that the phase 
transition occurs when the squared effective mass, $M^2 (x)$, of the 
quasiparticles becomes zero, we can say that the formalism may be 
used in the region, where $|M^2 (x)|$ is not small as compared to 
the typical scale parameter(s) of the system. 

Microscopic reactions discussed above cause change in the density 
matrix, through which the number densities of quasiparticles 
change with macroscopic spacetime $X^\mu$. Dealing with this is the 
subject of the \lq\lq next stage,'' where (weak) $X^\mu$-dependence 
of $\Delta (X; P)$, $\Sigma (X; P)$, and $\vec{\varphi} (X)$ are 
explicitly taken into account. 

As mentioned above, much work has already been devoted to the issues 
settled above. However, most work devoted to the first issue (to be 
dealt with in the first part of this paper) assume (explicitly or 
implicitly) special forms for the density matrix. Also as mentioned 
above, the second issue has so far been studied under various 
approximations. In this paper, starting from first principles, we 
proceed with discussing matters in a context which is as general as 
possible. 

The plan of the paper is as follows: In Section II, a brief 
introduction is given to the CTP formalism for out-of-equilibrium 
quantum field theories. In Sections III~-~V, taking up a 
self-interacting relativistic complex-scalar quantum field theory, 
perturbative calculational schemes are constructed. In Section III, 
standard representation of the CTP Green functions in the 
interaction picture is derived from first principles. In the course 
of derivation, we clarify what assumption is necessary in arriving 
at the representation. In a standard perturbative scheme, the $n \, 
(\geq 4)$-point initial correlation functions are also usually 
discarded. We discuss to what extent they can really be discarded. 
Whenever necessary, they can be incorporated into the scheme (cf. 
\cite{chou}). In Section IV, we deal with the bare propagator 
(two-point function), which is a building block of perturbation 
theory. The quasiparticle fields are introduced through a sort of 
Bogoliubov transformation, which is a generalization of the 
so-called thermal Bogoliubov transformation in equilibrium thermal 
field theory \cite{lan,ume,le-b}. Quasiparticle fields describe, in 
a sense, stable modes in the system. With the aid of the 
quasiparticle picture, we obtain a $2 \times 2$ matrix 
representation for the bare propagator. One of the advantages of 
this picture is that the elementary-reaction part and the relaxation 
part, mentioned above, are clearly identified. At the final part of 
Section IV, we propose two perturbative schemes, which are 
equivalent to each other. In Section V, through analyzing the 
structure of $2 \times 2$ self-energy matrix, we find the form for 
the self-energy-part resummed or full propagator. Then, after 
discussing how the two perturbative schemes work, we show that both 
schemes are equivalent and lead to a generalized relativistic 
kinetic or Boltzmann equation. Sections VI~-~VIII are devoted to the 
analysis of a fermionless $O (N)$ linear $\sigma$ model, with the 
aid of the effective action. [The case of $N = 4$ is of practical 
interest.] Here, the condensate or order-parameter fields and the 
quantum fields come in. The latter describe quantum fluctuations 
around the former. The condensate fields, which define the internal 
reference frames for the quantum fields, change in spacetime. Taking 
this fact into account, in Section VI, we construct consistent 
perturbative loop-expansion scheme for studying how the chiral phase 
transition proceeds through disoriented chiral condensates. Besides 
the assumption that the transition proceed slowly, no further 
assumption is introduced. In Section VII, we compute the effective 
action to the lowest one-loop order and derive the system of coupled 
equations that governs the evolution of the condensate fields, the 
equations which consist of the generalized relativistic Boltzmann 
equation, the self-consistent gap equation, and the equation of 
motion for the condensate fields. At the region where the curvature 
of the \lq\lq potential'' is negative, the effective action $\Gamma$ 
develops imaginary part. It is shown that $Im \, \Gamma$ can be 
dealt with by introducing the random-force fields that act on the 
system. The random-force fields cause negative correlation between 
the condensate fields. In Section VIII, we set up simple situations, 
for which approximate analytic calculations may be carried out. 
Section IX is devoted to concluding remarks. 
\subsubsection*{Remarks on notations and the gradient approximation} 
Let $G (x, y)$ be (a component of) a generic two-point function. We 
introduce two kinds of spacetime coordinates: 
\begin{eqnarray*} 
& & x - y: \mbox{relative or microscopic spacetime coordinates,} \\ 
& & X \equiv \frac{x + y}{2}: \mbox{center of mass or macroscopic 
spacetime coordinates.} 
\end{eqnarray*} 
As mentioned above, $X$-dependence of $G (x, y)$ is assumed to be 
weak. Fourier transform of $G (x, y)$ with respect to $x - y$ reads 
(cf. (\ref{Fourier}))
\[ 
G (x, y) = \int \frac{d^{\, 4} P}{(2 \pi)^4} \, e^{ - i P \cdot (x - 
y)} G (X; P) . 
\] 
Throughout this paper, we use upper-case letters, $P$, etc., to 
denote a four-momentum vectors, $P = (P^\mu) = (p_0, {\bf p})$, etc. 
The magnitude of three vector ${\bf p}$ will be written as $p = 
|{\bf p}|$. As to the $X$-dependence of $G (X; P)$, throughout this 
paper, we keep only up to a first-order-derivative term (gradient 
approximation), 
\begin{equation} 
G (X; P) \simeq G (Y; P) + (X - Y) \cdot \partial_Y G (Y; P) . 
\label{kinji0} 
\end{equation} 

Functions of the type, $B ({\bf x}, {\bf y}; z_0)$ also enter in 
Sections IV and VI. Here ${\bf X}$ $(= ({\bf x} + {\bf y}) / 2)$ and 
$z_0$ are macroscopic spacetime coordinates, i.e., $B ({\bf x}, {\bf 
y}; z_0)$ depends on ${\bf X}$ and on $z_0$ only weakly. Fourier 
transform of $B ({\bf x}, {\bf y}; z_0)$ on ${\bf x} - {\bf y}$ 
reads 
\[ 
B ({\bf x}, {\bf y}; z_0) = \int \frac{d^{\, 3} p}{(2 \pi)^3} \, e^{ 
i {\bf p} \cdot ({\bf x} - {\bf y})} B ({\bf X}, z_0; {\bf p}) . 
\] 
Whenever necessary, we expand $B$ within the gradient approximation: 
\begin{equation} 
B ({\bf X}, z_0; {\bf p})  \simeq B ({\bf Y}, x_0; {\bf p}) + (z_0 - 
x_0) \frac{\partial B ({\bf Y}, x_0; {\bf p})}{\partial x_0}  + 
({\bf X} - {\bf Y}) \cdot \nabla_{\bf Y} B ({\bf Y}, x_0; {\bf p}) . 
\label{kinji1} 
\end{equation} 
For the system of our concern, $|\partial B / \partial x_0| \lesssim 
|B| / L_0$, $|\partial B / \partial x^i| \lesssim |B| / L^i$ with 
$L^\mu$ as in Section I. 

In Sections VI~-~VIII, there comes in the condensate fields 
$\vec{\varphi}$. Although the argument of $\vec{\varphi}$ is always 
macroscopic spacetime coordinates, we write $\vec{\varphi} (x)$ or 
$\vec{\varphi} (X)$, depending on the contexts. 
\setcounter{equation}{0} 
\setcounter{section}{1} 
\def\theequation{\mbox{\arabic{section}.\arabic{equation}}} 
\section{Closed-time-path formalism} 
Let us start with a generic Lagrangian density 
\[ 
{\cal L} \left( \phi^{(\alpha)} (x), \phi^{(\beta)} (x), ... \right) 
\] 
that governs the quantum-field systems. Here the superscripts label 
collectively the field type and internal degrees of freedom. 
Following standard procedure, we introduce an oriented  closed-time 
path $C$ $(= C_1 \oplus C_2)$ in a complex time plane, which goes 
from $- \infty$ to $+ \infty$ ($C_1$) and then returns back from $+ 
\infty$ to $- \infty$ ($C_2$). The time argument $x_0$ of the 
fields, $\phi^{(\alpha)} (x)$'s, is on the time path $C$. Along this 
closed time path, the classical action is defined as 
\begin{eqnarray} 
{\cal A} & = & \int_C d x_0 \int d^{\, 3} x \, {\cal L} \left( 
\phi^{(\alpha)} (x), \phi^{(\beta)} (x), ... \right) \nonumber \\ 
& \equiv & \int_C {\cal L} \left( \phi^{(\alpha)} (x), 
\phi^{(\beta)} (x), ... \right) . 
\label{ac} 
\end{eqnarray} 
Define a path-ordered product of field operators along $C$: 
\[ 
T_C \left( \phi^{(\alpha_1)} (x_1), ...,  \phi^{(\alpha_n)} (x_n) 
\right) . 
\] 
Here $T_C$ means to rearrange the field operators as follows: When 
$x_{j 0}$ of $\phi^{(\alpha_j)} (x_j)$ is nearer to the end point of 
$C$, $- \infty \in C_2$, than $x_{i 0}$ of $\phi^{(\alpha_i)} 
(x_i)$, $\phi^{(\alpha_j)} (x_j)$ is placed on the left of 
$\phi^{(\alpha_i)} (x_i)$ with the sign due to the statistics of the 
relevant fields taken into account. Path-ordered Green function is 
defined as an average over ensemble prepared at an initial time $t = 
- \infty$, which is characterized by a density matrix $\rho$: 
\begin{eqnarray} 
G & \equiv & (- i)^{n - 1} \mbox{Tr} \left[ T_C \left( 
\phi^{(\alpha_1)} (x_1) ...  \phi^{(\alpha_n)} (x_n) \right) \rho 
\right] \nonumber \\ 
& \equiv & (- i)^{n - 1} \langle T_C \left( \phi^{(\alpha_1)} (x_1) 
...  \phi^{(\alpha_n)} (x_n) \right) \rangle . 
\label{green} 
\end{eqnarray} 
\noindent In particular, the two-point function is defined as 
\[ 
G^{(\alpha \beta)} (x, y) = - i \, \langle T_C \left( 
\phi^{(\alpha)} (x) \, \phi^{(\beta)} (y) \right) \rangle . 
\] 
Let us introduce \cite{chou} matrix notation $\hat{G}^{(\alpha 
\beta)} (x, y)$, whose ($i, j$) component $G_{i j}^{(\alpha \beta)} 
(x, y)$ stands for $G^{(\alpha \beta)} (x, y)$ with $x_0 \in C_i$ 
and $y_0 \in C_j$: 
\begin{eqnarray} 
G_{1 1}^{(\alpha \beta)} (x, y) & = & - i \langle T \left( 
\phi^{(\alpha)} (x) \phi^{(\beta)} (y) \right) \rangle , 
\nonumber \\ 
G_{1 2}^{(\alpha \beta)} (x, y) & = & - i \sigma \langle 
\phi^{(\beta)} (y) \phi^{(\alpha)} (x) \rangle , \nonumber \\ 
G_{2 1}^{(\alpha \beta)} (x, y) & = & - i \langle 
\phi^{(\alpha)} (x) \phi^{(\beta)} (y) \rangle , \nonumber \\ 
G_{2 2}^{(\alpha \beta)} (x, y) & = & - i \langle \overline{T} 
\left( \phi^{(\alpha)} (x) \phi^{(\beta)} (y) \right) \rangle , 
\label{shu-pa} 
\end{eqnarray} 
where $T$ ($\overline{T}$) is a time- (an anti-time-) ordering 
symbol and $\sigma = -1$ when both $\phi^{(\alpha)}$ and 
$\phi^{(\beta)}$ are fermionic field and $\sigma = +1$ otherwise. In 
writing down (\ref{shu-pa}), the \lq\lq limit'' $\phi^{(\alpha)} 
(x_0 \in C_1, {\bf x}) = \phi^{(\alpha)} (x_0 \in C_2, {\bf x})$ has 
been taken. Throughout this paper, we freely take this 
limit.\footnote{In this respect, cf. the argument given below in 
conjunction with (\ref{trad}).} 

Let us break up the time path $C$ into two parts: 
\begin{equation} 
\int_C d x_0 = \int_{C_1} d x_0 + \int_{C_2} d x_0 = \int_{- 
\infty}^\infty d x_0 \, \rule[-3mm]{.14mm}{8.5mm} 
\raisebox{-2.85mm}{\scriptsize{$\; x_0 \in C_1$}} - \int_{- 
\infty}^\infty d x_0 \, \rule[-3mm]{.14mm}{8.5mm} 
\raisebox{-2.85mm}{\scriptsize{$\; x_0 \in C_2$}} . 
\label{wakeru}
\end{equation} 
Then, in conformity with the matrix notation, we may rewrite the 
classical action in (\ref{ac}) as 
\begin{eqnarray} 
{\cal A} & = & \int_{- \infty}^{+ \infty} d x_0 \int d^{\, 3} x \, 
\hat{\cal L} , \nonumber \\
\hat{\cal L} & = & {\cal L} \left( \phi^{(\alpha)}_1 (x), 
\phi^{(\beta)}_1 (x), ... \right) - {\cal L} \left( 
\phi^{(\alpha)}_2 (x), \phi^{(\beta)}_2 (x), ... \right) , 
\label{ac1} 
\end{eqnarray} 
where $\phi^{(\alpha)}_i (x)$ stands for the field whose time 
argument $x_0$ is on $C_i$ ($i = 1, 2$). The field 
$\phi^{(\alpha)}_1$ is called a type-1 or physical field while 
$\phi^{(\alpha)}_2$ is called a type-2 field. We call $\hat{\cal L}$ 
in (\ref{ac1}) the hat-Lagrangian density \cite{ume,ume1}. Let us 
write (\ref{shu-pa}) collectively 
\begin{equation} 
\hat{G}^{(\alpha \beta)} (x, y) = - i \langle \hat{T} \left( 
\hat{\phi}^{(\alpha)} (x) \; \displaystyle{ 
\raisebox{0.9ex}{\scriptsize{$t$}}} \mbox{\hspace{-0.1ex}} 
\hat{\phi}^{(\beta)} (y) \right) \rangle , 
\label{abr} 
\end{equation} 
where 
\[ 
\hat{\phi}^{(\alpha)} (x) = \left( 
\begin{array}{c} 
\phi^{(\alpha)}_1 (x) \\ 
\phi^{(\alpha)}_2 (x) 
\end{array} 
\right), \;\;\;\;\;\; \displaystyle{ 
\raisebox{0.9ex}{\scriptsize{$t$}}} \mbox{\hspace{-0.1ex}} 
\hat{\phi}^{(\beta)} (y) = (\phi^{(\beta)}_1 (y), 
\phi^{(\beta)}_2 (y)) . 
\] 

Four $G_{i j}^{(\alpha \beta)}$ ($i, j = 1, 2$) in (\ref{shu-pa}) 
are not mutually independent. As a matter of fact, it can readily be 
found from (\ref{shu-pa}) that 
\[ 
\sum_{i, \, j = 1}^2 (-)^{i + j} G_{i j}^{(\alpha \beta)} = 0 . 
\] 

Retarded, advanced, and correlation functions are related, in 
respective order, to $G_{i j}^{(\alpha, \beta)}$ through 
\begin{eqnarray} 
G_R^{(\alpha \beta)} (x, y) & = & - i \theta (x_0 - y_0) \langle 
\left[ \phi^{(\alpha)} (x), \phi^{(\beta)} (y) \right]_{- \sigma} 
\rangle \nonumber \\ 
& = & G_{1 1}^{(\alpha \beta)} (x, y) - G_{1 2}^{(\alpha \beta)} (x, 
y) , \nonumber \\ 
G_A^{(\alpha \beta)} (x, y) & = & i \theta (y_0 - x_0) \langle 
\left[ \phi^{(\alpha)} (x), \phi^{(\beta)} (y) \right]_{- \sigma} 
\rangle \nonumber \\ 
& = & G_{1 1}^{(\alpha \beta)} (x, y) - G_{2 1}^{(\alpha \beta)} (x, 
y) , \nonumber \\ 
G_c^{(\alpha \beta)} (x, y) & = & - i \langle \left[ \phi^{(\alpha)} 
(x), \phi^{(\beta)} (y) \right]_\sigma \rangle \nonumber \\ 
& = & G_{1 1}^{(\alpha \beta)} (x, y) + G_{2 2}^{(\alpha \beta)} (x, 
y) , 
\label{phys} 
\end{eqnarray} 
where $[A, B]_\sigma \equiv A B + \sigma B A$ $(\sigma = \pm)$. 
\setcounter{equation}{0} 
\setcounter{section}{2} 
\def\theequation{\mbox{\arabic{section}.\arabic{equation}}} 
\section{Complex scalar field} 
We take up self-interacting relativistic complex-scalar field theory 
with a conserved charge, whose Lagrangian density is 
\begin{eqnarray} 
{\cal L} & = & {\cal L}_0 + {\cal L}_{int} , \nonumber \\ 
{\cal L}_0 & = & \partial_\mu \phi^\dagger_B \partial^\mu \phi_B - 
m^2_B \phi^\dagger_B \phi_B , \nonumber \\ 
{\cal L}_{int} & = & - \frac{\lambda_B}{4} (\phi^\dagger_B \phi_B)^2 
, 
\label{Lag} 
\end{eqnarray} 
where the suffices \lq\lq $B$'' stand for bare quantity. Aiming at 
constructing perturbation theories, we go to an interaction picture. 
One can choose any value for the time $t = t_i$, when the 
interaction picture and the Heisenberg picture coincide. As in, 
e.g., \cite{chou,lan,IZ}, we choose $t_i = - \infty$. For dealing 
with ultraviolet (UV) divergences, throughout this paper, we adopt 
the modified minimal-subtraction ($\overline{\mbox{MS}}$) scheme 
\cite{muta}, which is a variant of mass-independent renormalization 
schemes. We introduce UV-renormalized field $\phi$, mass $m$, and 
coupling constant $\lambda$ through  
\begin{equation} 
\phi_B = \sqrt{Z} \phi, \;\;\;\;\; m_B^2 = Z_m m^2,  \;\;\;\;\; 
\lambda_B = Z_\lambda \lambda .
\label{kuri} 
\end{equation} 
Note that, although the same letter $\phi$ is used, $\phi$ here is 
the UV-renormalized field in the interaction picture, while $\phi$'s 
in Section II are the field in the Heisenberg picture. ${\cal L}$ in 
(\ref{Lag}) may now be written as 
\begin{eqnarray} 
{\cal L} & = & {\cal L}_0 (\phi^\dagger, \phi) + {\cal L}_{int} 
(\phi^\dagger, \phi) + {\cal L}_{rc} (\phi^\dagger, \phi) , 
\nonumber \\ 
{\cal L}_0 (\phi^\dagger, \phi) & = & \partial_\mu \phi^\dagger 
\partial^\mu \phi - m^2 \phi^\dagger \phi , \nonumber \\ 
{\cal L}_{int} (\phi^\dagger, \phi) & = & - \frac{\lambda}{4} 
(\phi^\dagger \phi)^2 , \nonumber \\ 
{\cal L}_{r c} (\phi^\dagger, \phi) & = & (Z - 1) \partial_\mu 
\phi^\dagger \partial^\mu \phi - (Z_m Z - 1) m^2 \phi^\dagger \phi 
- (Z_\lambda Z^2 - 1) \frac{\lambda}{4} (\phi^\dagger \phi)^2 . 
\label{Lag-1} 
\end{eqnarray} 
We shall treat ${\cal L}_0$ as the nonperturbative part and ${\cal 
L}_{int} + {\cal L}_{rc}$ as the perturbative part. The UV 
renormalizability of the theory is demonstrated at the end of this 
section. 

The hat-Lagrangian density (cf. (\ref{ac1})) reads 
\[ 
\hat{\cal L} = {\cal L} (\phi_1^\dagger, \phi_1) - {\cal L} 
(\phi_2^\dagger, \phi_2) . 
\] 
From this with (\ref{Lag-1}) we can read off the vertex matrices: 
${\cal L}_{int}$ yields the $(\phi^\dagger \phi)^2$ vertex matrix, 
\begin{equation} 
i \hat{V}_4 = - i \lambda \hat{\tau}_3 
\label{bun} 
\end{equation} 
with $\hat{\tau}_3$ the third Pauli matrix and ${\cal L}_{r c}$ 
yields 
\begin{eqnarray} 
i \hat{V}_2 & = & - i \left[ (Z - 1) \partial^2 + (Z_m Z - 1) m^2 
\right] \hat{\tau}_3 \;\;\;\;\;\;\;\; (\phi^\dagger \phi \; 
\mbox{vertex}) , \nonumber \\  
i \hat{V}_4' & = & - i (Z_\lambda Z^2 - 1) \lambda \hat{\tau}_3 
\;\;\;\;\;\;\;\; ((\phi^\dagger \phi)^2 \; \mbox{vertex}) . 
\label{bun-1} 
\end{eqnarray} 
\subsection{2n-point Green function} 
We assume that the density matrix $\rho$ commutes with the charge 
operator, 
\begin{eqnarray} 
\left[ \rho, Q \right] & = & 0 , 
\nonumber \\ 
Q & = & i \int d^{\, 3} x \, (\phi^\dagger \dot{\phi} - 
\dot{\phi}^\dagger \phi) ,  
\label{den-Q} 
\end{eqnarray} 
where $\dot{\phi} \equiv \partial \phi (x) / \partial x_0$, etc. 
(cf. below for undoing this assumption). Then, with obvious 
notation, nonvanishing path-ordered Green functions (cf. 
(\ref{green})) are 
\[ 
G (1, ..., n; 1', ..., n') = i (-)^n \langle T_C \left\{ \phi (1) 
... \phi (n) \phi^\dagger (1') ... \phi^\dagger (n') \, e^{i \int_C 
{\cal L}'} \right\} \rangle , 
\] 
where ${\cal L}' \equiv {\cal L}_{int} + {\cal L}_{rc}$. Here, it is 
convenient to introduce a generating functional: 
\begin{eqnarray} 
Z_C [J] & = & \langle T_C \mbox{exp} \left( i \int_C \left[ {\cal 
L}' (\phi^\dagger (x), \phi (x)) + J^* (x) \phi (x) + \phi^\dagger 
(x) J (x) \right] \right) \rangle 
\label{gene-0} 
\\ 
& = & e^{i \int_C {\cal L}' (- i \delta / \delta J (x), - i \delta / 
\delta J^* (x))} \langle T_C \mbox{exp} \left( i \int_C \left[ J^* 
(x) \phi (x) + \phi^\dagger (x) J (x) \right] \right) \rangle , 
\nonumber 
\end{eqnarray} 
where $J (x)$ and $J^* (x)$ are c-number external-source functions. 
$G$ is obtained from (\ref{gene-0}) through 
\begin{equation} 
G (1, ..., n; 1', ..., n') = i \frac{\delta^n Z_C [J]}{\delta J^* 
(1) ... \delta J^* (n) \delta J (1') ... \delta J (n')} 
\rule[-3mm]{.14mm}{8.5mm} \raisebox{-2.85mm}{\scriptsize{$\; J^* = 
J = 0$}} .  
\label{gene-de} 
\end{equation} 
Hereafter we refer $Z_C [J]$ to as the CTP generating functional. It 
should be understood \cite{chou} that at the end of calculation we 
set $J_1 (x) = J_2 (x)$, where $J_i (x) = J (x)$ with $x_0 \in C_i$ 
($i = 1, 2$). 

As a generalization of the Wick theorem in vacuum theory, we obtain 
\cite{chou,le-b,fau} 
\begin{eqnarray} 
& & T_C \, \mbox{exp} \left( i \int_C \left[ J^* (x) \phi (x) + 
\phi^\dagger (x) J (x) \right] \right) \nonumber \\ 
& & \mbox{\hspace*{4ex}} = e^{- i \int_C \int_C J^* (x) \Delta_{0 C} 
(x - y) J (y)} : T_C \mbox{exp} \left( i \int_C \left[ J^* (x) \phi 
(x) + \phi^\dagger (x) J (x) \right] \right) : , 
\label{kohon} 
\end{eqnarray} 
where the symbol $: ... :$ indicates to take the normal ordering 
with respect to the creation and annihilation operators in vacuum 
theory. In (\ref{kohon}), $\Delta_{0 C} (x - y)$ takes the form, in 
matrix notation, 
\begin{eqnarray} 
\hat{\Delta}_0 (x - y) & = & \int \frac{d^{\, 4} P}{(2 \pi)^4} e^{ - 
i P \cdot (x - y)} \hat{\Delta}_0 (P) \nonumber \\ 
\hat{\Delta}_0 (P) & = & \left( 
\begin{array}{cc} 
\Delta_F (P) & \;\;\; 
\theta (- p_0) [\Delta_F (P) - \Delta_F^* (P)] \\ 
\theta (p_0) [\Delta_F (P) - \Delta_F^* (P)] & \;\;\; - \Delta^*_F 
(P) 
\end{array} 
\right) . 
\label{G0} 
\end{eqnarray} 
where 
\[ 
\Delta_F (P) = \frac{1}{P^2 - m^2 + i \eta} \;\;\;\;\;\; (\eta = 
0^+) 
\] 
is the Feynman propagator in vacuum theory. Simple manipulation 
yields 
\[ 
\langle \, : T_C \, e^{i \int_C \left[ J^* (x) \phi (x) + 
\phi^\dagger (x) J (x) \right] } : \, \rangle = e^{i {\cal W}_C} 
\] 
with 
\begin{eqnarray} 
{\cal W}_C & = & \sum_{n = 1}^\infty \frac{i^{n - 1}}{n !} \langle 
\, : T_C \left\{ \int_C \left[ J^* (x) \phi (x) + \phi^\dagger (x) 
J (x) \right] \right\}^n : \, \rangle_c 
\label{W} 
\\ 
& = & - \sum_{n = 1}^\infty \frac{1}{(n !)^2} \int_C \cdot \cdot 
\cdot \int_C {\cal W}_{2 n} (x_1, ..., x_n; y_1, ..., y_n) J^* (x_1) 
... J^* (x_n) J (y_1) ... J (y_n) . 
\label{WW} 
\end{eqnarray} 
Here the subscript \lq $c$' denotes connected part and 
\begin{equation} 
{\cal W}_{2 n} (x_1, ..., x_n; y_1, ..., y_n) = i (-)^n \langle \, : 
\phi (x_1) ... \phi (x_n) \phi^\dagger (y_1) ... \phi^\dagger (y_n) 
: \, \rangle_c 
\label{sora} 
\end{equation} 
represents initial correlation. In obtaining (\ref{WW}) from 
(\ref{W}), we have used (\ref{den-Q}), which leads to 
\[ 
\langle \, : \phi (x_1) ... \phi (x_m) \phi^\dagger (y_1) ... 
\phi^\dagger (y_n) : \, \rangle_c = 0 \;\;\;\;\;\; \mbox{if} \; m 
\neq n . 
\] 
If the condition (\ref{den-Q}) does not hold, the following argument 
in this paper should be generalized accordingly. Each of the $2n$ 
time arguments in (\ref{sora}) is on either $C_1$ or $C_2$. Then 
(\ref{sora}) consists of $2^{2n}$ components, all of which are 
identical. In the interaction representation adopting here, 
$(\partial^2 + m^2) \phi (x) = 0$, from which we obtain 
\begin{eqnarray} 
& & (\partial^2_{x_j} + m^2) {\cal W}_{2 n} (x_1, ..., x_n; y_1, 
..., y_n) = (\partial^2_{y_j} + m^2) {\cal W}_{2 n} (x_1, ..., x_n; 
y_1, ..., y_n) = 0 \nonumber \\ 
& & \mbox{\hspace*{40ex}} (j = 1, ..., n) . 
\label{constraint} 
\end{eqnarray} 

The bare two-point function or propagator, $\hat{\Delta} (x, y)$, is 
obtained from (\ref{gene-0}) with ${\cal L}' = 0$ and 
(\ref{gene-de})~-~(\ref{WW}): 
\begin{eqnarray} 
\hat{\Delta} (x, y) & = & \hat{G} (x, y) \, 
\rule[-3mm]{.14mm}{8.5mm} \raisebox{-2.85mm}{\scriptsize{$\; {\cal 
L}' = 0$}} = \hat{\Delta}_0 (x - y) + {\cal W}_2 (x; y) \hat{A}_+ , 
\label{hadaka-2} \\ 
\hat{A}_\pm & \equiv & \left( 
\begin{array}{cc} 
1 & \pm 1 \\ 
\pm 1 & 1 
\end{array} 
\right) . 
\label{Apm} 
\end{eqnarray} 
Thus we have learned that, in perturbation theory, the 
nonperturbative part consists of the bare two-point function $i 
\hat{\Delta} (x, y)$ and the multi-point initial-correlation 
functions $i (-)^{n - 1} {\cal W}_{2 n}$ $(n \geq 2)$. From 
(\ref{constraint}), we see that ${\cal W}_{2 n}$ does not contribute 
to the on-shell amplitudes. Throughout this paper, we shall assume 
that ${\cal W}_{2 n}$'s $(n \geq 2)$ can be treated as 
(nonamputated) \lq\lq perturbative parts'' and, whenever necessary, 
we include \lq\lq $2 n$-point propagator'' 
\begin{equation} 
- i (-)^n {\cal W}_{2 n} (x_1, ..., x_n; y_1, ..., y_n) 
\;\;\;\;\;\;\; (n \geq 2) 
\label{tsuika} 
\end{equation} 
into the Feynman rules. We shall discuss, in the following 
subsection C, to what extent can one discard ${\cal W}_{2 n}$ $(n 
\geq 2)$. 
\subsection{Bare two-point function} 
We analyze in detail the bare two-point function (\ref{hadaka-2}), 
which is a building block of the perturbation theory: 
\begin{eqnarray} 
\hat{\Delta} (x, y) & = & - i \langle \hat{T} \left( \hat{\phi} (x) 
\hat{\phi}^\dagger (y) \right) \rangle \nonumber \\ 
& = & \int \frac{d^{\, 4} P}{(2 \pi)^4} e^{- i P \cdot (x - y)} 
\hat{\Delta}_0 (P) + {\cal W} (x; y) \hat{A}_+ , 
\label{del} 
\\ 
(\partial^2_x + m^2) \hat{\Delta} (x, y) & = & (\partial^2_y + m^2) 
\hat{\Delta} (x, y) = - \hat{\tau}_3 \, \delta^4 (x - y) . 
\label{const} 
\end{eqnarray} 
Here $\hat{\Delta}_0$ is as in (\ref{G0}) and 
\begin{eqnarray} 
\hat{\phi} & = & \displaystyle{ \raisebox{0.9ex}{\scriptsize{$t$}}} 
\mbox{\hspace{-0.1ex}} (\phi_1, \phi_2) , \;\;\;\;\; 
\hat{\phi}^\dagger = (\phi_1^\dagger, \phi_2^\dagger) , 
\nonumber \\ 
{\cal W} (x; y) & \equiv & {\cal W}_2 (x; y) = - i \langle \, : \phi 
(x) \phi^\dagger (y) : \, \rangle_c , 
\label{W2} 
\\ 
(\partial^2_x + m^2) {\cal W} (x; y) & = & (\partial^2_y + m^2) 
{\cal W} (x; y) = 0 . 
\label{W-yo} 
\end{eqnarray} 
Let us find an explicit form for ${\cal W} (x; y)$ in terms of the 
density matrix $\rho$. For the purpose of later use, we consider a 
system enclosed in a rectangular parallelepiped. It is natural to 
take the sizes $L^j$ $(j = 1, 2, 3)$, introduced in Section I, as 
the lengths of edges. As discussed above, when we deal with 
microscopic reactions, $L^j$ is regarded as large. For simplicity of 
presentation, we assume that $L^1 = L^2 = L^3$ $(\equiv L)$. For a 
set of single-particle wave functions, we employ the periodic 
boundary condition, on the basis of which $\phi (x)$ may be expanded 
as 
\begin{equation} 
\phi (x) = \sum_{\bf p} \frac{1}{\sqrt{2 E_p V}} \left[ a_{\bf p} 
e^{ - i P \cdot x} + b^\dagger_{\bf p} e^{i P \cdot x} \right] , 
\label{tenkai} 
\end{equation} 
where $V = L^3$, $E_p = \sqrt{p^2 + m^2}$ is the energy of $\phi$ 
with momentum ${\bf p}$, 
\[ 
{\bf p} = \frac{2 \pi}{L} \, {\bf n}, \;\;\; n^i = 0, \pm1, \pm2, 
... \;\;\;\; (i = 1, 2, 3) , 
\] 
and $a_{\bf p}$ ($b^\dagger_{\bf p}$) is an annihilation (a 
creation) operator of a particle (an antiparticle) of momentum ${\bf 
p}$: 
\begin{eqnarray} 
[a_{\bf p}, a_{{\bf p}'}^\dagger] & = & [b_{\bf p}, b_{{\bf 
p}'}^\dagger] = \delta_{{\bf p} {\bf p}'}, \nonumber \\ 
a_{\bf p} | 0 \rangle & = & b_{\bf p} | 0 \rangle = 0 . 
\label{aya} 
\end{eqnarray} 

Substituting (\ref{tenkai}) and its hermitian conjugate 
$\phi^\dagger$ to (\ref{W2}), we obtain, 
\[ 
{\cal W} (x; y) = - \frac{i}{2 V} \sum_{{\bf p}, \, {\bf q}} 
\frac{1}{\sqrt{E_p E_q}} \left[ \langle a_{\bf q}^\dagger a_{\bf p} 
\rangle e^{- i (P \cdot x - Q \cdot y)} + \langle b_{\bf q} b_{\bf 
p}^\dagger \rangle e^{i (P \cdot x - Q \cdot y)} \right] , 
\] 
where $p^0 = E_p$ and $q^0 = E_q$. Taking the limit $V$ $(\equiv 
L^3)$ $\to \infty$ and Fourier transforming on $x - y$, we obtain, 
after some manipulations, 
\begin{eqnarray} 
{\cal W} (X; P) & = & \int d^{\, 4} (x - y) \, e^{i P \cdot (x - y)} 
{\cal W} (x; y) \nonumber \\ 
& = & - 2 \pi i \int d^{\, 4} Q \, e^{- i Q \cdot X} {\cal F} (P, Q) 
, 
\label{dai1} 
\\ 
{\cal F} (P, Q) & \equiv & \delta (2 P \cdot Q) \, \delta (P^2 - m^2 
+ Q^2 / 4) \nonumber \\ 
& & \times \sum_{\tau = \pm} \theta (\tau p_0 + \tau q_0 / 2) \, 
\theta (\tau p_0 - \tau q_0 / 2) \, F_\tau (\tau {\bf p} - {\bf q} / 
2, \tau {\bf p} + {\bf q} / 2) , 
\label{dai11} 
\end{eqnarray} 
where 
\begin{equation} 
F_\pm ({\bf p}, {\bf q}) \equiv \langle a^\dagger_\pm ({\bf p}) 
a_\pm ({\bf q}) \rangle . 
\label{dai31} 
\end{equation} 
with 
\begin{equation} 
a_+ ({\bf p}) = \lim_{V \rightarrow \infty} \sqrt{\frac{2 E_p V}{(2 
\pi)^3}} \, a_{\bf p}, \;\;\;\;\; a_- ({\bf p}) = \lim_{V \rightarrow 
\infty} \sqrt{\frac{2 E_p V}{(2 \pi)^3}} \, b_{\bf p} . 
\label{kan-01} 
\end{equation} 

In the case where the density matrix $\rho$ is diagonal in Fock 
space, $F_\pm ({\bf p}, {\bf q}) \propto \delta ({\bf p} - {\bf 
q})$. Substitution of this into (\ref{dai11}) yields $\delta ({\bf 
q})$, with which we have $\delta (2 P \cdot Q) \to \delta (q_0) / (2 
|p_0|)$, so that ${\cal F} (P, Q) \propto \delta^4 (Q)$. Thus, we 
see from (\ref{dai1}) that ${\cal W} (X; P)$ is $X$ independent or 
${\cal W} (x; y)$ $(= {\cal W} (x - y))$ is translation invariant. 
The density matrix dealt with in, e.g., \cite{hu} is of this type. 
The case of equilibrium system is also of this type: 
\begin{equation} 
F_\pm ({\bf p}, {\bf q}) = 2 E_p \, \delta ({\bf p} -{\bf q}) N_\pm 
(E_p) , 
\label{simplest} 
\end{equation} 
where $N_{+ (-)}$ is a particle (an antiparticle) number density. 
Equation (\ref{simplest}) leads to 
\begin{eqnarray} 
{\cal W}_{equ} (X; P) & = & {\cal W}_{equ} (P) \nonumber \\ 
& = & - 2 \pi i \, \delta (P^2 - m^2) \left[ \theta (p_0) N_+ (E_p) 
+ \theta (- p_0) N_- (E_p) \right] , 
\label{equ1} 
\end{eqnarray} 
from which we have for the bare two-point function, 
\begin{equation} 
\hat{\Delta}_{equ} (P) = \hat{\Delta}_0 (P) - 2 \pi i \, \delta (P^2 
- m^2) \left[ \theta (p_0) N_+ (E_p) + \theta (- p_0) N_- (E_p) 
\right] , 
\label{equ2} 
\end{equation} 
a well-known form in equilibrium thermal field theory 
\cite{lan,le-b}. 

Let us turn back to the present out-of-equilibrium case. As 
discussed in Section I, ${\cal W} (X; P)$ does not change 
appreciably in a single spacetime cell. Then, we see from 
(\ref{dai1}) that, for $|Q^\mu| \gtrsim 1 / L^\mu$, ${\cal F} (P, Q) 
\simeq 0$, which means 
\begin{equation} 
F_\pm (\pm {\bf p} - {\bf q} / 2, \pm {\bf p} + {\bf q} / 2) \simeq 
0 \;\;\; \mbox{for} \; q \gtrsim 1 / L . 
\label{ryouiki}
\end{equation} 
Here let us make following observation: 
\begin{description} 
\item{1)} $P$ is the momentum that is conjugate to the relative 
spacetime coordinates $x - y$, where $x$ and $y$ are in a single 
spacetime cell. 
\item{2)} In computing an amplitude in perturbation theory, ${\cal 
W} (X; P)$ is to be multiplied by some function ${\cal F} (P)$ and 
is to be integrated out over $P^\mu$.  
\end{description} 
The point 1) means that ${\cal W} (X; P)$ is meaningful only for 
$|p^0| \gtrsim 1 / L^0$ and $p \gtrsim 1 / L$. Then, the computation 
of the amplitude is reliable only when 
\[ 
\int_0^\infty d^{\, 4} P \, {\cal W} (X; P) {\cal F} (P) \simeq 
\int_{1/ L^0}^\infty d p^0 \int_{1 / L}^\infty d^{\, 3} p \, {\cal 
W} (X; P) {\cal F} (P) . 
\] 
Let us start from analyzing the \lq\lq hard-$p$ region,'' $p > O (1 
/ L)$. Substituting (\ref{ryouiki}) into (\ref{dai11}), we see that 
$P^\mu$ in (\ref{dai11}) satisfies 
\[ 
p_0^2 = p^2 + m^2 +O (1 / L^2) . 
\] 
To take care of the $O (1 / L^2)$ term, we expand $\delta (P^2 - m^2 
+ Q^2 / 4)$ in (\ref{dai11}) in powers of $Q^2$, 
\[ 
\delta (P^2 - m^2 + Q^2 / 4) = \delta (P^2 - m^2) + \frac{Q^2}{4} \, 
\frac{d \delta (P^2 - m^2)}{d P^2} + ... . 
\] 
Substituting this back into (\ref{dai11}), we obtain for ${\cal W} 
(X; P)$ in (\ref{dai1}), with obvious notation, 
\begin{eqnarray} 
{\cal W} (X; P) & = & - 2 \pi i \left[ \delta (P^2 - m^2) - 
\frac{1}{4} \, \frac{d \delta (P^2 - m^2)}{d P^2} \, \partial^2_X + 
... \right] \int \frac{d^{\, 3} q}{2 |p_0|} e^{- i {\bf p} \cdot 
{\bf q} X_0 / p_0 + i {\bf q} \cdot {\bf X}} \nonumber \\ 
& & \times \sum_{\tau = \pm} \left[ \theta (\tau p_0 + \tau |{\bf p} 
\cdot {\bf q}| / 2 p_0) F_\tau (\tau {\bf p} - {\bf q} / 2; \tau 
{\bf p} + {\bf q} / 2) \right] . 
\label{exact} 
\end{eqnarray} 
In conformity with the gradient approximation we are adopting (cf. 
(\ref{kinji0})), we neglect the term with $\partial^2_X$ and with 
\lq\lq ...' in (\ref{exact}). Using $\theta (\pm p_0 \pm q_0 / 2) 
\simeq \theta (\pm p_0)$ in (\ref{dai11}), we finally obtain 
\begin{equation} 
{\cal W} (X; P) \simeq - 2 \pi i \delta (P^2 - m^2) \left[ \theta 
(p_0) N_+ (X; E_p, \hat{\bf p}) + \theta (- p_0) N_- (X; E_p, - 
\hat{\bf p}) \right] , 
\label{aha} 
\end{equation} 
where $\hat{\bf p} \equiv {\bf p} / p$ and 
\[ 
N_\pm (X; E_p, \pm \hat{\bf p}) = \int \frac{d^{\, 3} q}{2 |p_0|} \, 
e^{- i {\bf p} \cdot {\bf q} X_0 / p_0 + i {\bf q} \cdot {\bf X}} 
F_\pm (\pm {\bf p} - {\bf q} / 2; \pm {\bf p} + {\bf q} / 2) . 
\] 
Equation (\ref{aha}) is of the standard form, which is used in the 
literature. 

Now we turn to the analysis of the region $p \sim 1 / L$. It can 
readily be seen that the condition for ${\cal W} (X; P)$ to take the 
standard form (\ref{aha}) is 
\begin{equation} 
F_\pm (\pm {\bf p} - {\bf p} / 2; \, \pm {\bf p} + {\bf p} / 2) 
\simeq 0 \;\; \mbox{for} \; \; q \gtrsim |\vec{\delta}_1| < O ( 1 / 
L) . 
\label{100} 
\end{equation} 
This condition is much stronger than (\ref{ryouiki}). However, in 
most practical cases, requiring the weaker condition (\ref{ryouiki}) 
is sufficient. This can be seen as follows. Let ${\cal P}$ be a 
typical scale(s) of the system under consideration. In the case of 
thermal-equilibrium system, ${\cal P}$ is the temperature of the 
system. As will be seen later, there emerges an effective mass 
$M_{ind} (X)$. In the case of $m >> \sqrt{\lambda} {\cal P}$, 
$M_{ind} (X)$ is not much different from $m$ and, for $m \lesssim 
\sqrt{\lambda} {\cal P}$, $M_{ind} (X)$ is of $O (\sqrt{\lambda} 
{\cal P})$: 
\begin{equation} 
M_{ind} (X) = \mbox{Max} (m, \sqrt{\lambda} {\cal P}) . 
\label{y-mass} 
\end{equation} 
Most amplitudes, when computed in perturbation theory to be 
constructed below, are insensitive to the region $|P^\mu| << 
\mbox{Max} (m, \sqrt{\lambda} {\cal P})$ of ${\cal W} (X; P)$, and 
then the precise form of ${\cal W} (X; P)$ in this region is 
irrelevant. Here, it should be noted that $m$ and $\sqrt{\lambda} 
{\cal P}$ are the scales that are characteristic to microscopic 
correlations. Then, from the setup in Section I, we assume that 
$\mbox{Max} (m, \sqrt{\lambda} {\cal P}) > O (1 / L)$. Thus, for 
dealing with the amplitudes of the above-mentioned type, requiring 
the condition (\ref{ryouiki}) is sufficient. 

Throughout in the sequel, for the piece ${\cal W} (X; P)$ of the 
propagator (cf. (\ref{del}) and (\ref{dai1})), we use (\ref{aha}) 
for all $P^\mu$ regions $- \infty < P^\mu < + \infty$. As pointed 
out above, in computing some quantity, if infrared divergence arises 
at $p = 0$, we should reanalyze the quantity using the fundamental 
form, (\ref{dai1}) with (\ref{dai11}). 

After all this, we find for the Fourier transform of $\hat{\Delta} 
(x, y)$: 
\begin{equation} 
\hat{\Delta} (X; P) \simeq \hat{\Delta}_0 (P) + {\cal W} (X; P) 
\hat{A}_+ , 
\label{aha1} 
\end{equation} 
with $\hat{\Delta}_0 (P)$ as in (\ref{G0}) and ${\cal W} (X; P)$ as 
in (\ref{aha}). We refer the representation (\ref{aha1}) to as the 
$F \overline{F}$ representation, where $F$ ($\overline{F}$) stands 
for $\Delta_F = (\Delta_0)_{1 1}$ ($\Delta_F^* = - (\Delta_0)_{2 
2}$) in $\hat{\Delta}_0 (P)$, Eq.~(\ref{G0}). 

Infrared region in massless complex-scalar field theory, ${\cal 
R}_{IR}$, is defined in momentum space as ${\cal R}_{IR} = \{ P^\mu; 
|P^\mu| \leq O (\sqrt{\lambda} {\cal P}) \}$. In quantum 
electrodynamics (QED) and QCD, ${\cal R}_{IR} = \{ P^\mu; O (g^2 
{\cal P}) < |P^\mu| \leq O (g {\cal P}) \}$ with $g$ the gauge 
coupling constant. It is well known that, in such theories as QED 
and QCD, the infrared region is full of rich structures and contains 
interesting physics \cite{le-b,pis-f,thoma}. Some quantities in QED 
or QCD is sensitive \cite{le-b} to a yet \lq\lq deeper'' infrared 
region, ${\cal R}_{IR}' = \{ P^\mu; O (g^3 {\cal P}) < |P^\mu| \leq 
O (g^2 {\cal P}) \}$. If the inverse size of a spacetime cell, 
${\cal R}_{cell} = \{ P^\mu; |P^\mu| \leq 1 / L^\mu \}$, is much 
smaller than the region ${\cal R}_{IR}$ or ${\cal R}_{IR}'$, then 
the above-mentioned interesting physics may be treated within our 
perturbative schemes. In the opposite case, the interesting physics 
is hidden outside of our schemes. In more precise, the region ${\cal 
R}_{IR}$ or ${\cal R}_{IR}'$ is outside of the region, where the 
representation (\ref{aha1}) is meaningful. 
\subsection{Multi-point initial correlations} 
In this subsection, we discuss to what extent ${\cal W}_{2 n}$ $(n 
\geq 2)$, Eq.~(\ref{sora}), may be discarded. First of all, we 
recall that all the $2^{2 n}$ components of ${\cal W}_{2 n}$ are 
identical. Then, from (\ref{phys}) with (\ref{del}), we see that 
${\cal W}_2$ $(= {\cal W})$ does not appear in the retarded and 
advanced Green functions. This means that ${\cal W}_2$ does not 
appear in the two-point correlation in the linear response theory. 
With the aid of the standard formulae (cf. Sections 2 and 5 of 
\cite{chou}), one can also show that ${\cal W}_{2 n}$ does not 
appear in multi-point correlations in the nonlinear as well as the 
linear response theory. 

We restrict our concern to ${\cal W}_4$. The argument below may be 
generalized to the case of general ${\cal W}_{2 n}$ $(n \geq 3)$. 
Straightforward computation using (\ref{tenkai}) yields 
\begin{eqnarray} 
{\cal W}_4 (x_1, x_2; y_1, y_2) & = & i \sum_{{\bf p}_1, \, {\bf 
q}_1} \sum_{{\bf p}_2, \, {\bf q}_2} \left[ \prod_i \frac{1}{\sqrt{2 
E_{p_i} V}} \frac{1}{\sqrt{2 E_{q_i} V}} \right] {\cal C}_{{\bf q}_1 
{\bf q}_2; {\bf p}_1 {\bf p}_2} \nonumber \\ 
& & \times \left[ e^{- i \sum_{j = 1}^2 (P_j \cdot x_j - Q_j \cdot 
y_j)} + e^{i \sum_{j = 1}^2 (Q_j \cdot x_j - P_j \cdot y_j)} \right] 
, 
\label{soso-1} 
\end{eqnarray} 
where $p_1^0 = E_{p_1}$, etc., and 
\begin{equation} 
{\cal C}_{{\bf q}_1 {\bf q}_2; {\bf p}_1 {\bf p}_2} \equiv \langle 
a_{{\bf q}_1}^\dagger a_{{\bf q}_2}^\dagger a_{{\bf p}_1} a_{{\bf 
p}_2} \rangle - \langle a_{{\bf q}_1}^\dagger a_{{\bf p}_1} \rangle 
\langle a_{{\bf q}_2}^\dagger a_{{\bf p}_2} \rangle - \langle 
a_{{\bf q}_2}^\dagger a_{{\bf p}_1} \rangle \langle a_{{\bf 
q}_1}^\dagger a_{{\bf p}_2} \rangle . 
\label{soso-2} 
\end{equation} 
Here we have assumed for simplicity that $\langle a_{{\bf 
q}_1}^\dagger a_{{\bf q}_2}^\dagger a_{{\bf p}_1} a_{{\bf p}_2} 
\rangle = \langle b_{{\bf q}_1}^\dagger b_{{\bf q}_2}^\dagger 
b_{{\bf p}_1} b_{{\bf p}_2} \rangle$, etc. The first exponential 
function in (\ref{soso-1}) may be written as, with obvious notation, 
\[ 
e^{- i \sum_{j = 1}^2 (P_j \cdot x_j - Q_j \cdot y_j)} 
= e^{- i X \cdot \sum_{j = 1}^2 (P_j - Q_j)} 
e^{- i \sum_{j = 1}^2 (P_j \cdot x_j' - Q_j \cdot y_j')} 
\] 
where $X = (x_1 + x_2 + x_3 + x_4) / 4$ is the center-of-mass or 
the macroscopic spacetime coordinates and $x_j'$ and $y_j'$ are the 
microscopic spacetime coordinates. Similar decomposition can be made 
for the second exponential factor in (\ref{soso-1}). As setted up in 
Section I, ${\cal W}_4$ in (\ref{soso-1}) depends weakly on $X^\mu$, 
which means that, for $|p_1^0 + p_2^0 - q_1^0 - q_2^0| \gtrsim 1 / 
L^0$ and/or $|{\bf p}_1 + {\bf p}_2 - {\bf q}_1 - {\bf q}_2| \gtrsim 
1 / L$, 
\begin{equation} 
{\cal C}_{{\bf q}_1 {\bf q}_2; {\bf p}_1 {\bf p}_2} \simeq 0 . 
\label{hoho} 
\end{equation} 
Thus, the momentum conservation holds approximately in a single 
spacetime cell. Note that 
\begin{equation} 
{\cal C}_{{\bf q}_1 {\bf q}_2; {\bf p}_1 {\bf p}_2} = O (1) \;\;\; 
\mbox{for} \; \rule[-3mm]{.14mm}{8.5mm} \, \sum_{j = 1}^2 (p_j^0 - 
q_j^0) \, \rule[-3mm]{.14mm}{8.5mm} \lesssim 1 / L^0 \; \mbox{and} 
\; \rule[-3mm]{.14mm}{8.5mm} \, \sum_{j = 1}^2 ({\bf p}_j - {\bf 
q}_j) \, \rule[-3mm]{.14mm}{8.5mm} \lesssim 1 / L . 
\label{hoho-10} 
\end{equation} 

Now we take the limit $V \to \infty$ and go to the momentum space: 
\begin{eqnarray} 
G_4 & \equiv & - i \int d x_1 \, d x_2 \, d y_1 \, d y_2 \, e^{i 
\sum_{j = 1}^2 (P_j \cdot x_j - Q_j \cdot y_j)} {\cal W}_4 (x_1, 
x_2; y_1, y_2) \nonumber \\ 
& = & \frac{(2 \pi)^4}{4} \, \frac{V^2}{\sqrt{E_{p_1} E_{p_2} 
E_{q_1} E_{q_2}}} \left[ {\cal C}_{{\bf q}_1 {\bf q}_2; {\bf p}_1 
{\bf p}_2} \prod_{j =1}^2 \left( \delta (p_{j 0} - E_{p_j}) \, 
\delta (q_{j 0} - E_{q_j}) \right) \right. \nonumber \\ 
& & \left. + {\cal C}_{- {\bf p}_1 - {\bf p}_2; - {\bf q}_1 - {\bf 
q}_2} \prod_{j =1}^2 \left( \delta (p_{j 0} + E_{p_j}) \, \delta 
(q_{j 0} + E_{q_j}) \right) \right] . 
\label{kob} 
\end{eqnarray} 
A transition probability or a rate of a microscopic reaction is 
related to an on-shell amplitude. (In the case of equilibrium 
thermal field theories, this relation is settled in \cite{nie10}.) 
Let us now analyze the on-shell $(p_0^2 = E_p^2)$ amplitudes. As 
mentioned above and can be seen from (\ref{kob}), $G_4$ {\em per se} 
does not contribute to the on-shell amplitudes. Then, ${\cal W}_4$ 
or $G_4$ appears as a part(s) of an on-shell amplitude. Let us see 
the structure of a connected amplitude $S_4$, which includes a 
single $G_4$. We may write, with obvious notation, 
\begin{equation} 
S_4 = (2 \pi)^4 \int \left[ \prod_{j = 1}^2 \frac{d^{\, 4} P_j}{(2 
\pi)^4} \frac{d^{\, 4} Q_j}{(2 \pi)^4} \right] G_4 \, H \, \delta^4 
\left( \sum_l P_l + \sum_{j = 1}^2 (Q_j - P_j) \right).  
\label{hika} 
\end{equation} 
Here $\sum_l P_l + \sum_{j = 1}^2 (Q_j - P_j)$ is the sum of the 
momenta that flow in $H$. As mentioned above $P_1 + P_2 \simeq Q_1 
+ Q_2$, we may replace the delta function in (\ref{hika}) with 
$\delta^4 (\sum_l P_l)$: 
\begin{eqnarray} 
S_4 & \simeq & (2 \pi)^4 \delta^4 (\sum_l P_l) \, \tilde{S}_4 , 
\nonumber \\ 
\tilde{S}_4 & = & 
\int \left[ \prod_{j = 1}^2 \frac{d^{\, 4} P_j}{(2 
\pi)^4} \frac{d^{\, 4} Q_j}{(2 \pi)^4} \right] G_4 \, H . 
\label{hika-10} 
\end{eqnarray} 

Let $G_4'$ be the Fourier-transformed $2 n$-point Green function 
computed in perturbative scheme constructed below, which is the 
counterpart of $G_4$. Note that $G_4'$, being of $O (\lambda)$ with 
$\lambda$ the coupling constant, takes the form 
\[ 
G_4' = (2 \pi)^4 \delta^4 (P_1 + P_2 - Q_1 - Q_2) \, \tilde{G}_4' . 
\] 
The contribution (\ref{hika}) with $G_4'$ for $G_4$ reads 
\begin{eqnarray} 
S_4' & = & (2 \pi)^4 \delta^4 (\sum_l P_l) \, \tilde{S}_4' , 
\nonumber \\ 
\tilde{S}_4' & = & \int \frac{d^{\, 4} P_1}{(2 \pi)^4} \frac{d^{\, 
4} P_2}{(2 \pi)^4} \frac{d^{\, 4} Q_1}{(2 \pi)^4} \, \tilde{G}_4' \, 
H .  
\label{hika1} 
\end{eqnarray} 
Here we recall that the volume of the box, $V = L^3$, can be 
regarded as large. Let us see the dependence of (\ref{hika-10}) and 
(\ref{hika1}) on (the large) $V$. $\tilde{G}_4'$ and then 
$\tilde{S}_4'$ are independent of $V$. Owing to the properties 
(\ref{hoho}) and (\ref{hoho-10}), integration over ${\bf q}_2$ in 
(\ref{hika-10}) yields a factor of $1 / V$. Then, at this stage, 
$\tilde{S}_4$ is $O (V)$ larger than $\tilde{S}_4'$. Let us consider 
possible properties of $\langle a^\dagger_{{\bf q}_1} a_{{\bf 
q}_2}^\dagger a_{{\bf p}_1} a_{{\bf p}_2} \rangle$. 
\begin{description} 
\item{P1)} {\em Strong factorizability}. There exist scale 
parameters $\vec{\delta}_2$ $(|\vec{\delta}_2| > 1 / L)$, so that, 
for $|{\bf p}_1 - {\bf p}_2| \gtrsim |\vec{\delta}_2|$ {\em or} 
$|{\bf q}_1 - {\bf q}_2| \gtrsim |\vec{\delta}_2|$, 
\begin{equation} 
\langle a^\dagger_{{\bf q}_1} a_{{\bf q}_2}^\dagger a_{{\bf p}_1} 
a_{{\bf p}_2} \rangle \simeq \langle a^\dagger_{{\bf q}_1} a_{{\bf 
p}_1} \rangle \langle a^\dagger_{{\bf q}_2} a_{{\bf p}_2} \rangle + 
\langle a^\dagger_{{\bf q}_1} a_{{\bf p}_2} \rangle \langle 
a^\dagger_{{\bf q}_2} a_{{\bf p}_1} \rangle, 
\label{facto} 
\end{equation} 
which leads to ${\cal C}_{{\bf q}_1 {\bf q}_2; {\bf p}_1 {\bf p}_2} 
\simeq 0$. 
\item{P1')} {\em Weak factorizability}. There exist scale 
parameters $\vec{\delta}_2$ $(|\vec{\delta}_2| > 1 / L)$, so that, 
for $|{\bf p}_1 - {\bf p}_2| \gtrsim |\vec{\delta}_2|$ {\em and} 
$|{\bf q}_1 - {\bf q}_2| \gtrsim |\vec{\delta}_2|$, (\ref{facto}) 
holds. 
\item{P2)} {\em Strong \lq\lq short-range correlation''}. There 
exist scale parameters $\vec{\delta}_3$ $(|\vec{\delta}_3| > 1 / 
L)$, so that, for $|{\bf p}_1 - {\bf q}_1| \gtrsim |\vec{\delta}_3|$ 
{\em or} $|{\bf p}_1 - {\bf q}_2| \gtrsim |\vec{\delta}_3|$, 
\begin{equation} 
\langle a^\dagger_{{\bf q}_1} a_{{\bf q}_2}^\dagger a_{{\bf p}_1} 
a_{{\bf p}_2} \rangle \simeq 0 . 
\label{van} 
\end{equation} 
\item{P2')} {\em Weak \lq\lq short-range correlation''}. There 
exist scale parameters $\vec{\delta}_3$ $(|\vec{\delta}_3| > 1 / 
L)$, so that, for $|{\bf p}_1 - {\bf q}_1| \gtrsim |\vec{\delta}_3|$ 
{\em and} $|{\bf p}_1 - {\bf q}_2| \gtrsim |\vec{\delta}_3|$, 
(\ref{van}) holds. From (\ref{hoho}) and (\ref{hoho-10}), this 
condition implies $|{\bf p}_2 - {\bf q}_1| \gtrsim |\vec{\delta}_3|$ 
{\em and} $|{\bf p}_2 - {\bf q}_2| \gtrsim |\vec{\delta}_3|$. 
\end{description} 

Incidentally, P1) and P2) are not compatible with each other. In 
fact, when ${\bf p}_1 \simeq {\bf q}_1$ and $|{\bf p}_1 - {\bf q}_2| 
>> Max (|\vec{\delta}_2|, |\vec{\delta}_3|)$, P2) yields $\langle 
a^\dagger_{{\bf q}_1} a_{{\bf q}_2}^\dagger a_{{\bf p}_1} a_{{\bf 
p}_2} \rangle \simeq 0$. On the other hand, P1) or P1') leads to 
(\ref{facto}). Owing to (\ref{100}) with (\ref{dai31}) and 
(\ref{kan-01}), the second term on the right-hand side (RHS) of 
(\ref{facto}) vanishes but the first term does not, which 
contradicts $\langle a^\dagger_{{\bf q}_1} a_{{\bf q}_2}^\dagger 
a_{{\bf p}_1} a_{{\bf p}_2} \rangle \simeq 0$ above. 

Let us enumerate the cases, where $\tilde{S}_4$ and then ${\cal 
W}_4$ may possibly be ignored. 
\begin{description} 
\item{Property P1)}: 
\begin{equation} 
\frac{\tilde{S}_4}{\tilde{S}_4'} = O \left( \frac{|\vec{\delta}_2|^6 
V}{\lambda {\cal P}^3}\right) . 
\label{pq-1} 
\end{equation} 
\item{Property P2)}: 
\begin{equation} 
\frac{\tilde{S}_4}{\tilde{S}_4'} = O \left( \frac{|\vec{\delta}_3|^6 
V}{\lambda {\cal P}^3}\right) . 
\label{pq-2} 
\end{equation} 
\item{Properties P1') and P2)}: 
\begin{equation} 
\frac{\tilde{S}_4}{\tilde{S}_4'} = O \left( 
\frac{|\vec{\delta}_3|^3 \, \mbox{Min} (|\vec{\delta}_2|^3, \, 
|\vec{\delta}_3|^3) \, V}{\lambda {\cal P}^3}\right) . 
\label{pq-3} 
\end{equation} 
\item{Properties P1) and P2')}: 
\begin{equation} 
\frac{\tilde{S}_4}{\tilde{S}_4'} = O \left( \frac{|\vec{\delta}_2|^3 
\, \mbox{Min} (|\vec{\delta}_2|^3, \, |\vec{\delta}_3|^3) \, 
V}{\lambda {\cal P}^3}\right) . 
\label{pq-5} 
\end{equation} 
\item{Properties P1') and P2')}: 
\begin{equation} 
\frac{\tilde{S}_4}{\tilde{S}_4'} = O \left( \frac{|\vec{\delta}_2|^3 
\, |\vec{\delta}_3|^3) \, V}{\lambda {\cal P}^3}\right) . 
\label{pq-4} 
\end{equation} 
\end{description} 
Let us inspect (\ref{pq-1}). If $|\vec{\delta}_2|^6 < O (\lambda 
{\cal P}^3 / V)$ or $|\vec{\delta}_2| < O (\lambda^{1 / 6} 
\sqrt{{\cal P} / L})$, $G_4$ or ${\cal W}_4$ may be discarded. An 
exceptional case is where $H$ in (\ref{hika}) is as large as 
$\propto \lambda / (|\vec{\delta}_2|^6 V)$ at $P_1 \simeq P_2 \simeq 
Q_1 \simeq Q_2$ with $P_1^2 = P_2^2 = Q_1^2 = Q_2^2 = m^2$. If 
$|\vec{\delta}_2| = O (1 / L)$, (\ref{pq-1}) becomes 
\begin{equation} 
\frac{\tilde{S}_4}{\tilde{S}_4'} = O \left(\frac{1}{\lambda V 
{\cal P}^3} \right) , 
\label{pq-10} 
\end{equation} 
and then ${\cal W}_4$ may generally be discarded. Similar 
inspection may be made for (\ref{pq-2}) - (\ref{pq-4}). 

Let us analyze the special case where the density matrix $\rho$ is 
diagonal in Fock space: 
\begin{eqnarray*} 
\langle n_{\bf k}^{(a)} | \rho | n_{{\bf k}'}^{(a)} \rangle & = & 
\delta_{{\bf k} {\bf k}'} \rho_{n_{\bf k}^{(a)}} , \;\;\;\;\;\; 
\langle n_{\bf k}^{(b)} | \rho | n_{{\bf k}'}^{(b)} \rangle = 
\delta_{{\bf k} {\bf k}'} \rho_{n_{\bf k}^{(b)}} , \\ 
a_{\bf k}^\dagger a_{\bf k} | n_{\bf k}^{(a)} \rangle & = & n_{\bf 
k}^{(a)} | n_{\bf k}^{(a)} \rangle , \;\;\;\;\;\; b_{\bf k}^\dagger 
b_{\bf k} | n_{\bf k}^{(b)} \rangle = n_{\bf k}^{(b)} | n_{\bf 
k}^{(b)} \rangle , 
\end{eqnarray*} 
where $n_{\bf k}^{(a) / (b)} = 0, 1, 2, ...$. In this case, Green 
functions are translation invariant and independent of the 
center-of-mass or macroscopic spacetime coordinates (cf. above after 
(\ref{kan-01})). Equation (\ref{soso-1}) becomes 
\begin{eqnarray*} 
{\cal W}_4 & = & \sum_{\bf p} \frac{i}{(2 E_p V)^2} \left[ \langle 
a_{\bf p}^{\dagger 2} a_{\bf p}^2 \rangle - 2 \langle a_{\bf 
p}^\dagger a_{\bf p} \rangle^2 \right] 
\nonumber \\ 
& & \times \left[ e^{- i P \cdot (x_1 + x_2 - y_1 - y_2)} + 
\mbox{c.c.} \right] , 
\end{eqnarray*} 
where \lq c.c.' stands for the complex conjugate and (\ref{kob}) 
becomes 
\begin{eqnarray} 
G_4 & = & - i \int d x_1 d x_2 d y_1 d y_2 e^{i \sum_{j = 1}^2 (P_j 
\cdot x_j - Q_j \cdot y_j)} {\cal W}_4 
\label{henkan} \\ 
& = & (2 \pi)^{1 2} \delta^4 \left( \sum_{j = 1}^2 (P_j - Q_j) 
\right) \frac{1}{V} \frac{2 \pi \delta (P_1^2 - m^2)}{2 E_{p_1}} 
\left[ \langle a_{{\bf p}_1}^{\dagger 2} a_{{\bf p}_1}^2 \rangle - 2 
\langle a_{{\bf p}_1}^\dagger a_{{\bf p}_1} \rangle^2 \right] 
\nonumber \\ 
& & \times \delta^4 (P_2 - P_1) \delta^4 (Q_1 - P_1) , 
\label{ef} 
\end{eqnarray} 
where, for simplicity, we have assumed $N_{\bf p} = N_{- {\bf p}}$. 
This leads to 
\begin{equation} 
\frac{S_4}{\tilde{S}_4} = O \left( \frac{1}{\lambda V {\cal P}^3} 
\right) , 
\label{futa} 
\end{equation} 
which is of the same order of magnitude as (\ref{pq-10}). Thus, 
$G_4$ or ${\cal W}_4$ may generally be ignored. Incidentally, for 
${\cal W}_{2 n}$, we obtain, in place of (\ref{futa}), $S_{2 n} / 
\tilde{S}_{2 n} = O (1 / (\lambda V {\cal P}^3)^{n - 1})$. 
 
In the sequel of this paper, we ignore throughout the multi-point 
correlations, ${\cal W}_{2 n}$ $(n \geq 2)$. 
\subsection{UV renormalizability} 
Here, a comment is made on the UV renormalizability. We have 
completed the construction of the building blocks of Feynman rules. 
The vertices are given by (\ref{bun}) and (\ref{bun-1}) while the 
propagator is given by (\ref{aha1}). As discussed at the end of 
Section IIIA, it should be stressed that, if the \lq\lq $2 n$-point 
propagator'' $(n \geq 2)$ ${\cal W}_{2 n}$ cannot be discarded, we 
should include it into the rules. We assume that, as in equilibrium 
thermal field theories, ${\cal W} (X; P)$ in (\ref{aha1}) and ${\cal 
W}_{2 n}$ $(n \geq 2)$ in (\ref{tsuika}) are \lq\lq soft.'' Here, by 
\lq\lq soft'' we mean that, in the UV limit $|P^\mu| \to \infty$, 
${\cal W} (X; P)$ $(\in \hat{\Delta} (X; P))$ tends to zero more 
quickly than the vacuum part $\hat{\Delta}^{(0)} (P)$ and ${\cal 
W}_{2 n}$ tends to zero more quickly than the (leading contribution 
to the) $2 n$-point function. Then, genuine UV divergences emerge 
only from the vacuum part $\hat{\Delta}^{(0)} (P)$ of $\hat{\Delta} 
(X; P)$. This means that Green functions are UV finite if the vacuum 
theory is renormalizable and has been renormalized, which is in fact 
the case for vacuum theory of complex-scalar fields. For more 
elaborate argument, we refer the reader to \cite{lan,mou}. 
\setcounter{equation}{0} 
\setcounter{section}{3} 
\def\theequation{\mbox{\arabic{section}.\arabic{equation}}} 
\section{Quasiparticle representation} 
\subsection{Introduction of quasiparticle} 
It is convenient to rewrite $\hat{\Delta} (x, y)$, Eq.~(\ref{del}) 
or (\ref{aha1}), using the so-called \lq\lq Re\-tard\-ed/Advanced 
basis'' \cite{aure}: 
\begin{eqnarray} 
\hat{\Delta} (x, y) & = & \hat{\Delta}_{R A} (x - y) - i F (x, y) 
\hat{A}_+ , 
\label{imp} 
\\ 
\hat{\Delta}_{R A} (x - y) & = & \hat{\Delta}_{diag} (x - y) + 
(\Delta_R (x - y) - \Delta_A (x - y)) \left( 
\begin{array}{cc} 
0 & \; 0 \\ 
1 & \; 0 
\end{array} 
\right) , 
\label{1-11} 
\\ 
F (x, y) & = & i {\cal W} (x; y) + \int \frac{d^{\, 4} P}{(2 \pi)^3} 
\, e^{- i P \cdot (x - y)} \theta (- p_0) \, \delta (P^2 - m^2) . 
\label{1-1} 
\end{eqnarray} 
Here 
\begin{eqnarray} 
\hat{\Delta}_{diag} (x - y) & = & \mbox{diag} \left( \Delta_R (x - 
y), \, - \Delta_A (x - y) \right) , 
\label{shin} 
\\ 
\Delta_{R (A)} (x - y) & = & \int \frac{d^{\, 4} P}{(2 \pi)^4} e^{- 
i P \cdot (x - y)} \Delta_{R (A)} (P) 
\label{6.2-d} 
\end{eqnarray} 
with 
\begin{equation} 
\Delta_{R (A)} (P) = \frac{1}{P^2 - m^2 \pm i \epsilon (p_0) \eta} 
\label{6.2-e} 
\end{equation} 
the retarded (advanced) Green function. We refer (\ref{imp}) to as 
the $R / A$ representation. When compared to the $F \overline{F}$ 
representation (\ref{aha1}), the $R / A$ representation makes the 
subsequent expressions compact and convenient for handling. A cost 
to be paid is that the $R / A$ representation generates less 
intuitive expressions. 

In the case of equilibrium system (cf. (\ref{equ1}) and 
(\ref{equ2})), we have 
\begin{eqnarray*} 
\hat{\Delta}_{equ} (x - y) & = & \hat{\Delta}_{R A} (x - y) \\ 
& & + \int \frac{d^{\, 4} P}{(2 \pi)^4} \, e^{- P \cdot (x - y)} 
\epsilon (p_0) \left[ \theta (- p_0) + N (E_p) \right] \left[ 
\Delta_R (P) - \Delta_A (P) \right] \hat{A}_+ , 
\end{eqnarray*} 
where a chemical potential (being conjugate to the charge) has been 
assumed to vanish, so that $N_+ (E_p) = N_- (E_p)$ $(\equiv N 
(E_p))$. We note that, in momentum space, $\hat{\Delta}_{equ}$ may 
be diagonalized as 
\begin{equation} 
\hat{\Delta}_{equ} (P) = \left( 
\begin{array}{cc} 
1 & \;\;\; f_{equ} (p_0, E_p) \\ 
1 & \;\;\; 1 + f_{equ} (p_0, E_p) 
\end{array} 
\right) 
\hat{\Delta}_{diag} (P) \left( 
\begin{array}{cc} 
1 + f_{equ} (p_0, E_p) & \;\;\; f_{equ} (p_0, E_p) \\ 
1 & \;\;\; 1 
\end{array} 
\right) , 
\label{equ4} 
\end{equation} 
where 
\[ 
f_{equ} (p_0, E_p) = \epsilon (p_0) [\theta (- p_0) + N (E_p)] , 
\] 
In configuration space, (\ref{equ4}) reads 
\begin{equation} 
\hat{\Delta}_{equ} (x - y) = \int d^{\, 3} u \, d^{\, 3} v 
\sum_{\tau = \pm} \hat{B}_L^{(\tau)} ({\bf x} - {\bf u}) \, 
\hat{\Delta}_{diag}^{(\tau)} ({\bf u} - {\bf v}; x_0 - y_0) \, 
\hat{B}_R^{(\tau)} ({\bf v} - {\bf y}) , 
\label{conf} 
\end{equation} 
where $\hat{\Delta}_{diag}^{(+)}$ $[\hat{\Delta}_{diag}^{(-)}$] 
stands for a positive [negative] frequency part of 
$\hat{\Delta}_{diag}$ and 
\begin{equation} 
\hat{B}_L^{(\tau)} ({\bf x}) = \left( 
\begin{array}{cc}
\delta ({\bf x}) & \;\; f^{(\tau)}_{equ} ({\bf x}) \\ 
\delta ({\bf x}) & \;\; \delta ({\bf x}) + f^{(\tau)}_{equ} ({\bf 
x}) 
\end{array}
\right) , \;\;\;\; 
\hat{B}_R^{(\tau)} ({\bf x}) = \left( 
\begin{array}{cc}
\delta ({\bf x}) + f^{(\tau)}_{equ} ({\bf x}) & \;\; 
f^{(\tau)}_{equ} ({\bf x}) \\ 
\delta ({\bf x}) & \;\; \delta ({\bf x}) 
\end{array}
\right) 
\label{Bo-0} 
\end{equation} 
with 
\[ 
f^{(\tau)}_{equ} ({\bf x}) = \epsilon (\tau) \int \frac{d^{\,3} 
p}{(2 \pi)^3} \, e^{i {\bf p} \cdot {\bf x}} [\theta (- \tau) + N 
(E_p)] . 
\] 
It is to be noted that the form for the matrices $\hat{B}_{L 
(R)}^{(\tau)}$, which leads to (\ref{equ4}) with diagonal 
$\hat{\Delta}_{diag}$, is not unique and the choice here is the 
simplest one \cite{ume,ume1}. 

The representation (\ref{conf}) may be interpreted in terms of 
quasiparticle picture. In fact, (\ref{conf}) naturally leads us to 
introduce quasiparticle fields, $\hat{\phi}_{R A}$ and 
$\hat{\phi}^\dagger_{R A}$, through 
\begin{eqnarray} 
\hat{\phi} (x) & = & \int d^{\, 3} u \sum_{\tau = \pm} 
\hat{B}_L^{(\tau)} ({\bf x} - {\bf u}) \hat{\phi}_{R A}^{(\tau)} 
({\bf u}, x_0) \nonumber \\ 
\hat{\phi}^\dagger (y) & = & \int d^{\, 3} v \sum_{\tau = \pm} 
\hat{\phi}_{R A}^{(\tau) \dagger} ({\bf v}, y_0) \hat{B}_L^{(\tau)} 
({\bf v} - {\bf y}) , 
\label{qu-0} 
\end{eqnarray} 
where $\hat{\phi}^{(+)}_{R A}$ [$\hat{\phi}^{(-) \dagger}_{R A}$] 
denotes positive frequency part of $\hat{\phi}_{R A}$ 
[$\hat{\phi}^\dagger_{R A}$] and $\hat{\phi}^{(-)}_{R A}$ 
[$\hat{\phi}^{(+) \dagger}_{R A}$] denotes negative frequency part. 
The transformation (\ref{qu-0}) is called thermal Bogoliubov 
transformation. The statistical average of $\hat{T} \hat{\phi}_{R A} 
\hat{\phi}^\dagger_{R A}$ (cf. (\ref{abr})) assumes the form 
\begin{equation} 
\langle \hat{T} \hat{\phi}_{R A} (x) \, \hat{\phi}^\dagger_{R A} (y) 
\rangle = i \hat{\Delta}_{daig} (x - y) . 
\label{qu-1} 
\end{equation} 
Thus the fields $\hat{\phi}_{R A}$ and $\hat{\phi}^\dagger_{R A}$ 
describe well-defined propagating modes in the system, which we call 
retarded/advanced quasiparticles. It is to be noted that 
$\hat{\phi}^\dagger_{R A}$ is not a hermitian conjugate of 
$\hat{\phi}_{R A}$, which is a characteristic feature of the 
theory of this type \cite{ume,ume1}. 

With (\ref{qu-0}) and (\ref{qu-1}), we can readily show that 
$\hat{\Delta}_{equ} (x - y)$ turns out to (\ref{conf}). From 
(\ref{Bo-0}), we see that 
\begin{equation} 
\int d^{\, 3} u \, \hat{B}_L^{(\tau)} ({\bf x} - {\bf u}) \, 
\hat{\tau}_3 \, \hat{B}_R^{(\tau)} ({\bf u} - {\bf y}) = 
\hat{\tau}_3 \, \delta ({\bf x} - {\bf y}) , 
\label{comm} 
\end{equation} 
which shows that, under the transformation (\ref{qu-0}), the 
canonical commutation relation is preserved: 
\begin{eqnarray} 
\left[ \hat{\phi} (x), \, \dot{\hat{\phi}}^\dagger (y) \right] 
\delta (x_0 - y_0) & = & \left[ \hat{\phi}_{R A} (x), \, 
\dot{\hat{\phi}}^\dagger_{R A} (y) \right] \delta (x_0 - y_0) 
\nonumber \\ 
& = & i \hat{\tau}_3 \, \delta^4 (x - y) , 
\label{comm-2} 
\end{eqnarray} 
where $\dot{\hat{\phi}}^\dagger (y) \equiv \partial 
\hat{\phi}^\dagger (y) / \partial y_0$. Furthermore, the 
nonperturbative or bilinear part of the hat-Lagrangian density is 
form invariant, 
\begin{eqnarray} 
\hat{\cal L}_0 & = & - \phi_1^\dagger \, (\partial^2 + m^2) 
\phi_1 + \phi_2^\dagger \, (\partial^2 + m^2) \phi_2 \nonumber \\ 
& = & - \hat{\phi}^\dagger \, \hat{\tau}_3 \, (\partial^2 + m^2) 
\hat{\phi} \nonumber \\ 
& = & - \hat{\phi}^\dagger_{R A} \, \hat{\tau}_3 \, (\partial^2 + 
m^2) \, \hat{\phi}_{R A} . 
\label{form} 
\end{eqnarray} 

Let us turn back to the out-of-equilibrium case. As a generalization 
of (\ref{qu-0}), we assume that the fields, $\hat{\phi}$ and 
$\hat{\phi}^\dagger$, are related to the retarded/advanced 
quasiparticle fields, $\hat{\phi}_{R A}$ and 
$\hat{\phi}^\dagger_{R A}$, through the transformation, 
\begin{eqnarray} 
\hat{\phi} (x) & = & \int d^{\, 3} u \sum_{\tau = \pm} 
\hat{B}_L^{(\tau)} ({\bf x}, {\bf u}; x_0) \, \hat{\phi}_{R 
A}^{(\tau)} ({\bf u}, x_0) \nonumber \\ 
\hat{\phi}^\dagger (y) & = & \int d^{\, 3} v \sum_{\tau = \pm} 
\hat{\phi}_{R A}^{(\tau) \dagger} ({\bf v}, y_0) \, 
\hat{B}_L^{(\tau)} ({\bf v}, {\bf y}; y_0) , 
\label{qu-2} 
\end{eqnarray} 
where 
\begin{eqnarray} 
\hat{B}_L^{(\tau)} ({\bf x}, {\bf u}; x_0) & = & \left( 
\begin{array}{cc}
\delta ({\bf x} - {\bf u}) & \;\;\; f^{(\tau)} ({\bf x}, {\bf u}; 
x_0) \\ 
\delta ({\bf x} - {\bf u}) & \;\;\; \delta ({\bf x} - {\bf u}) + 
f^{(\tau)} ({\bf x}, {\bf u}; x_0) 
\end{array}
\right) , \nonumber \\ 
\hat{B}_R^{(\tau)} ({\bf v}, {\bf y}; y_0) & = & \left( 
\begin{array}{cc}
\delta ({\bf v} - {\bf y}) + f^{(\tau)} ({\bf v}, {\bf y}; y_0) & 
\;\;\; f^{(\tau)} ({\bf v}, {\bf y}; y_0) \\ 
\delta ({\bf v} - {\bf y}) & \;\;\; \delta ({\bf v} - {\bf y}) 
\end{array}
\right) . 
\label{Bogo-R} 
\end{eqnarray} 
We assume that $\hat{B}_{L (R)}^{(\tau)} ({\bf x}, {\bf u}; x_0)$ 
depends on $({\bf x} + {\bf u}) / 2$ and on $x_0$ only weakly. 
Corresponding to (\ref{comm}), we have 
\begin{equation} 
\int d^{\, 3} u \, \hat{B}_L^{(\tau)} ({\bf x}, {\bf u}; x_0) \, 
\hat{\tau}_3 \, \hat{B}_R^{(\tau)} ({\bf u}, {\bf y}; x_0) = 
\hat{\tau}_3 \, \delta ({\bf x} - {\bf y}) , 
\label{Bo-k} 
\end{equation} 
so that, as in (\ref{comm-2}), the transformation (\ref{qu-2}) 
preserves the canonical commutation relation.  

Incidentally, the so-called nonequilibrium thermo field dynamics 
\cite{ume,ume1} is formulated by taking (\ref{qu-2}) as a starting 
hypothesis. 
\subsection{Propagator} 
Using (\ref{qu-2}) and (\ref{qu-1}), we obtain for the propagator 
$\hat{\Delta}$, 
\begin{eqnarray} 
\hat{\Delta} (x, y) & = & - i \langle \hat{T} \hat{\phi} (x) 
\hat{\phi}^\dagger (y) \rangle \nonumber \\ 
& = & - i \int d^{\, 3} u \, d^{\, 3} v \, \hat{B}_L^{(\tau)} ({\bf 
x}, {\bf u}; x_0) \langle \hat{T} \hat{\phi}_{R A}^{(\tau)} ({\bf 
u}, x_0) \, \hat{\phi}^{(\tau) \dagger}_{R A} ({\bf v}; y_0) \rangle 
\hat{B}_R^{(\tau)} ({\bf v}, {\bf y}; y_0) \nonumber \\ 
& = & \int d^{\, 3} u \, d^{\, 3} v \sum_{\tau = \pm} 
\hat{B}_L^{(\tau)} ({\bf x}, {\bf u}; x_0) \, 
\hat{\Delta}_{diag}^{(\tau)} ({\bf u} - {\bf v}; x_0 - y_0) \, 
\hat{B}_R^{(\tau)} ({\bf v}, {\bf y}; y_0) , 
\label{imple} 
\end{eqnarray} 
which leads to (\ref{imp}) with 
\begin{eqnarray} 
F (x, y) & = & \sum_{\tau = \pm} F^{(\tau)} (x, y) \nonumber \\ 
& = & i \int d^{\, 3} u \sum_{\tau = \pm} \left[ \Delta_R^{(\tau)} 
({\bf x} - {\bf u}; x_0 - y_0) \, f^{(\tau)} ({\bf u}, {\bf y}; y_0) 
\right. \nonumber \\ 
& & \left. - f^{(\tau)} ({\bf x}, {\bf u}; x_0) \, \Delta_A^{(\tau)} 
({\bf u} - {\bf y}; x_0 - y_0) \right] \nonumber \\ 
& = & i {\cal W} (x; y) + \int \frac{d^{\, 4} P}{(2 \pi)^3} \, 
e^{- i P \cdot (x - y)} \, \theta (- p_0) \, \delta (P^2 - m^2) . 
\label{F} 
\end{eqnarray} 
The last line is the definition of $F (x, y)$, Eq.~(\ref{1-1}). 
Equation (\ref{F}) is a determining equation for $f^{(\tau)}$. 

Straightforward manipulation using (\ref{kinji1}) in (\ref{imple}) 
yields 
\begin{eqnarray} 
\hat{\Delta} (X; P) & \equiv & \int d^{\, 4} (x - y) \, e^{i P \cdot 
(x - y)} \hat{\Delta} (x, y) \nonumber \\ 
& \simeq & \hat{\Delta}^{(0)} (X; P) + \hat{\Delta}^{(1)} (X; P) 
\;\;\;\;\;\; (X \equiv (x + y)/ 2) , 
\label{non-e} 
\\ 
\hat{\Delta}^{(0)} (X; P) & = & \hat{\Delta}_{R A} (P) + 
\hat{\Delta}^{(0) '} (X; P) , \nonumber 
\end{eqnarray} 
where 
\begin{eqnarray} 
\hat{\Delta}_{R A} (P) & = & \left( 
\begin{array}{cc} 
\Delta_R (P) & \;\;\; 0 \\ 
\Delta_R (P) - \Delta_A (P) & \;\;\; - \Delta_A (P) 
\end{array} 
\right) 
\label{vac} 
\\ 
\hat{\Delta}^{(0) '} (X; P) & = & - 2 \pi i \epsilon (p_0) \, 
f^{(\epsilon (p_0))} (X; {\bf p}) \, \delta (P^2 - m^2) \, \hat{A}_+ 
\nonumber \\ 
\hat{\Delta}^{(1)} (X; P) & = & \frac{i}{2} \frac{\partial 
f^{(\epsilon (p_0))} (X; {\bf p})}{\partial X^\mu} \, 
\frac{\partial}{\partial P_\mu} (\Delta_R (P) + \Delta_A (P)) 
\hat{A}_+ \\ 
& = & - i P \cdot \partial_X f^{(\epsilon (p_0))} (X; {\bf p}) 
\left[ \Delta_R^2 (P) + \Delta_A^2 (P) \right] \hat{A}_+ . 
\label{prop1} 
\end{eqnarray} 

This is not the end of the story. We should impose the condition 
(\ref{const}), which leads to a constraint on $f^{(\tau)} (X; P)$. 
The most convenient way is to start from (\ref{W-yo}) with 
(\ref{F}), 
\begin{equation} 
(\partial^2_x + m^2) F^{(\tau)} (x, y) = (\partial^2_y + m^2) 
F^{(\tau)} (x, y) = 0 . 
\label{free} 
\end{equation} 
However, for the purpose of later use, we analyze (\ref{const}) 
directly. From (\ref{imple}), we obtain 
\begin{eqnarray} 
(\partial^2_x + m^2) \, \hat{\Delta} (x, y) & = & (\partial^2_x + 
m^2) \int d^{\, 3} u \, d^{\, 3} v \sum_{\tau = \pm} 
\hat{B}_L^{(\tau)} ({\bf x}, {\bf u}; x_0) \nonumber \\ 
& & \times \hat{\Delta}_{diag}^{(\tau)} ({\bf u} - {\bf v}; x_0 - 
y_0) \, \hat{B}_R^{(\tau)} ({\bf v}; {\bf y}, y_0) \nonumber \\ 
& \simeq & - \hat{\tau}_3 \, \delta^4 (x - y) + \int d^{\, 3} u \, 
d^{\, 3} v \sum_{\tau = \pm} \left[ 2 \, \frac{\partial 
\hat{B}_L^{(\tau)} ({\bf x}, {\bf u}; x_0)}{\partial x_0} 
\frac{\partial}{\partial x_0} \right. \nonumber \\ 
& & \left. - \left\{ (\nabla^2_{\bf x} - \nabla^2_{\bf u}) \, 
\hat{B}_L^{(\tau)} ({\bf x}, {\bf u}; x_0) \right\} \right] \, 
\hat{\tau}_3 \, \hat{B}_R^{(\tau)} ({\bf u}, {\bf v}; x_0) \, 
\hat{\tau}_3 \, \hat{\Delta}^{(\tau)} ({\bf v}, {\bf y}; x_0, y_0) , 
\nonumber 
\\ 
\label{kita} 
\end{eqnarray} 
where $\hat{\Delta}^{(+)}$ [$\hat{\Delta}^{(-)}$] is a positive 
[negative] frequency part of $\hat{\Delta}$. Comparing this result 
with (\ref{const}), we see that the second term should vanish, which 
reads, in the momentum space, 
\begin{equation} 
- 2 i \int \frac{d^{\, 4} P}{(2 \pi)^4} \, e^{- i P \cdot (x - y)} 
\sum_{\tau = \pm} \theta (\tau p_0) \left[ P \cdot \partial_X 
f^{(\tau)} (X; {\bf p}) \right] \hat{\Delta}^{(\tau)} (X; P) = 0 , 
\label{evo-pre} 
\end{equation} 
where $X = (x + y) / 2$. Since $p_0$ and $p$ are mutually 
independent, we obtain 
\begin{equation} 
\partial_{X_0} f^{(\tau)} (X; {\bf p}) = {\bf p} \cdot \nabla_{\bf 
X} f^{(\tau)} (X; {\bf p}) = 0 \;\;\;\;\; (\tau = \pm) , 
\label{evo-1} 
\end{equation} 
as it should be, since, interaction is absent here. Thus 
$f^{(\tau)} (X; {\bf p})$ is $X_0$ independent and has vanishing 
gradient along the \lq\lq flow lines.'' It can easily be shown that 
(\ref{evo-1}) guarantees also $(\partial^2_y + m^2) F^{(\tau)} = 0$, 
Eq.~(\ref{free}). Thanks to the relation (\ref{evo-1}), 
$\hat{\Delta}^{(1)}$ in (\ref{prop1}) vanishes, and then 
\begin{eqnarray} 
\hat{\Delta} (X; P) & \simeq & \hat{\Delta}_{R A} (P) - 2 \pi i 
\epsilon (p_0) \, f^{(\epsilon (p_0))} (X; {\bf p}) \, \delta (P^2 - 
m^2) \, \hat{A}_+ \nonumber \\ 
& \simeq & \hat{\Delta}_{R A} (P) + [{\cal W} (P; X) - 2 \pi i 
\theta (- p_0) \, \delta (P^2 - m^2)] \, \hat{A}_+ . 
\label{free-p} 
\end{eqnarray} 
\subsubsection*{$|p_0|$ prescription} 
Comparison between (\ref{free-p}) and (\ref{aha}) yields 
\begin{eqnarray} 
f^{(+)} (X; {\bf p}) & = & N_+ (X; E_p, \hat{\bf p}) \nonumber \\ 
f^{(-)} (X; {\bf p}) &=& - 1 - N_- (X; E_p, - \hat{\bf p}) , 
\label{ryuushi} 
\end{eqnarray} 
where $N_+$ $(N_-)$ is the particle- (antiparticle)-number density 
function. 

In the case of equilibrium system with vanishing chemical potential 
(cf. (\ref{equ2})), 
\begin{eqnarray*} 
f^{(+)}_{equ} (E_p) & = & - 1 - f^{(-)}_{equ} (E_p) \\ 
& = & \frac{1}{e^{E_p / T} - 1} . 
\end{eqnarray*} 
$f^{(\tau)}_{equ} (E_p)$ always appears in combination with $\delta 
(P^2 - m^2) = \delta (p_0^2 - E_p^2)$. Then, at first sight, it 
seems that no difference arises between $f^{(\tau)}_{equ} (|p_0|)$ 
and $f^{(\tau)}_{equ} (E_p)$. This is, however, not the case, since, 
in general, $\hat{\Delta} (P)$ is to be multiplied by the functions 
that are singular at $|p_0| = E_p$. The correct choice has been 
known \cite{nie} for some time now to be $f^{(\tau)}_{equ} (|p_0|)$ 
--- the \lq\lq $|p_0|$ prescription.'' 

Let us turn back to the present out-of-equilibrium case. On the 
basis of the above observation, we assume that the $|p_0|$ 
prescription should be adopted: 
\begin{eqnarray} 
f^{(+)} (X; {\bf p}) & = & N_+ (X; E_p, \hat{{\bf p}}) \to N_+ (X; 
p_0, \hat{\bf p}) \equiv f^{(+)} (X; P) \nonumber \\ 
f^{(-)} (X; {\bf p}) & = & - 1 - N_- (X; E_p, - \hat{{\bf p}}) \to 
- 1 - N_- (X; - p_0, - \hat{\bf p}) \equiv f^{(-)} (X; P) , 
\nonumber \\ 
f (X; P) & = & \theta (p_0) \, f^{(+)} (X; P) + \theta (- p_0) \, 
f^{(-)} (X; P) . 
\label{p0f} 
\end{eqnarray} 
In what follows, we simply refer $f$ to as {\em the number density 
(function)}. An argument for the necessity of adopting the $|p_0|$ 
prescription is given in Appendix A. 

It is to be noted that the translation into the $|p_0|$ prescription 
is {\em formally} achieved by replacing (\ref{imple}) and 
(\ref{Bogo-R}) with, in respective order, 
\begin{eqnarray} 
\hat{\Delta} (x, y) & = & \int d^{\, 4} u \, d^{\, 4} v \, \hat{B}_L 
(x, u) \, \hat{\Delta}_{diag} (u - v) \, \hat{B}_R (v, y) \nonumber 
\\ 
& \equiv & \left[ \hat{B}_L \cdot \hat{\Delta}_{diag} \cdot 
\hat{B}_R \right] (x, y) , 
\label{imple-p0} 
\end{eqnarray} 
and 
\begin{eqnarray} 
\hat{B}_L (x, u) & = & \left( 
\begin{array}{cc} 
\delta^4 (x - u) & \;\;\; f (x, u) \\ 
\delta^4 (x - u) & \;\;\; \delta^4 (x - u) + f (x, u) 
\end{array} 
\right) , \nonumber \\ 
\hat{B}_R (v, y) & = & \left( 
\begin{array}{cc}
\delta^4 (v - y) + f (v, y) & \;\;\; f (v, y) \\ 
\delta^4 (v - y) & \;\;\; \delta^4 (v - y) 
\end{array}
\right) . 
\label{Bogo-p0} 
\end{eqnarray} 
The function $f (x, y)$ here is an inverse Fourier transform of the 
number-density function $f (X; P)$ in (\ref{p0f}): 
\[ 
f (X; P) = \int d^{\, 4} (x - y) \, e^{i P \cdot (x - y)} f (x, y) 
\;\;\;\;\;\;\; (X = (x + y)/ 2) . 
\] 
Now (\ref{evo-pre}) turns out to 
\[ 
- 2 i \int \frac{d^{\, 4} P}{(2 \pi)^4} \, e^{- i P \cdot (x - y)} 
\left[ P \cdot \partial_X f (X; P) \right] \hat{\Delta} (X; P) = 0 , 
\] 
and then, in place of (\ref{evo-1}), we have 
\begin{equation} 
P \cdot \partial_X f (X; P) = 0 . 
\label{cont-f} 
\end{equation} 
This is a continuity equation for $f$ along the \lq\lq flow line'' 
in a four-dimensional space. Equation (\ref{cont-f}) may be solved 
as 
\begin{equation} 
f (X; P) = {\cal F} ({\bf X} - (X_0 - T_{in}) {\bf p} / p_0; P) , 
\label{toke} 
\end{equation} 
where ${\cal F}$ is an arbitrary function and $T_{in}$ is an 
(arbitrary) \lq\lq initial'' time. Given the initial data $f (X_0 = 
T_{in}, {\bf X}; P)$ that characterizes the ensemble of the systems 
at $X_0 = T_{in}$, (\ref{toke}) fixes the form of ${\cal F}$ and we 
obtain 
\[ 
f (X; P) = f (X_0 = T_{in}, {\bf X} - (X_0 - T_{in}) {\bf p} / p_0 ; 
P) . 
\] 
The propagator $\hat{\Delta} (X; P)$ with this $f(X; P)$ still takes 
the form (\ref{free-p}): 
\begin{equation} 
\hat{\Delta} (X; P) = \hat{\Delta}_{R A} (P) + f (X; P) \, (\Delta_R 
(P) - \Delta_A (P)) \, \hat{A}_+ . 
\label{free-fin} 
\end{equation} 

In order to see the physical meaning of $f$ to be the \lq\lq number 
density,'' let us compute a statistical average of the current 
density: 
\begin{eqnarray} 
\langle j^\mu (x) \rangle & \equiv & \frac{i}{2} \left[ \langle 
\phi^\dagger (x) \stackrel{\leftrightarrow}{\partial^\mu} \phi (x) 
\rangle - \langle \phi (x) \stackrel{\leftrightarrow}{\partial^\mu} 
\phi^\dagger (x) \rangle \right] \nonumber \\ 
& = & \frac{1}{2} (\partial_{x_\mu} - \partial_{y_\mu}) \left[ 
\Delta_{1 2} (y, x) + \Delta_{2 1} (y, x) \right]_{y = x} \nonumber 
\\ 
& = & i \int \frac{d^{\, 4} P}{(2 \pi)^4} \, P^\mu \left[ \Delta_{1 
2} (x; P) + \Delta_{2 1} (x; P) \right] , 
\label{current} 
\end{eqnarray} 
where $\stackrel{\leftrightarrow}{\partial} \equiv \partial - 
\stackrel{\leftarrow}{\partial}$. Straightforward manipulation using 
(\ref{free-fin}) with (\ref{p0f}) yields 
\begin{eqnarray} 
\langle j^0 (x) \rangle & \simeq & \int \frac{d^{\, 3} p}{(2 \pi)^3} 
\, \left[ \tilde{N}_+ (x; {\bf p}) - \tilde{N}_- (x; - {\bf p}) 
\right] \nonumber \\ 
\langle {\bf j} (x) \rangle & \simeq & \int \frac{d^{\, 3} p}{(2 
\pi)^3} \, \frac{\bf p}{E_p} \left[ \tilde{N}_+ (x; {\bf p}) + 
\tilde{N}_- (x; - {\bf p}) \right] , 
\label{curr1} 
\end{eqnarray} 
where 
\begin{eqnarray} 
\tilde{N}_+ (x; {\bf p}) & \equiv & N_+ (x; P) \, 
\rule[-3mm]{.14mm}{8.5mm} \raisebox{-2.85mm}{\scriptsize{$\; p_0 = 
E_p$}} \nonumber , \\ 
\tilde{N}_- (x; - {\bf p}) & \equiv & N_- (x; - P) \, 
\rule[-3mm]{.14mm}{8.5mm} \raisebox{-2.85mm}{\scriptsize{$\; p_0 = - 
E_p$}} . 
\label{num-den} 
\end{eqnarray} 
Thus $\tilde{N}_+ (x; {\bf p})$ [$\tilde{N}_- (x; - {\bf p})$] is a 
number density of particles with momentum ${\bf p}$ [antiparticles 
with momentum $- {\bf p}$]. It is to be noted that (\ref{num-den}) 
is valid in the gradient approximation, i.e., the terms with first 
derivative $\partial N_\pm (x; \pm P) / \partial x$ do not appear in 
(\ref{num-den}). Also to be noted is that the argument $x$ here is a 
macroscopic spacetime coordinates. 
\subsection{Two perturbative schemes} 
In this subsection, we introduce two perturbative schemes of 
calculating various quantities. The first scheme, which we call the 
\lq\lq bare$\,$-$f$ scheme,'' is the naive one introduced above. In 
this scheme, perturbative calculation goes with (\ref{bun}) and 
(\ref{bun-1}) for vertices and (\ref{free-fin}) for propagators. We 
designate the number-density function in (\ref{free-fin}) in this 
scheme as $f_B$. 

The second scheme, which we call the \lq\lq physical$\,$-$f$ 
scheme,'' is defined as follows. We {\em redefine} the fields, 
$\hat{\phi}$ and $\hat{\phi}^\dagger$, by (\ref{qu-2}) with 
(\ref{Bogo-R}), where $f^{(\tau)}$ is different from $f_B^{(\tau)}$. 
Then, the fields in this scheme are different from the fields in the 
bare-$f$ scheme. Changing the definition of fields leads to 
emergence of a counter term in the Lagrangian. In order to extract 
it, we first note that (\ref{cont-f}) does not hold now and 
(\ref{kita}) may be written as 
\begin{eqnarray} 
& & \int d^{\, 3} u \, d^{\, 3} v \sum_{\tau = \pm} D_\tau (x; {\bf 
u}, {\bf v}) \hat{\Delta}^{(\tau)} ({\bf v}, {\bf y}; x_0, 
y_0) \simeq - \hat{\tau}_3 \, \delta^4 (x - y) , 
\nonumber \\ 
& & D_\tau (x; {\bf u}, {\bf v}) \equiv 
(\partial^2_x + m^2) \, \delta ({\bf x} - {\bf u}) \, \delta ({\bf 
u} - {\bf v}) - \left[ 2 \, \frac{\partial \hat{B}_L^{(\tau)} ({\bf 
x}, {\bf u}; x_0)}{\partial x_0} \frac{\partial}{\partial x_0} 
\right. \nonumber \\ 
& & \left. \mbox{\hspace*{13ex}} - \left\{ (\nabla^2_{\bf x} - 
\nabla^2_{\bf u}) \hat{B}_L^{(\tau)} ({\bf x}, {\bf u}; x_0) 
\right\} \right] \hat{\tau_3} \, \hat{B}_R^{(\tau)} ({\bf u}, {\bf 
v}; x_0) \hat{\tau_3} . 
\label{kita1} 
\end{eqnarray} 
This means that, as a free hat-Lagrangian density, we should take 
\begin{equation} 
\hat{\cal L}_0 = - \int d^{\, 3} u \, d^{\, 3} v \sum_{\tau = \pm} 
\hat{\phi}^{(\tau) \dagger} ({\bf x}; x_0) \hat{\tau}_3 D_\tau (x; 
{\bf u}, {\bf v}) \hat{\phi}^{(\tau)} ({\bf v}; x_0) , 
\label{ff} 
\end{equation} 
where $\hat{\phi}^{(\tau) \dagger} = (\phi_1^{(\tau) \dagger}, 
\phi_2^{(\tau) \dagger})$ and $\hat{\phi}^{(\tau)} = \, ^t 
\mbox{\hspace{-.2mm}} (\phi_1^{(\tau)}, \phi_2^{(\tau)})$ with 
$\phi^{(+)}_i$ $(\phi^{(-)}_i)$ a positive (negative) frequency part 
of $\phi_i$, and $\phi^{(\pm) \dagger}_i$ a Hermitian conjugate of 
$\phi^{(\pm)}_i$. Instead of (\ref{kita}), starting with 
$(\partial^2_y + m^2) \hat{\Delta} (x, y)$, we are led to the same 
$\hat{\cal L}_0$ as (\ref{ff}) above, as it should be. Equation 
(\ref{ff}) tells us that, for compensating the difference between 
(\ref{ff}) and the original free hat-Lagrangian density, the counter 
term should be introduced in the hat-Lagrangian density: 
\begin{eqnarray} 
\hat{\cal L}_c & = & - \int d^{\, 3} u \, d^{\, 3} v \sum_{\tau = 
\pm} \hat{\phi}^{(\tau) \dagger} ({\bf x}; x_0) \, \hat{\tau}_3 
\left[ 2 \, \frac{\partial \hat{B}_L^{(\tau)} ({\bf x}, {\bf u}; 
x_0)}{\partial x_0} \frac{\partial}{\partial x_0} \right. \nonumber 
\\ 
& & \left. - (\nabla_{\bf x}^2 - \nabla_{\bf u}^2) \, 
\hat{B}_L^{(\tau)} ({\bf x}, {\bf u}; x_0) \right] 
\hat{\tau}_3 \, \hat{B}_R^{(\tau)} ({\bf u}, {\bf v}; x_0) \, 
\hat{\tau}_3 \, \hat{\phi}^{(\tau)} ({\bf v}; x_0) , 
\label{count} 
\end{eqnarray} 
which leads to a two-point vertex function, 
\begin{eqnarray*} 
i \hat{{\cal V}}_c (x, y) & = & - i \hat{\tau}_3 \int d^{\, 3} u 
\sum_{\tau = \pm} \left[ 2 \, \frac{\partial \hat{B}_L^{(\tau)} 
({\bf x}, {\bf u}; x_0)}{\partial x_0} \frac{\partial}{\partial x_0} 
- (\nabla_{\bf x}^2 - \nabla_{\bf u}^2) \, \hat{B}_L^{(\tau)} ({\bf 
x}, {\bf u}; x_0) \right] \nonumber \\ 
& & \times \hat{\tau}_3 \, \hat{B}_R^{(\tau)} ({\bf u}, {\bf y}; 
x_0) \, \hat{\tau}_3 \, \delta (x_0 - y_0) \nonumber \\ 
& \simeq & i \hat{A}_- \sum_{\tau = \pm} \left[ 2 \, \frac{\partial 
f ({\bf x}, {\bf y}, x_0)}{\partial x_0} \frac{\partial}{\partial 
x_0} - (\nabla_{\bf x}^2 - \nabla_{\bf y}^2) \, f^{(\tau)} ({\bf x}, 
{\bf y}, x_0) \right] \delta (x_0 - y_0) \\ 
& = & 2 \int \frac{d^{\, 4} P}{(2 \pi)^4} \, e^{- i P \cdot (x - y)} 
\sum_{\tau = \pm} \theta (\tau p_0) \left[ P \cdot \partial_X 
f^{(\tau)} (X; {\bf p}) \right] \hat{A}_- . 
\end{eqnarray*} 
Here $X = (x + y) / 2$ and $\hat{A}_-$ is defined as in (\ref{Apm}). 
Going to the $|p_0|$ prescription, we have (cf. (\ref{p0f})),
\begin{equation} 
i \hat{{\cal V}}_c (x, y) \to 2 \int \frac{d^{\, 4} P}{(2 \pi)^4} \, 
e^{- i P \cdot (x - y)} \left[ P \cdot \partial_X f (X; P) \right] 
\hat{A}_- . 
\label{V} 
\end{equation} 
The form for $f (X; P)$ will be determined later so as to be the 
physical \lq\lq number density.'' If we impose (\ref{cont-f}), 
(\ref{V}) vanishes as it should be. 

The \lq\lq free form'' of the propagator (\ref{free-p}) should be 
replaced by (\ref{non-e}) - (\ref{prop1}) with the $|p_0|$ 
prescription: 
\begin{eqnarray} 
\hat{\Delta} (x, y) & = & \int \frac{d^{\, 4} P}{(2 \pi)^4} \, e^{- 
i P \cdot (x - y)} \hat{\Delta} (X; P) , 
\label{prop20} 
\\ 
\hat{\Delta} (X; P) & = & \hat{\Delta}^{(0)} (X; P) + 
\hat{\Delta}^{(1)} (X; P) 
\label{prop2} 
\\ 
\hat{\Delta}^{(0)} (X; P) & = & \hat{B}_L (X; P) \, 
\hat{\Delta}_{diag} (P) \, \hat{B}_R (X; P) = \hat{\Delta}_{R A} (P) 
+ \hat{\Delta}^{(0) '} (X; P) , 
\label{prop3} 
\\ 
\hat{B}_L (X; P) & = & \left( 
\begin{array}{cc} 
1 & \;\;\; f (X; P) \\ 
1 & \;\;\; 1 + f (X; P) 
\end{array}
\right) \, , \;\;\;\;\;\;\; 
\hat{B}_R (X; P) = \left( 
\begin{array}{cc} 
1 + f (X; P) & \;\;\; f (X; P) \\ 
1 & \;\;\; 1 
\end{array}
\right) , 
\label{Bog-yo} 
\end{eqnarray} 
where $\hat{\Delta}_{R A} (P)$ is as in (\ref{vac}) and (cf. 
(\ref{p0f})) 
\begin{eqnarray} 
\hat{\Delta}^{(0) '} (X; P) & = & f (X; P) \, (\Delta_R (P) - 
\Delta_A (P)) \, \hat{A}_+ , 
\label{prop5} \\ 
\hat{\Delta}^{(1)} (X; P) & = & \frac{i}{2} \frac{\partial f (X; 
P)}{\partial X^\mu} \, \frac{\partial}{\partial P_\mu} (\Delta_R (P) 
+ \Delta_A (P)) \, \hat{A}_+ \\ 
& = & - i P \cdot \partial_X f (X; P) \left[ \Delta_R^2 (P) + 
\Delta_A^2 (P) \right] \hat{A}_+ . 
\label{prop4} 
\end{eqnarray} 

Let us discuss two perturbative schemes in analogy with mass 
renormalization in vacuum theory. We take the Lagrangian density 
\begin{eqnarray*} 
{\cal L} & = & {\cal L}_0 + {\cal L}_{int} , \\ 
{\cal L}_0 & = & \partial_\mu \phi^\dagger \, \partial^\mu \phi - 
m_B^2 \, \phi^\dagger \phi .  
\end{eqnarray*} 
The \lq\lq bare scheme'' in vacuum theory takes ${\cal L}_0$ as a 
nonperturbative part. Then, the propagator reads, 
\[ 
\Delta (P) = \frac{1}{P^2 - m_B^2 + i \eta} , 
\] 
which corresponds to (\ref{free-p}) in the bare-$f$ scheme here. The 
\lq\lq renormalized scheme'' in vacuum theory adopts (cf. 
(\ref{kuri}) and (\ref{Lag-1})) 
\begin{equation} 
{\cal L}_0' = \partial_\mu \phi^\dagger \, \partial^\mu \phi - m^2 
\phi^\dagger \phi , 
\label{mogi} 
\end{equation} 
as the nonperturbative part. Here $m = Z_m^{- 1 / 2} m_B$ is a 
renormalized mass. Equation (\ref{mogi}) leads to 
\[ 
\Delta (P) = \frac{1}{P^2 - m^2 + i \eta} , 
\] 
for the propagator, which corresponds to 
(\ref{prop20})~-~(\ref{prop4}) in the physical-$f$ scheme here. To 
compensate the difference ${\cal L}_0 - {\cal L}_0'$, a counter term 
should be introduced into the Lagrangian density, ${\cal L}_c = - 
(Z_m Z - 1) \, m^2 \, \phi^\dagger \phi$, which corresponds to 
(\ref{count}) here. 

To summarize, the bare-$f$ scheme is constructed in terms of 
original bare number density $f_B (X; P)$ [respect. bare mass 
$m_B$]. On the other hand, the physical-$f$ scheme is constructed in 
terms of physical number density $f (X; P)$ [respect. renormalized 
mass $m$]. Both perturbative schemes are equivalent. The first 
scheme starts with the \lq\lq bare quantity'' and the \lq\lq 
renormalization'' is done at the end, while in the second scheme, 
the \lq\lq renormalization'' is done at the beginning by introducing 
the counter hat-Lagrangian $\hat{L}_c$ [respect. ${\cal L}_c$]. It 
is worth recalling that the renormalized mass $m$ is defined so that 
$m$ is free from UV divergences. However, there is arbitrariness in 
the definition of the \lq\lq finite part'' of $m$. $m$ is determined 
by imposing some condition. (In this paper, we are adopting the 
$\overline{\mbox{MS}}$ scheme.) Thus, $m$ is determined order by 
order in perturbation series. As will be seen in the next section, 
this also applies to the present physical-$f$ scheme. Namely, there 
is arbitrariness in defining $f$. In Section VC, we shall impose the 
condition for determining $f$ so as to be the physical number 
density, under which $P \cdot \partial_X f (X; P)$ in 
$\hat{\Delta}^{(1)} (X; P)$, Eq.~(\ref{prop4}), and then also $f (X; 
P)$ turn out to be determined order by order in perturbation series. 
\setcounter{equation}{0} 
\setcounter{section}{4} 
\def\theequation{\mbox{\arabic{section}.\arabic{equation}}} 
\section{Interacting field} 
Interactions among fields give rise to reactions taking place in the 
system, which, in turn, causes a nontrivial change in number 
density of quasiparticles. In this section, we analyze full 
propagator in two perturbative schemes presented at the last section 
and demonstrate their equivalence. 
\subsection{Self-energy part} 
The self-energy part $\hat{\Sigma}$ takes the form 
\begin{eqnarray} 
\hat{\Sigma} (x, y) & = & \hat{\Sigma}_{int} (x, y) + \hat{\Sigma}_c 
(x, y) , \nonumber \\ 
\hat{\Sigma}_c (x, y) & = & - \hat{\cal V}_c (x, y) \equiv - {\cal 
V}_c (x, y) \hat{A}_- , 
\label{self0} 
\end{eqnarray} 
where $\hat{\Sigma}_{int}$ is a loop-diagram contribution, 
$\hat{\Sigma}_c$ comes from the counter hat-Lagrangian density 
$\hat{\cal L}_c$ in (\ref{count}), and $\hat{\cal V}_c$ is as in 
(\ref{V}). 

It can be shown that (cf. Appendix B) 
\begin{equation} 
\sum_{i, \, j = 1}^2 \Sigma_{i j} = \sum_{i, \, j = 1}^2 \left( 
\Sigma_{int} \right)_{i j} = 0 , 
\label{iden-self} 
\end{equation} 
with which we obtain, in the configuration space, 
\begin{equation} 
\hat{\underline{\Sigma}} \equiv \hat{B}_R \cdot \hat{\Sigma} \cdot 
\hat{B}_L = \left( 
\begin{array}{cc}
\Sigma_R & \;\; \Sigma_{off} \\ 
0 & \;\; - \Sigma_A 
\end{array}
\right) , 
\label{self1} 
\end{equation} 
where $\hat{B}_{R (L)}$ is as in (\ref{Bogo-p0}) and 
\begin{eqnarray} 
\Sigma_R & = & \Sigma_{1 1} + \Sigma_{1 2} = (\Sigma_{int})_{1 1} + 
(\Sigma_{int})_{1 2} , \nonumber \\ 
\Sigma_A & = & - (\Sigma_{2 2} + \Sigma_{1 2}) = - \{ 
(\Sigma_{int})_{2 2} + (\Sigma_{int})_{1 2} \} \; (= \Sigma_R^*) , 
\nonumber \\ 
\Sigma_{off} & = & \Sigma_{1 2} + \Sigma_{1 2} \cdot f - \Sigma_{2 
1} \cdot f + \Sigma_A \cdot f - f \cdot \Sigma_A + {\cal V} . 
\label{self2}
\end{eqnarray} 
$\Sigma_R$ and $\Sigma_A$ are called the retarded and advanced 
self-energy parts, respectively. 

Going to momentum space, we obtain, after some manipulations, 
\begin{equation} 
\Sigma_{off} (X; P) \simeq - i \left\{ f(X; P), \, P^2 - m^2 - Re 
\Sigma_R (X; P) \right\} + i \tilde{\Gamma}^{(p)} (X; P) , 
\label{off} 
\end{equation} 
where 
\begin{eqnarray} 
\left\{ A (X; P), \, B (X; P) \right\} & \equiv & \frac{\partial A 
(X; P)}{\partial X_\mu} \frac{\partial B (X; P)}{\partial P^\mu} - 
\frac{\partial A (X; P)}{\partial P^\mu} \frac{\partial B (X; 
P)}{\partial X_\mu} , 
\label{teigi} 
\\ 
i \tilde{\Gamma}^{(p)} (X; P) & \equiv & (1 + f (X; P)) \Sigma_{1 2} 
(X; P) - f (X; P) \Sigma_{2 1} (X; P) . 
\label{net-sei} 
\end{eqnarray} 
In deriving (\ref{off}), use has been made of the relation 
$\Sigma_A (X; P) = [\Sigma_R (X; P)]^*$, a proof of which is given 
in Appendix B. 
\subsection{Self-energy-part resummed propagator} 
In this subsection, we compute a self-energy-part resummed full 
propagator trough the Dyson equation, 
\[ 
\hat{G} (x, y) = \hat{\Delta} (x, y) + \left[ \hat{\Delta} \cdot 
\hat{\Sigma} \cdot \hat{G} \right] (x, y) .  
\] 
We would like to obtain $\hat{G} (X; P)$ ($X = (x + y) / 2$), the 
Fourier transform of $\hat{G} (x, y)$ on $x - y$. Using the 
transformation (\ref{imple-p0}), we obtain 
\begin{eqnarray*} 
\hat{\underline{G}} (x, y) & = & \left[ \hat{B}_L^{- 1} \cdot 
\hat{G} \cdot \hat{B}_R^{- 1} \right] (x, y) \nonumber \\ 
& = & \hat{\Delta}_{diag} (x, y) + \left[ \hat{\Delta}_{diag} \cdot 
\hat{\underline{\Sigma}} \cdot \hat{\underline{G}} \right] (x, y) . 
\end{eqnarray*} 
\noindent Fourier transforming on $x - y$ using (\ref{al}) in 
Appendix C, we obtain 
\begin{eqnarray} 
\hat{\underline{G}} (X; P) & \simeq & \hat{\Delta}_{diag} (P) + 
\hat{\Delta}_{diag} (P)\hat{\underline{\Sigma}} (X; P) 
\hat{\underline{G}} (X; P) \nonumber \\ 
& & + \frac{i}{2} \left[ \frac{\partial 
\hat{\Delta}_{diag}}{\partial P^\mu} \frac{\partial 
\hat{\underline{\Sigma}}}{\partial X_\mu} \hat{\underline{G}} - 
\hat{\Delta}_{diag} \frac{\partial 
\hat{\underline{\Sigma}}}{\partial X_\mu} \frac{\partial 
\hat{\underline{G}}}{\partial P^\mu} + \frac{\partial 
\hat{\Delta}_{diag} \hat{\underline{\Sigma}}}{\partial P^\mu} 
\frac{\partial \hat{\underline{G}}}{\partial X_\mu} \right] , 
\label{sum0} 
\end{eqnarray} 
where (cf. (\ref{self1})) 
\begin{eqnarray} 
\hat{\underline{\Sigma}} & = & \hat{\Sigma}_{diag} + 
\hat{\Sigma}_{off} \left( 
\begin{array}{cc} 
0 & 1 \\ 
0 & 0 
\end{array} 
\right) , \nonumber \\ 
\hat{\Sigma}_{diag} & = & \mbox{diag} \left( \Sigma_R, - \Sigma_A 
\right) . 
\label{sum1} 
\end{eqnarray} 
The terms in the square brackets in (\ref{sum0}) involve 
$X^\mu$-derivatives and are small when compared to the first term on 
the RHS. 

Let us solve (\ref{sum0}) iteratively. In doing so, we assume that 
the leading terms on the RHS of (\ref{sum0}) are 
$\hat{\Delta}_{diag} + \hat{\Delta}_{diag} \hat{\Sigma}_{diag} 
\hat{\underline{G}}$ and $\hat{\underline{\Sigma}}$'s in the square 
brackets may be approximated as $\hat{\underline{\Sigma}} \simeq 
\hat{\Sigma}_{diag}$. This assumption will be justified a posteriori 
(cf. the next subsection). Straightforward manipulation yields 
\begin{equation} 
\hat{\underline{G}} (X; P) = \hat{G}_{diag} (X; P) - \Sigma_{off} 
(X; P) G_R (X; P) G_A (X; P) \left( 
\begin{array}{cc} 
0 & \;\; 1 \\ 
0 & \;\; 0 
\end{array} 
\right) , 
\label{sum2} 
\end{equation} 
where 
\begin{eqnarray} 
\hat{G}_{diag} (X; P) & = & \mbox{diag} \left( G_R (X; P), \, - G_A 
(X; P) \right) , \nonumber \\ 
G_{R (A)} (X; P) & = & \frac{1}{P^2 - m^2 - \Sigma_{R (A)} (X; P) 
\pm i \epsilon (p_0) \eta} . 
\label{re-re} 
\end{eqnarray} 
It is to be noted that (\ref{sum2}) is exact to the derivative 
approximation, i.e., no term including the first derivative (with 
respect to $X^\mu$) arises. Then, using again (\ref{al}) in Appendix 
C, we obtain 
\begin{eqnarray} 
\hat{G} (X; P) & = & \hat{G}^{(0)} (X; P) + \hat{G}^{(1)} (X; P) , 
\label{sum3} 
\\ 
\hat{G}^{(0)} (X; P) & = & \hat{B}_L (X; P) \hat{G}_{diag} (X; P) 
\hat{B}_R (X; P) \nonumber \\ 
& = & \hat{G}_{R A} (X; P) + \hat{G}^{(0) '} (X; P) , 
\label{sum33} 
\end{eqnarray} 
where 
\begin{eqnarray} 
\hat{G}_{R A} (X; P) & = & \left( 
\begin{array}{cc} 
G_R (X; P) & \;\;\; 0 \\ 
G_R (X; P) - G_A (X; P) & \;\;\; - G_A (X; P) 
\end{array}
\right) , 
\label{sum34} 
\\ 
\hat{G}^{(0) '} (X; P) & = & f (X; P) \left( G_R (X; P) - G_A (X; P) 
\right) \hat{A}_+ , 
\label{sum38} 
\\ 
\hat{G}^{(1)} (X; P) & = & - \Sigma_{off} (X; P) G_R (X; P) G_A (X; 
P) \hat{A}_+ 
\label{sum35} 
\\ 
& & - \frac{i}{2} \left[ \left\{ f (X; P), \, P^2 - m^2 - \Sigma_R 
(X; P) \right\} G_R^2 (X; P) \right. \nonumber \\ 
& & \left. + \left\{ f (X; P), \; P^2 - m^2 - \Sigma_A (X; P) 
\right\} G_A^2 (X; P) \right] \hat{A}_+ \nonumber \\ 
& = & \frac{1}{2} \Sigma_{off} (X; P) \left( G_R (X; P) - G_A (X; P) 
\right)^2 \hat{A}_+ \nonumber \\ 
& & - \frac{1}{2} \left\{ f (X; P), \, Im \Sigma_R (X; P) \right\} 
\left( G_R^2 (X; P) - G_A^2 (X; P) \right) \hat{A}_+ \nonumber \\ 
& & - \frac{1}{2} \tilde{\Gamma}^{(p)} (X; P) \left( G_R^2 (X; P) + 
G_A^2 (X; P) \right) \hat{A}_+ . 
\label{sum8} 
\end{eqnarray} 
In narrow-width approximation, $Im \Sigma_R = - Im \Sigma_A \to - 
\epsilon(p_0) 0^+$, the first term on the RHS of (\ref{sum8}) 
develops pinch singularity, while the rest turns out to the 
well-defined distributions. 

For clarifying the physical meaning of (\ref{sum8}), we compute the 
contribution to the physical number density through analyzing the 
contribution to the statistical average of the current density 
$\langle j^\mu \rangle$, Eq.~(\ref{current}). The second and third 
terms on the RHS of (\ref{sum8}) leads to perturbative corrections 
to $\langle j^\mu \rangle$ in (\ref{curr1}) due to quantum and 
medium effects, while the first term yields a large contribution, 
\begin{equation} 
\langle \delta j^\mu (x) \rangle = i \int \frac{d^{\, 4} P}{(2 
\pi)^4} \, P^\mu \, \Sigma_{off} (x; P) \left( G_R (x; P) - G_A (x; 
P) \right)^2 , 
\label{div} 
\end{equation} 
which diverges in the narrow-width approximation. (Note again that 
the argument $x$ here is a macroscopic spacetime coordinates.) In 
the next subsection, we inspect this large contribution more closely 
in the bare-$f$ and physical-$f$ schemes. 
\subsection{Two perturbative schemes revisited} 
\subsubsection{Bare-$f$ scheme} 
Let us estimate (\ref{div}). All the formulae displayed above may be 
used with $f \to f_B$ and with imposition of (\ref{cont-f}). 
Important region of integration is where $Re (G_R^{- 1} (x; P)) = 
P^2 - m^2 - Re \Sigma_R (x; P) \simeq 0$. We define \lq\lq on the 
mass-shell'' $p_0 = \pm \omega_\pm (X; \pm {\bf p})$ $(\equiv \pm 
\omega_\pm)$ through 
\begin{equation} 
[P^2 - m^2 - Re \Sigma_R (X; P))] \, \rule[-3mm]{.14mm}{8.5mm} 
\raisebox{-2.85mm}{\scriptsize{$\; p_0 = \pm \omega_\pm$}} = 0 . 
\label{on-shell} 
\end{equation} 
We make following approximations; 
\begin{eqnarray} 
P^\mu \Sigma_{off} & \simeq & \theta (p_0) \, (\omega_+, {\bf 
p})^\mu \, \Sigma_{off}^{(bare) (+)} + \theta (- p_0) \, (- 
\omega_-, {\bf p})^\mu \, \Sigma_{off}^{(bare) (-)} , \nonumber \\ 
\frac{- i}{2} \left[ G_R (x; P) - G_A (x; P) \right] & \simeq & 
\sum_{\tau \pm} \frac{\theta (\tau p_0) \, Im 
\Sigma_R^{(\tau)}}{\left[ 2 \omega_\tau Z_\tau^{- 1} (p_0 - \tau 
\omega_\tau) \right]^2 + (Im \Sigma_R^{(\tau)})^2} , 
\label{koto} 
\end{eqnarray} 
where 
\begin{eqnarray} 
\Sigma^{(bare) (\pm)}_{off} & = & \Sigma_{off} (X; p_0 = \pm 
\omega_\pm, {\bf p}), \;\;\; \Sigma^{(\pm)}_R = \Sigma_R (X; p_0 = 
\pm \omega_\pm, {\bf p}) , 
\label{hati} 
\\ 
Z_\pm^{- 1} & \equiv & 1 \mp \frac{\partial Re \Sigma_R}{\partial 
p_0} \rule[-3mm]{.14mm}{8.5mm} \raisebox{-2.85mm}{\scriptsize{$\; 
p_0 = \pm \omega_\pm$}} . 
\label{Z} 
\end{eqnarray} 
It is convenient to carry out the wave-function renormalization such 
that $Z_\pm = 1$, which we assume to be done for a while. Using 
(\ref{koto}) in (\ref{div}), we obtain 
\begin{equation} 
\langle \delta j^\mu (x) \rangle \simeq \frac{1}{2} \int \frac{d^{\, 
3} p}{(2 \pi)^3} \left[ \left( 1, {\bf p} / \omega_+ \right)^\mu 
\frac{- i \Sigma_{off}^{(bare) (+)}}{|Im \Sigma_R^{(+)}|} - \left( 
1, - {\bf p} / \omega_- \right)^\mu \frac{- i \Sigma_{off}^{(bare) 
(-)}}{|Im \Sigma_R^{(-)}|} \right] . 
\label{ii} 
\end{equation} 
In deriving this, we have assumed that $|Im \Sigma_R^{(\pm)}| << 
\omega_\pm$, which is the case in equilibrium case. As in 
(\ref{current})~-~(\ref{num-den}), we can read off from (\ref{ii}) 
the contribution to the physical number density: 
\begin{eqnarray} 
\delta \tilde{N}_\pm (x, \pm {\bf p}) & = & \frac{1}{2} \frac{- i 
\Sigma_{off}^{(bare) (\pm)}}{|Im \Sigma_R^{(\pm)}|} , \nonumber \\ 
& = & \frac{1}{2} \tau_\pm \left[ - i \Gamma_\pm^{(p)} + 
\frac{1}{2 \omega_\pm} \left\{ f, \, Re \Sigma_R \right\} 
\rule[-3mm]{.14mm}{8.5mm} \raisebox{-2.85mm}{\scriptsize{$\; p_0 = 
\pm \omega_\pm$}} \right] . 
\label{iii} 
\end{eqnarray} 
Here $\tau_+ \equiv 2 \omega_+ / |Im \Sigma_R^{(+)}|$ [$\tau_- 
\equiv 2 \omega_- / |Im \Sigma_R^{(-)}|$] is a characteristic time 
during which a particle with momentum ${\bf p}$ [an antiparticle 
with momentum $- {\bf p}$] damps: 
\begin{eqnarray*} 
G_R (X; x_0 - y_0; {\bf p}) & = & \int \frac{d^{\, 3} p}{(2 \pi)^3} 
\, e^{i {\bf p} \cdot ({\bf x} - {\bf y})} \, G_R (x, y) \nonumber 
\\ 
& \propto & e^{- (x_0 - y_0) / \tau_+} \;\;\;\;\;\;\; \mbox{for a 
particle mode} , \nonumber \\ 
& \propto & e^{- (x_0 - y_0) / \tau_-} \;\;\;\;\;\;\; \mbox{for an 
antiparticle mode} . 
\end{eqnarray*} 
In (\ref{iii}), 
\begin{equation} 
\Gamma^{(p)}_\pm = \frac{- i}{2 \omega_\pm} [(1 + N_\pm) \Sigma_{1 2 
(2 1)} - N_\pm \Sigma_{2 1 (1 2)}] \, \rule[-3mm]{.14mm}{8.5mm} 
\raisebox{-2.85mm}{\scriptsize{$\; p_0 = \pm \omega_\pm$}} , 
\label{net-pro} 
\end{equation} 
which comes from $\tilde{\Gamma}^{(p)}$ $\in - i \Sigma_{off}$ in 
(\ref{net-sei}), is a net production rate of a particle (an 
antiparticle) of momentum ${\bf p}$ ($- {\bf p}$). In fact, 
$\Gamma^{(p)}_\pm$, Eq.~(\ref{net-pro}), is a difference between the 
production rate and the decay rate, so that $\Gamma^{(p)}_\pm$ is 
the net production rate. In the case of equilibrium system, 
$\Gamma^{(p)}_\pm = 0$ (detailed balance formula). 

To disclose the physical meaning of (\ref{iii}), for the time being, 
we put aside the second term in the square brackets, whose physical 
interpretation will become clear later at (\ref{mawasi}). Then, the 
RHS of (\ref{iii}) is the change in the physical number density, 
during the time interval $\tau_\pm / 2$, due the the net production 
rate. In Section I, we have introduced spacetime cells, whose size 
is $L^\mu$ $(\mu = 0, 1, 2, 3)$. It is quite natural to take 
$\tau_\pm = L_0$, where $L_0$ is the size of a time direction of the 
spacetime cell (including the spacetime point $x^\mu$ in 
(\ref{iii})). [Strictly speaking, in general, $\tau_+ \neq \tau_-$. 
However, in the present crude argument, we ignore this difference.] 
It is interesting to note that (\ref{iii}) is {\em half} of the net 
production probability, due to the reaction, during the time 
interval $\tau_\pm = L_0$. [In this respect, cf. \cite{nie-pl}.] 

Noticing that we are concerned about the particle mode with momentum 
${\bf p}$ [antiparticle mode with momentum $- {\bf p}$], we see that 
(\ref{iii}) leads to the relations for the physical number density, 
\begin{eqnarray*} 
\tilde{N}_+ (x_0 + L_0, {\bf x} + L_0 {\bf p} / \omega_+; {\bf p}) 
- \tilde{N}_+ (x_0, {\bf x}; {\bf p}) & \simeq & L_0 \frac{- i 
\Sigma_{off}^{(bare)} (x; \omega_+, {\bf p})}{2 \omega_+} , \\ 
\tilde{N}_- (x_0 + L_0, {\bf x} - L_0 {\bf p} / \omega_-; - {\bf p}) 
- \tilde{N}_- (x_0, {\bf x}; - {\bf p}) & \simeq & L_0 \frac{- i 
\Sigma_{off}^{(bare)} (x; - \omega_-, {\bf p})}{2 \omega_-} , 
\end{eqnarray*} 
from which we obtain 
\begin{equation} 
\frac{\partial \tilde{N}_\pm (x; \pm {\bf p})}{\partial x_0} \pm 
\frac{{\bf p}}{\omega_\pm} \cdot \frac{\partial 
\tilde{N}_\pm}{\partial {\bf x}} \simeq  \frac{- i 
\Sigma_{off}^{(bare)} (x; \omega_\pm, {\bf p})}{2 \omega_\pm} . 
\label{bare-evo} 
\end{equation} 
Physical interpretation of these relations will be discussed below. 

In calculating some quantity in the bare-$f$ scheme, one carries out 
perturbative calculation using the propagator $\hat{\Delta} (X; P)$ 
given in (\ref{free-fin}). The calculated quantity is written in 
terms of the bare number density $f_B (X; P)$ or $N_{B \pm} (X; E_p, 
\pm {\bf p})$. Using the solution to (\ref{bare-evo}), one can 
rewrite the quantity in terms of the \lq\lq physical number 
density'' $\tilde{N}_\pm (X; \pm {\bf p})$. 
\subsubsection{Physical-$f$ scheme} 
As defined in Section IVC, we change the definition of the 
number-density function $f$. Then, in this physical-$f$ scheme, the 
number density function $f$ does not obey (\ref{cont-f}). It is 
convenient to define $f$, such that the number densities 
\begin{eqnarray*} 
N_+ (X; E_{\bf p}, \hat{\bf p}) & = & f (X; p_0 = E_{\bf p}, {\bf 
p}) , \\ 
N_- (X; E_{\bf p}, - \hat{\bf p}) & = & - 1 - f (X; p_0 = - E_{\bf 
p}, {\bf p}) , 
\end{eqnarray*} 
are as close as the physical number densities $\tilde{N}_\pm (X; \pm 
{\bf p})$. As seen above, the first term on the RHS of (\ref{sum8}), 
which is proportional to $\Sigma_{off} (X; P)$, yields a large 
contribution to the physical number density. Thus, we demand that $f 
(X; P)$ changes in time $X_0$ so that $\Sigma_{off} (X; P)$ 
vanishes: 
\begin{eqnarray} 
i \Sigma_{off} (X; P) & = & \left\{ f (X; P), \, P^2 - m^2 - Re 
\Sigma_R (X; P) \right\} - \tilde{\Gamma}^{(p)} (X; P) \nonumber \\ 
& = & 0 . 
\label{r-cond} 
\end{eqnarray} 
Given an initial data at the \lq\lq initial'' time $X_0 = T_{in}$, 
$f (X_0 = T_{in}, {\bf X}; P)$, (\ref{r-cond}) serves as determining 
$f (X_0, {\bf X}; P)$ at later time $X_0$. Note that $Re \Sigma_R$ 
and $\tilde{\Gamma}^{(p)}$ are evaluated order by order in 
perturbation series. Then, through (\ref{r-cond}), $f (X; P)$ is 
also determined order by order. (In this relation, see the argument 
at Section IVC for the mass 
renormalization in vacuum theory.) We refer (\ref{r-cond}) to as the 
\lq\lq renormalization condition'' for the number density, which 
replaces the condition (\ref{cont-f}) in the bare scheme. In the 
present scheme, transformed self-energy part 
$\hat{\underline{\Sigma}}$, Eq.~(\ref{self1}), is diagonal: 
\begin{equation} 
\hat{B}_R \hat{\Sigma} \hat{B}_L = \left( 
\begin{array}{cc} 
\Sigma_R & 0 \\ 
0 & - \Sigma_A 
\end{array} 
\right) . 
\label{ref3} 
\end{equation} 
It is to be noted that, with the condition (\ref{r-cond}), $\hat{G} 
(X; P)$ in (\ref{sum3})~-~(\ref{sum8}) is free from singular 
contribution in the narrow-width limit (cf. above after 
(\ref{sum8})) and the difference between $\hat{G}$ and the bare 
propagator $\hat{\Delta}$ leads to a perturbative correction to the 
physical number density. 

In the sequel of this paper, we adopt the physical-$f$ scheme. 

Let us inspect the physical implication of the \lq\lq 
renormalization condition'' (\ref{r-cond}): 
\begin{equation} 
2 P \cdot \partial_{X} f (X; P) - \left\{ f (X; P), \; Re \Sigma_R 
(X; P) \right\} = \tilde{\Gamma}^{(p)} (X; P) . 
\label{kik} 
\end{equation} 
Here let us recall the definition (\ref{on-shell}) of the \lq\lq 
mass-shell'' and find the form of (\ref{kik}) on the mass-shell. 
From (\ref{on-shell}), we have 
\begin{eqnarray} 
& & \left[ \left\{ 2 p_0 - \frac{\partial Re 
\Sigma_R}{\partial p_0} \right\} {\bf v}_\pm - 2 {\bf p} - 
\frac{\partial Re \Sigma_R}{\partial {\bf p}} \right] 
\rule[-3mm]{.14mm}{8.5mm} \raisebox{-2.85mm}{\scriptsize{$\; p_0 
= \pm \omega_\pm (X; \pm {\bf p})$}} = 0 , \nonumber \\ 
& & \frac{\partial Re \Sigma_R}{\partial X} = \pm \frac{\partial 
\omega_\pm}{\partial X} \left( 2 p_0 - \frac{\partial Re 
\Sigma_R}{\partial p_0} \right) \rule[-3mm]{.14mm}{8.5mm} 
\raisebox{-2.85mm}{\scriptsize{$\; p_0 = \pm \omega_\pm (X; \pm {\bf 
p})$}} , 
\label{yaha} 
\end{eqnarray} 
where ${\bf v}_\pm \equiv \pm \partial \omega_\pm (X; \pm {\bf p}) / 
\partial {\bf p}$ is the velocity of the $\pm$ mode. We also recall 
that $f = \theta (p_0) N_+ - \theta (- p_0) (1 + N_-)$, where $N_+ = 
N_+ (X; p_0 = \omega_+, \hat{\bf p})$ [$N_- = N_- (X; - p_0 = 
\omega_-, - {\bf p})$] is the number density of a (quasi)particle 
[an anti(quasi)particle] with momentum ${\bf p}$ [$- {\bf p}$]. It 
is straightforward to show that, on the mass-shell, $p_0 = \pm 
\omega_\pm (X; \pm {\bf p})$ ($= \pm \omega_\pm$), (\ref{kik}) 
becomes 
\begin{eqnarray} 
& & \frac{\partial N_\pm}{\partial X_0} + {\bf v}_\pm \cdot \nabla_X 
N_\pm \pm \frac{\partial \omega_\pm}{\partial X_\mu} \frac{\partial 
N_\pm}{\partial P^\mu} \rule[-3mm]{.14mm}{8.5mm} 
\raisebox{-2.85mm}{\scriptsize{$\; p_0 = \pm \omega_\pm$}} \nonumber 
\\ 
& & \mbox{\hspace*{5ex}} = \frac{d N_{\pm} (X; \omega_\pm (X; \pm 
{\bf p}), \pm \hat{\bf p})}{d X_0} + \frac{d \omega_\pm}{d {\bf p}} 
\cdot \frac{d N_\pm}{d {\bf X}} - \frac{\partial 
\omega_\pm}{\partial {\bf X}} \cdot \frac{d N_\pm}{d {\bf p}} 
\nonumber \\ 
& & \mbox{\hspace*{5ex}} = Z_\pm \Gamma_\pm^{(p)} 
\rule[-3mm]{.14mm}{8.5mm} \raisebox{-2.85mm}{\scriptsize{$\; p_0 = 
\pm \omega_\pm$}} , 
\label{evo}
\end{eqnarray} 
where $Z_\pm$ is the wave-function renormalization factor (cf. 
(\ref{Z})) and, as defined in (\ref{net-pro}), $\Gamma^{(p)}_\pm$ 
is the net production rate of a particle (an antiparticle) with 
momentum ${\bf p}$ [$- {\bf p}$]. If we employ the wave-function 
renormalization condition $Z_\pm = 1$, the RHS turns out to the net 
production rate $\Gamma^{(p)}_\pm$. The quasiparticle distribution 
function $N_\pm = N_\pm (X; p_0 = \omega_\pm, \hat{\bf p})$ here is 
nothing but the relativistic Wigner function and (\ref{evo}) is the 
generalized kinetic or Boltzmann equation (cf. \cite{Hei}) for the 
relativistic complex-scalar-field system. 

Here we make a comment on a role of the second term on the left-hand 
side (LHS) of (\ref{kik}). The second term represents the effect due 
to the change of \lq\lq mass;'' 
\begin{equation} 
\left\{ f (X; P), \; Re \Sigma_R (X; P) \right\} = \frac{\partial f 
(X; P)}{\partial X_\mu} \frac{\partial Re \Sigma_R (X; P)}{\partial 
P^\mu} - \frac{\partial f (X; P)}{\partial P^\mu} \frac{\partial Re 
\Sigma_R (X; P)}{\partial X_\mu} . 
\label{mawasi} 
\end{equation} 
The first term on the RHS together with the first term on the LHS of 
(\ref{kik}) yields the term being proportional to the first two 
terms on the LHS of (\ref{evo}), while the second term on the RHS 
yields the term being proportional to the third term on the LHS of 
(\ref{evo}). 

One can show using (\ref{yaha}) that (\ref{bare-evo}) is the same in 
form as (\ref{evo}). [Recall that we have set $Z_\pm = 1$ in 
(\ref{bare-evo}).] This observation gives a support for the 
qualitative argument made above in conjunction with 
(\ref{bare-evo}). 
\subsection{Resummation of the lowest-order $\hat{\Sigma}$} 
To lowest order in $\lambda$ (cf. (\ref{Lag-1})), a tadpole diagram 
contributes to $\hat{\Sigma}_{int}$: 
\begin{eqnarray} 
\hat{\Sigma}_{int} (X) & = & M^2 (X) \hat{\tau}_3 , \\ 
M^2 (X) & \simeq & i \lambda \int \frac{d^{\, D} P}{(2 \pi)^D} \, 
\Delta_{1 1}^{(0)} (X; P) + (Z_m Z - 1) \, m^2 \nonumber \\ 
& = & i \lambda \int \frac{d^{\, D} P}{(2 \pi)^D} \left[ 
\frac{1}{P^2 - m^2 + i \eta} \right. \nonumber \\ 
& & \left.- 2 \pi i \left\{ \theta (p_0) N_+ (X; p_0, \hat{\bf p}) + 
\theta (- p_0) N_- (X; - p_0, - \hat{\bf p}) \right\} \right] \delta 
(P^2 - m^2) \nonumber \\ 
& & + (Z_m Z - 1) \, m^2 . 
\label{mass-in0} 
\end{eqnarray} 
Here use has been made of the dimensional regularization: $D$ is the 
spacetime dimension and $\epsilon = 4 - D$ works as a regulator for 
the UV divergence. For the purpose of later use, we have adopted 
here the $F \overline{F}$ representation (cf. (\ref{aha1})) for 
$\Delta_{1 1}^{(0)}$, (\ref{prop2}) with (\ref{prop3}) and 
(\ref{prop5}). $N_\pm$ in (\ref{mass-in0}) is as in (\ref{p0f}). 
Note that $\Delta_{1 1}^{(1)}$ in (\ref{prop2}) with (\ref{prop4}) 
leads to a contribution, which is of higher order in $\lambda$. It 
should be kept in mind that, in dealing with $O (\lambda^2)$ 
contributions, (\ref{mass-in0}) with $\Delta_{1 1}^{(1)}$ for 
$\Delta_{1 1}^{(0)}$, as well as the two-loop contribution, should 
be taken into account. 

Standard manipulation yields 
\begin{eqnarray} 
M^2 (X) & = & M_0^2 + M^2_{induced} (X) , \nonumber \\ 
M_0^2 & = & - \frac{\lambda}{(4 \pi)^2} m^2 \left[ 
\frac{2}{\overline{\epsilon}} + 1 - \ln \frac{m^2}{\mu_d^2} \right] 
, 
\label{vacc} 
\\ 
M^2_{induced} (X) & = & \lambda \int \frac{d^{\, 3} p}{(2 \pi)^3} 
\frac{1}{2 E_p} \left[ N_+ (X; E_p, \hat{\bf p}) + N_- (X; E_p, - 
\hat{\bf p}) \right] , 
\label{mass-in}
\end{eqnarray} 
where 
\begin{equation} 
\frac{1}{\overline{\epsilon}} \equiv \frac{1}{4 - D} - 
\frac{\gamma_E}{2} + \frac{1}{2} \ln (4 \pi) . 
\label{dim}
\end{equation} 
Here $M_0^2$ is the squared self mass in vacuum theory, $\gamma_E$ 
is the Euler constant, and $\mu_d$ is an arbitrary scale parameter. 
As an UV renormalization scheme in vacuum theory, we are adopting 
the $\overline{\mbox{MS}}$ scheme. Thus, at the present lowest 
order, we have 
\[ 
Z_m = 1 + \frac{\lambda}{(4 \pi)^2} \, \frac{2}{\overline{\epsilon}} 
, \;\;\;\;\;\; Z = 1 , 
\] 
and then 
\begin{equation} 
M_0^2 = \frac{\lambda}{(4 \pi)^2} m^2 \ln \frac{m^2}{e \mu_d^2} . 
\label{aso} 
\end{equation} 
If we choose $\mu_d = m / \sqrt{e}$, we have $M_0^2 = 0$. 

In passing, for a locally thermal equilibrium system, we have 
\[ 
N_+ (X; E_p, \hat{\bf p}) = N_- (X; E_p, - \hat{\bf p}) = 
\frac{1}{e^{E_p / T (X)} - 1} , 
\] 
where $T (X)$ is the local temperature of the system. In this case, 
we obtain 
\[ 
M^2_{induced} (X) = \frac{\lambda}{2 \pi^2} m^2 \int_1^\infty d u \, 
\frac{\sqrt{u^2 - 1}}{e^{m u / T (X)} - 1} . 
\] 
In the high-temperature regime, $T (X) >> m$, we have 
\begin{equation} 
M^2_{induced} (X) \simeq \frac{\lambda}{1 2} \, T^2 (X) \left( 1 - 
\frac{3}{\pi} \frac{m}{T (X)} \right) \;\;\;\;\;\;\;\; (T (X) >> m) 
, 
\label{high-T} 
\end{equation} 
while, at low temperature, we have 
\begin{equation} 
M^2_{induced} (X) \simeq \frac{\lambda \, m^2}{(2 \pi)^{3 / 2}} 
\left( \frac{T (X)}{m} \right)^{3 / 2} e^{- m / T (X)} \left( 1 + 
\frac{3}{8} \frac{T (X)}{m} \right) \;\;\;\;\;\;\;\; (T (X) << m) . 
\label{low-T} 
\end{equation} 

Let us construct a $\hat{\Sigma}$-resummed propagator. To the 
present lowest order, $\Sigma_{1 2} = \Sigma_{2 1} = 0$ and then, 
from (\ref{net-sei}), $\tilde{\Gamma}^{(p)}$ vanishes. Then, the 
renormalization condition (\ref{kik}), $\Sigma_{off} = 0$, reads 
\begin{equation} 
i \Sigma_{off} (X; P) = 2 P \cdot \partial_{X} f (X; P) + 
\frac{\partial f (X; P)}{\partial P^\mu} \frac{\partial M^2 
(X)}{\partial X_\mu} = 0 , 
\label{resmm-1}
\end{equation} 
whose approximate solution is given by 
\begin{equation} 
f (X; P) \simeq {\cal F} (\underline{\bf X}; P^2 - M^2 (X), 
\underline{\bf p}) \simeq G (\underline{\bf X}; \underline{P}) . 
\label{resmm-sol} 
\end{equation} 
Here, ${\cal F}$ or $G$ is an arbitrary function and 
\begin{eqnarray} 
\underline{p_0} & \equiv & p_0 - \frac{\partial M^2 (X)}{\partial 
X^0} \frac{(X_0 - T_{in})}{2 p_0} , \nonumber \\ 
\underline{p_i} & \equiv & p_i - \frac{\partial M^2 (X)}{\partial 
X^i} \frac{(X_0 - T_{in})}{2 p_0} \;\;\;\;\;\; (\mbox{no summation 
over} \; i) , \nonumber \\ 
\underline{X_i} & \equiv & X_i - \frac{p_i}{p_0} (X_0 - T_{in}) + 
\frac{p_0 \partial M^2 (X) / \partial X^i - p_i \partial M^2 (X) / 
\partial X^0}{4 (p_0)^3} (X_0 - T_{in})^2 . 
\label{kai-tame} 
\end{eqnarray} 
Given an initial data $f (X_0 = T_{in}, {\bf X}; P) = G ({\bf X}; 
P)$, (\ref{resmm-sol}) gives $f (X_0, {\bf X}; P)$ at later time 
$X_0$ to $O (\lambda)$. 

Since $Im \Sigma_R = \tilde{\Gamma}^{(p)} = 0$, $\hat{G}^{(1)}$ in 
(\ref{sum8}) vanishes due to the condition (\ref{resmm-1}). Then, 
the $\hat{\Sigma}$-resummed propagator $\hat{G} (X; P)$ reads 
\begin{eqnarray} 
\hat{G} (X; P) & \simeq & \hat{G}^{(0)} (X; P) \nonumber \\ 
& = & G_{R A} + f (X; P) \, (G_R - G_A) \, \hat{A}_+ , 
\label{resmm0} 
\end{eqnarray} 
where 
\begin{eqnarray} 
G_{R A} & = & \left( 
\begin{array}{cc} 
G_R & \;\;\; 0 \\ 
G_R - G_A & \;\;\; - G_A 
\end{array} 
\right) , \nonumber \\ 
G_{R (A)} & = & G_{R (A)} (X; P) = \frac{1}{P^2 - m^2 - M^2 (X) \pm 
i \epsilon(p_0) \eta} . 
\label{resmm1}
\end{eqnarray} 
\subsection{Changing the mass parameter} 
The analysis in the last subsection shows that an efficient 
perturbative scheme is obtained as follows. Rewrite the Lagrangian 
density (\ref{Lag-1}) by introducing a new \lq\lq mass function'' $M 
(x)$, which is assumed to depend weakly on $x$: 
\begin{eqnarray} 
{\cal L} & = & {\cal L}_0 + {\cal L}_{int} + {\cal L}_{r c} , 
\nonumber \\ 
{\cal L}_0 & = & \partial_\mu \phi^\dagger (x) \, \partial^\mu \phi 
(x) - M^2 (x) \, \phi^\dagger (x) \, \phi (x) , \nonumber \\ 
{\cal L}_{int} & = & - \frac{\lambda}{4} \left( \phi^\dagger (x) \, 
\phi (x) \right)^2 , \nonumber \\ 
{\cal L}_{r c} & = & \chi (x) \, \phi^\dagger (x) \, \phi (x) + (Z - 
1) \partial_\mu \phi^\dagger (x) \, \partial^\mu \phi (x) - (Z_m Z - 
1) (M^2 (x) - \chi (x) ) \, \phi^\dagger (x) \, \phi (x) \nonumber 
\\ 
& & - \frac{\lambda}{4} \, (Z_\lambda Z^2 - 1) \left( \phi^\dagger 
(x) \, \phi (x) \right)^2 , 
\label{Lag-2}
\end{eqnarray} 
where 
\begin{equation} 
\chi (x) \equiv M^2 (x) - m^2 . 
\label{suri} 
\end{equation} 
Note that $M^2$ here is not $M^2$ in (\ref{mass-in0}). As discussed 
in \cite{chiku}, perturbation theory based on (\ref{Lag-2}) is 
UV-renormalizable. We give an simple argument for this in Appendix 
D. 

It should be emphasized \cite{chiku} that the perturbation theory 
based on (\ref{Lag-2}) is defined by taking ${\cal L}_0$ as a 
nonperturbative or \lq\lq free'' part and ${\cal L}_{int} +{\cal 
L}_{r c}$ as perturbative part. Then, $\chi (x) \phi^\dagger \phi$ 
in ${\cal L}_{r c}$ is a part of the perturbative terms. Thus, in 
traditional perturbation expansion, we assign 
\begin{equation} 
M^2 (x) = O (\lambda^0) \, , \;\;\;\;\;\; \chi (x) \; (= M^2 (x) - 
m^2) = O (\lambda) , 
\label{kan} 
\end{equation} 
and, in loop or $\delta$ expansion (cf. the next section), we assign 
\begin{equation} 
M^2 (x) = O (\delta^0) \, , \;\;\;\;\;\; \chi (x) \; (= M^2 (x) - 
m^2) = O (\delta) , 
\label{kan-1} 
\end{equation} 

It is obvious from the argument in the above subsection that, in the 
bare-$f$ scheme in perturbation theory based on (\ref{Lag-2}), the 
bare propagator is given by (\ref{resmm0}), 
\begin{equation} 
\hat{G} (X; P) = G_{R A} + f (X; P) \, (G_R - G_A) \, \hat{A}_+ , 
\label{yayo} 
\end{equation} 
where 
\begin{equation} 
G_{R (A)} = G_{R (A)} (X; P) = \frac{1}{P^2 - M^2 (X) \pm i 
\epsilon (p_0) \eta} . 
\label{kota} 
\end{equation} 
The number-density function $f (X; P)$ in (\ref{yayo}) obeys 
(\ref{resmm-1}). 

This result may also be derived through the same procedure as in 
Section IV. This time we start with the $|p_0|$-prescription-formula 
(\ref{imple-p0}) from the beginning, 
\begin{equation} 
\hat{G} (x, y) = \int d^{\, 4} u \, d^{\, 4} v \, \hat{B}_L (x, u) 
\, \hat{G}_{diag} (u, v) \, \hat{B}_R (v, y) . 
\label{yuri} 
\end{equation} 
Here $\hat{G}_{diag} = \mbox{diag} (G_R, - G_A)$ with $G_{R (A)} (x, 
y)$ a solution to 
\begin{equation} 
\left( \partial^2_x + M^2 (x) \right) G_{R (A)} (x, y) = \left( 
\partial^2_y + M^2 (y) \right) G_{R (A)} (x, y) = - \delta^4 (x - y) 
, 
\label{yosiki} 
\end{equation} 
with retarded (advanced) boundary condition. It can easily be shown 
that the solution to (\ref{yosiki}) is (\ref{kota}). Note that no 
first $X^\mu$-derivative term appears there. In place of 
(\ref{kita}), we have 
\begin{eqnarray} 
( \partial^2_x + M^2 (x)) \hat{G} (x, y) & = & - \hat{\tau}_3 \, 
\delta^4 (x - y) + \int d^{\, 4} u \, d^{\, 4} v \left[ \left\{ 
(\partial^2_x - \partial^2_u) \right. \right. \nonumber \\ 
& & \left. \left. + M^2 (x) - M^2 (u) \right\} \hat{B}_L (x, u) 
\right] \hat{\tau}_3 \, \hat{B}_R (u, v) \, \hat{\tau}_3 \, \hat{G} 
(v, y) , 
\label{new-1} 
\end{eqnarray} 
and in place of (\ref{count}), we have 
\[ 
\hat{\cal L}_c = - \int d^{\, 4} u \, d^{\, 4} v \, 
\hat{\phi}^\dagger (x) \, \hat{\tau}_3 \left[ \left\{ (\partial^2_x 
- \partial^2_u) + M^2 (x) - M^2 (u) \right\} \hat{B}_L (x, u) 
\right] \hat{\tau}_3 \, \hat{B}_R (u, v) \, \hat{\tau}_3 \hat{\phi} 
(v) .  
\] 
This yields a two-point vertex function 
\begin{eqnarray} 
i \hat{\cal V}_c (X; P) & = & 2 P \cdot \partial_X f (X; P) + 
\frac{d M^2 (X)}{d X_\mu} \frac{\partial f (X; P)}{\partial P^\mu} 
\hat{A}_- \nonumber \\ 
& = & \left\{f (X; P), \, P^2 - M^2 (X) \right\} \, \hat{A}_- . 
\label{new-2} 
\end{eqnarray} 
Since we are to obtain the bare propagator, $\hat{\Sigma}_{int} = 
\tilde{\Gamma}^{(p)} = 0$. Then, the \lq\lq renormalization 
condition'' (\ref{r-cond}) turns out to 
\begin{equation} 
\left\{ f (X; P) , \; P^2 - M^2 (X) \right\} = 0 , 
\label{kuri-j} 
\end{equation} 
with which we have $\hat{\cal V}_c (X; P) = 0$. Equation 
(\ref{kuri-j}) is nothing but (\ref{resmm-1}). Using this 
condition, we can derive (\ref{yayo}) from (\ref{yuri}), which is 
the bare propagator in the bare-$f$ scheme. The physical-$f$ scheme 
may be constructed in a similar manner as in Section IVC. 

In this theory, $\hat{\Sigma}_{int}$ to $O (\lambda)$ reads 
\[ 
\hat{\Sigma}_{int} (X) = \hat{\tau}_3 \left[ \frac{\lambda}{(4 
\pi)^2} M^2 (X) \ln \frac{M^2 (X)}{e \mu_d^2} + M^2_{induced} (X) - 
M^2 (X) + m^2 \right] \nonumber , 
\] 
where $M^2_{induced} (X)$ is as in (\ref{mass-in}) with $M^2 (X)$ 
for $m^2$ and $- \hat{\tau}_3 [M^2 (X) - m^2]$ has come from $\chi 
\phi^\dagger \phi \in {\cal L}_{r c}$, Eq. (\ref{Lag-2}). 

Up to this point, the function $M^2 (X)$ $(\geq 0)$ is arbitrary. An 
efficient perturbation theory is obtained by demanding, e.g., 
$\hat{\Sigma}_{int} (X) = 0$; 
\begin{equation} 
M^2 (X) \left[ 1 - \frac{\lambda}{(4 \pi)^2} \, \ln \frac{M^2 (X)}{e 
\mu_d^2} \right] = m^2 +  M^2_{induced} (X) . 
\label{nasi} 
\end{equation} 
The scheme with this $M (X)$ may be called \lq\lq presummation 
scheme.'' Equation (\ref{nasi}) is the gap equation, which serves as 
determining $M^2 (X)$ in terms of $\mu_d$, which is still an 
arbitrary parameter. 

It is worth mentioning that, in the case where the condition $[\rho, 
Q] = 0$, Eq.~(\ref{den-Q}), is not fulfilled, we may generalize 
(\ref{Lag-2}), although not necessary, by changing ${\cal L}_0$ to 
\[ 
{\cal L}_0 = \partial^\mu \phi^\dagger (x) \partial_\mu \phi (x) - 
M_+ (x) \phi^{(+) \dagger} (x) \phi^{(+)} (x) - M_- (x) \phi^{(-) 
\dagger} (x) \phi^{(-)} (x) , 
\] 
and changing ${\cal L}_{r c}$ accordingly. Here the superscript 
\lq\lq $(+)$'' [\lq\lq $(-)$] refers to the positive [negative] 
frequency part, and then $M_+ (x)$ [$M_- (x)$] is an effective mass 
of a quasiparticle [anti-quasiparticle]. 
\setcounter{equation}{0} 
\setcounter{section}{5} 
\def\theequation{\mbox{\arabic{section}.\arabic{equation}}} 
\section{The $O (N)$ linear $\sigma$ model} 
\subsection{Preliminary} 
In this section, we deal with a fermionless $O (N)$ linear $\sigma$ 
model, in which $N$ real-scalar fields constitute a vector 
representation $\vec{\phi} = \displaystyle{ 
\raisebox{0.9ex}{\scriptsize{$t$}}} \mbox{\hspace{-0.1ex}} (\phi_1, 
..., \phi_n)$ of $O (N)$. The case of $N = 4$ is of practical 
interest. Specifically, through computing the effective action, we 
analyze how the phase transition proceeds. The Lagrangian of the 
theory reads 
\begin{equation} 
{\cal L} = \frac{1}{2} \left[ (\partial \vec{\phi}_B )^2 - \mu^2_B 
\vec{\phi}^{\, 2}_B \right]  - \frac{\lambda_B}{4 !} ( 
\vec{\phi}^{\, 2}_B)^2 + H_B (\displaystyle{ 
\raisebox{0.9ex}{\scriptsize{$t$}}} \mbox{\hspace{-0.1ex}} 
\vec{e} \cdot \vec{\phi}_B ) , 
\label{on-lag} 
\end{equation} 
where $\mu^2_B < 0$, so that the theory describes the system whose 
ground state is in a broken phase in the classical limit. In 
(\ref{on-lag}), we have introduced the term $H_B (\displaystyle{ 
\raisebox{0.9ex}{\scriptsize{$t$}}} \mbox{\hspace{-0.1ex}} \vec{e} 
\cdot \vec{\phi}_B )$, with $\vec{e}$ an unit vector, which 
explicitly breaks $O (N)$ symmetry. Noticing the fact that an 
UV-renormalization scheme for the symmetric phase ($\mu_B^2 > 0$) 
works \cite{lee} as it is for the symmetry-broken phase 
($\mu_B^2 < 0$), we introduce UV-renormalized quantities, 
\[ 
\vec{\phi}_B = \sqrt{Z} \vec{\phi}, \;\;\;\; \mu_B^2 = Z_\mu \mu^2, 
\;\;\;\; \lambda_B = Z_\lambda \lambda . 
\] 
In terms of renormalized quantities, ${\cal L}$ reads 
\begin{eqnarray} 
{\cal L} (\vec{\phi}) & = & \frac{1}{2} \left[ (\partial \vec{\phi} 
)^2 - \mu^2 \vec{\phi}^{\, 2} \right] - \frac{\lambda}{4 !} ( 
\vec{\phi}^{\, 2})^2 + H (\displaystyle{ 
\raisebox{0.9ex}{\scriptsize{$t$}}} \mbox{\hspace{-0.1ex}} \vec{e} 
\cdot \vec{\phi}) + \frac{1}{2} (Z - 1) \left[ (\partial \vec{\phi} 
)^2 - \mu^2 \vec{\phi}^{\, 2} \right] \nonumber \\ 
& & - (Z_\lambda Z^2 - 1) \frac{\lambda}{4 !} ( \vec{\phi}^{\, 2})^2 
+ A \mu^4 , 
\label{ki-ki} 
\end{eqnarray} 
where $H = \sqrt{Z} H_B$. In (\ref{ki-ki}), according to 
\cite{chiku}, we have introduced the c-number counter term $A 
\mu^4$, which is necessary to make the effective action finite. 

For the purpose of later use, we here construct 
\begin{equation} 
{\cal L} (\vec{\phi} (x); \vec{\varphi} (x)) \equiv {\cal L} 
(\vec{\phi} (x) + \vec{\varphi} (x)) - \frac{\partial {\cal L} 
(\vec{\varphi} (x))}{\partial \vec{\varphi} (x)} \cdot  \vec{\phi} 
(x) . 
\label{shift} 
\end{equation} 
To avoid too many notations, for ${\cal L} (\vec{\phi} (x); 
\vec{\varphi} (x))$, we have used the same letter \lq\lq ${\cal 
L}$'' as in (\ref{on-lag}) and (\ref{ki-ki}). As will be discussed 
below, $\vec{\varphi} (x)$ in (\ref{shift}) is the (classical) 
condensate or order-parameter fields and $\vec{\phi} (x)$ is the 
quantum fields, which describes the fluctuation around 
$\vec{\varphi} (x)$. Straightforward manipulation yields 
\begin{eqnarray} 
{\cal L} (\vec{\phi} (x); \vec{\varphi} (x)) & = & {\cal L}_{cond} + 
{\cal L}_{cond}' + {\cal L}_q' + {\cal L}_{r c}' , \nonumber \\ 
{\cal L}_{cond} & = & \frac{1}{2} \left[ (\partial \vec{\varphi})^2 
- \mu^2 \vec{\varphi}^{\, 2} \right] - \frac{\lambda}{4 !} ( 
\vec{\varphi}^2)^2 + H (\displaystyle{ 
\raisebox{0.9ex}{\scriptsize{$t$}}} \mbox{\hspace{-0.1ex}} \vec{e} 
\cdot \vec{\varphi} ) , 
\label{cla} 
\\ 
{\cal L}_{cond}' & = & \frac{1}{2} (Z - 1) (\partial 
\vec{\varphi})^2 - \frac{1}{2} (Z_\mu Z - 1) \mu^2 
\vec{\varphi}^{\, 2} - (Z_\lambda Z^2 - 1) \frac{\lambda}{4 !} ( 
\vec{\varphi}^{\, 2})^2 + A \mu^4 , 
\label{cla-1} 
\\ 
{\cal L}_q' & = & \frac{1}{2} \left[ ( \partial \vec{\phi})^2 - 
\mu^2 \vec{\phi}^{\, 2}  \right] - \frac{\lambda}{12} \left[ 
\vec{\varphi}^2 \vec{\phi}^{\, 2} + 2 ( 
\displaystyle{ \raisebox{0.9ex}{\scriptsize{$t$}}} 
\mbox{\hspace{-0.1ex}} \vec{\varphi} \cdot \vec{\phi})^2 \right] - 
\frac{\lambda}{3 !} ( \displaystyle{ 
\raisebox{0.9ex}{\scriptsize{$t$}}} \mbox{\hspace{-0.1ex}} 
\vec{\varphi} \cdot \vec{\phi}) \, \vec{\phi}^{\, 2} - 
\frac{\lambda}{4 !} ( \vec{\phi}^{\, 2})^2 , \nonumber \\ 
{\cal L}_{r c}' & = & \frac{1}{2} (Z - 1) ( \partial \vec{\phi})^2 - 
\frac{1}{2} (Z_\mu Z - 1) \mu^2 \vec{\phi}^{\, 2} - (Z_\lambda Z^2 - 
1) \left[ \frac{\lambda}{12} \left\{ \vec{\varphi}^{\, 2} 
\vec{\phi}^{\, 2} + 2 ( \displaystyle{ 
\raisebox{0.9ex}{\scriptsize{$t$}}} \mbox{\hspace{-0.1ex}} 
\vec{\varphi} \cdot \vec{\phi})^2 \right\} \right. \nonumber \\ 
& & \left. + \frac{\lambda}{3 !} ( \displaystyle{ 
\raisebox{0.9ex}{\scriptsize{$t$}}} \mbox{\hspace{-0.1ex}} 
\vec{\varphi} \cdot \vec{\phi}) \, \vec{\phi}^{\, 2} + 
\frac{\lambda}{4 !} ( \vec{\phi}^{\, 2})^2 \right] . 
\label{L-q} 
\end{eqnarray} 
It should be noted that ${\cal L}$ enjoys $O (N - 1)$ symmetry. 

For obtaining an efficient or rather physically-sensible 
(perturbative) scheme, as in Section VE, it is necessary 
\cite{chiku,wein,fen,law}) to introduce weakly $x$-dependent masses, 
$M_\pi(x)$ and $M_\sigma (x)$, for which we assume that 
\begin{equation} 
|\mu^2 - M_\pi^2 (x)|, \;\; |M_\pi^2 (x) - M_\sigma^2 (x)| = O 
(\delta) . 
\label{diff} 
\end{equation} 
How to determine $M_\pi^2 (x)$ and $M_\sigma^2 (x)$ will be 
discussed in subsection D. We then rewrite ${\cal L} (\vec{\phi} 
(x); \vec{\varphi} (x))$ in the form, 
\begin{equation} 
{\cal L} (\vec{\phi} (x); \vec{\varphi} (x)) = {\cal L}_{cond} + 
{\cal L}_{cond}' + {\cal L}_0 (\vec{\phi} (x); \vec{\varphi} (x)) + 
{\cal L}_{int} + {\cal L}_{r c} . 
\label{owari} 
\end{equation} 
Here ${\cal L}_{cond}$ and ${\cal L}_{cond}'$ are as in (\ref{cla}) 
and (\ref{cla-1}), respectively, and 
\begin{eqnarray} 
{\cal L}_0 (\vec{\phi} (x); \vec{\varphi} (x)) & = & \frac{1}{2} \, 
\displaystyle{ \raisebox{0.9ex}{\scriptsize{$t$}}} 
\mbox{\hspace{-0.1ex}} \vec{\phi} (x) \, {\bf \Delta}^{- 1} (x) \, 
\vec{\phi} (x) , 
\label{hi-setsu} \\ 
{\cal L}_{int} & = & - \frac{\lambda}{3 !} ( \displaystyle{ 
\raisebox{0.9ex}{\scriptsize{$t$}}} \mbox{\hspace{-0.1ex}} 
\vec{\varphi} \cdot \vec{\phi}) \vec{\phi}^{\, 2} - \frac{\lambda}{4 
!} ( \vec{\phi}^{\, 2})^2 , \nonumber \\ 
{\cal L}_{r c} & = & {\cal L}_{r c}' + \frac{1}{2} \displaystyle{ 
\raisebox{0.9ex}{\scriptsize{$t$}}} \mbox{\hspace{-0.1ex}} 
\vec{\phi} (x) \left[ \chi_\pi (x) {\bf P}_\pi (x) + \chi_\sigma (x) 
{\bf P}_\sigma (x) \right] \vec{\phi} (x) . 
\label{teigi-yo} 
\end{eqnarray} 
Here ${\cal L}_{r c}'$ is as in (\ref{L-q}) and 
\begin{eqnarray} 
{\bf \Delta}^{- 1} (x) & \equiv & - \left( \partial^2 + M_\pi^2 (x) 
+ \frac{\lambda}{6} \vec{\varphi}^{\, 2} (x) \right) {\bf P}_\pi 
(x) - \left( \partial^2 + M_\sigma^2 (x) + \frac{\lambda}{2} 
\vec{\varphi}^{\, 2} (x) \right) {\bf P}_\sigma (x) , \nonumber \\ 
\chi_\xi (x) & = & M_\xi^2 (x) - \mu^2 \;\;\;\;\;\; (\xi = \pi, \, 
\sigma) , \nonumber \\ 
{\bf P}_\pi (x) & = & {\bf I} - | \varphi (x) \rangle \langle 
\varphi (x) | , \nonumber \\ 
{\bf P}_\sigma (x) & = & | \varphi (x) \rangle \langle \varphi (x) | 
. 
\label{teigi-10} 
\end{eqnarray} 
In these equations, the boldface letters denote $N \times N$ 
matrices in the real vector space: ${\bf P}_\pi$ (${\bf P}_\sigma$) 
is the projection operator onto the $\pi$- $(\sigma)$-subspace; 
${\bf I}$ is an unit matrix. $| \varphi (x) \rangle$ is a unit 
vector along $\vec{\varphi} (x)$, and $\langle \varphi (x)|$ is an 
adjoint of $| \varphi (x) \rangle$. As has been noted above, when 
compared to $M_\xi^2 (x)$ $(\xi = \pi, \sigma)$, $\chi_\xi (x)$ is 
of higher order. 

As mentioned above, the quantum fields $\vec{\phi}$ describe the 
fluctuation around the condensate fields $\vec{\varphi} (x)$. The 
construction of perturbation theory based on (\ref{owari}) starts 
with constructing the Fock space of the quanta described by 
$\vec{\phi}$, which is defined \lq\lq on $\vec{\varphi} (x)$.'' As 
stated in Section I, dealing with the dynamical aspect of the 
system, we are assuming that microscopic or elementary reactions can 
be regarded as taking place in a single spacetime cell. While, in a 
single cell, relaxation phenomena are negligible. This means that 
the theory to be developed in the sequel is consistent only when 
$\vec{\varphi} (x)$ does not change appreciably in a single 
spacetime cell. We shall assume that this is the case and treat $x$ 
of $\vec{\varphi} (x)$ as macroscopic spacetime coordinates. 
\subsection{Effective action} 
Spacetime evolution of the condensate fields $\vec{\varphi} (x)$ is 
governed by the Euler-Lagrange equation, which is derived from the 
CTP effective action $\Gamma_C [\vec{\varphi}]$. If one restricts to 
the case where $\vec{\varphi} (x)$ is space-time independent, 
$\vec{\varphi} (x) \to \vec{\varphi}$, $\Gamma_C [\vec{\varphi}]$ 
reduces to minus the CTP effective potential, $\Gamma_C 
[\vec{\varphi}] = - V_C (\vec{\varphi})$. The {\em physical part} of 
the CTP effective potential (cf. below) $V_{phys} (\vec{\varphi})$ 
is the minimum free energy density of the system under the 
constraint, $\langle \phi (x) \rangle = \varphi$, the minimum free 
energy density with quantum and medium effects taken into account. A 
computational algorithm of effective potential in vacuum theory is 
established in \cite{jak}. On the basis of a generalization of this 
algorithm to the case of equilibrium thermal field theory, a 
calculational scheme of the effective potential and its {\em 
physical part} have been settled \cite{fuji} (see also \cite{eva}). 

We generalize this algorithm to the present case of out of 
equilibrium systems. Following standard procedure \cite{chou} of 
obtaining an expression for the CTP effective action, we start with 
the CTP generating functional for Green functions in the interaction 
picture (cf. (\ref{gene-0})), 
\begin{equation} 
Z_C [\vec{J}] \equiv e^{i W_C [\vec{J}]} = \langle T_C \mbox{exp} \, 
i \int_C \left[ {\cal L}' (\vec{\phi} (x)) + \vec{J} (x) \cdot 
\vec{\phi} (x) \right] \rangle , 
\label{moha} 
\end{equation} 
where, as in Section II, $C$ $(= C_1 \oplus C_2)$ is the closed 
time-path in the complex time-plane and ${\cal L}'$ stands for the 
perturbative part of the Lagrangian density, which is to be 
identified below in (\ref{rei}). In (\ref{moha}), $W_C [J]$ is the 
generating functional for connected Green functions. Expectation 
value of $\vec{\phi} (x)$, $\langle \vec{\phi} (x) \rangle_J$ 
$(\equiv \vec{\varphi} (x))$, in the presence of the external source 
$\vec{J} (x)$ is obtained through 
\begin{equation} 
\vec{\varphi} (x) = \frac{\delta W_C [\vec{J}]}{\delta \vec{J} (x)} 
. 
\label{p-1} 
\end{equation} 
The CTP effective action $\Gamma_C$ is defined as the functional 
Legendre transform, 
\begin{equation} 
\Gamma_C [\vec{\varphi}] = W_C [\vec{J}] - \int_C \vec{J} (x) \cdot 
\vec{\varphi} (x) , 
\label{Yu-eff} 
\end{equation} 
from which the Euler-Lagrange equation follows: 
\begin{equation} 
\frac{\delta \Gamma_C [\vec{\varphi}]}{\delta \vec{\varphi} (x)} = - 
\vec{J} (x) . 
\label{bibi} 
\end{equation} 
The CTP effective action $\Gamma_C$ is the generating functional for 
connected, one-particle irreducible, amputated Green functions. 
Substitution of (\ref{moha}) and (\ref{bibi}) into (\ref{Yu-eff}) 
yields 
\[ 
e^{i \Gamma_C [\vec{\varphi}]} = \langle T_C \mbox{exp} \left( i 
\int_C \left[ {\cal L}' (\vec{\phi} (x)) - \frac{\delta \Gamma_C [ 
\vec{\varphi}]}{\delta \vec{\varphi} (x)} \cdot \{ \vec{\phi} (x) - 
\vec{\varphi} (x) \} \right] \right) \rangle . 
\] 
Redefining $\vec{\phi} (x) - \vec{\varphi} (x)$ as new quantum 
fields $\vec{\phi} (x)$, we have 
\begin{equation} 
e^{i \Gamma_C [\vec{\varphi}]} = \langle T_C \mbox{exp} \left[ 
\frac{i}{\delta} \int_C \left\{ {\cal L}' (\vec{\phi} (x) + 
\vec{\varphi} (x)) - \frac{\delta \Gamma_C [\vec{\varphi}]}{\delta 
\vec{\varphi} (x)} \cdot \vec{\phi} (x) \right\} \right] \rangle . 
\label{eff-ac0} 
\end{equation} 
Thus, the new quantum fields $\vec{\phi} (x)$ describe the 
fluctuation around $\vec{\varphi} (x)$. Here, for the purpose 
stated below, we have inserted the factor $\delta$ $( = 1)$. Going 
to the two-component representation, we have 
\begin{equation} 
e^{i \hat{\Gamma} [\vec{\varphi}_1, \vec{\varphi}_2]} = \langle T_C 
\mbox{exp} \left[ \frac{i}{\delta} \int d^{\, 4} x \sum_{j = 1}^2 
\left\{ (-)^{j - 1} {\cal L}' (\vec{\phi}_j (x) + \vec{\varphi}_j 
(x)) - \frac{\delta \hat{\Gamma} [\vec{\varphi}_1, 
\vec{\varphi}_2]}{\delta \vec{\varphi}_j (x)} \cdot \vec{\phi}_j (x) 
\right\} \right] \rangle , 
\label{eff-ac2} 
\end{equation} 
where $\vec{\phi}_j$ and $\vec{\varphi}_j$ $(j = 1, 2)$ are the 
type-$j$ fields and 
\begin{equation} 
\frac{\delta \hat{\Gamma} [\vec{\varphi}_1, \vec{\varphi}_2]}{\delta 
\vec{\varphi}_j (x)} = (-)^{j - 1} \frac{\delta \Gamma_C 
[\vec{\varphi}]}{\delta \vec{\varphi} (x)} \;\; \mbox{with} \; x_0 
\in C_j \;\;\;\;\;\; (j = 1, 2) . 
\label{yayakosi} 
\end{equation} 
The minus sign on the RHS of (\ref{yayakosi}) with $j = 2$ comes 
from the fact that, on $C_2$, the time ticks away backward from $+ 
\infty$ to $- \infty$ (cf. (\ref{wakeru})). 

Equation (\ref{eff-ac2}) is of the form which allows for computing 
$\hat{\Gamma} [\vec{\varphi}_1, \vec{\varphi}_2]$ perturbatively. 
Perturbative calculation of (\ref{eff-ac2}) goes by taking suitable 
\begin{equation} 
\hat{\cal L}_0 \equiv \sum_{j = 1}^2 (-)^{j - 1} {\cal L}_0 
(\vec{\phi}_j (x), \, \vec{\varphi}_j (x)) \;\; (\in \hat{\cal L} 
\equiv \sum_{j = 1}^2 (-)^{j - 1} {\cal L} (\vec{\phi}_j (x) + 
\vec{\varphi}_j (x))) , 
\label{ato} 
\end{equation} 
being bilinear in quantum fields $\vec{\phi}_j (x)$, as the free 
hat-Lagrangian density, which defines $2 \times 2$ propagators of 
the fields $\hat{\vec{\phi}} (x) = \displaystyle{ 
\raisebox{0.9ex}{\scriptsize{$t$}}} \mbox{\hspace{-0.1ex}} ( 
\vec{\phi}_1 (x), \vec{\phi}_2 (x))$. The form of ${\cal L} 
(\vec{\phi} (x) + \vec{\varphi} (x))$ is already been obtained in 
the last subsection and ${\cal L}_0$ is defined in (\ref{hi-setsu}) 
with (\ref{teigi-10}). The $2 \times 2$ matrix propagators will be 
constructed in the next subsection. The vertex factors  can be reads 
off from 
\begin{equation} 
\hat{\cal L} - \sum_{j = 1}^2 (-)^{j - 1} {\cal L} (\vec{\varphi}_j 
(x)) - \hat{\cal L}_0 - \sum_{j = 1}^2 \frac{\delta \hat{\Gamma} [ 
\vec{\varphi}_1, \vec{\varphi}_2 ]}{\delta \vec{\varphi}_j (x)} 
\cdot \vec{\phi}_j (x) . 
\label{rei} 
\end{equation} 
Note that, to zeroth order (tree level), 
\begin{equation} 
\hat{\Gamma}^{(0)} [\vec{\varphi}_1, \vec{\varphi}_2] = \sum_{j = 
1}^2 (-)^{j - 1} \int d^{\, 4} x \, {\cal L} (\vec{\varphi}_j (x)) . 
\label{rei-11} 
\end{equation} 
Then, the exponent in (\ref{eff-ac2}) with $\hat{\Gamma}^{(0)}$ for 
$\hat{\Gamma}$ on the RHS is free from the term being linear in 
$\vec{\phi}$, which means that the \lq\lq one-point vertex'' is 
absent. [In constructing ${\cal L} (\vec{\phi} (x); \vec{\varphi} 
(x))$ above, Eq.~(\ref{shift}), this has already been taken into 
account.] Thus, in terms of Feynman diagrams, (\ref{eff-ac2}) 
consists of bubble or no-leg diagrams. The loop expansion is defined 
as an expansion by $\delta$ ($\delta$-expansion), which is set to 1 
at the end of calculation (see, e.g., \cite{hon}). Computation of 
$\hat{\Gamma}$ should be carried out successively: In computing 
$l$th loop-order contribution, the contributions to $\hat{\Gamma}$ 
up to $(l - 1)$th loop-order should be substituted for 
$\hat{\Gamma}$ on the RHS of (\ref{eff-ac2}). This part, $- i 
\delta^{- 1} \sum_{j = 1}^2 \vec{\phi}_j \cdot \delta \hat{\Gamma} / 
\delta \vec{\varphi}_j$, plays a roll of eliminating the 
contribution of one-particle reducible diagrams. 

The one-loop contribution to $\hat{\Gamma}$ will be dealt with in 
Section VII. Computation of a contribution $\hat{\Gamma}^{(n)}$ from 
a multi-loop diagram, which includes $n$ $(\geq 1)$ vertices, goes 
in standard manner. Each vertex consists of two types, the type-1 
and type-2. Then, $\hat{\Gamma}^{(n)}$ is obtained by summing over 
types of vertices, so that $\hat{\Gamma}^{(n)}$ consists of $2^n$ 
terms, which we can write, with obvious notation, as 
\begin{equation} 
\hat{\Gamma}^{(n)} [\vec{\varphi}_1, \vec{\varphi}_2] = 
\sum_{i_1, \, i_2, \, ..., \, i_n = 1}^2 \hat{\cal F}_{i_1, \, i_2, 
\, ..., \, i_n} [\vec{\varphi}_1, \vec{\varphi}_2] , 
\label{eff-gamm10} 
\end{equation} 
where $i_j$ $(j = 1, 2, ..., n)$ denotes the type of vertex. 

In the two-component representation, the Euler-Lagrange equation 
(\ref{bibi}) reads (cf. above after (\ref{yayakosi})), 
\begin{equation} 
\frac{\delta \hat{\Gamma} [\vec{\varphi}_1, \vec{\varphi}_2]}{\delta 
\vec{\varphi}_j (x)} = - (-)^{j - 1} \vec{J}_j (x) \;\;\;\;\;\;\; 
(j = 1, 2) . 
\label{maae} 
\end{equation} 
In the limit of absence of the external sources, $\vec{J}_1 (x) = 
\vec{J}_2 (x) = 0$, $\vec{\varphi}_1 (x) = \vec{\varphi}_2 (x)$ 
holds and (\ref{maae}) becomes 
\begin{equation} 
\frac{\delta \hat{\Gamma} [\vec{\varphi}_1, \vec{\varphi}_2]}{\delta 
\vec{\varphi}_j (x)} \rule[-3mm]{.14mm}{8.5mm} 
\raisebox{-2.85mm}{\scriptsize{$\; \vec{\varphi}_1 = \vec{\varphi}_2 
= \vec{\varphi}$}} = 0 \, , \;\;\;\;\;\;\; (j = 1, 2) . 
\label{maae-1} 
\end{equation} 
The equation with $j = 1$ is the equation of motion for the physical 
or type-1 condensate fields, while the one with $j = 2$ is for the 
unphysical or the type-2 condensate fields. The {\em physical} 
effective action $\Gamma_{phys}$ and the {\em unphysical} effective 
action $\Gamma_{unphys}$ are defined through  
\begin{eqnarray} 
\frac{\delta \Gamma_{phys} [\vec{\varphi}]}{\delta \vec{\varphi} 
(x)} & = & \frac{\delta \hat{\Gamma} [\vec{\varphi}_1, 
\vec{\varphi}_2]}{\delta \vec{\varphi}_1 (x)} 
\rule[-3mm]{.14mm}{8.5mm} \raisebox{-2.85mm}{\scriptsize{$\; 
\vec{\varphi}_1 = \vec{\varphi}_2 = \vec{\varphi}$}} , 
\nonumber \\ 
\frac{\delta \Gamma_{unphys} [\vec{\varphi}]}{\delta \vec{\varphi} 
(x)} & = & \frac{\delta \hat{\Gamma} [\vec{\varphi}_1, 
\vec{\varphi}_2]}{\delta \vec{\varphi}_2 (x)} 
\rule[-3mm]{.14mm}{8.5mm} \raisebox{-2.85mm}{\scriptsize{$\; 
\vec{\varphi}_1 = \vec{\varphi}_2 = \vec{\varphi}$}} . 
\label{maae-11} 
\end{eqnarray} 
With suitable (functional) integration constant, (\ref{maae-11}) 
determines $\Gamma_{phys}$ and $\Gamma_{unphys}$. In the case of 
thermally equilibrium system, the physical effective potential 
($\Gamma_{phys}$ with constant $\vec{\varphi}$) has been computed in 
\cite{fuji} up to two-loop order. 

At the previous subsection, in defining ${\cal L}_0$, we have 
introduced (arbitrary) \lq\lq mass functions'' $M_\xi^2 (x)$ $(\xi = 
\pi, \sigma)$. The free hat-Lagrangian density in the present CTP 
formalism is given by (\ref{ato}). Then, as in the fields, there are 
two types of $M_\xi^2 (x)$, the type-1 and type-2, which we write as 
$M_{\xi j}^2 (x)$ $(j = 1, 2)$. [$M_{\xi j}^2 (x) = M_\xi^2 (x)$ 
with $x_0 \in C_j$ $(j = 1, 2)$.]  As will be discussed in 
subsection D, various methods of determining $M_{\xi j}^2 (x)$ are 
available. All those methods are based on the notion of \lq\lq 
optimization'' of some physical quantity. The determined $M_{\xi 
j}^2 (x)$ generally depends on $\vec{\varphi}_1 (x)$ and 
$\vec{\varphi}_2 (x)$. Let $\hat{\Gamma} [\vec{\varphi}_1; 
\vec{\varphi}_2; M_{\pi 1}^2, M_{\pi 2}^2, M_{\sigma 1}^2, M_{\sigma 
2}^2]$ be the effective action computed to some order of loop 
expansion. Here a question arises: In computing the equation of 
motion (\ref{maae-1}), does the functional derivative $\delta / 
\delta \vec{\varphi}_j (x)$ $(j = 1, 2)$ act on $M_{\xi 1}^2 (x)$ 
and $M_{\xi 2}^2 (x)$?. The answer is as follows: If we intend to 
\lq\lq optimize'' the effective action, from which the equation of 
motion is derived, $\delta / \delta \vec{\varphi}_j (x)$ should act 
on $M_{\xi 1}^2 (x)$ and $M_{\xi 2}^2 (x)$. On the contrary, if we 
intend to \lq\lq optimize'' the equation of motion, $\delta / \delta 
\vec{\varphi}_j (x)$ should not act on $M_{\xi 1}^2 (x)$ and $M_{\xi 
2}^2 (x)$. We take the stand that the effective action is an 
intermediate device to derive an equation of motion for 
$\vec{\varphi}_j (x)$ $(j = 1, 2)$, etc., and then the equation of 
motion is of physical importance over the effective action. On the 
basis of this observation, following \cite{chiku}, we take the 
latter throughout in the sequel, i.e., $\delta / \delta 
\vec{\varphi}_j (x)$ does not act on $M_{\xi 1}^2 (x)$ and $M_{\xi 
2}^2 (x)$. 
\subsection{Propagator} 
\subsubsection{Retarded/advanced propagator} 
Here, we obtain a retarded/advanced propagator, $\hat{\bf 
\Delta}_{diag}$, which is an inverse of 
\begin{equation} 
\hat{\bf \Delta}^{- 1} (x) = {\bf \Delta}^{- 1} (x) \, \hat{\tau}_3, 
\label{moto} 
\end{equation} 
with ${\bf \Delta}^{- 1} (x)$ as in (\ref{teigi-10}), under the 
retarded/advanced boundary condition (cf. (\ref{1-11}) and 
(\ref{shin})). 

We start with finding an inverse of ${\bf \Delta}^{- 1} (x)$ under 
the retarded boundary condition: 
\begin{equation} 
{\bf \Delta}^{- 1} (x) {\bf \Delta}_R (x, y) = {\bf I} \, \delta^{4} 
(x - y) . 
\label{prop-on} 
\end{equation} 
To this end, we introduce 
\begin{eqnarray} 
{\bf \Delta}^{(0)}_R (x, y) & = & {\bf \Delta}^{(\pi)}_R (x, y) + 
{\bf \Delta}^{(\sigma)}_R (x, y) , \nonumber \\ 
{\bf \Delta}^{(\xi)}_R (x, y) & = & {\bf P}_\xi (x) \int \frac{d^{\, 
4} P}{(2 \pi)^4} \, e^{- i P \cdot (x - y)} \frac{1}{P^2 - {\cal 
M}^2_\xi (X)}  \;\;\;\;\;\;\; (\xi = \pi, \sigma) , 
\label{cal-MM} 
\end{eqnarray} 
where $X = (x + y) / 2$ and 
\begin{equation} 
{\cal M}_\pi^2 (X) = M_\pi^2 (X) + \frac{\lambda}{6} \, 
\vec{\varphi}^{\, 2} (X) , \;\;\;\;\;\; {\cal M}_\sigma^2 (X) = 
M_\sigma^2 (X) + \frac{\lambda}{2} \, \vec{\varphi}^{\, 2} (X) . 
\label{134} 
\end{equation} 
${\cal M}_\xi^2 (X)$ in (\ref{cal-MM}) should be understood to be 
${\cal M}^2_\xi (X) - i \epsilon (p_0) \eta$. 

Straightforward manipulation using (\ref{al}) in Appendix C yields 
\begin{eqnarray} 
{\bf \Delta}^{- 1} (x) {\bf \Delta}^{(0)}_R (x, y) & = & {\bf I} \, 
\delta^{4} (x - y) - 2 i \int \frac{d^{\, 4} P}{(2 \pi)^4} \, e^{- i 
P \cdot (x - y)} \, P^\mu \frac{\partial | \varphi (X) \rangle 
\langle \varphi (X) |}{\partial X^\mu} \nonumber \\ 
& & \times \left[ \frac{1}{P^2 - {\cal M}^2_\pi (X)} - \frac{1}{P^2 
- {\cal M}^2_\sigma (X)} \right] . 
\label{yotei} 
\end{eqnarray} 
With (\ref{yotei}) in hand, we are now ready to obtain ${\bf 
\Delta}_R (x, y)$ in (\ref{prop-on}) to the gradient approximation: 
\begin{eqnarray*} 
{\bf \Delta}_R (x, y) & \simeq & \sum_{\xi = \pi, \, \sigma} {\bf 
P}_\xi (x) \int \frac{d^{\, 4} P}{(2 \pi)^4} \, e^{- i P \cdot (x - 
y)} \frac{1}{P^2 - {\cal M}_\xi^2 (X)} \nonumber \\ 
& & + 2 i \sum_{\xi = \pi, \, \sigma} {\bf P}_\xi (X) \int 
\frac{d^{\, 4} P}{(2 \pi)^4} \, e^{- i P \cdot (x - y)} \frac{P 
\cdot \partial_X {\bf P}_\sigma (X)}{P^2 - {\cal M}^2_\xi (X)} 
\nonumber \\ 
& & \times \left[ \frac{1}{P^2 - {\cal M}^2_\pi (X)} - \frac{1}{P^2 
- {\cal M}^2_\sigma (X)} \right] . 
\end{eqnarray*} 
Using (\ref{al}) in Appendix C and Fourier transforming the 
resultant expression, we obtain 
\begin{eqnarray} 
{\bf \Delta}_R (X; P) & \simeq & \sum_{\xi = \pi, \, \sigma} 
\frac{{\bf P}_\xi (X)}{P^2 - {\cal M}^2_\xi (X)} \nonumber \\ 
& & - i P_\mu \left( | \varphi (X) \rangle 
\stackrel{\leftrightarrow}{\partial}_{X_\mu} \langle \varphi (X)| 
\right) \left[ \frac{1}{P^2 - {\cal M}^2_\pi (X)} - \frac{1}{P^2 - 
{\cal M}^2_\sigma (X)} \right]^2 . 
\label{R} 
\end{eqnarray} 
It can be shown that ${\bf \Delta}_R (x, y)$, inverse Fourier 
transform of ${\bf \Delta}_R (X; P)$, satisfies 
\[ 
{\bf \Delta}_R (x, y) \stackrel{\longleftarrow}{\bf \Delta^{- 1}} 
(y) = {\bf I} \, \delta^{4} (x - y) . 
\] 

Form for ${\bf \Delta}_A (x, y)$ may be obtained in a similar 
manner. Thus we obtain for $\hat{\bf \Delta}_{diag} (X; P)$, the 
counterpart to the Fourier transform of $\hat{\Delta}_{diag} (u - 
v)$ in (\ref{imple-p0}), 
\begin{eqnarray} 
\hat{\bf \Delta}_{diag} (X; P) & \simeq & \hat{\bf 
\Delta}_{diag}^{(0)} (X; P) + \hat{\bf \Delta}_{diag}^{(1)} (X; P) 
\label{on-d} 
\\ 
\hat{\bf \Delta}_{diag}^{(0)} (X; P) & = & \mbox{diag} \left( {\bf 
\Delta}_R^{(0)} (X; P) , \, - {\bf \Delta}_A^{(0)} (X; P) \right) , 
\nonumber \\ 
& = & \sum_{\xi = \pi, \, \sigma} {\bf P}_\xi (X) \, \mbox{diag} 
\left( \Delta_R^{(\xi)} (X; P) , \, - \Delta_A^{(\xi)} (X; P) 
\right) \nonumber \\ 
& \equiv & \sum_{\xi = \pi, \, \sigma} {\bf P}_\xi (X) \, 
\hat{\Delta}_{diag}^{(\xi)} (X; P) , 
\label{on-d1} 
\\ 
\hat{\bf \Delta}_{diag}^{(1)} (X; P) & = & \mbox{diag} \left( {\bf 
\Delta}_R^{(1)} (X; P) , \, - {\bf \Delta}_A^{(1)} (X; P) \right) , 
\label{on-d2} 
\end{eqnarray} 
where 
\begin{eqnarray} 
\Delta_{R (A)}^{(\xi)} (X; P) & \equiv & \frac{1}{P^2 - {\cal 
M}^2_\xi (X) \pm i \epsilon(p_0) \eta} , \nonumber \\ 
{\bf \Delta}_{R (A)}^{(1)} (X; P) & = & - i P_\mu \left( | \varphi 
(X) \rangle \stackrel{\leftrightarrow}{\partial}_{X_\mu} \langle 
\varphi (X)| \right) \nonumber \\ 
& & \times \left[ \frac{1}{P^2 - {\cal M}^2_\pi (X) \pm i 
\epsilon(p_0) \eta} - \frac{1}{P^2 - {\cal M}^2_\sigma (X) \pm i 
\epsilon(p_0) \eta} \right]^2 . 
\label{b-ret} 
\end{eqnarray} 
$\hat{\bf \Delta}_{diag} (X; P)$ corresponds to $\Delta_{R (A)} 
(P)$, Eq.~(\ref{6.2-e}), in the complex-scalar field theory. 

The following form for $\hat{\bf \Delta}_{diag} (x, y)$ is also 
useful for later analysis: 
\begin{eqnarray} 
\hat{\bf \Delta}_{diag} (x, y) & = & \sum_{\xi = \pi, \, \sigma} 
{\bf P}_\xi (x) \hat{\Delta}_{diag}^{(\xi)} (x, y) {\bf P}_\xi (y) 
+ \hat{\bf \Delta}^{(1) '}_{diag} (x, y) , \nonumber \\ 
\hat{\Delta}_{diag}^{(\xi)} (x, y) & = & \int \frac{d^{\, 4} P}{(2 
\pi)^4} \, e^{- i P \cdot (x - y)} \hat{\Delta}^{(\xi)}_{diag} (X; 
P) , \nonumber \\ 
\hat{\bf \Delta}^{(1) '}_{diag} (x, y) & = & 2 i \int \frac{d^{\, 4} 
P}{(2 \pi)^4} \, e^{- i P \cdot (x - y)} P_\mu \left( | \varphi (X) 
\rangle \stackrel{\leftrightarrow}{\partial}_{X_\mu} \langle \varphi 
(X) | \right) \nonumber \\ 
& & \times \mbox{diag} \left( \Delta_R^{(\pi)} (X; P) 
\Delta_R^{(\sigma)} (X; P), - \Delta_A^{(\pi)} (X; P) 
\Delta_A^{(\sigma)} (X; P) \right) . 
\label{ana} 
\end{eqnarray} 

Let us find the property of the operators $|\varphi (X) \rangle 
\partial_{X_\mu} \langle \varphi (X)|$ and $|\varphi (X) \rangle \! 
\stackrel{\leftarrow}{\partial}_{X_\mu} \! \langle \varphi (X)|$. As 
defined above, $|\varphi (X) \rangle$ is the unit vector along 
$\vec{\varphi} (X)$ in the $O (N)$ vector space, so that $\langle 
\varphi (X) | \varphi (X) \rangle = 1$. Then $\langle \varphi (X) 
| \partial_{X_\mu} | \varphi (X) \rangle = \langle \varphi (X) | 
\! \stackrel{\leftarrow}{\partial}_{X_\mu} \! | \varphi (X) \rangle 
= 0$. From these relations, we obtain 
\begin{eqnarray} 
{\bf P}_\xi (X) \left( |\varphi (X) \rangle \partial_{X_\mu} \langle 
\varphi (X)| \right) {\bf P}_\zeta (X) & = & \left\{ 
\begin{array}{ll} 
|\varphi (X) \rangle \partial_{X_\mu} \langle 
\varphi (X)| \; & \mbox{if} \; \xi = \sigma \; \mbox{and} \; 
\zeta = \pi , \\ 
0 \; & \mbox{otherwise} , 
\end{array}
\right. \nonumber \\ 
{\bf P}_\xi (X) \left( |\varphi (X) \rangle \! 
\stackrel{\leftarrow}{\partial}_{X_\mu} \! \langle \varphi (X)| 
\right) {\bf P}_\zeta (X) & = & \left\{ 
\begin{array}{ll} 
|\varphi (X) \rangle \! \stackrel{\leftarrow}{\partial}_{X_\mu} \! 
\langle \varphi (X)| \; & \mbox{if} \; \xi = \pi \; \mbox{and} \; 
\zeta = \sigma , \\ 
0 \; & \mbox{otherwise} . 
\end{array}
\right. 
\label{sei} 
\end{eqnarray} 
Thus, $\hat{\bf \Delta}^{(1)}_{diag}$ in (\ref{on-d}) and $\hat{\bf 
\Delta}^{(1) '}_{diag}$ in (\ref{ana}) represent mixing between the 
$\pi$ sector and the $\sigma$ sector. 
\subsubsection{$2 \times 2$ matrix propagator} 
Let us introduce \lq\lq Bogoliubov'' transformation (cf. 
(\ref{qu-2})), 
\begin{eqnarray*} 
\hat{\vec{\phi}} (x) & = & \int d^{\, 3} u \sum_{\tau = \pm} 
\hat{\bf B}^{(\tau)}_L ({\bf x}, {\bf u}; x_0) \hat{\vec{\phi}}_{R 
A}^{(\tau)} ({\bf u}, x_0) , \\ 
\displaystyle{ \raisebox{0.9ex}{\scriptsize{$t$}}} 
\mbox{\hspace{-0.1ex}} \hat{\vec{\phi}} (y) & = & \int d^{\, 3} v 
\sum_{\tau = \pm} \displaystyle{ \raisebox{0.9ex}{\scriptsize{$t$}}} 
\mbox{\hspace{-0.1ex}} \hat{\vec{\phi}}_{R A}^{(\tau)} ({\bf v}, 
y_0) \hat{\bf B}^{(\tau)}_R ({\bf v}, {\bf y}; y_0) . 
\end{eqnarray*} 
As seen above after (\ref{L-q}), ${\cal L}_q$ enjoys $O (N - 1)$ 
symmetry. We assume that the density matrix enjoys $O (N - 1)$ 
symmetry in the sense that the $N \times N$ matrices $\hat{\bf B}_{L 
(R)}^{(\tau)}$ take the form, 
\[ 
\hat{\bf B}_{L (R)}^{(\tau)} ({\bf x}, {\bf u}; x_0) = \sum_{\xi, \, 
\zeta = \pi, \, \sigma} {\bf P}_\xi ({\bf x}, x_0) \hat{B}_{L 
(R)}^{(\tau) ( \xi, \, \zeta )} ({\bf x}, {\bf u}; x_0) {\bf 
P}_\zeta ({\bf u}, x_0) . 
\] 
The condition that the canonical commutation relations are preserved 
(cf. (\ref{comm-2})) yields a solution, 
\[ 
\hat{\bf B}_{L (R)}^{(\tau)} ({\bf x}, {\bf u}; x_0) = \sum_{\xi = 
\pi, \, \sigma} {\bf P}_\xi ({\bf x}, x_0) \hat{B}_{L (R)}^{(\tau) 
(\xi)} ({\bf x}, {\bf u}; x_0) {\bf P}_\xi ({\bf u}, x_0) , 
\] 
where $\hat{B}_{L (R)}^{(\tau) (\xi)}$ is as in (\ref{Bogo-R}) with 
$f^{(\tau)} \to f^{(\tau)}_\xi$. Translated into the $|p_0|$ 
prescription (cf. (\ref{imple-p0})), we obtain for the propagator, 
\begin{eqnarray} 
\hat{\bf \Delta} (x, y) & = & \int d^{\, 4} u \, d^{\, 4} v \, 
\hat{\bf B}_L (x, u) \hat{\bf \Delta}_{diag} (u, v) \hat{\bf B}_R 
(v, y) \nonumber \\ 
& = & \int d^{\, 4} u \, d^{\, 4} v \sum_{\xi, \, \zeta = \pi, \, 
\sigma} {\bf P}_\xi (x) \hat{B}_L^{(\xi)} (x, u) {\bf P}_\xi (u) 
\hat{\bf \Delta}_{diag} (u, v) {\bf P}_\zeta (v) \hat{B}_R^{(\zeta)} 
(v, y) {\bf P}_\zeta (y) , \nonumber \\ 
\hat{B}_L^{(\xi)} (x, u) & = & \left( 
\begin{array}{cc} 
\delta^4 (x - u) & \;\;\; f_\xi (x, u) \\ 
\delta^4 (x - u) & \;\;\; \delta^4 (x - u) + f_\xi (x, u) 
\end{array}
\right) , \nonumber \\ 
\hat{B}_R^{(\xi)} (v, y) & = & \left( 
\begin{array}{cc} 
\delta^4 (v - y) + f_\xi (v, y) & \;\;\; f_\xi (v, y) \\ 
\delta^4 (v - y) & \;\;\; \delta^4 (v - y) 
\end{array}
\right) . 
\label{o-prop} 
\end{eqnarray} 
Substituting (\ref{ana}) into (\ref{o-prop}), we obtain 
\begin{eqnarray} 
\hat{\bf \Delta} (x, y) & = & \int d^{\, 4} u \, d^{\, 4} v 
\sum_{\xi = \pi, \, \sigma} {\bf P}_\xi (x) \hat{B}_L^{(\xi)} (x, u) 
{\bf P}_\xi (u) \hat{\Delta}_{diag}^{(\xi)} (u, v) {\bf P}_\xi (v) 
\hat{B}_R^{(\xi)} (v, y) {\bf P}_\xi (y) \nonumber \\ 
& & + \int d^{\, 4} u \, d^{\, 4} v \sum_{\xi, \, \zeta = \pi, \, 
\sigma} {\bf P}_\xi (x) \hat{B}_L^{(\xi)} (x, u) {\bf P}_\xi (u) 
\hat{\bf \Delta}^{(1) '}_{diag} (u, v) {\bf P}_\zeta (v) 
\hat{B}_R^{(\zeta)} (v, y) {\bf P}_\zeta (y) . 
\label{ii-o} 
\end{eqnarray} 
From the first term on the RHS, we pick out 
\begin{equation} 
{\bf P}_\xi (x) {\bf P}_\xi (u) {\bf P}_\xi (v) {\bf P}_\xi (y) . 
\label{pick} 
\end{equation} 
Using ${\bf P}_\xi (u) \simeq {\bf P}_\xi (x) + (u - x)_\mu 
\partial_{X_\mu} {\bf P}_\xi (x)$, etc., we obtain 
\begin{eqnarray*} 
\mbox{Eq.~(\ref{pick})} & = & {\bf P}_\xi (x) {\bf P}_\xi (y) + 
(u - x)_\mu {\bf P}_\xi (x) \left( \partial_{x_\mu} {\bf P}_\xi 
(x) \right) {\bf P}_\xi (x) \nonumber \\ 
& & + (v - y)_\mu {\bf P}_\xi (y) \left( \partial_{y_\mu} {\bf 
P}_\xi (y) \right) {\bf P}_\xi (y) . 
\end{eqnarray*} 
In the second (third) term on the RHS, we have set ${\bf P}_\xi (v) 
\simeq {\bf P}_\xi (y) \simeq {\bf P}_\xi (x)$  (${\bf P}_\xi (x) 
\simeq {\bf P}_\xi (u) \simeq {\bf P}_\xi (y))$, which is allowed 
since the term already contains the $X_\mu$-derivative. The 
relations (\ref{sei}) tell us that the second and third terms on the 
RHS vanish. Thus, we obtain, for the first term on the RHS of 
(\ref{ii-o}), 
\begin{equation} 
\sum_{\xi = \pi, \, \sigma} {\bf P}_\xi (x) \left[ \int d^{\, 4} u 
\, d^{\, 4} v \, \hat{B}_L^{(\xi)} (x, u) 
\hat{\Delta}_{diag}^{(\xi)} (u, v) \hat{B}_R^{(\xi)} (v, y) \right] 
{\bf P}_\xi (y) . 
\label{chukan} 
\end{equation} 
The quantity in the square brackets is the same as in complex-scalar 
theory dealt with in Section IV. Fourier transformation of 
(\ref{chukan}) may be carried out using (\ref{o-prop}), (\ref{ana}) 
and (\ref{al}) - (\ref{al1}). The last term in (\ref{ii-o}) may also 
be computed by using (\ref{ana}) and (\ref{sei}). Fourier 
transforming the resultant expression, we have 
\begin{eqnarray*} 
& & - 2 i P_\mu \frac{\partial | \varphi (X) \rangle}{\partial 
X_\mu} \langle \varphi (X) | \left[ \left( 
\begin{array}{cc} 
\Delta_R^{\pi} \Delta_R^{\sigma} & \;\;\; 0 \\ 
\Delta_R^{\pi} \Delta_R^{\sigma} - \Delta_A^{\pi} \Delta_A^{\sigma} 
& \;\;\; - \Delta_A^{\pi} \Delta_A^{\sigma} 
\end{array} 
\right) + ( f_\sigma \Delta_R^{\pi} \Delta_R^\sigma - f_\pi 
\Delta_A^{\pi} \Delta_A^\sigma ) \hat{A}_+ \right] \nonumber \\ 
& & + 2 i P_\mu | \varphi (X) \rangle \frac{\partial \langle \varphi 
(X) |}{\partial X_\mu} \left[ \pi \longleftrightarrow \sigma \right] 
, 
\end{eqnarray*} 
where $\Delta_R^\pi = \Delta_R^\pi (X; P)$, $f_\sigma = f_\sigma 
(X)$, etc. 

After all this, we finally obtain for the Fourier transform of 
$\hat{\bf \Delta} (x, y)$ on $x - y$: 
\begin{eqnarray} 
\hat{\bf \Delta} (X; P) & = & \hat{\bf \Delta}^{(0)} (X; P) + 
\hat{\bf \Delta}^{(1)} (X; P) , 
\label{yaya} 
\\ 
\hat{\bf \Delta}^{(0)} (X; P) & = & \sum_{\xi = \pi, \, \sigma} {\bf 
P}_\xi (X) \hat{B}_L^{(\xi)} (X; P) \hat{\Delta}_{diag}^{(\xi)} (X; 
P) \hat{B}_R^{(\xi)} (X; P) \nonumber \\ 
& = & \sum_{\xi = \pi, \, \sigma} {\bf P}_\xi (X) 
\hat{\Delta}^{(\xi)} (X; P) , 
\label{yaya-1} 
\\ 
\hat{B}_L^{(\xi)} (X; P) & = & \left( 
\begin{array}{cc} 
1 & \;\;\; f_\xi (X; P) \\ 
1 & \;\;\; 1 + f_\xi (X; P) 
\end{array}
\right) \, , \;\;\;\;\;\;\; 
\hat{B}_R^{(\xi)} (X; P) = \left( 
\begin{array}{cc} 
1 + f_\xi (X; P) & \;\;\; f_\xi (X; P) \\ 
1 & \;\;\; 1 
\end{array}
\right) , \nonumber \\ 
\hat{\Delta}^{(\xi)} (X; P) & = & \hat{\Delta}_{R A}^{(\xi)} (X; P) 
+ \hat{\Delta}^{(\xi) '} (X; P) , 
\label{yaya-12} 
\end{eqnarray} 
where 
\begin{eqnarray} 
\hat{\Delta}_{R A}^{(\xi)} (X; P) & = & \left( 
\begin{array}{cc} 
\Delta_R^{(\xi)} & \;\;\; 0 \\ 
\Delta_R^{(\xi)} - \Delta_A^{(\xi)} & \;\;\; - \Delta_A^{(\xi)} 
\end{array} 
\right) , 
\label{yaya-22} \\ 
\Delta_{R (A)}^{(\xi)} (X; P) & = & \frac{1}{P^2 - {\cal M}_\xi^2 
(X) \pm i \epsilon (p_0) \eta} , 
\label{yaza} 
\\ 
\hat{\Delta}^{(\xi) '} (X; P) & = & f_\xi (X; P) \, [ 
\Delta_R^{(\xi)} (X; P) - \Delta_A^{(\xi)} (X; P) ] \, \hat{A}_+ , 
\label{yaya-2} 
\\ 
\hat{\bf \Delta}^{(1)} (X; P) & = & - \frac{i}{2} \sum_{\xi = \pi, 
\, \sigma} {\bf P}_\xi (X) \left\{ f_\xi (X; P), \, P^2 - {\cal 
M}^2_\xi (X) \right\} \left[ (\Delta_R^{(\xi)} )^2 + 
(\Delta_A^{(\xi)} )^2 \right] \hat{A}_+ \nonumber \\ 
& & + i P^\mu \frac{\partial |\varphi (X) \rangle}{\partial X^\mu} 
\langle \varphi (X) | \, \hat{\Omega} - i P^\mu |\varphi (X) \rangle 
\frac{\partial \langle \varphi (X)|}{\partial X^\mu} \, \hat{\Omega} 
\, \rule[-3mm]{.14mm}{8.5mm} \raisebox{-2.85mm}{\scriptsize{$\; \pi 
\leftrightarrow \sigma$}} \nonumber \\  
& & + \frac{i}{2} \left[ |\varphi (X) \rangle 
\stackrel{\leftrightarrow}{\partial}_{X_\mu} \langle \varphi (X)| 
\right] \sum_{\xi = \pi, \, \sigma} \frac{\partial f_\xi (X; 
P)}{\partial P_\mu} (\Delta^{(\xi)}_R - \Delta_A^{(\xi)}) \hat{A}_+ 
. 
\label{yayaya} 
\end{eqnarray} 
Here 
\begin{eqnarray} 
\hat{\Omega} & = & \hat{\Omega}_0 + \left( f_\sigma 
[(\Delta_R^{(\sigma)})^2 - 2 \Delta_R^{(\pi)} \Delta_R^{(\sigma)} - 
(\Delta_A^{(\sigma)})^2] - f_\pi [(\Delta_A^{(\pi)})^2 - 2 
\Delta_A^{(\pi)} \Delta_A^{(\sigma)} - (\Delta_R^{(\pi)})^2] \right) 
\hat{A}_+ , \nonumber \\ 
& & 
\label{yaya-3} \\ 
\hat{\Omega}_0 & = & \left( 
\begin{array}{cc} 
(\Delta_R^{(\pi)} - \Delta_R^{(\sigma)})^2 & \;\;\; 0 \\ 
(\Delta_R^{(\pi)} - \Delta_R^{(\sigma)})^2 - (\Delta_A^{(\pi)} - 
\Delta_A^{(\sigma)})^2 & \;\;\; - (\Delta_A^{(\pi)} - 
\Delta_A^{(\sigma)})^2 
\end{array} 
\right) . 
\label{yaya-4} 
\end{eqnarray} 
Note that $f_\xi$ is related to the \lq\lq genuine'' number density 
$N^{(\xi)}$ through (cf. (\ref{p0f})) 
\[ 
f_\xi (X; P) = \theta (p_0) N^{(\xi)} (X; E_p^{(\xi)}; \hat{\bf p}) 
- ( 1 + \theta (- p_0)) N^{(\xi)} (X; E_p^{(\xi)}; - \hat{\bf p}) . 
\] 
\subsubsection{Counter Lagrangian and renormalized scheme} 
Following the procedure in Section VE (cf. (\ref{new-1}) to 
(\ref{new-2})), we can extract the counter hat-Lagrangian density 
$\hat{\cal L}_c$ through analyzing $\hat{\bf \Delta}^{- 1} (x) 
\hat{\bf \Delta} (x, y)$ and $\hat{\bf \Delta} (x, y) 
\stackrel{\longleftarrow}{\hat{\bf \Delta}^{- 1}} (y)$. 
$\hat{\cal L}_c$ thus found gives a two-point vertex function 
$\hat{\bf V}_c (X; P)$. Straightforward manipulation yields 
\begin{eqnarray} 
i \hat{\bf V}_c (X; P) & \simeq & \sum_{\xi = \pi, \, \sigma} 
\left\{ {\bf P}_\xi (X) f_\xi (X; P), \, P^2 - {\cal M}^2_\xi (X) 
\right\} \hat{A}_- 
\label{v-n} \\ 
& = & \sum_{\xi = \pi, \, \sigma} {\bf P}_\xi (X) \left\{ f_\xi 
(X; P), \, P^2 - {\cal M}^2_\xi (X) \right\} \hat{A}_- \nonumber \\ 
& & - 2 P_\mu [f_\pi (X; P) - f_\sigma (X; P)] \, \frac{\partial | 
\varphi (X) \rangle \langle \varphi (X)|}{\partial X_\mu} \, 
\hat{A}_- . 
\label{V-on} 
\end{eqnarray} 
Let us adopt the physical-$f$ scheme. In the case of complex-scalar 
field, we have adopted the \lq\lq renormalization condition'' 
$\hat{\cal V}_c (X; P) = 0$ (cf. (\ref{new-2}) and (\ref{kuri-j})). 
In the present case, we cannot set $\hat{\bf V}_c (X; P) = 0$. Then, 
we adopt the condition, 
\begin{equation} 
\left\{ f_\xi (X; P), \, P^2 - {\cal M}^2_\xi (X) \right\} = 0 , 
\label{ashi} 
\end{equation} 
which is the same as in the complex-scalar-field case. Then, in 
$\hat{\bf V}_c$ in (\ref{V-on}), the second term on the RHS 
survives. Let us compute the $\hat{\bf V}_c$-resummed propagator 
$\hat{\bf \Delta}_{c-res}$. Since $\hat{\bf V}_c$ contains the 
$X_\mu$-derivative, one can directly write down the 
Fourier-transformed form: 
\begin{equation} 
\hat{\bf \Delta} (X; P) \left[ 1 + \sum_{n = 1} \left\{ - 
\hat{\bf V}_c (X; P) \, \hat{\bf \Delta} (X; P) \right\}^n \right] . 
\label{resum-ro} 
\end{equation} 
In $2 \times 2$ matrix space, $\hat{\bf V}_c$ is proportional to 
$\hat{A}_-$, Eq.~(\ref{Apm}), which shows that, in (\ref{resum-ro}), 
the contributions with $n \geq 2$ vanish. Thus, after some 
manipulations, we obtain, 
\begin{eqnarray*} 
\hat{\bf \Delta}_{c-resum} (X; P) & \simeq & - 2 i (f_\pi - 
f_\sigma) P_\mu \nonumber \\ 
& & \times \left[ | \varphi (X) \rangle 
\stackrel{\leftarrow}{\partial}_{X_\mu} \langle \varphi (X) | \, 
\Delta_R^\pi \, \Delta_A^\sigma + | \varphi (X) \rangle 
\partial_{X_\mu} \langle \varphi (X) | \, \Delta_R^\sigma \, 
\Delta_A^\pi \right] \, \hat{A}_+ . 
\end{eqnarray*} 
Summing this to $\hat{\bf \Delta} (X; P)$ in (\ref{yaya}), we obtain 
(\ref{yaya})~-~(\ref{yayaya}) with (\ref{ashi}), provided that 
$\hat{\Omega}$ in (\ref{yaya-3}) is changed to 
\begin{eqnarray} 
\hat{\Omega} \to \hat{\Omega}_{renorm} & = & \hat{\Omega}_0 + \left( 
f_\sigma \, [ \Delta_R^{(\sigma)} - \Delta_A^{(\sigma)}] 
[\Delta_R^{(\sigma)} + \Delta_A^{(\sigma)} - 2 \Delta_R^{(\pi)} ] 
\right. \nonumber \\ 
& & \left. - f_\pi \, [\Delta_A^{(\pi)} - \Delta_R^{(\pi)} ] 
[\Delta_A^{(\pi)} + \Delta_R^{(\pi)} - 2 \Delta_A^{(\sigma)} ] 
\right) \hat{A}_+ . 
\label{complete} 
\end{eqnarray} 
This completes the deduction of bare propagators in the bare-$f$ 
scheme. The physical-$f$ scheme may be constructed in a similar 
manner as in Section IVC. In place of (\ref{r-cond}) or 
(\ref{kik}), we have, with obvious notation, 
\begin{equation} 
\Sigma_{off}^{(\xi)} (X; P) = 0 \;\;\;\;\;\; (\xi = \pi, \sigma) , 
\label{iyya} 
\end{equation} 
which lead to the generalized Boltzmann equations. 
\subsection{Determining $M_\pi^2 (X)$ and $M_\sigma^2 
(X)$} 
In this subsection, we present a scheme for determining so far 
arbitrary $M_\xi^2 (X)$ ($\xi = \pi, \sigma$), introduced in 
(\ref{diff})~-~(\ref{teigi-10}). Various methods of fixing 
$M_\xi^2 (X)$ are available. Among those, we mention the following 
three methods. 

1) {\em Resummation method}. Argument in Section V shows that the 
quantities to be resummed are the real part of the retarded and 
advanced self-energy parts, 
\[ 
{\bf \Sigma}_R (X; R) = {\bf \Sigma}_{1 1} (X; R) + {\bf \Sigma}_{1 
2} (X; R) = \left( {\bf \Sigma}_A (X; R) \right)^* . 
\] 
Determine $M_\xi^2 (X)$ ($\xi = \pi, \sigma$) such that the real 
part of ${\bf \Sigma}_R (X; R)$ computed up to required $L$th loop 
order, $Re \, \hat{\bf \Sigma}^{(L)}_R (X; R)$, vanishes at $R^\mu = 
R_0^\mu$ with $R_0^\mu$ a suitable four momentum: 
\begin{equation} 
Re \, \hat{\bf \Sigma}^{(L)}_R (X; R = R_0) = 0 . 
\label{gap} 
\end{equation} 

2) {\em The principles of minimal sensitivity} \cite{ste}. Total 
Lagrangian density ${\cal L}$ in (\ref{owari}) is independent of 
arbitrary functions $M_\pi^2 (X)$ and $M_\sigma^2 (X)$. Let $\Xi$ be 
a quantity, in which we are interested. Then, if we could evaluate 
$\Xi$ to all orders of perturbation series or loop expansion, it 
is independent of $M_\xi^2 (X)$ $(\xi = \pi, \sigma)$. On the other 
hand, $\Xi$ computed on the basis of ${\cal L}$ in (\ref{owari}) up 
to the $L$th order, $\Xi^{(L)}$, depends on $M_\xi^2 (X)$. This 
observation leads one to {\em impose} the condition 
\[ 
\frac{\delta \Xi^{(L)}}{\delta M_\xi^2 (X)} = 0 \;\;\;\;\;\;\;\; 
(\xi = \pi, \sigma) , 
\] 
which serves as a determining equation for $M_\xi^2 (X)$. Note that, 
for $\Xi = \Gamma_{phys}$ or $\Xi = \delta \Gamma_{phys} / \delta 
\vec{\varphi} (X)$, this method works for $L \geq 2$. 

3) {\em The criterion of the fastest apparent convergence} 
\cite{grunberg}. Compute $\Xi$ up to $L$th order, $\Xi^{(L)}$ $(L 
\geq 2)$. Then, determine all the parameters of the theory, such as 
the coupling constant, mass functions $M_\xi^2 (X)$ $(\xi = \pi, 
\sigma)$, etc., so that all the terms but the first term of the 
series vanish. 

A number of works, which employ the methods 2) and 3) above for 
analyzing the equilibrium effective potential, is cited in 
\cite{chiku}. 

In this paper, we follow the method 1). The requirement in 1) gives 
the gap equation. 
\subsection{Summary of this section} 
In this section, we have found that the spacetime evolution of the 
condensate fields $\vec{\varphi} (X)$ is governed by the following 
system of coupled equations: 
\begin{description} 
\item{E1)} Generalized relativistic kinetic or Boltzmann equation 
(\ref{iyya}). 
\item{E2)} Gap equation (\ref{gap}). 
\item{E3)} Equation of motion for $\vec{\varphi} (X)$, 
(\ref{maae-1}) with $j = 1$. 
\end{description} 

As seen in Section VC, (\ref{ashi}) in E1) is a determining equation 
for the number-density function $f_\xi (X; P)$ $(\xi = \pi, \sigma)$ 
under a given initial data $f_\xi (X_0 = T_{in}, {\bf X}; P)$. This 
equation involves yet unknown functions $M_\xi^2 (X)$ $(\xi = \pi, 
\sigma)$. Equation (\ref{gap}) determines $M_\xi^2 (X)$ in terms of 
$f_\xi (X; P)$. Thus, (\ref{ashi}) and (\ref{gap}) interplay to 
determine $M_\xi^2 (X)$ and $f_\xi (X; P)$. Through computing the 
effective action $\hat{\Gamma}$ using $M_\xi^2 (X)$ and $f_\xi (X; 
P)$ thus determined, the equation of motion (\ref{maae-11}) for 
$\vec{\varphi} (X)$ is obtained. Solving it under a given initial 
data determines the spacetime evolution of $\vec{\varphi} (X)$, 
which describes the progress of phase transition. 
\setcounter{equation}{0}
\setcounter{section}{6}
\def\theequation{\mbox{\arabic{section}.\arabic{equation}}}
\section{Lowest-order effective action in the $\sigma$ model} 
In this section, we compute the CTP effective action to the lowest 
nontrivial order, which comes\footnote{The difference between 
$\hat{\Gamma}$ in (\ref{saa}) and $\hat{\Gamma}$ computed using the 
physical-$f$ scheme, whose form has not been displayed explicitly, 
is of higher order.} from an one-loop diagram with $\hat{\cal L}_0$ 
as in (\ref{hi-setsu}). The contribution is 
\begin{equation} 
\hat{\Gamma} = \frac{i}{2} \, \mbox{Tr} \, \ln \left( \frac{1}{2} \, 
\hat{\tau}_3 \left\{ {\bf I} + \sum_{\xi = \pi, \, \sigma} 
\hat{\bf P}_\xi \, \hat{\cal M}_\xi^2 \, \partial^{- 2} \right\} 
\right) , 
\label{saa} 
\end{equation} 
where \lq Tr' acts on matrices in spacetime, $2 \times 2$ matrices 
in the \lq\lq complex-time plane,'' and $N \times N$ matrices in 
the $O (N)$ vector space. In (\ref{saa}), 
\[ 
\left( \hat{\tau}_3 \, \partial^{- 2} \right) (x, y) = - 
\hat{\Delta} (x, y) \, \rule[-3mm]{.14mm}{8.5mm} 
\raisebox{-2.85mm}{\scriptsize{$\; m = 0$}} , 
\] 
with $\hat{\Delta} (x, y)$ as in (\ref{imple-p0}) and 
\begin{eqnarray*} 
\hat{\cal M}^2_\xi (x) & = & \mbox{diag} \left[ 
{\cal M}_{\xi 1}^2 (x), \; - 
{\cal M}_{\xi 2}^2 (x) \right] 
\;\;\;\;\;\;\;\; (\xi = \pi, \sigma) \nonumber \\ 
\hat{\vec{\varphi}}^{\, 2} & = & \mbox{diag} \left( 
\vec{\varphi}_1^{\, 2} (x), \; - \vec{\varphi}_2^{\, 2} (x) \right) 
, \nonumber \\ 
\hat{\bf P}_\xi (x) & = & \mbox{diag} \left( {\bf P}_{\xi 1} 
(x) , \; - {\bf P}_{\xi 2} (x) \right) . 
\end{eqnarray*} 
Here $\vec{\varphi}_i$ $(i = 1, 2)$ are the type-$i$ condensate 
fields, ${\cal M}_{\xi i}^2$ and ${\bf P}_{\xi i}$ $(i = 1, 2)$ are 
as in (\ref{134}) and (\ref{teigi-10}), respectively, with 
$\vec{\varphi}_i$ in place of $\vec{\varphi}$. 

For extracting the {\em physical} effective action from 
(\ref{saa}), we use a method \cite{fuji} of computing {\em 
physical} effective potential, in which the tadpole method is 
employed . The method starts with taking the derivatives, 
\begin{eqnarray} 
\frac{\delta \hat{\Gamma}}{\delta {\cal M}_{\xi 1}^2 (x)} & = & 
\frac{i}{2} \, \mbox{tr} \, {\bf P}_{\xi 1} (x) \left[ {\bf I} \, 
\partial^{- 2} (x, x) - \left\{ \partial^{- 2} \left( \hat{\bf 
P}_\pi \hat{\cal M}_\pi^2 + \hat{\bf P}_\sigma \hat{\cal M}_\sigma^2 
\right) \partial^{- 2} \right\} (x, x) + ... \right]_{1 1} \nonumber 
\\ 
& & \mbox{\hspace*{57ex}} (\xi = \pi, \sigma) , 
\label{mine-01} 
\end{eqnarray} 
where \lq\lq tr'' acts on $2 \times 2$ matrices and $N \times N$ 
matrices. Then, we set $\vec{\varphi}_1^{\, 2} (x) = 
\vec{\varphi}_2^{\, 2} (x) \equiv \vec{\varphi}^{\, 2} (x)$. 
Equation (\ref{mine-01}) now becomes (cf. (\ref{maae-11})) 
\begin{eqnarray} 
\frac{\delta \hat{\Gamma}}{\delta {\cal M}_{\xi 1}^2 (x)} \, 
\rule[-3mm]{.14mm}{8.5mm} \raisebox{-2.85mm}{\scriptsize{$\; 
\vec{\varphi}_1 = \vec{\varphi}_2 = \vec{\varphi}$}} & = & - 
\frac{i}{2} \, \mbox{tr} \, {\bf P}_\xi (x) {\bf G}_{1 1} (x, x) 
\nonumber \\ 
& = & \frac{\delta \Gamma_{phys}}{\delta {\cal M}_\xi^2 (x)} , 
\label{miyo-1} 
\end{eqnarray} 
where 
\begin{equation} 
\hat{\bf G} = - \hat{\tau}_3 \left[ {\bf I} \, \partial^2 + 
\hat{\bf P}_\pi \hat{\cal M}_\pi^2 + \hat{\bf P}_\sigma \hat{\cal 
M}_\sigma^2 \right]^{- 1} \, \rule[-3mm]{.14mm}{8.5mm} 
\raisebox{-2.85mm}{\scriptsize{$\; \vec{\varphi}_1 = \vec{\varphi}_2 
= \vec{\varphi}$}} . 
\label{G} 
\end{equation} 
The one-loop contribution to the {\em physical} effective action 
$\Gamma_{phys}$ is obtained \cite{fuji} by integrating 
(\ref{miyo-1}) over ${\cal M}_\xi^2  (x)$. 

Incidentally, the \lq\lq unphysical part'' of $\Gamma$, which comes 
from $\delta \hat{\Gamma} / \delta {\cal M}_{\xi 2}^2 (X)$, becomes 
\begin{equation} 
\Gamma_{unphys} = - \left[ \Gamma_{phys} \right]^* . 
\label{hi-b} 
\end{equation} 

So far, we have made no assumption on the $x$-dependence of ${\cal 
M}_\xi^2 (x)$ and $\vec{\varphi} (x)$. Here we recall that ${\cal 
M}_\xi^2 (x)$ and $\vec{\varphi} (x)$ weakly depend on $x$. We 
first note that 
\[ 
\mbox{Sp} [\partial_{X_\mu} {\bf P}_\xi (X)] = \mbox{Sp} [{\bf 
P}_\xi \partial_{X_\mu} {\bf P}_\zeta (X)] = 0 \;\;\;\;\;\; (\xi, 
\zeta = \pi, \sigma) , 
\] 
where \lq Sp' stands for the trace operation that acts on an $N 
\times N$ matrix. Then, the term with $\partial_{X_\mu} {\bf P}_\xi 
(X)$ does not appear. This means that, within the gradient 
approximation, we can assign common spacetime coordinates for all 
$\hat{\bf P}_\xi$ in (\ref{G}). Thus, including the zeroth-order 
contribution (cf. (\ref{rei-11})), we obtain 
\begin{eqnarray} 
\Gamma_{phys} & = &  \int d^{\, 4} X \, \gamma_{phys} (X) , 
\label{gam-tot} 
\\ 
\gamma_{phys} (X) & = & \frac{1}{2} \left[ (\partial \vec{\varphi} 
(X))^2 - \mu^2 \vec{\varphi}^2 (X) \right] - \frac{\lambda}{4 !} ( 
\vec{\varphi}^2 (X))^2 + H ( \displaystyle{ 
\raisebox{0.9ex}{\scriptsize{$t$}}} \mbox{\hspace{-0.1ex}} \vec{e} 
\cdot \vec{\varphi} (X)) \nonumber \\ 
& & + \frac{1}{2} \left[ (N - 1) \gamma_\pi (X) + \gamma_\sigma (X) 
\right] , \label{gam} \\ 
\gamma_\xi (X) & = & - \int d {\cal M}_\xi^2 (X) \, {\cal I}_\xi (X) 
\;\;\;\;\;\; (\xi = \pi, \sigma) . 
\label{gam-1} 
\end{eqnarray} 
Here 
\begin{eqnarray} 
{\cal I}_\xi (X) & = & i \int \frac{d^{\, 4} P}{(2 \pi)^4} 
\Delta_{1 1}^{(\xi)} (X; P) = {\cal I}_\xi^{(0)} (X) + {\cal 
I}_\xi^{(\beta)} (X) , 
\label{yoko} \\ 
{\cal I}_\xi^{(0)} (X) & \equiv & \int \frac{d^{\, 4} P}{(2 
\pi)^4} \frac{i}{P^2 - {\cal M}_\xi^2 + i \eta} , 
\label{yoko-10} 
\\ 
{\cal I}_\xi^{(\beta)} (X) & \equiv & \int \frac{d^{\, 3} p}{(2 
\pi)^3} \, \frac{N^{(\xi)} (X; {\cal E}_p^{(\xi)} (X) , \hat{\bf 
p})}{{\cal E}_p^{(\xi)} (X)} , 
\label{yoko-11} 
\end{eqnarray} 
where ${\cal E}_p^{(\xi)} (X) = \sqrt{p^2 + {\cal M}_\xi^2 (X)}$. 
Thus the physical effective action $\Gamma_{phys}$ is given by an 
integral of the \lq\lq local effective action'' $\gamma_{phys} (X)$ 
over the macroscopic spacetime coordinates $X^\mu$. Computation of 
$\gamma_{phys} (X)$ goes essentially in the the same manner as that 
of the physical effective potential. 

In the following two subsections, we separately analyze the 
positive-mass-square region and negative-mass-square region. As 
mentioned in Section I, our scheme does not apply to the region near 
the transition point between the two regions. 
\subsection{The region ${\cal M}_\pi^2 (X), \, {\cal 
M}_\sigma^2 (X) > 0$.} 
The lowest-order contribution to ${\cal I}_\xi (X)$, 
Eq.~(\ref{yoko}), has already been computed in Section VD (see 
(\ref{mass-in0})~-~(\ref{dim})): 
\begin{eqnarray} 
{\cal I}_\xi^{(0)} (X) & = & - \frac{{\cal M}_\xi^2 (X)}{(4 \pi)^2} 
\left[ \frac{2}{\overline{\epsilon}} - \ln \frac{{\cal M}_\xi^2 
(X)}{e \mu_d^2} \right] \nonumber \\ 
{\cal I}_\xi^{(\beta)} (X) & = & 
\frac{1}{2 \pi^2} \int_{{\cal M}_\xi (X)}^\infty d {\cal E} \, 
\sqrt{{\cal E}^2 -{\cal M}_\xi^2 (X)} \; \langle N^{(\xi)} (X; 
{\cal E}) \rangle , 
\label{tad-1} 
\end{eqnarray} 
where 
\begin{equation} 
\langle N^{(\xi)} (X ; {\cal E}) \rangle \equiv \int \frac{d 
\Omega_{\bf p}}{4 \pi} \, N^{(\xi)} (X; {\cal E}, \hat{\bf p}) 
\label{are} 
\end{equation} 
with $d \Omega_{\hat{\bf p}}$ a solid-angle element in a ${\bf 
p}$-space. Substituting these into (\ref{gam-1}), we obtain 
\begin{equation} 
\gamma_\xi (X) = \gamma_\xi^{(0)} (X) + \gamma_\xi^{(\beta)} (X) 
\label{sekibun} 
\end{equation} 
with 
\begin{eqnarray} 
\gamma_\xi^{(0)} (X) & = & i \int \frac{d^{\, 4} P}{(2 \pi)^4} \, 
\ln \left( P^2 - {\cal M}_\xi^2 (X) + i \eta \right) 
\label{T-xi10} \\ 
& = & \frac{1}{3 2 \pi^2} \, {\cal M}_\xi^4 
(X) \left[ \frac{2}{\overline{\epsilon}} + \frac{3}{2} - \ln 
\frac{{\cal M}_\xi^2 (X)}{\mu_d^2} \right] ,
\label{T-xi0} 
\\ 
\gamma_\xi^{(\beta)} (X) & = & \frac{1}{3 \pi^2} \int_{{\cal 
M}_\xi (X)}^\infty d {\cal E} \, ({\cal E}^2 - {\cal M}_\xi^2 
(X))^{3 / 2} \, \langle N^{(\xi)} (X ; {\cal E}) \rangle . 
\label{T-xi} 
\end{eqnarray} 
$2 / \overline{\epsilon}$ in (\ref{T-xi0}) is canceled by the 
counter Lagrangian ${\cal L}_{cond}'$, Eq.~(\ref{cla-1}). We note 
that, to the present one-loop order, we can replace $- (Z_\mu Z - 1) 
\mu^2 \vec{\varphi}^{\, 2} / 2$ in (\ref{cla-1}) as 
\begin{equation} 
- (Z_\mu Z - 1) \mu^2 \vec{\varphi}^{\, 2} / 2 \to - \frac{1}{2 (N + 
2)} (Z_\mu Z - 1) \left[(N - 1) M_\pi^2 + 3 M_\sigma^2 \right] 
\vec{\varphi}^{\, 2} . 
\label{kae} 
\end{equation} 
The difference between the LHS and the RHS, which is proportional to 
the UV-diverging factor $1 / \overline{\epsilon}$, is of higher 
order (cf. above after (\ref{teigi-10})). This difference is 
canceled by the higher-order one-loop diagrams, one of which 
includes the vertex coming from $\displaystyle{ 
\raisebox{0.9ex}{\scriptsize{$t$}}} \mbox{\hspace{-0.1ex}} 
\vec{\phi} \chi_\pi {\bf P}_\pi \vec{\phi} / 2$ $(\in {\cal L}_{r 
c})$ and the other one includes the vertex coming from 
$\displaystyle{ \raisebox{0.9ex}{\scriptsize{$t$}}} 
\mbox{\hspace{-0.1ex}} \vec{\phi} \chi_\sigma {\bf P}_\sigma 
\vec{\phi} / 2$ $(\in {\cal L}_{r c})$. Since we are choosing the 
$\overline{\mbox{MS}}$ scheme, we have 
\begin{equation} 
Z = 1 \, , \;\;\;\;\;\; Z_\mu = 1 + \lambda \frac{N + 2}{48 
\pi^2} \frac{1}{\overline{\epsilon}} \, , \;\;\;\;\;\; 
Z_\lambda = 1 + \lambda \frac{N + 8}{48 \pi^2} 
\frac{1}{\overline{\epsilon}} . 
\label{Z-fac} 
\end{equation} 
In a similar manner, we obtain for $A$ in (\ref{cla-1}), 
\[ 
A = - \frac{N}{32 \pi^2} \frac{1}{\overline{\epsilon}} . 
\] 
After all this, we obtain for the local effective action, to 
one-loop order, 
\begin{equation} 
\gamma_{phys} (X) = \gamma_{phys}^{(0)} (X) + 
\gamma_{phys}^{(\beta)} (X) , 
\label{yuukou} 
\end{equation} 
where 
\begin{eqnarray} 
\gamma_{phys}^{(0)} (X) & = & \frac{1}{2} \left[ (\partial 
\vec{\varphi} (X))^2 - \mu^2 \vec{\varphi}^2 (X) \right] - 
\frac{\lambda}{4 !} ( \vec{\varphi}^2 (X))^2 + H ( \displaystyle{ 
\raisebox{0.9ex}{\scriptsize{$t$}}} \mbox{\hspace{-0.1ex}} 
\vec{e} \cdot \vec{\varphi} (X)) \nonumber \\ 
& & - \frac{1}{(8 \pi)^2} \left[ (N - 1) {\cal M}_\pi^4 (X) \ln 
\frac{{\cal M}_\pi^2 (X)}{e^{3 / 2} \mu_d^2} + {\cal M}_\sigma^4 (X) 
\ln \frac{{\cal M}_\sigma^2 (X)}{e^{3 / 2} \mu_d^2} \right] , 
\label{yuukou-0} \\ 
\gamma_{phys}^{(\beta)} (X) & = & \frac{1}{2} \left[ (N - 1) \, 
\gamma^{(\beta)}_\pi (X) + \gamma^{(\beta)}_\sigma (X) 
\right] . 
\label{yuukou-1} 
\end{eqnarray} 
It should be emphasized that the factor of the \lq\lq 
$\vec{\varphi}^{\, 2}$ term'' in (\ref{yuukou-0}) is neither ${\cal 
M}_\pi^2 (X)$ nor ${\cal M}_\sigma^2 (X)$ but $\mu^2$. 

In the case of thermal equilibrium system with constant 
$\vec{\varphi}$, $- \gamma_{phys}$ reduces to the effective 
potential (density) of the system, $V (\vec{\varphi}) = - 
\gamma_{phys} (\vec{\varphi})$. Besides the terms due to the 
different UV-renormalization scheme, $V (\vec{\varphi})$ obtained 
above agrees with the result of \cite{chiku}, where ${\cal M}_\pi^2 
(X) = {\cal M}_\sigma^2 (X)$, being constant, is assumed. 

Now we are ready to write down the equation of motion for the 
condensate fields (cf. Section VIB), 
\begin{eqnarray}  
\frac{\delta \Gamma_{phys}}{\delta \vec{\varphi} (X)} & = & - 
\left[ \partial^2 + \Omega (\vec{\varphi}^{\, 2} (X)) \right] \, 
\vec{\varphi} (X) + H \vec{e} = 0 , 
\label{tititi} 
\\ 
\Omega (\vec{\varphi}^{\, 2} (X)) & = & - 2 \frac{\partial 
\gamma_{phys} (X)}{\partial \vec{\varphi}^{\, 2} (X)} 
\label{haha-pre} 
\\ 
& = & \mu^2 + \frac{\lambda}{6} \vec{\varphi}^{\, 2} (X) + 
\frac{\lambda}{6} \left[ (N - 1) {\cal I}_\pi (X) + 3 {\cal 
I}_\sigma (X) \right] 
\label{haha} 
\\ 
& = & \Omega_0 (\vec{\varphi}^{\, 2} (X)) + \Omega_\beta 
(\vec{\varphi}^{\, 2} (X)) , 
\label{Ome-11} 
\end{eqnarray}  
where 
\begin{eqnarray}  
\Omega_0 (\vec{\varphi}^{\, 2} (X)) & = & \mu^2 + \frac{\lambda}{6} 
\vec{\varphi}^{\, 2} (X) \nonumber \\ 
& & + \frac{\lambda}{96 \pi^2} \left[ (N - 1) \, {\cal M}_\pi^2 (X) 
\, \ln \frac{{\cal M}_\pi^2 (X)}{e \mu_d^2} + 3 {\cal M}_\sigma^2 
(X) \, \ln \frac{{\cal M}_\sigma^2 (X)}{e \mu_d^2} \right] , 
\label{Ome-0} 
\\ 
\Omega_\beta (\vec{\varphi}^{\, 2} (X)) & = & \frac{\lambda}{6} 
\left[ (N - 1) {\cal I}_\pi^{(\beta)} (X) + 3 \, {\cal 
I}_\sigma^{(\beta)} (X) \right] . 
\label{eq-m} 
\end{eqnarray}  
Here ${\cal I}_\xi^{(\beta)} (X)$ $(\xi = \pi, \sigma)$ is as in 
(\ref{tad-1}). 

It is worth mentioning in passing that the equation of motion for 
unphysical or type-2 condensate fields are obtained from 
(\ref{hi-b}), which is of the same form as (\ref{tititi}). Since the 
time argument of the type-2 field is on $C_2$, the type-2 field 
\lq\lq lives'' in the \lq\lq time-reversed world.'' Then, one can 
say that the equation of motion is time-reversal invariant. 

Since we are assuming that $\vec{\varphi} (X)$ does not appreciably 
change in a single spacetime cell (cf. Section I), (\ref{tititi}) 
applies only in the region of $\vec{\varphi} (X)$ where $|\Omega 
(\vec{\varphi}^{\, 2} (X)) | \lesssim 1 / (L^\mu)^2$ $(\mu = 0, 1, 
2, 3)$ with $L^\mu$ the size of the spacetime cell. 

Let us study the small oscillation of $\vec{\varphi} (X)$ at the 
vicinity of the quasiminimum point $\vec{\varphi} (X) = \varphi_0 
(X) \vec{e}$: 
\begin{equation} 
\Omega (\varphi_0^2 (X)) \, \varphi_0 (X) = H . 
\label{q-min} 
\end{equation} 
Setting $\vec{\varphi} (X) = \varphi_0 (X) \vec{e} + \vec{\epsilon} 
(X)$, we obtain, 
\begin{eqnarray}  
\left[ \partial^2 + \Omega^{(\sigma)} (\vec{\varphi}_0^{\, 2} (X)) + 
\frac{H}{\varphi_0 (X)} \right] (\displaystyle{ 
\raisebox{0.9ex}{\scriptsize{$t$}}} \mbox{\hspace{-0.1ex}} \vec{e} 
\cdot \vec{\epsilon} (X)) + \partial^2 \varphi_0 (X) & \simeq & 0 , 
\label{Gold-11}  \\ 
\left[ \partial^2 + \frac{h}{\varphi_0 (X)} \right] \, 
\vec{\epsilon}_\perp (X) & \simeq & 0 , 
\label{Gold} 
\end{eqnarray}  
where $\vec{\epsilon}_\perp (X) \equiv \vec{\epsilon} (X) -  
(\displaystyle{ \raisebox{0.9ex}{\scriptsize{$t$}}} 
\mbox{\hspace{-0.1ex}} \vec{e} \cdot \vec{\epsilon} (X)) \, \vec{e}$ 
and 
\[ 
\Omega^{(\sigma)} (\vec{\varphi}_0^{\, 2} (X)) = \Omega_0^{(\sigma)} 
(\vec{\varphi}_0^{\, 2} (X)) + \Omega_\beta^{(\sigma)} 
(\vec{\varphi}_0^{\, 2} (X)) 
\] 
with 
\begin{eqnarray}  
\Omega_0^{(\sigma)} (\vec{\varphi}_0^{\, 2} (X)) & = & 
\frac{\lambda \varphi_0^2 (X)}{3} \left[ 1 + \frac{\lambda}{96 
\pi^2} \left\{ (N - 1) \, \ln \frac{{\cal M}_\pi^2 (X)}{\mu_d^2} + 
9 \, \ln \frac{{\cal M}_\sigma^2 (X)}{\mu_d^2} 
\right\}_{\scriptsize{\vec{\varphi} (X) = \varphi_0 (X) \vec{e}}} 
\right] , \nonumber \\ 
& & 
\label{Om-0} \\ 
\Omega_\beta^{(\sigma)} (\vec{\varphi}_0^{\, 2} (X)) & = & 
\frac{\lambda^2 \varphi_0^2 (X)}{18} \left[ (N - 1) \frac{\partial 
\, {\cal I}_\pi (X)}{\partial {\cal M}_\pi^2 (X)} + 9 
\frac{\partial \, {\cal I}_\sigma (X)}{\partial {\cal M}_\sigma^2 
(X)} \right]_{\scriptsize{\vec{\varphi} (X) = \varphi_0 (X) 
\vec{e}}} . 
\label{Om-1} 
\end{eqnarray}  
$\vec{\epsilon}_\perp (X)$ describes quasi Goldstone mode. Equation 
(\ref{Gold-11}) may be used only in the region of $\varphi_0 (X)$ 
where $|\Omega^{(\sigma)} (\vec{\varphi}_0^{\, 2} (X))| \lesssim 1 / 
(L^\mu)^2$. 
\subsubsection{\lq\lq High-temperature'' expansion} 
The system is characterized by a scale parameter or several scale 
parameters of mass dimension. Let us assume that, for small ${\cal 
E}$, $\langle N^{(\xi)} (X ; {\cal E}) \rangle$ in (\ref{are}) 
behaves as 
\begin{equation} 
\langle N^{(\xi)} (X; {\cal E}) \rangle \simeq \frac{{\cal T}_\xi 
(X)}{\cal E} + {\cal C}_\xi (X) \;\;\;\; \mbox{for} \;\; {\cal 
T}_\xi (X) >> {\cal E} , 
\label{takai} 
\end{equation} 
where ${\cal T}_\xi (X)$, which we call the temperature, is 
proportional to the scale parameter or to some combination of the 
scale parameters. In the case of locally thermal equilibrium system, 
the averaged number-density function reads 
\begin{equation} 
\langle N^{(\xi)} (X; {\cal E}) \rangle = \frac{1}{e^{{\cal E} / T 
(X)} - 1} , 
\label{bunbun} 
\end{equation} 
where $T (X)$ is the local temperature of the system. Then, we have 
\[ 
{\cal T}_\xi (X) = T (X) \, , \;\;\;\;\; {\cal C}_\xi (X) = - 
\frac{1}{2} . 
\] 
Let us analyze $\gamma_{phys} (X)$ in the high-temperature region. 
We start with rewriting (\ref{T-xi}) as 
\begin{eqnarray} 
\gamma_\xi^{(\beta)} (X) & = & \gamma_{\xi 1}^{(\beta)} (X) + 
\gamma_{\xi 2}^{(\beta)} (X) \nonumber \\ 
\gamma_{\xi 1}^{(\beta)} (X) & = & \frac{1}{3 \pi^2} \int_0^\infty d 
{\cal E} \left[ {\cal E}^3 - \frac{3}{2} {\cal E} {\cal M}_\xi^2 (X) 
\right] \langle N^{(\xi)} (X; {\cal E}) \rangle \nonumber \\ 
& \equiv & \frac{\pi^2}{45} b_\xi (X) {\cal T}_\xi^4 (X) - 
\frac{1}{1 2} c_\xi (X) {\cal M}_\xi^2 (X) {\cal T}_\xi^2 (X) , 
\label{takai-1} \\ 
\gamma_{\xi 2}^{(\beta)} (X) & = & \frac{1}{3 \pi^2} \int_0^\infty d 
{\cal E} \left[ \theta \left( {\cal E} - {\cal M}_\xi (X) \right) \, 
({\cal E}^2 - {\cal M}_\xi^2 (X) )^{3 / 2} - {\cal E}^3 + 
\frac{3}{2} {\cal E} {\cal M}_\xi^2 (X) \right] \nonumber \\ 
& & \times \langle N^{(\xi)} (X; {\cal E}) \rangle . 
\label{takai-2} 
\end{eqnarray} 
In the case of locally equilibrium system, $b_\xi (X) = c_\xi (X) = 
1$. At \lq\lq high temperature,'' $\gamma_{\xi 1}^{(\beta)}$ 
dominates, $|\gamma_{\xi 2}^{(\beta)}| << \gamma_{\xi 1}^{(\beta)}$. 
$\gamma_{\xi 2}^{(\beta)}$ may be estimated by substituting 
(\ref{takai}) into (\ref{takai-2}): 
\[ 
\gamma_{\xi 2}^{(\beta)} (X) \simeq \frac{1}{6 \pi} \, {\cal 
M}_\xi^3 (X) \, {\cal T}_\xi (X) . 
\] 

Putting all this together, we obtain 
\begin{equation} 
\gamma_\xi^{(\beta)} (X) \simeq \frac{\pi^2}{4 5} b_\xi (X) {\cal 
T}_\xi^4 (X) - \frac{1}{1 2} c_\xi (X) {\cal M}_\xi^2 (X) {\cal 
T}_\xi^2 (X) + \frac{1}{6 \pi} {\cal M}_\xi^3 (X) \, {\cal 
T}^{(\xi)} (X) . 
\label{koou} 
\end{equation} 
The local effective action $\gamma_{phys} (X)$, Eq.~(\ref{yuukou}), 
now becomes 
\[ 
\gamma_{phys} (X) \simeq \gamma^{(0)}_{phys} (X) + 
\gamma^{(\beta)}_{phys} (X) , 
\] 
where $\gamma^{(0)}_{phys} (X)$ is as in (\ref{yuukou-0}), and 
\begin{eqnarray*} 
\gamma^{(\beta)}_{phys} (X) & \simeq & \frac{1}{9 0} \left[ (N - 
1) \left\{ \pi^2 b_\pi (X) {\cal T}_\pi^4 (X) - \frac{15}{4} c_\pi 
(X) {\cal M}_\pi^2 (X) \, {\cal T}_\pi^2 (X) \right\} \right. 
\nonumber \\ 
& & \left. + \left\{ \pi^2 b_\sigma (X) {\cal T}_\sigma^4 (X) - 
\frac{15}{4} c_\sigma (X) {\cal M}_\sigma^2 (X) \, {\cal T}_\sigma^2 
(X) \right\} \right] \nonumber \\ 
& & + \frac{1}{12 \pi} \left[ (N - 1) {\cal M}_\pi^3 (X) \, {\cal 
T}^{(\pi)} (X) + {\cal M}_\sigma^3 (X) \, {\cal T}^{(\sigma)} (X) 
\right] . 
\end{eqnarray*} 
In the case where ${\cal M}_\pi^2 (X) << {\cal T}_\pi (X)$ and 
${\cal M}_\sigma^2 (X) >> {\cal T}_\sigma (X)$, the \lq\lq 
$\sigma$-terms'' here and below are dropped. 

High-temperature expansion of $\Omega_\beta (\vec{\varphi}^{\, 2} 
(X))$, Eq.~(\ref{eq-m}), and $\Omega_\beta^{(\sigma)} 
(\vec{\varphi}_0^{\, 2} (X))$, Eq.~(\ref{Om-1}), reads, in 
respective order, 
\begin{eqnarray*}  
\Omega_\beta (\vec{\varphi}^{\, 2} (X)) & \simeq & \frac{\lambda}{7 
2} \left[ (N - 1) \left\{ c_\pi {\cal T}_\pi^2 (X) - \frac{3}{\pi} 
\, {\cal M}_\pi (X) \, {\cal T}_\pi (X) \right\} \right. \nonumber \\ 
& & \left. + 3 \left\{ c_\sigma {\cal T}_\sigma^2 (X) - \frac{3}{\pi} 
\, {\cal M}_\sigma (X) \, {\cal T}_\sigma (X) \right\} \right] 
\end{eqnarray*}  
and 
\[ 
\Omega_\beta^{(\sigma)} (\vec{\varphi}_0^{\, 2} (X)) \simeq - 
\frac{\lambda^2 \varphi_0^2 (X)}{144 \pi} \left[ (N - 1) \frac{{\cal 
T}_\pi (X)}{{\cal M}_\pi (X)} + 9 \frac{{\cal T}_\sigma (X)}{{\cal 
M}_\sigma (X)} \right] . 
\] 
\subsubsection{\lq\lq Low-temperature'' expansion} 
Here we consider locally thermal equilibrium system, which is 
characterized by (\ref{bunbun}). In the low-temperature limit, $T 
(X) << {\cal M}_\xi^2 (X)$, $\gamma_\xi^{(\beta)} (X)$ in 
(\ref{T-xi}) may be computes as 
\begin{eqnarray} 
\gamma_\xi^{(\beta)} (X) & \simeq & \frac{1}{3 \pi^2} \int_{{\cal 
M}_\xi (X)}^\infty d E \, \left( E^2 - {\cal M}_\xi^2 (X) \right)^{3 
/ 2} \, \frac{1}{e^{E / T (X)} - 1} \nonumber \\ 
& \simeq & \frac{1}{3 \pi^2} \left( {\cal M}_\xi^2 (X) \right)^{3 / 
4} \, e^{- {\cal M}_\xi^2 (X) / T (X)} \int_0^\infty d x \, x^{3 / 
2} \, e^{- x / T (X)} \nonumber \\ 
& = & \frac{1}{\sqrt{2} \pi} \left( {\cal M}_\xi^2 (X) \right)^{3 / 
4} \left( T (X) \right)^{5 / 2} e^{- {\cal M}_\xi (X) / T (X)} . 
\label{Low-pos} 
\end{eqnarray} 
Then, $\gamma_{phys}^{(\beta)} (X)$ in (\ref{yuukou-1}) is 
approximated as 
\begin{eqnarray*} 
\gamma_{phys}^{(\beta)} (X) & \simeq & \frac{1}{2^{3 / 2} \pi} 
\left( T (X) \right)^{5 / 2} 
\left[ (N - 1) \left( {\cal M}_\pi^2 (X) \right)^{3 / 4} e^{- {\cal 
M}_\pi (X) / T (X)} \right. \nonumber \\ 
& & \left. + \left( {\cal M}_\sigma^2 (X) \right)^{3 / 4} e^{- 
{\cal M}_\sigma (X) / T (X)} \right] .
\end{eqnarray*} 
$\Omega_\beta (\vec{\varphi}_0^{\, 2} (X))$ in (\ref{eq-m}) becomes 
\begin{eqnarray*} 
\Omega_\beta (\vec{\varphi}_0^{\, 2} (X)) & \simeq & 
\frac{\lambda}{12 \sqrt{2} \pi} \left( T (X) \right)^{3 / 2} \left[ 
(N - 1) \left( {\cal M}_\pi^2 (X) \right)^{1 / 4} e^{- {\cal M}_\pi 
(X) / T (X)} \right. \nonumber \\ 
& & \left. + 3 \left( {\cal M}_\sigma^2 (X) \right)^{1 / 4} e^{- 
{\cal M}_\sigma (X) / T (X)} \right] 
\end{eqnarray*} 
and $\Omega_\beta^{(\sigma)} (\vec{\varphi}_0^{\, 2} (X))$ in 
(\ref{Om-1}) becomes 
\begin{eqnarray*} 
\Omega_\beta^{(\sigma)} (\vec{\varphi}_0^{\, 2} (X)) & \simeq & - 
\frac{\lambda^2 \varphi_0^2 (X)}{72 \sqrt{2} \pi} \left( T (X) 
\right)^{1 / 2} \left[ (N - 1) \left( {\cal M}_\pi^2 (X) \right)^{- 
1 / 4} e^{- {\cal M}_\pi (X) / T (X)} \right. \nonumber \\ 
& & \left. + 9 \left( {\cal M}_\sigma^2 (X) \right)^{- 1 / 4} e^{- 
{\cal M}_\sigma (X) / T (X)} \right] . 
\end{eqnarray*} 
\subsection{The region ${\cal M}^2_\pi (X)$ and/or ${\cal 
M}^2_\sigma (X) < 0$} 
It can be seen that, through analytically continuing the one-loop 
effective action obtained above to the region where ${\cal M}_\pi^2 
(X)$ and/or ${\cal M}^2_\sigma (X) < 0$, an imaginary part emerges 
and then we face the issue of interpreting it. In the case of 
effective potential in vacuum theory, this issue has been studied, 
e.g., in \cite{weinberg}. Let us summarize the observation in 
\cite{weinberg}. The quantity analyzed there is essentially 
(\ref{T-xi10}). In (\ref{T-xi10}), the quantum modes with $p^2 \geq 
|{\cal M}_\xi^2 (X)|$ remain harmonic oscillator as in the case of 
${\cal M}_\xi^2 (X) \geq 0$. However, the long-wavelength modes, 
$p^2 < |{\cal M}_\xi^2 (X)|$, have imaginary frequencies and thus 
unstable. These unstable modes turn out to develop imaginary part of 
$\gamma_\xi^{(0)} (X)$, $Im \, \gamma_\xi^{(0)} (X)$, the emergence 
of which is a signal of appearance of perturbative instability. The 
$Im \, \gamma_\xi^{(0)} (X)$ may be interpreted as half the decay 
rate per unit volume of a particular quantum state. More precisely, 
$Im \, \gamma_\xi^{(0)} (X)$ is the sum of half the decay rate per 
unit volume over all unstable modes, $\phi_{long} (x)$, which are 
localized at $\langle \phi_{long} (x_0 = (x_0)_{in}, {\bf x}) 
\rangle \simeq 0$ at the initial time $x_0 = (x_0)_{in}$. Here 
$\phi_{long} (x)$ stands for the long-wavelength components of the 
quantum field $\phi (x)$. The perturbative instability arises due to 
the growth of $\langle \phi_{long} (x) \rangle$. Owing to this 
instability, the original quasiuniform configuration beaks up into 
many uncorrelated domains. Within a domain the quantum field is 
correlated and $\langle \phi_{long} (x) \rangle$ is increasing in 
time. From one domain to the next, $\langle \phi_{long} (x) 
\rangle$'s vary randomly from positive to negative. These domains is 
expected to have size at least as great as $|{\cal M}_\xi^2 (X)|^{- 
1 / 2}$, the minimum wavelength for an unstable mode.  

As seen above, leading one-loop contribution to the  effective 
action is given by an integral of the \lq\lq local effective 
action'' $\gamma_{phys} (X)$ over the whole macroscopic spacetime 
region. We recall that $\gamma_{phys} (X)$ is a function of $X$ that 
labels the spacetime cells. Then, in a single cell, $X$ of 
$\gamma_{phys} (X)$ is approximately constant. Moreover 
$\gamma_{phys} (X)$ is the same in form as the effective potential 
(density). From this observation, together with the conclusion 
\cite{weinberg} for the zero-temperature effective potential, we 
assume that the imaginary part of the effective action in the 
present case also reflects the proper physical situation. 

For the purpose of performing analytic continuation, we return back 
to (\ref{sekibun})~-~(\ref{T-xi}) with ${\cal M}_\xi^2 (X) \geq 0$. 
Equations (\ref{mass-in0}) and (\ref{T-xi10}) tell us that ${\cal 
M}_\xi^2 (X)$ in (\ref{T-xi}) should be understood to be ${\cal 
M}_\xi^2 (X) - i \eta$ with $\eta$ a positive infinitesimal 
constant. Now let us analytically continue the function $\gamma_\xi 
= \gamma_\xi^{(0)} + \gamma_\xi^{(\beta)}$ from the region ${\cal 
M}_\xi^2 (X) \geq 0$ to the region ${\cal M}_\xi^2 (X) < 0$. The 
UV-renormalized vacuum part $\gamma_\xi^{(0)}$, Eq.~(\ref{T-xi0}), 
goes to 
\begin{equation} 
\gamma_\xi^{(0)} = - \frac{1}{32 \pi^2} {\cal M}_\xi^4 (X) \, \ln 
\frac{- {\cal M}_\xi^2 (X)}{e^{3 / 2} \mu_d^2} + \frac{i}{32 \pi} 
{\cal M}_\xi^4 (X) \;\;\;\;\;\; ({\cal M}_\xi^2 (X) < 0) . 
\label{neg-yo} 
\end{equation} 
It is worth mentioning that a \lq \lq direct calculation'' of 
(\ref{yoko-10}) yields the correct result for $Im \, 
\gamma_\xi^{(0)}$:  
\begin{eqnarray*} 
i \, Im \, {\cal I}_\xi^{(0)} (X) & = & i \, Im \int \frac{d^{\, 4} 
P}{(2 \pi)^4} \theta \left( P^2 - {\cal M}_\xi^2 (X) \right) \, 
\frac{i}{P^2 - {\cal M}_\xi^2 (X) + i \eta} \nonumber \\ 
& = & i \int \frac{d p}{4 \pi^2} \frac{p^2}{\sqrt{- p^2 - 
{\cal M}_\xi^2 (X)}} \nonumber \\ 
& = & - \frac{i}{16 \pi} \, {\cal M}_\xi^2 (X) , 
\end{eqnarray*} 
from which we obtain 
\begin{eqnarray} 
i \, Im \, \gamma_\xi^{(0)} & \equiv & - \int d {\cal M}_\xi^2 
(X) \, i \, Im \, {\cal I}_\xi^{(0)} (X) \nonumber \\ 
& = & \frac{i}{3 2 \pi} \, {\cal M}_\xi^4 (X) , 
\label{kyo-zero} 
\end{eqnarray} 
in accord with (\ref{neg-yo}). 

As to $\gamma_\xi^{(\beta)}$, we rewrite (\ref{T-xi}) as 
\begin{eqnarray} 
\gamma_\xi^{(\beta)} & = & \frac{i}{6 \pi^2} \int_{C_{\cal E}} d 
{\cal E} \, ({\cal E}^2 - {\cal M}_\xi^2 (X) + i \eta) 
({\cal M}_\xi^2 (X) - {\cal E}^2 - i \eta)^{1 / 2} \, \langle 
N^{(\xi)} (X; {\cal E}) \rangle \;\;\;\;\; ({\cal M}_\xi^2 (X) > 0) 
, \nonumber \\ 
& & 
\label{hyouji} 
\end{eqnarray} 
where $\langle N^{(\xi)} (X; {\cal E}) \rangle$ is as in 
(\ref{are}). In a complex-${\cal E}$ plane, the function $({\cal 
M}_\xi^2 (X) - {\cal E}^2 - i \eta)^{1 / 2}$ has branch points at 
${\cal E} = \pm {\cal E}_{b r}$ with ${\cal E}_{b r} = ({\cal 
M}_\xi^2 (X) - i \eta)^{1 / 2}$. We take a branch cut associated 
with $+ {\cal E}_{b r}$ to be a straight line $[{\cal E}_{b r}, + 
\infty - i \alpha)$, where $\alpha = - Im ({\cal M}_\xi^2 (X) - i 
\eta)^{1 / 2}$. The integration contour $C_{\cal E}$ in 
(\ref{hyouji}) is defined so as to enclose the branch cut $[{\cal 
E}_{b r}, + \infty - i \alpha)$ in a clock-wise direction, 
\begin{eqnarray*} 
C_{\cal E} : & & + \infty - i (\alpha + \epsilon) \to 
{\cal M}_\xi (X) - \epsilon - i (\alpha + \epsilon) \nonumber \\ 
& & \to {\cal M}_\xi (X) - \epsilon + i (\epsilon - \alpha) \to 
+ \infty + i (\epsilon - \alpha) , 
\end{eqnarray*} 
where $0 < \epsilon < \alpha$. 

Continuing to the negative ${\cal M}_\xi^2 (X)$ region, the branch 
points $\pm E_{b r}$ go to 
\begin{equation} 
\pm {\cal E}_{b r} \to \pm \left( - i (- {\cal M}_\xi^2 (X))^{1 / 2} 
+ \alpha \right) \;\;\;\;\; (\alpha = Im (- {\cal M}_\xi^2 (X) + i 
\eta)^{1 / 2}) 
\label{keiro} 
\end{equation} 
and the branch cut associated with ${\cal E}_{b r}$ goes to $[{\cal 
E}_{b r}, (1 - i) \alpha] \oplus [(1 - i) \alpha, + \infty - i 
\alpha)$. The integration contour $C_{\cal E}$ is deformed 
accordingly. We assume that $\langle N^{(\xi)} (X; {\cal E}) 
\rangle$ is analytic in the region 
\begin{eqnarray*} 
\alpha - \epsilon < Re {\cal E} < (- {\cal M}_\xi^2 (X))^{1 / 2} + 
\alpha + \epsilon , \nonumber \\ 
- (- {\cal M}_\xi^2 (X))^{1 / 2} - \epsilon < Im {\cal E} < \epsilon 
- \alpha . 
\end{eqnarray*} 
Then, we can deform the branch cut associated with ${\cal E}_{b r}$ 
to 
\begin{eqnarray*} 
& & [ {\cal E}_{b r}, {\cal E}_{b r} + (- {\cal M}_\xi^2 (X))^{1 / 
2} ] \oplus [ {\cal E}_{b r} + ( - {\cal M}_\xi^2 (X))^{1 / 2}, \; (
- {\cal M}_\xi^2 (X))^{1 / 2} - i \alpha ] \nonumber \\ 
& & \mbox{\hspace*{4ex}} \oplus [(- {\cal M}_\xi^2 (X))^{1 / 2} - i 
\alpha, \; + \infty - i \alpha ) . 
\end{eqnarray*} 
The integration contour $C_{\cal E}$ is deformed accordingly. 

Then, we obtain, after some manipulations, 
\begin{equation} 
\gamma_\xi^{(\beta)} = \frac{1}{3 \pi^2} \int_{C'} d z \, (z^2 - 
{\cal M}_\xi^2 (X))^{3 / 2} \, \langle N^{(\xi)} (X; z) \rangle , 
\label{shimai} 
\end{equation} 
where $C' = C_1' \oplus C_2' \oplus C_3'$ with $C_1' = - i (- {\cal 
M}_\xi^2 (X))^{1 / 2}$ $\to$ $(1 - i) (- {\cal M}_\xi^2 (X))^{1 / 
2}$, $C_2' = (1 - i) (- {\cal M}_\xi^2 (X))^{1 / 2}$ $\to$ $(- 
{\cal M}_\xi^2 (X))^{1 / 2}$, and $C_3' = (- {\cal M}_\xi^2 (X))^{1 
/ 2}$ $\to$ $+ \infty$. 

After all this, we obtain, for the local effective action, 
\begin{equation} 
\gamma_{phys} (X) = \gamma^{(0)}_{phys} (X) + 
\gamma^{(\beta)}_{phys} (X) , 
\label{loc-neg0} 
\end{equation} 
where 
\begin{eqnarray} 
Re \, \gamma^{(0)}_{phys} (X) & = & \frac{1}{2} \left[ (\partial 
\vec{\varphi} (X))^2 - \mu^2 \vec{\varphi}^{\, 2} (X) \right] - 
\frac{\lambda}{4 !} \, (\vec{\varphi}^{\, 2} (X))^2 + h ( 
\displaystyle{ \raisebox{0.9ex}{\scriptsize{$t$}}} 
\mbox{\hspace{-0.1ex}} \vec{e} \cdot \vec{\varphi} (X)) 
\nonumber \\ 
& & - \frac{1}{(8 \pi)^2} \left[ (N - 1) {\cal M}_\pi^4 (X) 
\ln \frac{|{\cal M}_\pi^2 (X)|}{e^{2 / 3} \mu_d^2} + {\cal 
M}_\sigma^4 (X) \ln \frac{|{\cal M}_\sigma^2 (X)|}{e^{2 / 3} 
\mu_d^2} \right] 
\label{loc-neg1} 
\\ 
Im \, \gamma^{(0)}_{phys} (X) & = &  \frac{1}{6 4 \pi} \left[ 
(N - 1) \theta \left( - {\cal M}_\pi^2 (X) \right) {\cal M}_\pi^4 
(X) + \theta \left( - {\cal M}_\sigma^2 (X) \right) {\cal 
M}_\sigma^4 (X) \right] , 
\label{loc-neg2} 
\\ 
\gamma^{(\beta)}_{phys} (X) & = & \frac{1}{2} \left[ (N - 1) 
\gamma_\pi^{(\beta)} (X) + \gamma_\sigma^{(\beta)} (X) \right] . 
\label{loc-neg3} 
\end{eqnarray} 

As mentioned at the beginning of this subsection, $Im \, 
\Gamma_{phys} (X)$ causes the perturbative instability, which in 
turn leads to formation of a domain structure. In most cases 
analyzed so far in the literature, imaginary part appears in the CTP 
effective action $\Gamma$ in the form, 
\begin{equation} 
Im \Gamma = i \sum_{i, \, j = 1}^N \int d^{\, 4} x \, d^{\, 4} y \, 
(\varphi_1 (x) - \varphi_2 (x))_i \, \gamma_{i j} (x, y) (\varphi_1 
(y) - \varphi_2 (y))_j , 
\label{trad}  
\end{equation} 
where $\gamma_{i j}$ are real functions. Recalling that the limit 
$\vec{\varphi}_2 (x) \to\vec{\varphi}_1 (x)$ is taken at the final 
stage, we see that (\ref{trad}) describes small fluctuations. In the 
present case, (\ref{hi-b}) shows that $Im \Gamma$ under 
consideration is not of the form (\ref{trad}) and describes \lq\lq 
large'' fluctuations.\footnote{Even in the present case, the small 
fluctuations of the type (\ref{trad}) also appears, which we do not 
consider in this paper.} Here, for dealing with $Im \Gamma_{phys}$, 
we follow a standard procedure \cite{morikawa}, which starts with 
writing $\mbox{exp} \, i \Gamma_{phys}$ in the form, 
\begin{eqnarray} 
e^{i \Gamma_{phys}} & = & \mbox{exp} \, i \int d^{\, 4} X \, 
\gamma_{phys} (X) \nonumber \\ 
& = & \mbox{exp} \left[ i \int d^{\, 4} X \, Re \, \gamma_{phys} (X) 
- \int d^{\, 4} X \, Im \, \gamma_{phys} (X) \right] . \nonumber \\ 
\label{mori-0} 
\end{eqnarray} 
Provided that 
\begin{equation} 
\int [d \vec{\varphi}] \mbox{exp} \left[ - \int d^{\, 4} X \, Im \, 
\gamma_{phys} (X) \right] < \infty , 
\label{jyo-cond} 
\end{equation} 
we can make a functional Fourier transformation of $\mbox{exp} [- 
\int d^{\, 4} X \, Im \, \gamma_{phys} (X)]$: 
\begin{eqnarray} 
\mbox{exp} [ - \int d^{\, 4} X \, Im \, \gamma_{phys} (X) ] & = & 
\mbox{exp} \left[ - \int d^{\, 4} X \, Im \, \gamma_{phys} (X) 
\, \rule[-3mm]{.14mm}{8.5mm} \raisebox{-2.85mm}{\scriptsize{$\; 
\vec{\varphi} = 0$}} \right] \nonumber \\ 
& & \times \int [d \vec{\zeta}] \, \mbox{exp} [- \int d^{\, 4} X \, 
G (\vec{\zeta}^{\, 2} (X)) + i \int d^{\, 4} X \, \displaystyle{ 
\raisebox{0.9ex}{\scriptsize{$t$}}} \mbox{\hspace{-0.1ex}} 
\vec{\zeta} (X) \cdot \vec{\varphi} (X)] . \nonumber \\ 
& & 
\label{mori} 
\end{eqnarray} 
Substituting (\ref{mori}) into (\ref{mori-0}), we obtain 
\begin{eqnarray} 
e^{i \Gamma_{phys}} & = & \mbox{exp} \left[ - \int d^{\, 4} 
X \, Im \, \gamma_{phys} (X) \, \rule[-3mm]{.14mm}{8.5mm} 
\raisebox{-2.85mm}{\scriptsize{$\; \vec{\varphi} = 0$}} \right] 
\nonumber \\ 
& & \times \langle \mbox{exp}\left[ i \int d^{\, 4} X \left\{ Re \, 
\gamma_{phys} (X) + \displaystyle{ 
\raisebox{0.9ex}{\scriptsize{$t$}}} \mbox{\hspace{-0.1ex}} 
\vec{\zeta} (X) \cdot \vec{\varphi} (X) \right\} \right] 
\rangle_\zeta , 
\label{mori-2} 
\end{eqnarray} 
where the $\vec{\zeta}$-average $\langle ... \rangle_\zeta$ is 
defined as 
\[ 
\langle ... \rangle_\zeta \equiv \int [d \vec{\zeta}] \, e^{- \int 
d^{\, 4} X \, G (\vec{\zeta}^{\, 2} (X))} \, ... . 
\] 
Equation (\ref{mori-2}) tells us that $Re \, \gamma_{phys} (X) + 
\displaystyle{ \raisebox{0.9ex}{\scriptsize{$t$}}} 
\mbox{\hspace{-0.1ex}} \vec{\zeta} (X) \cdot \vec{\varphi} (X)$ may 
be interpreted as the local effective action, from which follows the 
equation of motion for $\vec{\varphi} (X)$, 
\begin{equation} 
\frac{\partial Re \, \gamma_{phys} (X)}{\partial \vec{\varphi} 
(X)} + \vec{\zeta} (X) = 0 . 
\label{sto} 
\end{equation} 
Thus, the auxiliary fields $\vec{\zeta} (X)$ represent random 
forces, whose character is determined by $G (\vec{\zeta}^{\, 2} 
(X))$. 

Taking the functional derivatives of (\ref{mori}) with respect to 
$\vec{\varphi}$ and then setting $\vec{\varphi} = 0$, we obtain the 
relations: 
\begin{eqnarray} 
\langle 1 \rangle_\zeta & = & 1 , \nonumber \\ 
\langle \vec{\zeta} (X) \rangle_\zeta & = & 0 , \nonumber \\ 
\langle \zeta_i (X) \, \zeta_j (Y) \rangle_\zeta & = & 2 
\delta_{i j} \, \delta^4 (X - Y) \, \frac{\partial Im  
\gamma_{phys} (X)}{\partial \vec{\varphi}^{\, 2} (X)} \, 
\rule[-3mm]{.14mm}{8.5mm} \raisebox{-2.85mm}{\scriptsize{$\; 
\vec{\varphi} = 0$}} \nonumber \\ 
& = & \delta_{i j} \, \delta^4 (X - Y) \left[ \frac{\lambda}{9 6 
\pi} \left\{ (N - 1) \theta \left( - M_\pi^2 (X) \right) \, M_\pi^2 
(X) \right. \right. \nonumber \\ 
& & \left. + 3 \, \theta \left( - M_\sigma^2 (X) \right) 
\, M_\sigma^2 (X) \right\} + (N- 1) \theta \left(- M_\pi^2 (X) 
\right) \frac{\partial \, Im \, \gamma_\pi^{(\beta)} (X)}{\partial 
\vec{\varphi}^{\, 2} (X)} \rule[-3mm]{.14mm}{8.5mm} 
\raisebox{-2.85mm}{\scriptsize{$\; \vec{\varphi} = 0$}} \nonumber \\ 
& & \left. + \theta \left( - M_\sigma^2 (X) \right) \frac{\partial 
\, Im \, \gamma_\sigma^{(\beta)} (X)}{\partial \vec{\varphi}^{\, 
2} (X)} \rule[-3mm]{.14mm}{8.5mm} \raisebox{-2.85mm}{\scriptsize{$\; 
\vec{\varphi} = 0$}} \right] , 
\label{2-mom} 
\end{eqnarray} 
\begin{eqnarray} 
& & \langle \zeta_i (X) \, \zeta_j (Y) \, \zeta_k (U) \, \zeta_l 
(V) \rangle_{\zeta c} \nonumber \\ 
& & \mbox{\hspace*{12.3ex}} \equiv \langle \zeta_i (X) \, \zeta_j 
(Y) \, \zeta_k (U) \, \zeta_l (V) \rangle_\zeta - \left[ \langle 
\zeta_i (X) \, \zeta_j (Y) \rangle_\zeta \; \langle \zeta_k (U) \, 
\zeta_l (V) \rangle_\zeta \right. \nonumber \\ 
& & \left. \mbox{\hspace*{15.3ex}} + \langle \zeta_i (X) \, \zeta_k 
(U) \rangle_\zeta \; \langle \zeta_j (Y) \, \zeta_l (V) 
\rangle_\zeta + \langle \zeta_i (X) \, \zeta_l (V) \rangle_\zeta \; 
\langle \zeta_j (Y) \, \zeta_k (U) \rangle_\zeta \right] \nonumber 
\\ 
& & \mbox{\hspace*{12.3ex}} = - 2 \, \delta^4 (X - U) \, \delta^4 (V 
- Y) \delta^4 (X - Y) \left[ \delta_{i k} \, \delta_{l j} + 
\delta_{k j} \, \delta_{l i} + \delta_{i j} \, \delta_{l k} \right] 
\nonumber \\ 
& & \mbox{\hspace*{15.3ex}} \times \left\{ \frac{\lambda^2}{576 \pi} 
\left[ (N - 1) \, \theta \left( - M_\pi^2 (X) \right) + 9 \, \theta 
\left( - M_\sigma^2 (X) \right) \right] \right. \nonumber \\ 
& & \mbox{\hspace*{15.3ex}} + (N - 1) \theta \left( - M_\pi^2 (X) 
\right) \frac{\partial^2 Im \, \gamma_\pi^{(\beta)} (X)}{\partial 
(\varphi^2 (X))^2} \rule[-3mm]{.14mm}{8.5mm} 
\raisebox{-2.85mm}{\scriptsize{$\; \vec{\varphi} = 0$}} 
\nonumber \\ 
& & \left. \mbox{\hspace*{15.3ex}} + \theta \left( - M_\sigma^2 (X) 
\right) \frac{\partial^2 Im \, \gamma_\sigma^{(\beta)} (X)}{\partial 
(\varphi^2 (X))^2} \rule[-3mm]{.14mm}{8.5mm} 
\raisebox{-2.85mm}{\scriptsize{$\; \vec{\varphi} = 0$}} 
\right\} , 
\label{4-pt} 
\end{eqnarray} 
etc. As mentioned in Section I, the macroscopic spacetime 
coordinates are the indices that labels the spacetime cells, whose 
size is $L^\mu$ $(\mu = 0, 1, 2, 3)$. Then, $\delta (X^\mu - Y^\mu)$ 
in (\ref{2-mom}) and (\ref{4-pt}) should be understood to be 
\[ 
\delta (X^\mu - Y^\mu) \simeq \frac{\delta_{i^\mu j^\mu}}{L^\mu} 
\;\;\;\;\; (\mbox{no summation over} \; \mu) , 
\]    
where $i^\mu$, etc., are the $\mu$-component of the label for the 
spacetime cells. 

Let $\vec{\varphi} (X) = \vec{\varphi} (X; [\vec{\zeta}])$ be a 
solution to (\ref{sto}) under a given initial data (at the initial 
time $X_0 = T_{in}$), $\vec{\varphi} (X_0 = T_{in}, {\bf X})$ and 
$\partial \vec{\varphi} (X_0, {\bf X}) / \partial X_0 \, 
\rule[-3mm]{.14mm}{8.5mm} \raisebox{-2.85mm}{\scriptsize{$\; X_0 = 
T_{in}$}}$. Then using (\ref{2-mom}), (\ref{4-pt}), etc., one can 
compute the $\zeta$-averaged correlation functions 
\begin{eqnarray*} 
& & \langle \varphi_i (X) \rangle_\zeta \, , \;\;\;\;\;  
\langle \varphi_i (X) \varphi_j (Y) \rangle_\zeta , \\ 
& & \langle \varphi_i (X) \varphi_j (Y) \varphi_k (Z) 
\rangle_\zeta \, , \;\;\;\;\; \langle \varphi_i (X) \varphi_j (Y) 
\varphi_k (U) \varphi_l (V) \rangle_\zeta , 
\end{eqnarray*} 
etc. 

A multi-point correlation function consists of \lq\lq 
connected'' and \lq\lq disconnected'' parts. For example, a 
two-point correlation function $\langle \varphi_i (X) \varphi_j (Y) 
\rangle_\zeta$ takes the form 
\[ 
\langle \varphi_i (X) \varphi_j (Y) \rangle_\zeta = \langle 
\varphi_i (X) \varphi_j (Y) \rangle_{\zeta c} + \langle \varphi_i 
(X) \rangle_\zeta \langle \varphi_j (Y) \rangle_\zeta , 
\]
where the first term on the RHS stands for \lq\lq connected'' part 
and the second term stands for \lq\lq disconnected'' part. It is 
convenient to introduce a generating functional 
\begin{equation} 
{\cal G} [\vec{y}, \vec{\varphi}] \equiv \langle e^{i \int d^{\, 4} 
Z \, \vec{y} (Z) \cdot \vec{\varphi} (Z)} \rangle_\zeta , 
\label{sei-i} 
\end{equation} 
which generates $\langle \varphi_{i_1} (X_1) ... \varphi_{i_n} (X_n) 
\rangle_\zeta$ through 
\[ 
\langle \varphi_{i_1} (X_1) ... \varphi_{i_n} (X_n) \rangle_\zeta = 
(- i)^n \prod_{j = 1}^n \frac{\delta}{\delta y_{i_j} (X_j)} 
{\cal G} [\vec{y}, \vec{\varphi}] \, \rule[-3mm]{.14mm}{8.5mm} 
\raisebox{-2.85mm}{\scriptsize{$\; \vec{y} = 0$}} . 
\] 
$\vec{\zeta}$-averaged \lq\lq connected'' correlation functions are 
obtained through 
\begin{equation} 
\langle \varphi_{i_1} (X_1) ... \varphi_{i_n} (X_n) \rangle_{\zeta 
c} = (- i)^n \prod_{j = 1}^n \frac{\delta}{\delta y_{i_j} (X_j)} 
\ln {\cal G} [\vec{y}, \vec{\varphi}] \, \rule[-3mm]{.14mm}{8.5mm} 
\raisebox{-2.85mm}{\scriptsize{$\; \vec{y} = 0$}} . 
\label{comu} 
\end{equation} 

It is worth mentioning here the equation of motion for type-2 or 
unphysical condensate fields. From (\ref{hi-b}), it is obvious that, 
for the unphysical condensate fields, all the above formulae apply 
with the substitutions, 
\[ 
Re \, \gamma_{phys} \to Re \, \gamma_{phys} , \;\;\;\;\;\; 
Im \, \gamma_{phys} \to - Im \, \gamma_{phys} . 
\] 
This means that (\ref{2-mom}) and (\ref{4-pt}), etc., change sign. 
Recalling again that the type-2 field \lq\lq lives'' in the \lq\lq 
time-reversed world,'' we see that the random forces that 
statistically upsetting the system act in completely opposite way in 
the time-reversed world. 
\subsubsection{\lq\lq High-temperature'' expansion} 
In a similar manner as in the case of ${\cal M}_\pi^2 (X), \, {\cal 
M}_\sigma^2 (X) \geq 0$, one can obtain a \lq\lq high-temperature'' 
expansion of $\gamma_{phys} (X)$. Using (\ref{takai}) in 
(\ref{shimai}), we obtain (cf. (\ref{takai-1})~-~(\ref{koou})) 
\begin{eqnarray} 
\gamma_\xi^{(\beta)} (X) & \simeq & \frac{1}{3 \pi^2} \int_0^\infty 
d {\cal E} \left[ {\cal E}^3 - \frac{3}{2} {\cal E} {\cal M}_\xi^2 
(X) \right] \langle N^{(\xi)} (X; {\cal E}) \rangle \nonumber \\ 
& & + \frac{1}{3 \pi^2} \left[ \int_{C'} d z \, (z^2 - {\cal 
M}_\xi^2 (X) )^{3 / 2} - \int_0^\infty d z \left( z^3 - \frac{3}{2} 
z {\cal M}_\xi^2 (X) \right) \right] \frac{{\cal T}_\xi (X)}{z} 
\nonumber \\ 
& \simeq & \frac{\pi^2}{4 5} b_\xi (X) \, {\cal T}_\xi^4 (X) - 
\frac{1}{1 2} c_\xi (X) \, {\cal M}_\xi^2 (X) \, {\cal T}_\xi^2 (X) 
+ \frac{i}{6 \pi} \left( - {\cal M}_\xi^2 (X) \right)^{3 / 2} {\cal 
T}_\xi (X) . 
\label{final-1} 
\end{eqnarray} 
It is to be noted that (\ref{final-1}) is directly obtained by 
analytically continuing the asymptotic form (\ref{koou}) from the 
region ${\cal M}_\xi^2 (X)$ $( > 0)$ to the region ${\cal M}_\xi^2 
(X) - i \eta$ (${\cal M}_\xi^2 (X) < 0)$. It is also worth 
mentioning that the following \lq\lq direct calculation'' reproduces 
(\ref{final-1}). We start from (\ref{takai-2}), 
\begin{eqnarray*} 
\gamma_{\xi 2}^{(\beta)} (X) & = & \frac{1}{3 \pi^2} \left[ 
\int_0^\infty d p \, \frac{p^4}{\sqrt{p^2 + {\cal M}_\xi^2 (X)}} \, 
\langle N^{(\xi)} (X; \sqrt{p^2 + {\cal M}_\xi^2 (X)} \, \rangle 
\right. \nonumber \\ 
& & \left. - \int_0^\infty d {\cal E} \left( {\cal E}^3 - 
\frac{3}{2} {\cal E} {\cal M}_\xi^2 (X) \right) 
\langle N^{(\xi)} (X; {\cal E}) \rangle \right] , 
\end{eqnarray*} 
which is valid for ${\cal M}_\xi^2 \geq 0$. At this stage, let us 
naively change ${\cal M}_\xi^2 (X)$ from positive to negative. Then 
we have 
\begin{eqnarray*} 
\gamma_{\xi 2}^{(\beta)} (X) & = & \frac{1}{3 \pi^2} \left[ 
\int_0^\infty d p \, \frac{p^4}{\sqrt{p^2 - |{\cal M}_\xi^2 (X)|}} 
\, \langle N^{(\xi)} (X; \sqrt{p^2 - |{\cal M}_\xi^2 (X)|} \, 
\rangle \right. \nonumber \\ 
& & \left. - \int_0^\infty d {\cal E} \left( {\cal E}^3 - 
\frac{3}{2} {\cal E} {\cal M}_\xi^2 (X) \right) 
\langle N^{(\xi)} (X; {\cal E}) \rangle \right] . 
\end{eqnarray*} 
Substituting (\ref{takai}) and using (\ref{takai-1}), we obtain 
\begin{eqnarray*} 
\gamma_{\xi 2}^{(\beta)} (X) & \simeq & \frac{{\cal T}_\xi (X)}{3 
\pi^2} \left[ \int_0^\infty d p \, \frac{p^4}{p^2 - |{\cal M}_\xi^2 
(X)| - i \eta} \right. \nonumber \\ 
& & \left. - \int_{\sqrt{|{\cal M}_\xi^2 (X)|}}^\infty d p \, p 
\left( \sqrt{p^2 - |{\cal M}_\xi^2 (X)|} + \frac{3}{2} \frac{|{\cal 
M}_\xi^2 (X)|}{\sqrt{p^2 - |{\cal M}_\xi^2 (X)|}} \right) \right] 
\nonumber \\ 
& = & \frac{i {\cal T}_\xi (X)}{3 \pi} \int_0^\infty d p \, p^4 \, 
\delta \left( p^2 - |{\cal M}_\xi^2 (X)| \right) \nonumber \\ 
& = & \frac{i}{6 \pi} |{\cal M}_\xi^2 (X)|^{3 / 2} {\cal T}_\xi (X) 
. 
\end{eqnarray*} 
This together with (\ref{takai-1}) reproduces (\ref{final-1}). 

Thus, $\gamma_{phys}^{(\beta)} (X)$ in (\ref{loc-neg3}) is 
approximated as 
\begin{eqnarray} 
\gamma^{(\beta)}_{phys} (X) & \simeq & \frac{1}{9 0} \left[ 
(N - 1) \left( \pi^2 b_\pi (X) {\cal T}_\pi^4 (X) - \frac{15}{4} 
c_\pi (X) {\cal M}_\pi^2 (X) \, {\cal T}_\pi^2 (X) \right) \right. 
\nonumber \\ 
& & \left. + \left( \pi^2 b_\sigma (X) {\cal T}_\sigma^4 (X) - 
\frac{15}{4} c_\sigma (X) {\cal M}_\sigma^2 (X) \, {\cal T}_\sigma^2 
(X) \right) \right] \nonumber \\ 
& & + \frac{1}{12 \pi} \left[ (N - 1) \theta \left( {\cal M}_\pi^2 
(X) \right) \, {\cal M}_\pi^3 (X) \, {\cal T}^{(\pi)} (X) + \theta 
\left( {\cal M}_\sigma^2 (X) \right) \, {\cal M}_\sigma^3 (X) \, 
{\cal T}^{(\sigma)} (X) \right] \nonumber \\ 
& & + \frac{i}{12 \pi} \left[ (N - 1) \theta \left( - {\cal M}_\pi^2 
(X) \right) \left( - {\cal M}_\pi^2 (X) \right)^{3 / 2} \, {\cal 
T}^{(\pi)} (X) \right. \nonumber \\ 
& & \left. + \theta \left( - {\cal M}_\sigma^2 (X) \right) \left( - 
{\cal M}_\sigma^2 (X) \right)^{3 / 2} \, {\cal T}^{(\sigma)} (X) 
\right] . 
\label{ukiyo} 
\end{eqnarray} 
The moment (\ref{2-mom}) and (\ref{4-pt}) are approximated, in 
respective order, as 
\begin{eqnarray*} 
\langle \zeta_i (X) \, \zeta_j (Y) \rangle_\zeta & = & 2 
\delta_{i j} \, \delta (X - Y) \, \frac{\partial Im  
\gamma_{phys} (X)}{\partial \vec{\varphi}^{\, 2} (X)} \, 
\rule[-3mm]{.14mm}{8.5mm} \raisebox{-2.85mm}{\scriptsize{$\; 
\vec{\varphi} = 0$}} \\ 
& \simeq & - \frac{\lambda}{2 4 \pi} \, \delta_{i j} 
\, \delta (X - Y) \left[ (N - 1) \theta \left( - M_\pi^2 (X) \right) 
\left(- M_\pi^2 (X) \right)^{1 / 2} \, {\cal T}_\pi (X) \right. 
\nonumber \\ 
& & + 3 \, \theta \left( - M_\sigma^2 (X) \right) \, \left(- 
M_\sigma^2 (X) \right)^{1 / 2} \, {\cal T}_\sigma (X) \\ 
& & \left. - \frac{1}{4} \, (N - 1) \theta \left( - M_\pi^2 (X) 
\right) \, M_\pi^2 (X) + \frac{3}{4} \, \theta \left( - M_\sigma^2 
(X) \right) \, M_\sigma^2 (X) \right] 
\end{eqnarray*} 
and 
\begin{eqnarray*} 
& & \langle \zeta_i (X) \, \zeta_j (Y) \, \zeta_k (U) \, \zeta_l 
(V) \rangle_{\zeta c} \nonumber \\ 
& & \mbox{\hspace*{4ex}} \simeq - \frac{\lambda^2}{288 \pi} \, 
\delta^4 (X - U) \, \delta^4 (V - Y) \delta^4 (X - Y) \left[ 
\delta_{i k} \, \delta_{l j} + \delta_{k j} \, \delta_{l i} + 
\delta_{i j} \, \delta_{l k} \right] \nonumber \\ 
& & \mbox{\hspace*{7ex}} \times \left\{ (N - 1) \, \theta \left( - 
M_\pi^2 (X) \right) + 9 \, \theta \left( - M_\sigma^2 (X) \right) 
\right. \\ 
& & \mbox{\hspace*{7ex}} + 2 \left[ (N - 1) \, \, \theta \left( - 
M_\pi^2 (X) \right) (- M_\pi^2 (X))^{- 1 / 2} \, {\cal T}_\pi (X) 
\right. \nonumber \\ 
& & \mbox{\hspace*{7ex}} \left. \left. + 9 \, \theta \left( - 
M_\sigma^2 (X) \right) \, (- M_\sigma^2 (X))^{- 1 / 2} \, {\cal 
T}_\sigma (X) \right] \right\} . 
\end{eqnarray*} 

$Im \, \gamma_{phys} (X) = Im \, \gamma_{phys}^{(0)} (X) + Im \, 
\gamma_{phys}^{(\beta)} (X)$ vanishes for ${\cal M}_\pi^2 (X) > 0$ 
{\em and} ${\cal M}_\sigma^2 (X) > 0$, i.e., for 
\[ 
\vec{\varphi}^{\, 2} (X) > \mbox{Max} \left( - \frac{6}{\lambda} \, 
M_\pi^2 (X), \; - \frac{2}{\lambda} \, M_\sigma^2 (X) \right) . 
\] 
Then, the condition (\ref{jyo-cond}) is met. 
\subsubsection{Low-temperature limit} 
We consider locally thermal equilibrium system, which is 
characterized by (\ref{bunbun}), and obtain $\gamma_{phys}^{(\beta)} 
(X)$ in the low-temperature limit, $T (X) << - {\cal M}_\xi^2 (X)$. 
The function (\ref{bunbun}) satisfy the assumption made above after 
(\ref{keiro}), we can use the representation (\ref{shimai}). 
From (\ref{shimai}), one can easily see that the contribution from 
the segment $C_1'$ dominates over others. Changing the integration 
variable from $z$ to $\rho$ through $z = (- {\cal M}_\xi^2 (X)) 
[\rho - i]$, we have 
\begin{eqnarray*} 
\gamma_\xi^{(\beta)} & = & \frac{1}{3 \pi^2} \left( - {\cal 
M}_\xi^2 (X) \right)^{2} \int_0 d \rho \, \frac{\rho^{3 / 2} (\rho 
- 2 i)^{3 / 2}}{\mbox{exp} \left( \frac{(- {\cal M}_\xi^2 (X))^{1 / 
2}}{T (X)} (\rho - i) \right) - 1} \nonumber \\ 
& \simeq & \frac{2 \sqrt{2}}{3 \pi^2} e^{- 3 \pi i / 4} e^{i (- 
{\cal M}_\xi^2 (X))^{1 / 2} / T (X)} \left( - {\cal M}_\xi^2 (X) 
\right)^{2} \int_0 d \rho \, \rho^{3 / 2} \, \mbox{exp} \left( - 
\frac{(- {\cal M}_\xi^2 (X))^{1 / 2}}{T (X)} \, \rho \right) 
\nonumber \\ 
& \simeq & \frac{1}{\sqrt{2} \pi} e^{- 3 \pi i / 4} e^{i (- 
{\cal M}_\xi^2 (X))^{1 / 2} / T (X)} (- {\cal M}_\xi^2 (X))^{3 / 4} 
\, (T (X))^{5 / 2} , 
\end{eqnarray*} 
from which, we obtain 
\begin{eqnarray} 
Re \, \gamma_\xi^{(\beta)} & = & \frac{1}{\sqrt{2} \pi} (- {\cal 
M}_\xi^2 (X))^{3 / 4} \, (T (X))^{5 / 2} \cos \left( \frac{(- {\cal 
M}_\xi^2 (X))^{1 / 2}}{T (X)} - \frac{3 \pi}{4} \right) , 
\nonumber \\ 
Im \, \gamma_\xi^{(\beta)} & = & \frac{1}{\sqrt{2} \pi} (- {\cal 
M}_\xi^2 (X))^{3 / 4} \, (T (X))^{5 / 2} \sin \left( \frac{(- {\cal 
M}_\xi^2 (X))^{1 / 2}}{T (X)} - \frac{3 \pi}{4} \right) . 
\label{Low-neg} 
\end{eqnarray} 
Incidentally, (\ref{Low-neg}) is directly obtained from 
(\ref{Low-pos}) through an analytic continuation, ${\cal M}_\xi^2 
(X)$ $(\geq 0)$ $\to$  ${\cal M}_\xi^2 (X) - i \eta$ (${\cal 
M}_\xi^2 < 0$). 

The form of $\gamma_{phys}^{(\beta)} (X)$ at low temperature is 
obtained by substituting (\ref{Low-neg}) into (\ref{loc-neg3}). It 
can be shown that the condition (\ref{jyo-cond}) holds. 
\subsection{\lq\lq Determining $M_\pi^2 (X)$ and $M_\sigma^2 
(X)$} 
In this subsection, following the method 1) in Section VID, we 
derive the gap equation (the determining equation for $M_\xi^2 (X)$ 
$(\xi = \pi, \sigma)$) to the leading one-loop order. Main task is 
to compute ${\bf \Sigma}_R (X; R)$. Let ${\cal D} \left[ {\cal L}_1, 
\, {\cal L}_2, \, ... \right]$ (${\cal L}_j \in {\cal L}_{int} + 
{\cal L}_{r c}$) be an one-loop diagram contributing to ${\bf 
\Sigma}_R (X; R)$, which includes the vertices coming from ${\cal 
L}_1$, ${\cal L}_2$, .... The relevant ones are 
\begin{description} 
\item{I.} ${\cal D} \left[ - \lambda (\vec{\phi}^{\, 2})^2 / 4 ! 
\right]\;\;\;$ (tadpole diagram). 
\item{II.} ${\cal D} \left[ - \lambda (\displaystyle{ 
\raisebox{0.9ex}{\scriptsize{$t$}}} \mbox{\hspace{-0.1ex}} 
\vec{\varphi} \cdot \vec{\phi}) \vec{\phi}^{\, 2} / 3 !, \, - 
\lambda (\displaystyle{ \raisebox{0.9ex}{\scriptsize{$t$}}} 
\mbox{\hspace{-0.1ex}} \vec{\varphi} \cdot \vec{\phi}) 
\vec{\phi}^{\, 2} / 3 ! \right]$. 
\end{description} 
Including the contributions coming from ${\cal L}_{r c}$ in 
(\ref{teigi-yo}), we write ${\bf \Sigma}_R (X; R)$, with obvious 
notation, in the form 
\begin{eqnarray} 
{\bf \Sigma}_R (X; R) & = & \left( {\bf \Sigma}_R (X) 
\right)_{\mbox{\scriptsize{I}}} + \left( {\bf \Sigma}_R (X; R) 
\right)_{\mbox{\scriptsize{II}}} + (Z_\mu Z - 1) \mu^2 \nonumber \\ 
& & + \frac{\lambda}{6} (Z_\lambda Z^2 - 1) \varphi^2 (X) \left( 
{\bf P}_\pi (X) + 3 {\bf P}_\sigma (X) \right) - \sum_{\xi = \pi, \, 
\sigma} {\bf P}_\xi (X) \chi_\xi (X) . 
\label{con-10} 
\end{eqnarray} 
The leading contribution to (\ref{con-10}) takes the form 
\begin{eqnarray} 
{\bf \Sigma}_R (X; R) & = & \sum_{\xi = \pi, \, \sigma} {\bf 
P}_\xi (X) \Sigma_R^{(\xi)} (X; R) \nonumber \\ 
& = & \sum_{\xi = \pi, \, \sigma} {\bf P}_\xi (X) \left[ \left( 
\Sigma_R^{(\xi)} (X) \right)_{\mbox{\scriptsize{I}}} + \left( 
\Sigma_R^{(\xi)} (X; R) \right)_{\mbox{\scriptsize{II}}} + (Z_\mu Z 
- 1) \mu^2 - \chi_\xi (X) \right. \nonumber \\ 
& & \left. + \frac{\lambda}{6} \, d_\xi (Z_\lambda Z^2 - 1) 
\varphi^2 (X) \right] ,  
\label{con-11} 
\end{eqnarray} 
with $d_\pi = 1$ and $d_\sigma = 3$. The resummation program 
completes by imposing the conditions 
\begin{equation} 
Re \, \Sigma_R^{(\xi)} (X; R^2 = {\cal M}_\xi^2 (X) ) = 0 \;\;\;\;
\;\;\ (\xi = \pi, \, \sigma) . 
\label{con-12} 
\end{equation} 
This is the gap equation, which serves as determining $M_\xi^2 (X)$ 
$(\xi = \pi, \, \sigma)$. 
\subsubsection{Computation of $\left( {\bf \Sigma}_R (X) 
\right)_{\mbox{\scriptsize{I}}}$} 
We start with computing the leading $O (\lambda)$ contribution from 
the tadpole diagram, $\left( {\bf \Sigma}_R (X) 
\right)_{\mbox{\scriptsize{I}}}$: 
\begin{eqnarray} 
\left( {\bf \Sigma}_R (X) \right)_{\mbox{\scriptsize{I}}} & = & 
\frac{i \lambda}{6} \int \frac{d^{\, 4} P}{(2 \pi)^4} \left[ 2 
\left\{ {\bf P}_\pi (X) \Delta_{1 1}^{(\pi)} (X; P) + {\bf P}_\sigma 
(X) \Delta_{1 1}^{(\sigma)} (X; P) \right\} \right. \nonumber \\ 
& & \left. + {\bf I} \left\{ (N - 1) \Delta_{1 1}^{(\pi)} 
(X; P) + \Delta_{1 1}^{(\sigma)} (X; P) \right\} \right] , 
\label{yareyare} 
\end{eqnarray} 
where $\Delta_{1 1}^{(\pi)}$ is the $(1, 1)$ component of 
$\hat{\Delta}^{(\pi)}$ in (\ref{yaya-12}), etc. Note that $\hat{\bf 
\Delta}^{(1)}$ in (\ref{yaya}) yields higher-order contribution to 
${\bf \Sigma}_R (X)$, which is pure imaginary and consists of terms 
that are proportional to $(\partial | \varphi (X) \rangle / \partial 
X^\mu) \langle \varphi (X) |$ and $| \varphi (X) \rangle ( \partial 
\langle \varphi (X) | / \partial X^\mu )$. When one performs 
two-loop calculation, this contribution should be taken into 
account. We rewrite (\ref{yareyare}) as 
\begin{eqnarray} 
\left( {\bf \Sigma}_{1 1} (X) \right)_{\mbox{\scriptsize{I}}} & = & 
{\bf P}_\pi (X) \left( \Sigma_R^{(\pi)} (X) 
\right)_{\mbox{\scriptsize{I}}} + {\bf P}_\sigma (X) \left( 
\Sigma_R^{(\sigma)} (X) \right)_{\mbox{\scriptsize{I}}} , 
\nonumber \\ 
\left( \Sigma_R^{(\xi)} (X) \right)_{\mbox{\scriptsize{I}}} & = & 
\frac{\lambda}{6} \left[ \alpha_\xi \, {\cal I}_\pi (X) + \beta_\xi 
\, {\cal I}_\sigma (X) \right] \;\;\;\;\;\; (\xi = \pi, \, \sigma) 
\nonumber \\ 
& & \mbox{\hspace*{5ex}} ((\alpha_\pi, \beta_\pi) = (N + 1, \, 1) , 
\; (\alpha_\sigma, \beta_\sigma) = (N - 1, \, 3)) , 
\label{tooi-yo} 
\end{eqnarray} 
where 
\begin{equation} 
{\cal I}_\xi (X) \equiv i \int \frac{d^{\, 4} P}{(2 \pi)^4} 
\Delta_{1 1}^{(\xi)} (X; P) . 
\label{yoko1-yo} 
\end{equation} 
The $O (\lambda)$ contribution to ${\cal I}_\xi (X)$ has already 
been computed in the above subsections. 
\subsubsection{Computation of $Re \left( {\bf \Sigma}_R (X; R) 
\right)_{\mbox{\scriptsize{II}}}$} 
Here we compute the leading $O (\lambda^2)$ contribution to $Re 
\left( {\bf \Sigma}_R (X; R) \right)_{\mbox{\scriptsize{II}}}$. 
Straightforward manipulation yields 
\begin{eqnarray*} 
\left( {\bf \Sigma}_R (X; R) \right)_{\mbox{\scriptsize{II}}} & = 
& {\bf P}_\pi (X) \left( \Sigma_R^{(\pi)} (X; R) 
\right)_{\mbox{\scriptsize{II}}} + {\bf P}_\sigma (X) \left( 
\Sigma_R^{(\sigma)} (X; R) \right)_{\mbox{\scriptsize{II}}} , 
\nonumber \\ 
\left( \Sigma_R^{(\pi)} (X; R) \right)_{\mbox{\scriptsize{II}}} & = 
& \frac{\lambda^2}{9} \varphi^2 (X) {\cal J}_{\pi \sigma} (X; R) , 
\nonumber \\ 
\left( \Sigma_R^{(\sigma)} (X; R) \right)_{\mbox{\scriptsize{II}}} 
& = & \frac{\lambda^2}{18} \varphi^2 (X) \left[ (N - 1) {\cal 
J}_{\pi \pi} (X; R) + 9 {\cal J}_{\sigma \sigma} (X; R) \right] . 
\nonumber 
\end{eqnarray*} 
Here 
\begin{eqnarray*} 
{\cal J}_{\xi \zeta} (X; R) & = & i \int \frac{d^{\, 4} P}{(2 
\pi)^4} \left[ \Delta^{(\xi)}_{1 1} (X; P) \, \Delta^{(\zeta)}_{1 1} 
(X; R - P) - \Delta^{(\xi)}_{1 2} (X; P) \, \Delta^{(\zeta)}_{1 2} 
(X; R - P) \right] \nonumber \\ 
& \equiv & {\cal J}_{\xi \zeta}^{(0)} (X; R) + {\cal J}_{\xi 
\zeta}^{(\beta)} (X; R) \mbox{\hspace{33ex}} (\xi, \zeta = \pi, 
\sigma) , 
\end{eqnarray*} 
with 
\begin{eqnarray*} 
{\cal J}_{\xi \zeta}^{(0)} (X; R) & = &i \int \frac{d^{\, D} P}{(2 
\pi)^D} \left[ \frac{1}{P^2 - {\cal M}_\xi^2 (X) + i \eta} \, 
\frac{1}{(R - P)^2 - {\cal M}_\zeta^2 (X) + i \eta} \right. 
\nonumber \\ 
& & \left. + 4 \pi^2 \theta (- p_0) \, \theta (p_0 - r_0) \, 
\delta (P^2 - {\cal M}_\xi^2 (X)) \, \delta ((R - P)^2 - {\cal 
M}_\zeta^2 (X)) \right] , \nonumber \\ 
{\cal J}_{\xi \zeta}^{(\beta)} (X; R) & = & \int \frac{d^{\, 4} 
P}{(2 \pi)^3} \left[ \frac{1}{P^2 - {\cal M}_\xi^2 (X) + i \epsilon 
(p_0) \eta} \, N^{(\zeta)} (X; |r_0 - p_0|; \hat{\bf p}) \, \delta 
((R - P)^2 - {\cal M}_\zeta^2 (X)) \right. \nonumber \\ 
& & \left. + \frac{1}{(R - P)^2 - {\cal M}_\zeta^2 (X) - i \epsilon 
(p_0 - r_0) \eta} \, N^{(\xi)} (X; |p_0|; \hat{\bf p}) \, \delta 
(P^2 - {\cal M}_\xi^2 (X)) \right] . \nonumber 
\end{eqnarray*} 
Here we have dropped the contributions that have to be dealt with in 
two-loop order (cf. above after (\ref{yareyare})). 

Calculation of ${\cal J}_{\xi \zeta}^{(0)}$ with ${\cal M}_\xi^2 
(X)$, ${\cal M}_\zeta^2 (X) \geq 0$ is straightforward. The 
resultant form for ${\cal J}_{\xi \zeta}^{(0)}$ may be analytically 
continued to other region than ${\cal M}_\xi^2 (X)$, ${\cal 
M}_\zeta^2 (X) \geq 0$: 
\begin{eqnarray} 
- (4 \pi)^2 {\cal J}_{\pi \sigma}^{(0)} (X; R) & = & 
\frac{2}{\overline{\epsilon}} - \ln \frac{{\cal M}_\sigma^2 (X)}{e 
\mu_d^2} + \frac{R^2 - {\cal M}_\sigma^2 (X) + {\cal M}_\pi^2 (X)}{2 
R^2} \, \ln \frac{{\cal M}_\sigma^2 (X)}{{\cal M}_\pi^2 (X)} 
\nonumber \\ 
& & + \frac{{\cal S}_1}{2 R^2} \left[ \ln \frac{{\cal M}_\sigma^2 
(X) - {\cal M}_\pi^2 (X) + R^2 - {\cal S}_1}{{\cal M}_\sigma^2 (X) - 
{\cal M}_\pi^2 (X) + R^2 + {\cal S}_1} \right. \nonumber \\ 
& & \left. + \ln \frac{{\cal M}_\sigma^2 
(X) - {\cal M}_\pi^2 (X) - R^2 + {\cal S}_1}{{\cal M}_\sigma^2 (X) - 
{\cal M}_\pi^2 (X) - R^2 - {\cal S}_1} \right] , 
\label{yoso} 
\end{eqnarray} 
where ${\cal M}_\xi^2 (X)$ is understood to be ${\cal M}_\sigma^2 
(X) - i \epsilon (r_0) \eta$ and 
\[ 
{\cal S}_1 \equiv \sqrt{(R^2 - {\cal M}_\pi^2 (X))^2 - 2 {\cal 
M}_\sigma^2 (X) (R^2 + {\cal M}_\pi^2 (X)) + {\cal M}_\sigma^4 (X)} 
. 
\] 
Equation (\ref{yoso}) is valid in the regions, 
\begin{description} 
\item{R1)} ${\cal M}_\sigma^2 (X) {\cal M}_\pi^2 (X) \geq 0$ and 
\begin{eqnarray*} 
& & R^2 \leq {\cal M}_\sigma^2 (X) + {\cal M}_\pi^2 (X) - 2 
\sqrt{{\cal M}_\sigma^2 (X) {\cal M}_\pi^2 (X)} \nonumber \\ 
& & \mbox{or} \nonumber \\ 
& & {\cal M}_\sigma^2 (X) + {\cal M}_\pi^2 (X) + 2 \sqrt{{\cal 
M}_\sigma^2 (X) {\cal M}_\pi^2 (X)} \leq R^2 . 
\end{eqnarray*} 
\item{R2)} ${\cal M}_\sigma^2 (X) {\cal M}_\pi^2 (X) < 0$. 
\end{description} 
In the region, 
\[ 
{\cal M}_\sigma^2 (X) - {\cal M}_\pi^2 (X) - |R^2| \geq 0 , 
\;\;\;\;\;\; {\cal M}_\sigma^2 (X), \; {\cal M}_\pi^2 (X) \geq 0 , 
\] 
Equation (\ref{yoso}) is as it is. The expression in the regions 
${\cal M}_\sigma^2 (X) - {\cal M}_\pi^2 (X) - |R^2| < 0$ and/or 
${\cal M}_\sigma^2 (X), \, {\cal M}_\pi^2 (X) < 0$ is obtained from 
(\ref{yoso}) through simple analytic continuation. 

Continuing (\ref{yoso}) to the region 
\begin{eqnarray*} 
& & {\cal M}_\sigma^2 (X) + {\cal M}_\pi^2 (X) - 2 \sqrt{{\cal 
M}_\sigma^2 (X) {\cal M}_\pi^2 (X)} < R^2 < {\cal M}_\sigma^2 (X) + 
{\cal M}_\pi^2 (X) + 2 \sqrt{{\cal M}_\sigma^2 (X) {\cal M}_\pi^2 
(X)} \nonumber \\ 
& & \mbox{\hspace*{50ex}} ({\cal M}_\sigma^2 (X) {\cal M}_\pi^2 (X) 
> 0) , 
\end{eqnarray*} 
we obtain 
\begin{eqnarray} 
- (4 \pi)^2 {\cal J}_{\pi \sigma}^{(0)} (X; R) & = & 
\frac{2}{\overline{\epsilon}} - \ln \frac{{\cal M}_\sigma^2 (X)}{e 
\mu_d^2} + \frac{R^2 - {\cal M}_\sigma^2 (X) + {\cal M}_\pi^2 (X)}{2 
R^2} \, \ln \frac{{\cal M}_\sigma^2 (X)}{{\cal M}_\pi^2 (X)} 
\nonumber \\ 
& & + \frac{{\cal S}_2}{R^2} \left[ \arctan 
\frac{{\cal S}_2}{{\cal M}_\sigma^2 (X)  - {\cal M}_\pi^2 (X) + R^2} 
\right. \nonumber \\ 
& & \left. - \arctan \frac{{\cal S}_2}{{\cal M}_\sigma^2 (X) - 
{\cal M}_\pi^2 (X) - R^2} \right] , 
\label{yoso-1} 
\end{eqnarray} 
where 
\[ 
{\cal S}_2 \equiv \sqrt{ 2 {\cal M}_\sigma^2 (X) (R^2 + {\cal 
M}_\pi^2 (X)) - (R^2 - {\cal M}_\pi^2 (X))^2 - {\cal 
M}_\sigma^4 (X)} . 
\] 
Equation (\ref{yoso-1}) is valid in the region 
\[ 
{\cal M}_\sigma^2 (X) - {\cal M}_\pi^2 (X) - |R^2| \geq 0 .  
\] 
If ${\cal M}_\sigma^2 (X) - {\cal M}_\pi^2 (X) + \tau R^2$ $(\tau = 
\pm)$ in (\ref{yoso-1}) is negative, the replacement 
\[ 
\arctan \frac{{\cal S}_2}{{\cal M}_\sigma^2 (X) - {\cal M}_\pi^2 (X) 
+ \tau R^2} \longrightarrow \pi - \arctan \frac{{\cal S}_2}{{\cal 
M}_\pi^2 (X) - {\cal M}_\sigma^2 (X) - \tau R^2} 
\] 
should be made. 

${\cal J}_{\xi \xi}^{(0)} (R)$ $(\xi = \pi, \sigma)$ is obtained 
from (\ref{yoso}) and (\ref{yoso-1}) by setting ${\cal M}_\pi^2 (X) 
= {\cal M}_\sigma^2 (X) = {\cal M}_\xi^2 (X)$. 

As seen in (\ref{con-12}), we need $Re \, \Sigma_R^{(\xi)} (X; R)$ 
at $R^2 = {\cal M}_\xi^2 (X)$. As to ${\cal J}^{(\beta)}_{\xi \zeta} 
(X; R)$, following \cite{chiku}, we compute it at $R = 0$ for 
simplicity. The difference between ${\cal J}^{(\beta)}_{\xi \zeta} 
(X; R = 0)$ and ${\cal J}^{(\beta)}_{\xi \zeta} (X; R)$ at $R^2 = 
{\cal M}_\xi^2 (X)$ is of higher order. Then, we have 
\begin{eqnarray*} 
Re \left( \Sigma_R^{(\pi)} \right)_{\mbox{\scriptsize{II}}} & \simeq 
& \frac{\lambda^2}{9} \, \varphi^2 (X) \left[ {\cal J}_{\pi 
\sigma}^{(0)} (X; R^2 = {\cal M}_\pi^2 (X)) + {\cal J}_{\pi 
\sigma}^{(\beta)} (X; R = 0) \right] , \nonumber \\ 
Re \left( \Sigma_R^{(\sigma)} \right)_{\mbox{\scriptsize{II}}} & 
\simeq & \frac{\lambda^2}{18} \, \varphi^2 (X) \left[ (N - 1) 
\left\{ {\cal J}_{\pi \pi}^{(0)} (X; R^2 = {\cal M}_\sigma^2 (X)) + 
{\cal J}_{\pi \pi}^{(\beta)} (X; R = 0) \right\} \right. \nonumber 
\\ 
& & \left. + 9 \left\{ {\cal J}_{\sigma \sigma}^{(0)} (X; R^2 = 
{\cal M}_\sigma^2 (X)) + {\cal J}_{\sigma \sigma}^{(\beta)} (X; R = 
0) \right\} \right] , \nonumber 
\end{eqnarray*} 
where 
\begin{eqnarray} 
{\cal J}_{\pi \sigma}^{(0)} (X; R^2 = {\cal M}_\pi^2 (X)) & = & - 
\frac{1}{(4 \pi)^2} \left[ \frac{2}{\overline{\epsilon}} - 
\frac{{\cal M}_\sigma^2 (X)}{2 \, {\cal M}_\pi^2 (X)}\,  \ln 
\frac{|{\cal M}_\sigma^2 (X)|}{|{\cal M}_\pi^2 (X)|} -  \ln 
\frac{|{\cal M}_\pi^2 (X)|}{e^2 \mu_d^2} + {\cal K}_{\pi 
\sigma}^{(\pi)} \right] , \nonumber \\ 
{\cal J}_{\pi \pi}^{(0)} (X; R^2 = {\cal M}_\sigma^2 (X)) & = & 
- \frac{1}{(4 \pi)^2} \left[ \frac{2}{\overline{\epsilon}} - 
\ln \frac{|{\cal M}_\pi^2 (X)|}{e^2 \mu_d^2} + {\cal K}_{\pi 
\pi}^{(\sigma)} \right] , \nonumber \\ 
{\cal J}_{\sigma \sigma}^{(0)} (X; R^2 = {\cal M}_\sigma^2 (X)) & 
= & - \frac{1}{(4 \pi)^2} \left[ \frac{2}{\overline{\epsilon}} - 
\frac{\pi}{\sqrt{3}} - \ln \frac{|{\cal M}_\sigma^2 (X)|}{e^2 
\mu_d^2} \right] 
\label{J} 
\end{eqnarray} 
with 
\begin{eqnarray} 
{\cal K}_{\pi \sigma}^{(\pi)} & = & \theta \left( {\cal M}_\sigma^2 
(X) ({\cal M}_\sigma^2 (X) - 4 {\cal M}_\pi^2 (X)) \right) \, 
\frac{\sqrt{{\cal M}_\sigma^2 (X) ({\cal M}_\sigma^2 (X) - 4 {\cal 
M}_\pi^2 (X))}}{2 \, {\cal M}_\pi^2 (X)} \nonumber \\ 
& & \times \left[ \ln \frac{|{\cal M}_\sigma^2 (X) - \sqrt{{\cal 
M}_\sigma^2 (X) ({\cal M}_\sigma^2 (X) - 4 {\cal M}_\pi^2 
(X))}|}{|{\cal M}_\sigma^2 (X) + \sqrt{{\cal M}_\sigma^2 (X) 
({\cal M}_\sigma^2 (X) - 4 {\cal M}_\pi^2 (X))}|} \right. \nonumber 
\\ 
& & \left. - \ln \frac{|{\cal M}_\sigma^2 (X) - 2 \, {\cal M}_\pi^2 
(X) - \sqrt{{\cal M}_\sigma^2 (X) ( {\cal M}_\sigma^2 (X) - 4 {\cal 
M}_\pi^2 (X))|}}{|{\cal M}_\sigma^2 (X) - 2 \, {\cal M}_\pi^2 (X) + 
\sqrt{{\cal M}_\sigma^2 (X) ( {\cal M}_\sigma^2 (X) - 4 {\cal 
M}_\pi^2 (X))|}} \right] , \nonumber \\ 
& & + \theta \left( {\cal M}_\sigma^2 (X) (4 {\cal M}_\pi^2 (X) - 
{\cal M}_\sigma^2 (X)) \right) \, \frac{\sqrt{{\cal M}_\sigma^2 (X) 
(4 {\cal M}_\pi^2 (X) - {\cal M}_\sigma^2 (X))}}{{\cal M}_\pi^2 (X)} 
\nonumber \\ 
& & \times \left[ \epsilon ({\cal M}_\sigma^2 (X) ) \arctan 
\frac{\sqrt{{\cal M}_\sigma^2 (X) (4 {\cal M}_\pi^2 (X) - {\cal 
M}_\sigma^2 (X))}}{|{\cal M}_\sigma^2 (X) |} \right. \nonumber \\ 
& & - \epsilon ({\cal M}_\sigma^2 (X) - 2 {\cal M}_\pi^2 (X) 
) \arctan \frac{\sqrt{{\cal M}_\sigma^2 (X) (4 {\cal M}_\pi^2 (X) - 
{\cal M}_\sigma^2 (X))}}{|{\cal M}_\sigma^2 (X) - 2 {\cal M}_\pi^2 
(X) |} \nonumber \\ 
& & \left. + \pi \left\{ \theta \left( - {\cal M}_\sigma^2 (X) 
\right) - \theta \left( 2 {\cal M}_\pi^2 (X) - {\cal M}_\sigma^2 (X) 
\right) \right\} \right] , 
\label{K} \\ 
{\cal K}_{\pi \pi}^{(\sigma)} & = & \theta \left( {\cal M}_\sigma^2 
(X) ({\cal M}_\sigma^2 (X) - 4 {\cal M}_\pi^2 (X)) \right) \nonumber 
\\ 
& & \times \frac{\sqrt{{\cal M}_\sigma^2 (X) ({\cal M}_\sigma^2 (X) 
- 4 {\cal M}_\pi^2 (X))}}{{\cal M}_\sigma^2 (X)} \, \ln \frac{|{\cal 
M}_\sigma^2 (X) - \sqrt{{\cal M}_\sigma^2 (X) ({\cal M}_\sigma^2 (X) 
- 4 {\cal M}_\pi^2 (X))|}}{|{\cal M}_\sigma^2 (X) + \sqrt{{\cal 
M}_\sigma^2 (X) ({\cal M}_\sigma^2 (X) - 4 {\cal M}_\pi^2 (X))|}} 
\nonumber \\ 
& & + \theta \left( {\cal M}_\sigma^2 (X) (4 {\cal M}_\pi^2 (X) - 
{\cal M}_\sigma^2 (X)) \right) \, \frac{\sqrt{{\cal M}_\sigma^2 (X) 
(4 {\cal M}_\pi^2 (X) - {\cal M}_\sigma^2 (X))}}{|{\cal M}_\sigma^2 
(X)|} \nonumber \\ 
& & \times \left[ 2 \arctan \frac{\sqrt{{\cal M}_\sigma^2 (X) (4 
{\cal M}_\pi^2 (X) - {\cal M}_\sigma^2 (X))}}{|{\cal M}_\sigma^2 
(X)|} - \pi \right] , 
\label{K-1} 
\end{eqnarray} 
and 
\begin{eqnarray} 
{\cal J}_{\pi \sigma}^{(\beta)} (X; R = 0) & = & \frac{{\cal 
I}_\sigma^{(\beta)} (X) - {\cal I}_\pi^{(\beta)} (X)}{
{\cal M}_\sigma^2 (X) - {\cal M}_\pi^2 (X)} , 
\label{yuugen-1} \\ 
{\cal J}_{\xi \xi}^{(\beta)} (X; R = 0) & = & \frac{\partial {\cal 
I}_\xi^{(\beta)} (X)}{\partial {\cal M}_\xi^2 (X)} . 
\label{yuugen} 
\end{eqnarray} 
Here ${\cal I}_\xi^{(\beta)} (X)$ is as in (\ref{yoko-11}) or 
(\ref{tad-1}). Expressions (\ref{J}), (\ref{K}), and (\ref{K-1}) do 
not diverge at ${\cal M}_\pi^2 (X) = 0$ and at ${\cal M}_\sigma^2 
(X) = 0$. Equations (\ref{yuugen-1}) and (\ref{yuugen}) are valid 
for ${\cal M}_\xi^2 (X) \geq 0$ ($\xi = \pi, \sigma$). Analytic 
continuation of them to the region ${\cal M}_\xi^2 (X) < 0$ may be 
performed as in subsection B: 
\begin{equation} 
{\cal I}_\xi^{(\beta)} (X) \to \frac{1}{2 \pi^2} Re \int_{C'} d z \, 
\sqrt{z^2 - {\cal M}_\xi^2 (X)} \langle N^{(\xi)} (X; z) \; \rangle 
, 
\label{setsu} 
\end{equation} 
where the integration contour $C'$ is as in (\ref{shimai}). 
\subsubsection{Gap equation} 
Let us turn back to the condition (\ref{con-12}), which may be 
written as 
\begin{eqnarray} 
\chi_\xi & = & M_\xi^2 (X) - \mu^2 \nonumber \\ 
& = & \left( \Sigma_R^{(\xi)} (X) \right)_{\mbox{\scriptsize{I}}} + 
\left( \Sigma_R^{(\xi)} (X) \right)_{\mbox{\scriptsize{II}}} + 
(Z_\mu Z - 1) \, \mu^2 + \frac{\lambda}{6} \, d_\xi (Z_\lambda Z^2 - 
1) \, \varphi^2 (X) \;\;\;\;\;\;\;\; (\xi = \pi, \, \sigma) . 
\nonumber \\ 
& & 
\label{cond-20} 
\end{eqnarray} 
The UV-diverging contributions in (\ref{cond-20}) cancels. In fact, 
substituting (\ref{Z-fac}) into (\ref{cond-20}), we obtain the 
UV-divergence free equation, 
\begin{equation} 
\left( \tilde{M}_\xi^2 (X) \right)_{\mbox{\scriptsize{I}}}  + \left( 
\tilde{M}_\xi^2 (X) \right)_{\mbox{\scriptsize{II}}} = M_\xi^2 (X) - 
\mu^2 \;\;\;\;\;\; (\xi = \pi, \, \sigma) . 
\label{cond-22} 
\end{equation} 
Here (cf. (\ref{tooi-yo})) 
\begin{equation} 
\left( \tilde{M}_\xi^2 (X) \right)_{\mbox{\scriptsize{I}}} = 
\frac{\lambda}{6} \left[ \alpha_\xi \tilde{\cal I}_\pi + \beta_\xi 
\tilde{\cal I}_\sigma \right] 
\label{soro} 
\end{equation} 
with 
\begin{equation} 
\tilde{\cal I}_\xi = \frac{1}{(4 \pi)^2} {\cal M}_\xi^2 (X) \, \ln 
\frac{|{\cal M}_\xi^2 (X)|}{e \mu_d^2} + Re \, {\cal 
I}_\xi^{(\beta)} (X) 
\label{ai} 
\end{equation} 
and 
\begin{eqnarray} 
\left( \tilde{M}_\pi^2 (X) \right)_{\mbox{\scriptsize{II}}} & = & 
\frac{\lambda^2}{9} \, \varphi^2 (X) \left[ \frac{1}{(4 \pi)^2} 
\left\{ \frac{{\cal M}_\sigma^2 (X)}{2 {\cal M}_\pi^2 (X)} \, \ln 
\frac{|{\cal M}_\sigma^2 (X)|}{|{\cal M}_\pi^2 (X)|} + \ln 
\frac{|{\cal M}_\pi^2 (X)|}{e^2 \mu_d^2} - {\cal K}_{\pi 
\sigma}^{(\pi)} \right\} \right. \nonumber \\ 
& & \left. + Re \, \frac{{\cal I}_\sigma^{(\beta)} (X) - {\cal 
I}_\pi^{(\beta)} (X)}{{\cal M}_\sigma^2 (X) - {\cal M}_\pi^2 (X)} 
\right] , 
\label{ai-1} \\ 
\left( \tilde{M}_\sigma^2 (X) \right)_{\mbox{\scriptsize{II}}} & = & 
\frac{ \lambda^2}{18} \, \varphi^2 (X) \left[ \frac{1}{(4 \pi)^2} 
\left\{ (N - 1) \left( \ln \frac{| {\cal M}_\pi^2 (X)|}{e^2 
\mu_d^2 } - {\cal K}_{\pi \pi}^{(\sigma)} \right) \right. \right. 
\nonumber \\ 
& & \left. \left. + 3 \sqrt{3} \pi + 9 \, \ln \frac{|{\cal 
M}_\sigma^2 (X)|}{e^2 \mu_d^2} \right\} + Re \left\{ (N - 1) \frac{ 
\partial {\cal I}_\pi^{(\beta)} (X)}{\partial {\cal M}_\pi^2 (X)} + 
9 \frac{\partial {\cal I}_\sigma^{(\beta)} (X)}{\partial {\cal 
M}_\sigma^2 (X)} \right\} \right] . 
\label{ai-2} 
\end{eqnarray} 
In the above equations, ${\cal I}_\xi^{(\beta)} (X)$ is as in 
(\ref{tad-1}) for ${\cal M}_\xi^2 (X) \geq 0$ and as in 
(\ref{setsu}) for ${\cal M}_\xi^2 (X) < 0$. ${\cal K}_{\pi 
\sigma}^{(\pi)}$ and ${\cal K}_{\pi \pi}^{(\sigma)}$ are as in 
(\ref{K}) and (\ref{K-1}), respectively. For the region, 
\begin{eqnarray*} 
& & {\cal M}_\sigma^2 (X) \geq 0 \, , \;\;\;{\cal M}_\sigma^2 (X) 
\geq 4 {\cal M}_\pi^2 (X) \nonumber \\ 
& & \mbox{or} \nonumber \\ 
& & {\cal M}_\sigma^2 (X) < 0 \, , \;\;\;{\cal M}_\sigma^2 (X) < 4 
{\cal M}_\pi^2 (X) , 
\end{eqnarray*} 
the first terms on the RHS of (\ref{K}) and of (\ref{K-1}) should be 
used and the second terms should be used for the region, 
\begin{eqnarray*} 
& & {\cal M}_\sigma^2 (X) \geq 0 \, , \;\;\;{\cal M}_\sigma^2 (X) < 
4 {\cal M}_\pi^2 (X) \nonumber \\ 
& & \mbox{or} \nonumber \\ 
& & {\cal M}_\sigma^2 (X) < 0 \, , \;\;\;{\cal M}_\sigma^2 (X) > 4 
{\cal M}_\pi^2 (X) . 
\end{eqnarray*} 
In the LHS of (\ref{cond-22}), we have dropped the higher-order 
term, 
\begin{equation} 
- \frac{\lambda}{48 \pi^2} \frac{1}{\overline{\epsilon}} [c_\xi 
\chi_\pi + d_\xi \chi_\sigma] , 
\label{cond-24} 
\end{equation} 
which should be dealt with when one computes next to the leading 
order. The one-loop diagram ${\cal D} \left[ \displaystyle{ 
\raisebox{0.9ex}{\scriptsize{$t$}}} \mbox{\hspace{-0.1ex}} 
\vec{\phi} \chi_\xi {\bf P}_\xi \vec{\phi} / 2, \, - \lambda 
(\vec{\phi}^{\, 2})^2 / 4 ! \right]$, which contributes at next to 
the leading order, cancels the diverging contribution 
(\ref{cond-24}). 

Equation (\ref{cond-22}) is the gap equation, by which $M_\xi^2 (X)$ 
($\xi = \pi, \sigma$) are determined self consistently. $M_\xi^2 
(X)$ thus determined depend on $\vec{\varphi}^{\, 2} (X)$. With 
$M_\xi^2 (X)$ in hand, one can judge if ${\cal M}_\xi^2 (X)$ in 
(\ref{134}) is positive or negative. 
\subsubsection{\lq\lq High-temperature'' expansion} 
\lq\lq High-temperature'' expansion of ${\cal I}_\xi^{(\beta)} (X)$, 
Eq.~(\ref{setsu}), is obtained as in subsections A and 
B, 
\begin{equation} 
Re \, {\cal I}_\xi^{(\beta)} (X) \simeq \frac{1}{1 2} \, 
c_\xi (X) \, {\cal T}_\xi^2 (X) - \frac{1}{4 \pi} \, \theta \left( 
{\cal M}_\xi^2 (X) \right) \left( {\cal M}_\xi^2 (X) \right)^{1 / 2} 
{\cal T}_\xi (X) . 
\label{soya} 
\end{equation} 
From this the \lq\lq high-temperature'' expansion of $\partial {\cal 
I}^{(\beta)}_\xi (X) / \partial {\cal M}_\xi^2 (X)$ $(\xi = \pi, 
\sigma)$ in (\ref{ai-2}) is obtained: 
\[ 
\frac{\partial \, {\cal I}_\xi^{(\beta)} (X)}{\partial {\cal 
M}_\xi^2 (X)} \simeq - \frac{1}{8 \pi} \, \theta \left( {\cal 
M}_\xi^2 (X) \right) \left( {\cal M}_\xi^2 (X) \right)^{- 1 / 2} \, 
{\cal T}_\xi (X) . 
\] 
When ${\cal M}_\xi^2 (X) < 0$, $\partial {\cal I}^{(\beta)}_\xi (X) 
/ \partial {\cal M}_\xi^2 (X)$ is negligibly small. 
\subsubsection{\lq\lq Low-temperature'' expansion} 
In the case of locally thermal equilibrium system at low 
temperature, we have, for ${\cal M}_\xi^2 (X) \geq 0$, 
\begin{equation} 
{\cal I}_\xi^{(\beta)} (X) \simeq 
\frac{1}{2^{3 / 2} \pi} \left( {\cal M}_\xi^2 (X) \right)^{1 / 4} 
\left( T (X) \right)^{3 / 2} e^{- {\cal M}_\xi^2 (X) / T (X)} 
\;\;\;\; ({\cal M}_\xi^2 (X) \geq 0) . 
\label{hikui-kawa} 
\end{equation} 
For ${\cal M}_\xi^2 (X) < 0$, computation may be carried out as in 
subsection B. The result agrees with that obtained from 
(\ref{hikui-kawa}) by an analytic continuation; 
\begin{equation} 
Re \, {\cal I}_\xi^{(\beta)} (X) \simeq \frac{1}{2^{3 / 2} \pi} 
\left( - {\cal M}_\xi^2 (X) \right)^{1 / 4} \left( T (X) \right)^{3 
/ 2} \cos \left( \frac{\left( - {\cal M}_\xi (X) \right)^{1 / 2}}{T 
(X)} - \frac{\pi}{4} \right) \;\;\;\; ({\cal M}_\xi^2 (X) < 0) . 
\label{hikui-kawa1} 
\end{equation} 
From (\ref{hikui-kawa}) and (\ref{hikui-kawa1}), the low-temperature 
expansion of $\partial {\cal I}^{(\beta)}_\xi (X) / \partial {\cal 
M}_\xi^2 (X)$ $(\xi = \pi, \sigma)$ in (\ref{ai-2}) is obtained: 
\begin{eqnarray*} 
Re \, \frac{\partial \, {\cal I}_\xi^{(\beta)} (X)}{\partial {\cal 
M}_\xi^2 (X)} & \simeq & - \frac{1}{4 \sqrt{2} \pi} \left[ \theta 
\left( {\cal M}_\xi^2 (X) \right) \, \left( {\cal M}_\xi^2 (X) 
\right)^{- 1 / 4} \left( T (X) \right)^{1 / 2} e^{- {\cal M}_\xi^2 
(X) / T (X)} \right. \nonumber \\ 
& & \left. - \theta \left( - {\cal M}_\xi^2 (X) \right) \, \sin 
\left( \frac{\left( - {\cal M}_\xi (X) \right)^{1 / 2}}{T (X)} - 
\frac{\pi}{4} \right) \, \left( - {\cal M}_\xi^2 (X) \right)^{- 1 / 
4} \left( T (X) \right)^{1 / 2} \right] . 
\end{eqnarray*} 
\setcounter{equation}{0}
\setcounter{section}{7}
\def\theequation{\mbox{\arabic{section}.\arabic{equation}}}
\section{Application --- Early stage of phase transition}
There are few cases, for which approximate analytic calculations 
yield reliable results. In this section, for the purpose of 
illustrating how the formalism constructed above works, we apply it 
to very simple cases, in which the system of coupled equations 
obtained in the previous section can approximately be solved 
analytically. 
\subsection{Spatially uniform system} 
In this subsection, we apply the formalism to a spatially uniform 
system. We suppose the following situation. At the initial time $X_0 
= 0$, i) the system is in a \lq\lq highly super-cooled state'' with 
$\vec{\varphi} (X_0 = 0) \simeq 0$ and ii) the temperature $T$ is 
still high and the system is in quasi thermal equilibrium, 
\begin{equation} 
N^{(\xi)} (X_0 = 0; |p_0|) = \frac{1}{e^{|p_0| / T} - 1} . 
\label{shoki} 
\end{equation} 
We are only interested in the small time interval, 
during which iii) the high-temperature expansion is applicable, so 
that keeping terms that \lq\lq explicitly'' depend on $T$ is a good 
approximation (cf. below) and iv) the condensate fields 
$\vec{\varphi} (X_0)$ remains small in the sense stated below. As 
mentioned above, our formalism applies to the system away from the 
critical region, which is the reason to include i) into the list.  

We start with the determining equation for $f^{(\xi)} (X_0; p_0) = 
\theta (p_0) N^{(\xi)} (X_0; p_0) - \theta (- p_0) (1 + N^{(\xi)} 
(X_0; - p_0))$ $(\xi = \pi, \sigma)$, Eq.~(\ref{ashi}), 
\begin{equation} 
2 p_0 \frac{\partial N^{(\xi)} (X_0; p_0)}{\partial X_0} + 
\frac{\partial N^{(\xi)} (X_0; p_0)}{\partial p_0} \frac{d {\cal 
M}_\xi (X_0)}{d X_0} = 0 . 
\label{kime} 
\end{equation} 
Since $N^{(\xi)} (X_0; p_0)$ in the propagator accompanies $\delta 
(P^2 - {\cal M}_\xi^2 (X_0))$, $p_0$ in (\ref{kime}) is $p_0 = 
\sqrt{p^2 + {\cal M}_\xi^2 (X_0)}$. Equation (\ref{kime}) may be 
solved as 
\begin{equation} 
N^{(\xi)} (X_0; p_0) = {\cal F} (P^2 - {\cal M}_\xi^2 (X_0)) , 
\label{kai} 
\end{equation} 
where ${\cal F}$ is an arbitrary function. For simplicity of 
presentation, we take $\vec{\varphi} (X_0 = 0) = d \vec{\varphi} 
(X_0) / d X_0 \, \rule[-3mm]{.14mm}{8.5mm} 
\raisebox{-2.85mm}{\scriptsize{$\; X_0 = 0$}} = 0$ in the sequel. In 
the case of $\vec{\varphi} (X_0 = 0) \neq 0$ and/or $d \vec{\varphi} 
(X_0) / d X_0 \, \rule[-3mm]{.14mm}{8.5mm} 
\raisebox{-2.85mm}{\scriptsize{$\; X_0 = 0$}} \neq 0$, no new 
feature arises, as long as they are small. Using (\ref{shoki}) in 
(\ref{kai}), we obtain, 
\begin{eqnarray} 
N^{(\xi)} (X_0; p_0) & = & \frac{1}{e^{\sqrt{p_0^2 - {\cal M}_\xi^2 
(X_0) + M_\xi^2 (X_0 = 0)} / T} - 1} \nonumber \\ 
& = & \frac{1}{e^{\sqrt{p^2 + M_\xi^2 (0)} / T} - 1} , 
\label{bunpupu} 
\end{eqnarray} 
where use has been made of the fact that $N^{(\xi)} (X_0; p_0)$ 
accompanies $\delta (P^2 - {\cal M}_\xi^2 (X_0))$ in the propagator. 

With (\ref{bunpupu}), let us compute the fundamental quantity ${\cal 
I}_\xi (X_0)$, Eq.~(\ref{yoko}). Computation goes in a similar 
manner as the \lq\lq direct calculation'' in Section VIIB: 
\begin{eqnarray} 
{\cal I}_\xi^{(\beta)} (X_0) & = & \frac{T^2}{1 2} + \frac{1}{2 
\pi^2} \left[ \int_0^\infty d p \frac{p^2}{\sqrt{p^2 + {\cal 
M}_\xi^2 (X_0)}} \frac{1}{e^{\sqrt{p^2 + M_\xi^2 (0)} / T} - 1} 
\right. \nonumber \\ 
& & \left. - \int_0^\infty d {\cal E} \, \frac{\cal E}{e^{{\cal 
E} / T} - 1}\right] \nonumber \\ 
& \simeq & \frac{T^2}{1 2} + \frac{T}{2 \pi^2} \left[ \int_0^\infty 
d p \, \frac{p^2}{\sqrt{p^2 - |{\cal M}_\xi^2 (X_0)|} \sqrt{p^2 - 
|M_\xi^2 (0)|}} \right. \nonumber \\ 
& & \left.- \int_{\sqrt{|M_\xi^2 (0)|}}^\infty d p 
\frac{p}{\sqrt{p^2 - |M_\xi^2 (0)|}} \right] , 
\label{momi} 
\end{eqnarray} 
where high-$T$ approximation has been used. Noticing that, 
\[ 
\sqrt{p^2 - |{\cal M}_\xi^2 (X_0)|} = - i \sqrt{|{\cal M}_\xi^2 
(X_0)| - p^2} \;\; \mbox{for} \;\;  p^2 < |{\cal M}_\xi^2 (X_0)|, 
\] 
etc., we obtain 
\begin{eqnarray} 
{\cal I}_\xi^{(\beta)} (X_0) & \simeq & \frac{T^2}{1 2} - 
\frac{\sqrt{|M_\xi^2 (0)|} \; T}{2 \pi^2} \; H \left( \frac{|{\cal 
M}_\xi^2 (X_0)|}{|M_\xi^2 (0)|} \right) \nonumber \\ 
& & + \frac{i T}{2 \pi^2} \sqrt{|M_\xi^2 (0)|} \; E \left( 
\frac{\pi}{2}, \, \sqrt{\frac{|M_\xi^2 (0)| - |{\cal M}_\xi^2 
(X_0)|}{|M_\xi^2 (0)|}} \right) , 
\label{owa-true} 
\end{eqnarray} 
where 
\begin{eqnarray*} 
H (\rho^2) & \equiv & \int_0^\rho d x \frac{x^2}{\sqrt{1 - x^2} 
\sqrt{\rho^2 - x^2}} - \int_1^\infty d x \, \frac{x}{\sqrt{x^2 - 1}} 
\left[ \frac{x}{\sqrt{x^2 - \rho^2}} - 1 \right] , \nonumber \\  
E (\pi / 2, \, k) & \equiv & \int_0^{\pi / 2} d \theta \sqrt{1 - 
k^2 \sin^2 \theta} . 
\end{eqnarray*} 
Here $E$ is an elliptic function of second kind. Note that $|M_\xi^2 
(0)| > |{\cal M}_\xi^2 (X_0)|$, so that $\rho < 1$. From the setup 
iv) above,  
\[
|M_\xi^2 (0)| - |{\cal M}_\xi^2 (X_0)| = \lambda e_\xi 
\vec{\varphi}^{\, 2} (X_0) < < |M_\xi^2 (0)| \;\;\;\;\; (e_\pi = 1 / 
6, \, e_\sigma = 1 / 2) , 
\] 
and then (\ref{owa-true}) may be approximated as 
\begin{eqnarray} 
{\cal I}_\xi^{(\beta)} (X_0) & \simeq & \frac{T^2}{1 2} - \frac{1}{4 
\pi^2} \frac{T}{\sqrt{|M_\xi^2 (0)|}} \left[ |M_\xi^2 (0)| - |{\cal 
M}_\xi^2 (X_0)| \right] \nonumber \\ 
& & + \frac{i T}{4 \pi} \left[ \sqrt{|{\cal M}_\xi^2 (X_0)|} + 
\frac{1}{4} \frac{|M_\xi^2 (0)| - |{\cal M}_\xi^2 
(X_0)|}{\sqrt{|M_\xi^2 (0)|}} \right] \nonumber \\ 
& = & \frac{T^2}{1 2} - e_\xi \frac{\lambda}{4 \pi^2} 
\frac{T}{\sqrt{|M_\xi^2 (0)|}} \vec{\varphi}^{\, 2} (X_0) + \frac{i 
T}{4 \pi} \left[ \sqrt{|{\cal M}_\xi^2 (X_0)|} + \frac{\lambda 
e_\xi}{4} \frac{\vec{\varphi}^{\, 2} (X_0)}{\sqrt{|M_\xi^2 (0)|}} 
\right] . 
\label{owa} 
\end{eqnarray} 

Let us turn to the determining equation for $M_\xi (X_0)$ $(\xi = 
\pi, \sigma)$, Eq.~(\ref{cond-22}). As mentioned above, we ignore 
the temperature-independent terms. Using (\ref{owa}) in 
(\ref{ai})~-~(\ref{ai-2}), we see that (\ref{cond-22}) yields 
\begin{eqnarray} 
M_\pi^2 (X_0) & \simeq & - \omega^2 - \left( \frac{\lambda}{12 
\pi} \right)^2 (N + 4) \frac{T}{\sqrt{|M_\pi^2 (0)|}} 
\vec{\varphi}^{\, 2} (X_0) \nonumber \\ 
& & - \frac{\lambda^3}{108 \pi^2} \frac{T}{\sqrt{|M_\pi^2 (0)|}} 
\frac{\left( \vec{\varphi}^{\, 2} (X_0) \right)^2}{M_\sigma^2 (X_0) 
- M_\pi^2 (X_0) + \lambda \vec{\varphi}^{\, 2} (X_0) / 3} , 
\label{pi} \\ 
M_\sigma^2 (X_0) & \simeq & - \omega^2 - 3 \left( 
\frac{\lambda}{12 \pi} \right)^2 (N + 8) \frac{T}{\sqrt{|M_\pi^2 
(0)|}} \vec{\varphi}^{\, 2} (X_0) , 
\label{sigma} 
\end{eqnarray} 
where 
\[ 
\omega^2 \equiv - \mu^2 - \frac{\lambda}{7 2} (N + 2) T^2 \; (> 0) . 
\] 
In obtaining (\ref{pi}) and (\ref{sigma}), use has been made of the 
approximation $(|M_\sigma^2 (0)| - |M_\pi^2 (0)|) < < |M_\pi^2 
(0)|$, which will be justified a posteriori. 

Approximate solution to (\ref{pi}) may be written in the form 
\begin{eqnarray} 
M_\pi^2 (X_0) & \simeq & - \omega^2 - \left( \frac{\lambda}{12 
\pi} \right)^2 (N + g) \frac{T}{\sqrt{|M_\pi^2 (0)|}} 
\vec{\varphi}^{\, 2} (X_0) , 
\label{pipi} 
\end{eqnarray} 
where 
\[ 
g = 4 \left[ 1 + \left\{ 1 - 3 \left( \frac{1}{12 \pi} \right)^2 
(2 N + 24 - g) \frac{\lambda T}{\sqrt{M_\pi^2 (0)}} \right\}^{- 1} 
\right] . 
\] 
For $\lambda T < O (\sqrt{|M_\pi^2 (0)|} \, )$, $g = 8$ and, for 
$\lambda T > O (\sqrt{|M_\pi^2 (0)|} \, )$, $g = 4$. 

Let us turn to deriving the equation of motion for $\vec{\varphi} 
(X)$. $\Omega (\vec{\varphi}^{\, 2} (X_0))$ in (\ref{haha-pre}) 
reads 
\begin{eqnarray*} 
\Omega (\vec{\varphi}^{\, 2} (X_0)) & \simeq & - \omega^2 + 
\frac{\lambda}{6} \vec{\varphi}^{\, 2} (X_0) \nonumber \\ 
& & - \frac{\lambda T}{12 \pi^2} \left[ (N - 1) \sqrt{|M_\pi^2 (0)|} 
\; H \left( \frac{ |{\cal M}_\pi^2 (X_0)|}{|M_\pi^2 (0)|} \right) + 
3 \sqrt{|M_\sigma^2 (0)|} \; H \left( \frac{ |{\cal M}_\sigma^2 
(X_0)|}{|M_\sigma^2 (0)|} \right) \right] . 
\end{eqnarray*} 
Up to $O (\vec{\varphi}^2)$ term, this becomes 
\[ 
\Omega (\vec{\varphi}^{\, 2} (X_0)) \simeq - \omega^2 + 
\frac{\lambda'}{6} \vec{\varphi}^{\, 2} (X_0) , 
\] 
where 
\begin{equation} 
\lambda' \equiv \lambda \left[ 1 - \frac{\lambda}{2 4 \pi^2} (N 
+ 8) \frac{T}{\omega} \right] . 
\label{lam} 
\end{equation} 
It should be emphasized that $\lambda' < \lambda$. Above equations 
are valid for $\lambda T < O (\omega)$ (cf. the setup i) at the 
beginning of this subsection). The equation of motion for 
$\vec{\varphi} (X_0)$, Eq.~(\ref{sto}), reads 
\begin{eqnarray}  
\ddot{\vec{\varphi}} (X_0) & = & \left[ \omega^2 - 
\frac{\lambda'}{6} \vec{\varphi}^{\, 2} (X_0) \right] \, 
\vec{\varphi} (X_0) + \vec{\zeta} (X_0) + H \vec{e} 
\label{eqevo} 
\\ 
& = & - \frac{\partial {\cal V}}{\partial \vec{\varphi} (X_0)} + 
\vec{\zeta} (X_0) , 
\label{eqevo1} 
\end{eqnarray} 
where $\ddot{\vec{\varphi}} (X_0) \equiv d^2 \vec{\varphi} (X_0) / d 
X_0^2$ and 
\begin{equation} 
{\cal V} \simeq \frac{\pi^2}{90} N T^4 - \frac{\omega^2}{2} \, 
\vec{\varphi}^{\, 2} (X_0) + \frac{\lambda'}{4 !} \, \left( 
\vec{\varphi}^{\, 2} (X_0) \right)^2 - H (\displaystyle{ 
\raisebox{0.9ex}{\scriptsize{$t$}}} \mbox{\hspace{-0.1ex}} 
\vec{e} \cdot \vec{\varphi} (X_0)) . 
\label{eqevo10} 
\end{equation} 
is related to $\gamma_{phys}$ through 
\[ 
{\cal V} = - Re \, \gamma_{phys} + \frac{1}{2} 
\dot{\vec{\varphi}}^{\, 2} (X_0) 
\] 
with $\dot{\vec{\varphi}} (X_0) \equiv d \vec{\varphi} (X_0) / d 
X_0$. In the following we simply call ${\cal V}$ the potential. From 
the setup i) at the beginning of this section, $\omega^2 \simeq - 
M_\pi^2 (0) \simeq - M_\sigma^2 (0)$ $(> 0)$ is quite large. The 
setup iv) means that we are interested in the time interval during 
which $\omega^2 >> \lambda' \vec{\varphi}^{\, 2} (X_0)$. [Recalling 
that $X_0$ of $\vec{\varphi} (X_0)$ is macroscopic time coordinate, 
$\omega$ should be $\omega \lesssim 1 / L_0$, where $L_0$ is the 
size of the time-direction of the spacetime cell (cf. Section I).] 
Above observation allows us to solve (\ref{eqevo}) iteratively. 
Ignoring the term $\lambda' \vec{\varphi}^{\, 2} (X_0) \vec{\varphi} 
(X_0) / 6$, we obtain the zeroth-order solution, 
$\vec{\varphi}^{(0)} (X_0)$, of (\ref{eqevo}), 
\begin{equation} 
\vec{\varphi}^{(0)} (X_0) = \frac{H}{\omega^2} \left[ \cosh 
(\omega X_0) - 1 \right] \vec{e} + \frac{1}{\omega} \int_0^{X_0} d 
X_0' \, \vec{\zeta} (X_0') \sinh (\omega (X_0 - X_0')) . 
\label{kai-yo} 
\end{equation} 
Substituting (\ref{kai-yo}) back into the term $\lambda' 
\vec{\varphi}^{\, 2} (X_0) \vec{\varphi} (X_0) / 6$ in 
(\ref{eqevo}), we obtain the first-order solution, 
\begin{eqnarray} 
\vec{\varphi} (X_0) & = & \frac{H}{\omega^2} \left[ \cosh 
(\omega X_0) - 1 \right] \vec{e} \nonumber \\ 
& & + \frac{1}{\omega} \int_0^{X_0} d X_0' \left[ \vec{\zeta} 
(X_0') - \frac{\lambda'}{6} \left(\vec{\varphi}^{(0)} (X_0') 
\right)^2 \vec{\varphi}^{(0)} (X_0') \right] \sinh (\omega (X_0 - 
X_0')) , 
\label{kai-yo1} 
\end{eqnarray} 
where $\vec{\varphi}^{(0)} (X_0')$ is as in (\ref{kai-yo}). 
More accurate solution may be obtained by further iterations. 

For the purpose of computing the $\zeta$-averaged correlation 
functions, we compute the partial derivatives of $Im \, 
\gamma_{phys} (X_0)$ with respect to $\vec{\varphi}^{\, 2} (X_0)$ 
using (\ref{haha-pre}) and (\ref{haha}) with (\ref{owa}): 
\begin{eqnarray} 
\frac{\partial \, Im \, \gamma_{phys} (X_0)}{\partial 
\vec{\varphi}^{\, 2} (X_0)} \rule[-3mm]{.14mm}{8.5mm} 
\raisebox{-2.85mm}{\scriptsize{$\; \vec{\varphi} = 0$}} 
& \simeq & - \frac{\lambda}{4 8 \pi} (N + 2) \omega T , 
\label{uni-1} \\ 
\frac{\partial^2 \, Im \, \gamma_{phys} (X_0)}{\partial 
(\vec{\varphi}^{\, 2} (X_0))^2} \rule[-3mm]{.14mm}{8.5mm} 
\raisebox{-2.85mm}{\scriptsize{$\; \vec{\varphi} = 0$}} 
& \simeq & \frac{\lambda^2}{1152 \pi} (N + 8) \frac{T}{\omega} . 
\label{uni-2} 
\end{eqnarray} 
Noticing that we are concerned about the spatially uniform system, 
(\ref{2-mom}) and (\ref{4-pt}) become, in respective order, 
\begin{equation} 
\langle \zeta_i (X_0) \zeta_j (Y_0) \rangle_\zeta \simeq - 
\frac{\lambda}{2 4 \pi} (N + 2) \frac{\omega T}{V} \, \delta_{i j} 
\, \delta (X_0 - Y_0) 
\label{iha-0} 
\end{equation} 
and 
\begin{eqnarray} 
\langle \zeta_i (X_0) \zeta_j (Y_0) \zeta_k (U_0) \zeta_l (V_0) 
\rangle_{\zeta c} & \simeq & - \frac{\lambda^2}{288 \pi} \frac{N + 
8}{V^3} \frac{T}{\omega} \left[ \delta_{i k} \delta_{l j} + 
\delta_{k j} \delta_{l i} + \delta_{i j}\delta_{l k} \right] 
\nonumber \\ 
& & \times \delta (X_0 - Y_0) \, \delta (X_0 - U_0 ) \, \delta (V_0 
- Y_0 ) , 
\label{iha} 
\end{eqnarray} 
where $V$ is the (large) volume of the system. 

We are now in a position to compute the $\zeta$-averaged \lq\lq 
connected'' correlation functions (\ref{comu}). From the 
zeroth-order solution (\ref{kai-yo}), we obtain using (\ref{iha-0}) 
and (\ref{iha}), 
\begin{eqnarray} 
\langle \vec{\varphi}^{\, (0)} (X_0) \rangle_\zeta & = & 
\frac{h}{\omega^2} \vec{e} \left[ \cosh (\omega X_0) - 1 \right] , 
\label{1-corr} \\ 
\langle \varphi_i^{(0)} (X_0) \varphi_j^{(0)} (Y_0) 
\rangle_{\zeta c} & = & \frac{1}{4 \omega^2} \sum_{\rho, \, 
\sigma = \pm} \rho \sigma e^{\omega (\rho X_0 - \sigma Y_0)} 
\nonumber \\ 
& & \times \int_0^{X_0} d X_0' \int_0^{Y_0} d Y_0' \, e^{- \omega 
(\rho X_0' - \sigma Y_0')} \langle \zeta_i (X_0') \zeta_j (Y_0') 
\rangle_\zeta \nonumber \\ 
& = & - \frac{\lambda}{4 8 \pi} \frac{T}{\omega^2} \frac{\delta_{i 
j}}{V} (N + 2) \left[ \cosh (\omega X_0) \sinh (\omega Y_0) - 
\omega_0 Y_0 \cosh (\omega (X_0 - Y_0)) \right] , \nonumber \\ 
& & 
\label{corr-2} 
\end{eqnarray} 
and 
\begin{eqnarray} 
& & \langle \varphi_i^{(0)} (X_0) \varphi_j^{(0)} (Y_0) 
\varphi_k^{(0)} (U_0) \varphi_l^{(0)} (V_0) \rangle_{\zeta c} 
\nonumber \\ 
& & \mbox{\hspace*{8ex}} \simeq - \frac{N + 8}{2 \pi} \left( 
\frac{\lambda}{4 8} \right)^2 \frac{T}{\omega^6 V^3} \left[ 
\delta_{i j} \delta_{k l} + \delta_{i k} \delta_{j l} + \delta_{i l} 
\delta_{j k} \right] \nonumber \\ 
& & \mbox{\hspace*{11ex}} \times \left[ \cosh (\omega (X_0 + Y_0 + 
U_0 - V_0) ) \sinh (2 \omega V_0) \right. \nonumber \\  
& & \mbox{\hspace*{11ex}} - 4 \sinh (\omega V_0) \left\{ \cosh ( 
\omega (X_0 + Y_0 - V_0) ) \cosh ( \omega (U_0 - V_0) ) \right. 
\nonumber \\ 
& & \mbox{\hspace*{11ex}} \left. + \cosh ( \omega (X_0 - Y_0) ) 
\cosh (\omega U_0) \right\} \nonumber \\ 
& & \left. \mbox{\hspace*{11ex}} + 2 \omega V_0 \left\{ 2 \cosh ( 
\omega (X_0 - V_0) ) \cosh ( \omega (Y_0 - U_0) ) + \cosh ( \omega 
(X_0 - Y_0 - U_0 + V_0) ) \right\} \right] . \nonumber \\ 
& & 
\label{corr-4} 
\end{eqnarray} 
In (\ref{corr-2}), $X_0 \geq Y_0$ has been assumed and in 
(\ref{corr-4}), $X_0, Y_0, U_0 \geq V_0$ has been assumed. Let us 
study (\ref{1-corr})~-~(\ref{corr-4}) in turn. 
\subsubsection{$\langle \vec{\varphi}^{\, (0)} (X_0) \rangle_\zeta$} 
Equation (\ref{1-corr}) shows that, as $X_0$ increases, the 
$\zeta$-averaged $\vec{\varphi}^{(0)} (X_0)$ blows up exponentially, 
\begin{equation} 
\langle \vec{\varphi}^{\, (0)} (X_0) \rangle_\zeta \sim 
\vec{\Phi} e^{\omega X_0} \;\;\;\;\;\;\; (\omega X_0 > > 1) , 
\label{1-ji} 
\end{equation} 
where $\vec{\Phi} \equiv H \vec{e} / (2 \omega^2)$. We should 
recall here that, for too large $X_0$, $\vec{\varphi} (X_0)$ is 
large and our approximation does not apply. For obtaining an 
improved expression for $\langle \vec{\varphi} (X_0) \rangle_\zeta$, 
we should use the improved solution (\ref{kai-yo1}), which yields, 
for large $e^{\omega X_0} > > 1$, 
\begin{equation} 
\langle \vec{\varphi} (X_0) \rangle_\zeta \simeq \vec{\Phi} 
e^{\omega X_0} - \frac{\lambda'}{48 \omega^2} \vec{\Phi} 
e^{3 \omega X_0} \left[ \vec{\Phi}^{\, 2} - \frac{\lambda}{1 9 2 
\pi} (N + 2)^2 \frac{1}{\omega^2} \frac{T}{V} \right] . 
\label{2-ji} 
\end{equation} 
Comparing (\ref{2-ji}) with (\ref{1-ji}), we see that (\ref{2-ji}) 
is reliable in the region 
\[ 
e^{2 \omega X_0} < O \left( \frac{\omega^2}{\lambda'} \, 
\rule[-3mm]{.14mm}{8.5mm} \, \vec{\Phi}^{\, 2} - \frac{\lambda}{192 
\pi} (N + 2)^2 \frac{T}{\omega^2 V} \, \rule[-3mm]{.14mm}{8.5mm}^{\, 
- 1} \right) ,  
\] 
which generally forces $\lambda$ to small values. When compared to 
the first term on the RHS of (\ref{2-ji}), the term with 
$\vec{\Phi}^{\, 2}$ in the second term prevents $|\langle 
\vec{\varphi} (X_0) \rangle|$ to become large, while the other term 
expedites it. The former can easily be understood from 
(\ref{eqevo}), since the term $\lambda' \vec{\varphi}^{(0) \, 2} 
(X_0) \vec{\varphi}^{(0)} (X_0) / 6$ in (\ref{eqevo}) acts as 
decreasing $\omega$. The latter comes from 
\begin{equation}  
\frac{\lambda'}{6} \left( \vec{\varphi}^{\, (0)} (X_0) \right)^2 
\varphi_i^{(0)} (X_0) \to \frac{\lambda'}{6} \langle \left( 
\vec{\varphi}^{\, (0)} (X_0) \right)^2 \rangle_\zeta \varphi_i^{(0)} 
(X_0) + \frac{\lambda'}{3} \langle \varphi_j^{\, (0)} (X_0) 
\varphi_i^{\, (0)} (X_0) \rangle_\zeta \varphi_j^{(0)} (X_0) 
\label{eff-ef} 
\end{equation}  
in (\ref{eqevo}). As seen from (\ref{corr-2}), the $\zeta$-averaged 
quantities in (\ref{eff-ef}) are negative so that, in (\ref{eqevo}), 
(\ref{eff-ef}) acts as increasing $|\vec{\varphi} (X_0)|$ as $X_0$ 
increases. 
\subsubsection{Two-point $\zeta$-correlation function} 
We now turn to (\ref{corr-2}). One can easily see that 
(\ref{corr-2}) is negative definite. Thus, $Im \, \gamma_{phys} 
(X_0)$ causes a negative correlation between the same component of 
$\vec{\varphi}$, $\varphi_i (X_0)$ and $\varphi_{j = i} (Y_0)$. If 
the term $\vec{\zeta} (X_0)$ in (\ref{eqevo}), which comes from $Im 
\, \gamma_{phys} (X_0)$, is absent, (\ref{kai-yo}) or 
(\ref{kai-yo1}) with $\vec{\zeta} = 0$ uniquely determines 
$\vec{\varphi} (X_0)$. $Im \, \gamma_{phys}$ works as stochastically 
upsetting $\vec{\varphi} (X_0)$ through the random forces 
$\vec{\zeta}$. Since $\langle \varphi_i^{(0)} (X_0) \varphi_i^{(0)} 
(Y_0) \rangle_\zeta$ (with no summation over $i$) in (\ref{iha-0}) 
is negative, $\langle \varphi_i (X_0) \varphi_{j = i} (Y_0) 
\rangle_\zeta$ tends to decrease when compared to the case where no 
random force acts, $\vec{\zeta} (X_0) = 0$. 

As $X_0$ increases, the $\zeta$-averaged two-point correlation 
function (\ref{corr-2}) blows up exponentially. For $X_0 = Y_0$ and 
$e^{\omega X_0} > > 1$, we have 
\begin{equation} 
\langle \varphi_i^{(0)} (X_0) \varphi_j^{(0)} (X_0) \rangle_{\zeta 
c} \simeq - \frac{\lambda}{1 9 2 \pi} (N + 2) \frac{T}{\omega^2} 
\frac{\delta_{i j}}{V} e^{2 \omega X_0} . 
\label{soukan} 
\end{equation} 
Using the improved solution (\ref{kai-yo1}), we may compute the 
improved $\zeta$-averaged two-point correlation function. Up to the 
first order with respect to the difference between (\ref{kai-yo1}) 
and (\ref{kai-yo}), we obtain 
\begin{equation} 
\langle \varphi_i (X_0) \varphi_j (X_0) 
\rangle_{\zeta c} \simeq \mbox{Eq.~(\ref{soukan})} 
+ {\cal C}_{i j}^{(1)} (X_0) + {\cal C}_{i j}^{(2)} (X_0) , 
\label{soukan-10} 
\end{equation} 
where 
\begin{eqnarray} 
& & {\cal C}_{i j}^{(1)} = \frac{\lambda \lambda'}{4608 \pi} (N + 2) 
\frac{1}{\omega^4} \frac{T}{V} e^{4 \omega X_0} \left( \delta_{i j} 
\vec{\Phi}^{\, 2} + 2 \Phi_i \Phi_j \right) , 
\label{soukan-11} \\ 
& & {\cal C}_{i j}^{(2)} = - \frac{\delta_{i j}}{\pi^2} \lambda^2 
\lambda' \left( \frac{N + 2}{96} \right)^3 \frac{1}{\omega^6} \left( 
\frac{T}{V} \right)^2 e^{4 \omega X_0} \left[1 - 2 \pi \frac{N + 
8}{(N + 2)^2} \frac{1}{\omega^2 T V} \right] . 
\label{soukan-12} 
\end{eqnarray} 
Let us write $\varphi_i (X_0)$, Eq.~(\ref{kai-yo1}), in the form 
\[ 
\varphi_i (X_0) = \sum_{l = 0}^3 \varphi_i^{[l]} (X_0) , 
\] 
where $\varphi_i^{[l]} (X_0)$ includes $l$ $\zeta$'s. ${\cal C}_{i 
j}^{(1)} (X_0)$, Eq.~(\ref{soukan-11}), comes from $\langle 
\varphi_i^{[1]} (X_0) \varphi_j^{[1]} (X_0) \rangle_\zeta$ and 
${\cal C}_{i j}^{(2)} (X_0)$ comes from $\langle \varphi_i^{[1]} 
(X_0) \varphi_j^{[3]} (X_0) \rangle_\zeta$ and $\langle 
\varphi_i^{[3]} (X_0) \varphi_j^{[1]} (X_0) \rangle_\zeta$. ${\cal 
C}_{i j}^{(2)} (X_0)$ consists of two terms. The term being 
proportional to $1 / V^2$ comes from (the square of) (\ref{iha-0}) 
while the term being proportional to $1 / V^3$ comes from 
(\ref{iha}). 

Here we summarize the characteristic features of 
(\ref{soukan-10})~-~(\ref{soukan-12}). 
\begin{itemize} 
\item Comparing (\ref{soukan-10})~-~(\ref{soukan-12}) with 
(\ref{soukan}) we see that (\ref{soukan-10}) with (\ref{soukan-11}) 
and (\ref{soukan-12}) is reliable in the region, 
\[ 
e^{2 \omega X_0} < O \left( \mbox{Min} \left( 
\frac{\omega^2}{\lambda' \vec{\Phi}^{\, 2}}, \, \frac{\omega^4 
V}{\lambda \lambda' T} , \, \frac{\omega^6 V^2}{\lambda \lambda'} 
\right) \right) . 
\] 
\item In contrast to (\ref{soukan}) and ${\cal C}_{i j}^{(2)} 
(X_0)$, ${\cal C}_{i j}^{(1)} (X_0)$ is not diagonal in 
$N$-dimensional vector space, which means that the correlation 
between different component of $\vec{\varphi} (X_0)$ exists. ${\cal 
C}_{i i}^{(1)} (X_0)$ (with no summation over $i$) is positive, 
while ${\cal C}_{i j (\neq i)}^{(1)} (X_0)$ can either be positive 
or negative depending on $\vec{\Phi}$. 
\item The $1 / V^2$ term in ${\cal C}_{i j}^{(2)} (X_0)$ is negative 
and then enhances (\ref{soukan}), while the $1 / V^3$ term is 
positive. 
\end{itemize} 
\subsubsection{Four-point $\zeta$-correlation function} 
Setting $X_0 = Y_0 = U_0 = V_0$ in (\ref{corr-4}), we obtain 
\begin{eqnarray} 
& & \langle \varphi_i^{(0)} (X_0) \varphi_j^{(0)} (Y_0) 
\varphi_k^{(0)} (U_0) \varphi_l^{(0)} (V_0) \rangle_{\zeta c} 
\nonumber \\ 
& & \mbox{\hspace*{8ex}} = - \frac{N + 8}{4 \pi} \left( 
\frac{\lambda}{4 8} \right)^2 \frac{T}{\omega^6 V^3} \left[ 
\delta_{i j} \delta_{k l} + \delta_{i k} \delta_{j l} + 
\delta_{i l} \delta_{j k} \right] {\cal H} (\omega X_0) , 
\label{4-ptpt} 
\end{eqnarray} 
where 
\[ 
{\cal H} (\omega X_0) = \sinh (4 \omega X_0) - 8 \sinh (2 \omega 
X_0) + 12 \omega X_0 . 
\] 
One can easily show that ${\cal H} (\omega X_0)$ $(X_0 > 0)$ is 
positive definite, so that the four-point correlation (\ref{4-ptpt}) 
is of negative. For $\omega X_0 << 1$ and $\omega X_0 > > 1$, we 
have 
\[ 
{\cal H} (\omega X_0) = \left\{ 
\begin{array}{ll} 
\frac{3 2}{5} (\omega X_0)^5 & \;\;\;\;\; (\omega X_0 < < 1) \\ 
\frac{1}{2} e^{4 \omega X_0} & \;\;\;\;\; (\omega X_0 > > 1) . 
\end{array}
\right. 
\]
\subsubsection{Features of the distribution function 
(\ref{bunpupu})} 
For the purpose of comparison, instead of (\ref{bunpupu}), we take 
the quasi equilibrium distribution function, 
\begin{equation} 
N^{(\xi)} (X_0; p_0) \, \rule[-3mm]{.14mm}{8.5mm} 
\raisebox{-2.85mm}{\scriptsize{$\;\mbox{\scriptsize{ref}}$}} = 
\frac{1}{e^{\sqrt{p^2 - |{\cal M}_\xi^2 (X_0)|} / T} - 1} . 
\label{kasou} 
\end{equation} 
Same approximations as above yield, 
\begin{eqnarray} 
{\cal I}_\xi^{(\beta)} (X_0) & \simeq & \frac{T^2}{12} + \frac{i 
T}{4 \pi} \sqrt{|{\cal M}_\xi^2 (X_0)|} , 
\label{moya} \\ 
\ddot{\vec{\varphi}} (X_0) & = & \left[ \omega^2 - \frac{\lambda}{6} 
\vec{\varphi}^{\, 2} (X_0) \right] \, \vec{\varphi} (X_0) + 
\vec{\zeta} (X_0) + H \vec{e} , 
\label{moya-1} \\ 
& & = - \frac{{\partial \cal V}}{\partial \vec{\varphi} (X_0)} + 
\vec{\zeta} (X_0) , 
\label{moya-21} \\ 
{\cal V} & \simeq & \frac{\pi^2}{90} N T^4 - \frac{\omega^2}{2} \, 
\vec{\varphi}^{\, 2} (X_0) + \frac{\lambda}{4 !} \, \left( 
\vec{\varphi}^{\, 2} (X_0) \right)^2 - H (\displaystyle{ 
\raisebox{0.9ex}{\scriptsize{$t$}}} \mbox{\hspace{-0.1ex}} 
\vec{e} \cdot \vec{\varphi} (X_0)) , \nonumber \\ 
& & 
\label{moya-3} \\ 
\langle \zeta_i (X_0) \zeta_j (Y_0) \rangle_\zeta & \simeq & - 
\frac{\lambda}{2 4 \pi} ( N + 2 ) \frac{\omega_0 T}{V} \, 
\delta_{i j} \, \delta (X_0 - Y_0) , 
\label{moya-5} \\ 
\langle \zeta_i (X_0) \zeta_j (Y_0) \zeta_k (U_0) \zeta_l (V_0) 
\rangle_{\zeta c} & \simeq & - \frac{\lambda^2}{144 \pi} \frac{N + 
8}{V^3} \frac{T}{\omega_0} \left[ \delta_{i k} \delta_{l j} + 
\delta_{k j} \delta_{l i} + \delta_{i j}\delta_{l k} \right] 
\nonumber \\ 
& & \times \delta (X_0 - Y_0) \, \delta (X_0 - U_0 ) \, \delta (V_0 
- Y_0 ) . 
\label{moya-10} 
\end{eqnarray} 
Let us summarize distinctive features of our results when compared 
to (\ref{moya})~-~(\ref{moya-10}). 
\begin{itemize} 
\item 
Change in $\lambda$, $\lambda \to \lambda'$ [Eq.~(\ref{lam})], does 
not occur in (\ref{moya})~-~(\ref{moya-10}). We emphasize again that 
$\lambda' < \lambda$. 
\item 
Equation (\ref{moya-10}) is two times (\ref{iha}), so that, besides 
the change $\lambda \to \lambda'$, (\ref{corr-4}), $1 / V^3$ part of 
${\cal C}_{i j}^{(2)}$ in (\ref{soukan-12}), and (\ref{4-ptpt}) are 
one-half of the corresponding formulae with the reference 
distribution (\ref{kasou}). 
\item 
We rewrite the potential ${\cal V}$, Eq.~(\ref{eqevo10}), in the 
form, 
\begin{equation} 
{\cal V} \simeq \frac{\pi^2}{90} N T^4 - \frac{\omega^2 + \Delta 
\omega^2}{2} \, \vec{\varphi}^{\, 2} (X_0) + \frac{\lambda}{4 !} 
\, \left( \vec{\varphi}^{\, 2} (X_0) \right)^2 - H 
(\displaystyle{ \raisebox{0.9ex}{\scriptsize{$t$}}} 
\mbox{\hspace{-0.1ex}} \vec{e} \cdot \vec{\varphi} (X_0)) . 
\label{bui} 
\end{equation} 
Here 
\begin{eqnarray*} 
\omega^2 + \Delta \omega^2 & = & \omega^2 + 2 (N + 8) \left( 
\frac{\lambda}{2 4 \pi} \right)^2 \frac{T}{\omega} \vec{\varphi}^{\, 
2} (X_0) \nonumber \\ 
& \simeq & - \mu^2 - \frac{\lambda}{7 2} (N + 2) (T + \Delta T)^2 
\end{eqnarray*} 
with 
\begin{eqnarray} 
\Delta T & \simeq & - \frac{\lambda}{8 \pi^2} \frac{N + 8}{N + 2} 
\frac{\vec{\varphi}^{\, 2} (X_0)}{\omega} . 
\label{moso-10} 
\end{eqnarray} 
Equation (\ref{bui}) is to be compared with (\ref{moya-3}). 
Comparing $\omega^2 + \Delta \omega^2$ with 
\begin{equation} 
\omega^2 = - \mu^2 - \frac{\lambda}{7 2} (N + 2) T^2 , 
\end{equation} 
we see that, as $\vec{\varphi} (X_0)$ is \lq\lq rolling down'' the 
potential hill, the temperature \lq\lq decreases,'' and thus the 
system \lq\lq cools.'' 
\item 
We write the distribution function (\ref{bunpupu}) in the form, 
\begin{eqnarray} 
N^{(\xi)} (X_0; p_0) & = & \frac{1}{e^{\sqrt{p^2 - |{\cal M}_\xi^2 
(X_0)| - e_\xi \lambda \vec{\varphi}^{\, 2} (X_0)} / T} - 1} 
\nonumber \\ 
& \simeq & \frac{1}{e^{\sqrt{p^2 - |{\cal M}_\xi^2 (X_0)|} / 
T_{eff} (X_0)} - 1} \;\;\;\;\; (e_\pi = 1 / 6 , \, e_\sigma = 1 /2) 
, 
\label{genjitsu} 
\end{eqnarray} 
where 
\begin{equation} 
\frac{T_{eff} (X_0)}{T} = \frac{\sqrt{p^2 - |{\cal M}_\xi^2 (X_0)|} 
}{\sqrt{p^2 - |{\cal M}_\xi^2 (X_0)| - e_\xi \lambda 
\vec{\varphi}^{\, 2} (X_0)} } . 
\label{eff-T} 
\end{equation} 
Let us compare this with the reference distribution (\ref{kasou}). 
We first note that the parts of the stable (oscillator) modes in 
(\ref{kasou}), 
\begin{equation} 
|{\cal M}_\xi^2 (X_0)| \leq p^2 \leq |{\cal M}_\xi^2 (X_0)| + 
e_\xi \lambda \vec{\varphi}^{\, 2} (X_0) , 
\label{huan} 
\end{equation} 
are the unstable modes in (\ref{genjitsu}) and thus the \lq\lq 
number'' of unstable modes in the latter are larger than that in the 
former. For the modes, $|{\cal M}_\xi^2 (X_0)| + e_\xi \lambda 
\vec{\varphi}^{\, 2} (X_0) \geq p^2$, (\ref{eff-T}) shows that the 
effective temperature increases when compared to the reference 
distribution (\ref{kasou}). This temperature increase is interpreted 
as follows. As seen above, while $\vec{\varphi} (X_0)$ becomes 
large, the potential ${\cal V}$ becomes small and the energy of the 
condensate fields $\vec{\varphi} (X_0)$ is released and converted to 
the kinetic energy of the quantum fields. On the other hand, for the 
unstable modes, $p^2 \leq |{\cal M}_\xi^2 (X_0)|$, (\ref{eff-T}) 
shows that the effective temperature decreases when compared to the 
reference distribution (\ref{kasou}). These unstable modes yield 
negative contribution to $Re \, {\cal I}_\xi^{(\beta)}$ in 
(\ref{owa-true}). All the effects mentioned above compete and $Re \, 
{\cal I}_\xi^{(\beta)}$ turns out to be smaller than the 
reference-distribution counterpart. 
\end{itemize} 
\subsection{Nonuniform system} 
In this subsection, we briefly analyze the spatially nonuniform 
system under the similar situation as in the last subsection. We 
assume that the initial distribution function is given by 
(\ref{shoki}): 
\begin{equation} 
N^{(\xi)} (X_0 = 0, {\bf X}; P) = \frac{1}{e^{|p_0| / T} - 1} . 
\label{i-data} 
\end{equation} 
The determining equation for $f^{(\xi)} (X; P) = \theta (p_0) 
N^{(\xi)} (X; P) - \theta (- p_0) (1 + N^{(\xi)} (X; - P))$, 
Eq.~(\ref{ashi}), may be solved as in (\ref{resmm-sol}): 
\begin{equation} 
N^{(\xi)} (X; P) \simeq {\cal F} (\underline{\bf X}; P^2 - {\cal 
M}_\xi^2 (X), \underline{\bf p}) , 
\label{kotae} 
\end{equation} 
where $\underline{\bf X}$ and $\underline{\bf p}$ are as in 
(\ref{kai-tame}) with $T_{in} = 0$. Substitution of the initial data 
(\ref{i-data}) into (\ref{kotae}) shows that ${\cal F}$ is 
independent of $\underline{\bf X}$ and $\underline{\bf p}$ and 
$N^{(\xi)} (X; P)$ becomes (\ref{bunpupu}): 
\[ 
N^{(\xi)} (X; P) = \frac{1}{e^{\sqrt{p^2 + M_\xi^2 (X_0 = 0, 
{\bf X})} / T} - 1} . 
\] 
At the initial time $X_0 = 0$, $|\vec{\varphi}|$ and $|\partial 
\vec{\varphi} / \partial X_\mu|$ are small, so that (cf. 
(\ref{pipi}) and ({\ref{sigma})) 
\[ 
M_\xi^2 (X_0 = 0, {\bf X}) \simeq - \omega^2 \;\;\;\;\;\; (\xi = 
\pi, \sigma) , 
\]  
which is independent of ${\bf X}$. In place of (\ref{iha-0}) and 
(\ref{iha}), we have, in respective order, 
\begin{eqnarray} 
\langle \zeta_i (X) \zeta_j (Y) \rangle_\zeta & \simeq & - 
\frac{\lambda}{2 4 \pi} (N + 2) \, T \omega \delta_{i j} \, 
\delta^4 (X - Y) , 
\label{iha-1} \\ 
\langle \zeta_i (X) \zeta_j (Y) \zeta_k (U) \zeta_l (V) 
\rangle_{\zeta c} & \simeq & - \frac{\lambda^2}{288 \pi} (N + 8) 
\frac{T}{\omega} \left[ \delta_{i j} \delta_{k l} + \delta_{i k} 
\delta_{j l} + \delta_{i l} \delta_{j k}  \right] \nonumber \\ 
& & \times \delta^4 (X - Y) \delta^4 (X - U) 
\delta^4 (V - Y) . 
\label{yon} 
\end{eqnarray} 
In this subsection, we are only concerned with a zeroth-order 
solution for $\vec{\varphi} (X)$ (cf. (\ref{kai-yo})), 
\begin{equation} 
\vec{\varphi}^{\, (0)} (X) = \vec{\varphi}^{\, (0)}_{pure} (X) 
+ \int \frac{d^{\, 3} p}{(2 \pi)^3} \frac{e^{i {\bf p} \cdot {\bf 
X}}}{\omega'} \int_0^{X_0} d X_0' \vec{\zeta} (X_0'; {\bf p}) 
\sinh \left( \omega' (X_0 - X_0') \right) , 
\label{non-u} 
\end{equation} 
where 
\begin{eqnarray} 
\omega' & = & \sqrt{\omega^2 - p^2} , 
\label{ome-dash} \\  
\vec{\zeta} (X_0'; {\bf p}) & = & \int d^{\, 3} X e^{- i {\bf p} 
\cdot {\bf X}} \vec{\zeta} (X_0', {\bf X}) . 
\end{eqnarray} 
For, $\omega^2 - p^2 < 0$, $\omega'$ should be understood to be 
$\omega' = i \sqrt{p^2 - \omega^2}$. In (\ref{non-u}), 
$\vec{\varphi}^{\, (0)}_{pure} (X)$ is the solution to the equation 
\[ 
(\partial_X^2 - \omega^2 ) \vec{\varphi}^{(0)}_{pure} (X) = H 
\vec{e} , 
\] 
under a given initial data. 
\subsubsection{Two-point $\zeta$-correlation function} 
Using (\ref{iha-1}), one can readily compute the two-point 
$\zeta$-correlation function: 
\begin{eqnarray} 
\langle \varphi_i^{(0)} (X) \varphi_j^{(0)} (Y) 
\rangle_{\zeta c} & = & - \frac{\lambda}{9 6 \pi} \delta_{i j} 
(N + 2) \omega T \int \frac{d^{\, 3} p}{(2 \pi)^3} e^{i {\bf p} 
\cdot ({\bf X} - {\bf Y})} \frac{1}{\omega^2 - p^2} \nonumber \\ 
& & \times \left[ \frac{\sinh (\sqrt{\omega^2 - p^2} (X_0 + Y_0)) - 
\sinh (\sqrt{\omega^2 - p^2} (X_0 - Y_0)}{\sqrt{\omega^2 - p^2}} 
\right. \nonumber \\ 
& & \left. - 2 Y_0 \cosh (\sqrt{\omega^2 - p^2} (X_0 - Y_0)) \right] 
\nonumber \\ 
& = & - \frac{\lambda}{192 \pi^3} \delta_{i j} (N + 2) 
\frac{\omega T}{|{\bf X} - {\bf Y}|} \int_0^\infty d p \, p 
\, \sin (p |{\bf X}  - {\bf Y}|) \frac{1}{\omega^2 - p^2} 
\nonumber \\ 
& & \times \left[ \theta (\omega^2 - p^2) \left\{ \frac{\sinh 
(\sqrt{\omega^2 - p^2} (X_0 + Y_0)) - \sinh (\sqrt{\omega^2 - p^2} 
(X_0 - Y_0))}{\sqrt{\omega^2 - p^2}} \right. \right. \nonumber \\ 
& & \left. - 2 Y_0 \cosh (\sqrt{\omega^2 - p^2} (X_0 - Y_0)) 
\right\} \nonumber \\ 
& & + \theta (p^2 - \omega^2) \left\{ \frac{\sin (\sqrt{p^2 - 
\omega^2} (X_0 + Y_0)) - \sin (\sqrt{p^2 - \omega^2} (X_0 - 
Y_0))}{\sqrt{p^2 - \omega^2}} \right. \nonumber \\ 
& & \left. \left. - 2 Y_0 \cos (\sqrt{p^2 - \omega^2} (X_0 - Y_0) ) 
\right\} \right] , 
\label{yoyo} 
\end{eqnarray} 
where $X_0 \geq Y_0$ has been assumed. Manipulation of (\ref{yoyo}) 
in Appendix E yields 
\begin{eqnarray} 
\langle \varphi_i^{(0)} (X) \varphi_j^{(0)} (Y) \rangle_{\zeta c} 
& = & - \frac{\lambda}{192 \pi^3} (N + 2) \delta_{i j} 
\frac{T}{|{\bf X} - {\bf Y}|} \nonumber \\ 
& & \times \left[ \theta (X_0 + Y_0 - |{\bf X} - {\bf Y}|) \, \theta 
(|{\bf X} - {\bf Y}| - (X_0 - Y_0)) \right. \nonumber \\ 
& & \times \left\{ \frac{\pi \omega}{2} (X_0 + Y_0) \cos (\omega 
|{\bf X} - {\bf Y}|) - \sin (\omega |{\bf X} - {\bf Y}|) \nonumber 
\right. \\ 
& & + \int_0^1 d \xi \frac{\xi}{(1 - \xi^2)^{3 / 2}} \left[ \sin 
(\omega |{\bf X} - {\bf Y}| \xi) \cosh \left( \sqrt{1 - \xi^2} 
\omega (X_0 + Y_0) \right) \right. \nonumber \\ 
& & \left. \left. - \sin (\omega |{\bf X} - {\bf Y}|) \right] 
\right\} \nonumber \\ 
& & + \theta (X_0 - Y_0 - |{\bf X} - {\bf Y}|) \nonumber \\ 
& & \times \left\{ \int_0^1 d \xi \frac{\xi \sin (\omega |{\bf X} - 
{\bf Y}| \xi)}{(1 - \xi^2)^{3 / 2}} \sum_{\tau = \pm} \tau \cosh 
\left( \sqrt{1 - \xi^2} \omega (X_0 + \tau Y_0) \right) \right. 
\nonumber \\ 
& & \left. \left. - 2 \omega Y_0 \int_0^1 d \xi \frac{\xi \sin 
(\omega |{\bf X} - {\bf Y}| \xi)}{1 - \xi^2} \sinh \left( \sqrt{1 - 
\xi^2} \omega (X_0 - Y_0) \right) \right\} \right] . 
\label{yoyoto} 
\end{eqnarray} 
First of all, we note that $\langle \varphi_i^{(0)} (X) 
\varphi_j^{(0)} (Y) \rangle_{\zeta c}$ vanishes for $|{\bf X} - {\bf 
Y}| > X_0 + Y_0$, which means that the correlation does not spread 
with super-light velocity. Several limiting cases are analyzed in 
Appendix E. 

1) $X_0 - Y_0 < |{\bf X} - {\bf Y}| < X_0 + Y_0$; $\, \omega 
|{\bf X} - {\bf Y}| < \omega (X_0 + Y_0) << 1$: 
\begin{eqnarray} 
\langle \varphi_i (X) \varphi_j (Y) \rangle_{\zeta c} & \simeq & - 
\frac{\lambda}{384 \pi^2} \, (N + 2) \delta_{i j} \frac{\omega 
T}{|{\bf X} - {\bf Y}| } \nonumber \\ 
& & \times \left[ X_0 + Y_0 - |{\bf X} - {\bf Y}| \left\{ 1 - 
\frac{\omega^2 (X_0 + Y_0)^2}{4} \right\} \right] . 
\label{chiisai-t} 
\end{eqnarray} 
This shows that the correlation is negative. For fixed $|{\bf X}|$ 
and $|{\bf Y}|$, the correlation is maximum at the smallest $|{\bf 
X} - {\bf Y}|$, $|{\bf X} - {\bf Y}| = \mbox{Max} (X_0 - Y_0, 
||{\bf X}| - |{\bf Y}||)$ and is minimum at the largest $|{\bf X} - 
{\bf Y}|$, $|{\bf X} - {\bf Y}| = \mbox{Min} (X_0 + Y_0, |{\bf X}| 
+ |{\bf Y}|)$. We are dealing with the zeroth-order solution to the 
evolution equation (the counterpart to (\ref{eqevo})), i.e., the 
nonlinear term, $- \lambda' \vec{\varphi}^{\, 2} \vec{\varphi} / 6$, 
in the evolution equation has been ignored. The condition that the 
ignored term is small restrict the domain, where (\ref{chiisai-t}) 
is valid, to 
\[ 
\lambda \lambda' \frac{X_0 + Y_0 - |{\bf X} - {\bf Y}|}{|{\bf X}- 
{\bf Y}|} < O \left(\frac{\omega}{T} \right) . 
\] 

2) $1 << \omega (X_0 + Y_0)$ and $\omega |{\bf X} - {\bf Y}| << 
\sqrt{\omega (X_0 + Y_0)}$: 
\begin{eqnarray} 
\langle \varphi_i (X) \varphi_j (Y) \rangle_{\zeta c} & \simeq & - 
\frac{\sqrt{2 \pi}}{768 \pi^3} \, \lambda (N + 2) \delta_{i j} 
\omega T \nonumber \\ 
& & \times \left[ \frac{1}{(\omega (X_0 + Y_0))^{3 / 2}} \, 
\mbox{exp} \left( \omega \left\{ X_0 + Y_0 - \frac{|{\bf X} - {\bf 
Y}|^2}{2 (X_0 + Y_0)} \right\} \right) \right. \nonumber \\ 
& & \times \left( 1 + O \left( \frac{1}{\omega (X_0 + Y_0)} \right) 
\right) \nonumber \\ 
& & + \sqrt{2 \pi} \theta (X_0 + Y_0 - |{\bf X} - {\bf Y}|) \theta 
(|{\bf X} - {\bf Y}| - (X_0 - Y_0)) \nonumber \\ 
& & \left. \times \frac{X_0 + Y_0}{|{\bf X} - {\bf Y}|} \cos (\omega 
|{\bf X} - {\bf Y}| ) \right] , 
\label{ooki-1} 
\end{eqnarray} 
which is valid for 
\[ 
\lambda \lambda' e^{\omega (X_0 + Y_0)} < O \left( \frac{\omega}{T} 
\{\omega (X_0 + Y_0) \}^{3 / 2}\right) \;\;\; \mbox{and} \;\;\; 
\lambda \lambda' \frac{X_0 + Y_0}{|{\bf X} - {\bf Y}|} < O \left( 
\frac{\omega}{T} \right) . 
\] 
Equation (\ref{ooki-1}) indicates that, for large $X_0 + Y_0$ with 
$\omega |{\bf X} - {\bf Y}| << \sqrt{\omega (X_0 + Y_0)}$, the 
(negative) correlation spreads over the distance, 
\[ 
|{\bf X} - {\bf Y}| \sim \sqrt{\frac{2 (X_0 + Y_0)}{\omega}} , 
\] 
and asymptotically spreads over whole space region. It should be 
emphasized, however, that, for too large $X_0 + Y_0$, our whole 
approximations in this section cease to hold. 

3) $|{\bf X} - {\bf Y}| < X_0 - Y_0$; $\, \omega (X_o + Y_0) << 1$: 
\[ 
\langle \varphi_i (X) \varphi_j (Y) \rangle_{\zeta c} \simeq - 
\frac{\lambda}{384 \pi^2} (N + 2) \delta_{i j} \omega T (\omega 
Y_0)^2 . 
\] 
\subsubsection{Four-point $\zeta$-correlation function} 
Using (\ref{non-u}) and (\ref{yon}), we obtain 
\begin{eqnarray} 
& & \langle \varphi_{i_1}^{(0)} (X_1) \varphi_{i_2}^{(0)} (X_2) 
\varphi_{i_3}^{(0)} (X_3) \varphi_{i_4}^{(0)} (X_4) \rangle_{\zeta 
c} \nonumber \\ 
& & \mbox{\hspace*{8ex}} \simeq - \frac{\lambda^2}{2 88 \pi} (N + 8) 
\frac{T}{\omega} \left[ \delta_{i_1 i_2} \delta_{i_3 i_4} + 
\delta_{i_1 i_3} \delta_{i_2 i_4} + \delta_{i_1 i_4} \delta_{i_2 
i_3} \right] \nonumber \\ 
& & \mbox{\hspace*{11ex}} \times \int_0^{X_0} d X_0' \int d {\bf X}' 
\prod_{j = 1}^4 \int \frac{d^{\, 3} p_j}{(2 \pi)^3} \frac{e^{i {\bf 
p}_j \cdot ({\bf x}_j - {\bf x}')}}{\sqrt{\omega^2 - p_j^2}} \sinh 
\left( \sqrt{\omega^2 - p_j^2} (X_{j 0} - X_0')  \right) , 
\label{sorayo} 
\end{eqnarray} 
where $X_{1 0}, X_{2 0}, X_{3 0} \geq X_{4 0}$ has been assumed. 
For equal times, $X_{1 0} = X_{2 0} = X_{3 0} = X_{4 0}$ $(\equiv 
X_0)$, computation of (\ref{sorayo}) in the limit, $\omega X_0 >> 1$ 
and $|{\bf X}_i - {\bf X}_j| << X_0$ $(i, j = 1, .., 4)$ is 
relatively simple. Similar calculation as in Appendix E yields 
\begin{eqnarray} 
& & \langle \varphi_{i_1}^{(0)} (X_0, {\bf X}_1) \varphi_{i_2}^{(0)} 
(X_0, {\bf X}_2) \varphi_{i_3}^{(0)} (X_0, {\bf X}_3) 
\varphi_{i_4}^{(0)} (X_0, {\bf X}_4) \rangle_{\zeta c} \nonumber \\ 
& & \mbox{\hspace*{8ex}} \simeq - \frac{\lambda^2}{288 \pi} (N + 8) 
\left[ \delta_{i_1 i_2} \delta_{i_3 i_4} + \delta_{i_1 i_3} 
\delta_{i_2 i_4} + \delta_{i_1 i_4} \delta_{i_2 i_3} \right] 
\nonumber \\ 
& & \mbox{\hspace*{11ex}} \times \omega^3 T \frac{e^{4 \omega 
X_0}}{(8 \pi \omega X_0)^{9 / 2}} \mbox{exp} \left( - 
\frac{\omega}{8 X_0}\sum_{i < j} ({\bf X}_i - {\bf X}_j)^2\right) , 
\label{soraho} 
\end{eqnarray} 
which is valid for 
\[ 
(\lambda \lambda')^2 e^{4 \omega X_0} < O \left( \frac{\omega}{T} 
(\omega X_0)^{9 / 2}\right) . 
\] 
It should be emphasized that, if we use the reference distribution 
(\ref{kasou}), we obtain two times (\ref{soraho}). 
\setcounter{equation}{0} 
\setcounter{section}{8} 
\def\theequation{\mbox{\arabic{section}.\arabic{equation}}} 
\section{Concluding remarks} 
In this paper, two related subjects are dealt with: 
\begin{description} 
\item{A)} Deduction from first principles of the perturbative 
framework for dealing with out-of-equilibrium relativistic 
complex-scalar-field systems. 
\item{B)} Deduction from first principles of the 
perturbative-loop-expansion framework for studying an $O (N)$ linear 
$\sigma$ model. 
\end{description} 
We have assumed the existence of two different spacetime scales, the 
microscopic and macroscopic. The first small scale characterizes the 
microscopic correlations and the second large scale is inherent in 
the relaxation phenomena. Besides this setup, in principle, we have 
not made use of any further approximations. Let us summarize what 
has been added and what has been clarified in this paper. Some 
additional comments are also given. 
\subsubsection{On the issue A)} 
\begin{description} 
\item{(1)} The assumption that is necessary for the propagator to 
take the standard form has been singled out. In most available 
work, among $2 n$-point initial correlation functions ${\cal W}_{2 
n}$ $(n \geq 1)$, Eq.~(\ref{sora}), ${\cal W}_{2 n}$ $(n \geq 2)$ 
are dropped and only ${\cal W}_2$ is kept into the propagator. This 
is so done as an approximation or by assuming some specific form for 
the density matrix. In this paper, we have seen that, to what 
extent, ${\cal W}_{2 n}$ $(n \geq 2)$ may be discarded. Whenever 
necessary, one can incorporate them into the perturbative framework 
(cf. \cite{chou}). 
\item{(2)} We have emphasized that the perturbative framework is 
effective only when the inverse size of a spacetime cell is much 
smaller than the infrared scale in the microscopic sector of the 
theory. In the present complex-scalar-field theory, this condition 
is $(L^\mu)^{- 1} \lesssim \sqrt{\lambda} {\cal P}$ $(\mu = 0, 1, 2, 
3)$ with $\lambda$ the coupling constant and ${\cal P}$ the typical 
parameter(s) (of mass dimension) of the system. 
\item{(3)} We have proposed mutually equivalent two perturbative 
schemes. Both of them lead to a generalized relativistic kinetic or 
Boltzmann equation. Derivation of it has proceeded in a physically 
transparent manner. In the physical-$f$ scheme, the equation is 
derived through the process of redefinition of the number-density or 
relativistic Wigner function, from the initial number-density 
function to the physically sensible one. Traditional derivation of 
the equation proceeds \cite{chou,hu} in a rather abstract way, using 
the Dyson equation. (For the case of nonrelativistic many-body 
theory, see, e.g., \cite{ram}.) As seen in Section V, the Dyson 
equation is the equation that serves as determining the full 
propagator in terms of the self-energy part. In other word, the 
Dyson equation simply serves as an efficient way of resumming the 
self-energy part. Then, by referring to the full propagator thus 
obtained, one introduces \cite{chou,hu,ram} a redefined distribution 
function. It is this function that subjects to the relativistic 
Boltzmann equation. 
\item{(4)} In massless theories, divergence due to infrared and/or 
mass (or collinear) singularities appear in some amplitudes. This 
occurs as a result of an interplay of {\em bare} massless 
propagators. By performing resummations of the self-energy part for 
such propagators (cf. (\ref{re-re})), one can get rid of this 
divergence disaster (cf., e.g., \cite{chou,alt1}. For a different 
calculational scheme, in which no divergence emerges, see 
\cite{nie-pl}. 
\end{description} 
\subsubsection{On the issue B)} 
\begin{description} 
\item{(1)} Through introducing effective masses, which depend on the 
macroscopic spacetime coordinates, we have derived self-consistent 
gap equation for them. Together with the generalized relativistic 
Boltzmann equation (cf. item (3) above) and the equation of motion 
for the condensate fields, this constitutes the system of coupled 
equations, which describes how the phase transition proceeds. It 
should be emphasized, however, that, as for other work of this sort, 
the present scheme applies only for the systems away from the 
critical region.  
\item{(2)} The negative curvature region of the \lq\lq potential'' 
has been dealt with by introducing random-force fields. To leading 
one-loop order, this leads to negative correlation between the 
condensate fields. 
\item{(3)} As in the issue A) above, within the gradient 
approximation, the propagator consists of two parts, $\hat{\bf 
\Delta} = \hat{\bf \Delta}^{(0)} + \hat{\bf \Delta}^{(1)}$, where 
$\hat{\bf \Delta}^{(0)}$ is the dominant part and the \lq\lq 
correction part'' $\hat{\bf \Delta}^{(1)}$ contains a derivative 
with respect to the macroscopic spacetime coordinates. In a 
$(\vec{\pi}, \sigma)$-space, $\hat{\bf \Delta}^{(0)}$ is diagonal, 
while $\hat{\bf \Delta}^{(1)}$ is not. This means that, in the level 
of $\hat{\bf \Delta}^{(1)}$, there occurs mixing between 
$\vec{\pi}$ and $\sigma$ sectors. Then, when one takes the part 
$\hat{\bf \Delta}^{(1)}$ into account, rediagonalization of the 
propagator in the $(\vec{\pi}, \sigma)$-space is necessary, which 
leads to \lq\lq physical'' $\vec{\pi}$ and $\sigma$. 
\end{description} 

Finally, we mention two related works \cite{ume,ume1,law}. In the 
theories developed there, a \lq\lq generalized mass (function)'' 
is introduced, which is $2 \times 2$ matrix in the \lq\lq 
complex-time plane'' and matrix elements are complex functions of 
spacetime. Due to the last fact, the theories provide for a 
treatment of dissipative effects. The \lq\lq generalized mass 
(function)'' is determined self consistently through renormalization 
\cite{ume,ume1} or through some optimization procedure \cite{law}. 
Through this process, the Boltzmann equation emerges. As applied to 
the complex-scalar field theory, the part of the hat-Lagrangian 
density that include the \lq\lq generalized mass (function)'' 
corresponds to (cf. (\ref{kita1})~-~(\ref{count}) and (\ref{Lag-2})) 
\[ 
- \sum_{j = 1}^2 (-)^{j - 1} \phi_j^* (x) M^2 (x) \phi_j (x) - 
 \hat{\cal L}_c 
\] 
(with $\hat{\cal L}_c$ as in (\ref{count})) in this paper. 
\section*{Acknowledgments}
This work was supported in part by the Grant-in-Aide for Scientific 
Research ((A)(1) (No.~08304024)) of the Ministry of Education, 
Science and Culture of Japan. 
\setcounter{equation}{0}
\setcounter{section}{1}
\section*{Appendix A: Mass-derivative formula} 
\def\theequation{\mbox{\Alph{section}\arabic{equation}}} 
Here an argument for necessity of adopting the $|p_0|$ prescription 
is given. We add ${\cal L}_m = - \delta m^2 \phi^\dagger \phi$ to 
the Lagrangian density (\ref{Lag-1}). The effect of ${\cal L}_m$ on 
the propagator $\hat{\Delta} (X; P)$ should be 
\begin{equation} 
\hat{\Delta} (X; P) \to \hat{\Delta} (X; P) \, 
\rule[-3mm]{.14mm}{8.5mm} \raisebox{-2.85mm}{\scriptsize{$\; m^2 
\to m^2 + \delta m^2$}} . 
\label{mass-der} 
\end{equation} 
Let us evaluate the $O (\delta m^2)$ correction, $\delta 
\hat{\Delta}$, 
to $\hat{\Delta}$ due to the perturbation ${\cal L}_m$: 
\begin{eqnarray*} 
\delta \hat{\Delta} (x, y) & = & \delta m^2 \int d^{\, 4} u \, 
\hat{\Delta} (x, u) \, \hat{\tau}_3 \, \hat{\Delta} (u, y) \\ 
& \simeq & \delta m^2 \int d^{\, 4} u 
\left[ \hat{\Delta} (X, x - u) + \frac{u - y}{2} \cdot \partial_X 
\hat{\Delta} (X, x - u) \right] \\ 
& & \times \hat{\tau}_3 
\left[ \hat{\Delta} (X, u - y) + \frac{u - x}{2} \cdot \partial_X 
\hat{\Delta} (X, u - y) \right] , 
\end{eqnarray*} 
where $X = (x + y) / 2$. Taking the Fourier transform with respect 
to $x - y$ and using $\hat{A}_+ \hat{\tau}_3 \hat{A}_+ = 0$ and 
(\ref{cont-f}), we obtain 
\begin{equation} 
\delta \hat{\Delta} (X; P) \simeq \delta m^2 \, \hat{\Delta} (X; P) 
\, \hat{\tau}_3 \, \hat{\Delta} (X; P) . 
\label{mass-d} 
\end{equation} 
It is to be noted that (\ref{mass-d}) is valid to the gradient 
approximation, i.e., the term with first derivative (with respect to 
$X$) is absent in (\ref{mass-d}). 

Substituting (\ref{free-p}), we finally obtain 
\begin{equation} 
\delta \hat{\Delta} (X; P) \simeq \delta m^2 \left[ 
\left( 
\begin{array}{cc}
\Delta_R^2 & \;\;\; 0 \\ 
\Delta_R^2 - \Delta_A^2 & \;\;\; - \Delta_A^2 
\end{array}
\right) 
+ f (\Delta_R^2 - \Delta_A^2) \hat{A}_+ \right] . 
\label{del-1} 
\end{equation} 
If we adopt the $|p_0|$ prescription, $f = f (X; P)$ is independent 
of $m^2$ (cf. (\ref{p0f})), so that $\delta \hat{\Delta}$ in 
(\ref{del-1}) may be written as 
\begin{equation} 
\delta \hat{\Delta} (X; P) = \delta m^2 \, \frac{\partial 
\hat{\Delta} (X; P)}{\partial m^2} , 
\label{mass-der-fin} 
\end{equation} 
which is in accord with (\ref{mass-der}). On the other hand, if we 
take original $f$ (Eq.~(\ref{p0f})), $f = \theta (p_0) N_+ (X; E_p, 
\hat{\bf p}) - \theta (- p_0) [1 + N_+ (X; E_p, - \hat{\bf p})]$, 
it depends on $m^2$ through $E_p$ and $\delta \hat{\Delta}$ in 
(\ref{del-1}) cannot be written in the \lq\lq mass-derivative form'' 
(\ref{mass-der-fin}). We have confirmed the consistency between 
the mass-derivative formula and the $|p_0|$ prescription at least up 
to the terms with second-order derivative with respect to $X$. 
\setcounter{equation}{0}
\setcounter{section}{2}
\section*{Appendix B: Properties of the self-energy part} 
\def\theequation{\mbox{\Alph{section}\arabic{equation}}}
\subsection*{B.1. Proof of $\Sigma_{1 1} + \Sigma_{1 2} + \Sigma_{2 
1} + \Sigma_{2 2} = 0$} 
Let $\Sigma_l (x, y; z_1, ..., z_n)$ be a contribution from a loop 
diagram to the self-energy part in configuration space, where $x$ 
and $y$ are the space-time coordinates of external vertices and 
$z_1, ..., z_n$ are those of internal vertices. Here an external 
vertex means a vertex to which an external leg is attached. The 
times $x_0, y_0, z_{1 0}, ..., z_{n 0}$ are on the closed time path 
$C_1 \oplus C_2$. 

Let us first show that, when one of the internal vertices, say 
$z_{i 0}$, is the largest time, 
\begin{equation} 
z_{i 0} > x_0, \, y_0, \; \mbox{other} \; z_0\mbox{'s} , 
\label{large-yo} 
\end{equation} 
$\Sigma_l$ vanishes \cite{kobes}. This can be proved as follows. 
From (\ref{bun}) and (\ref{bun-1}), we see that a vertex factor with 
$z_{i 0} \in C_1$ possesses an opposite sign relative to the 
corresponding vertex factor with $z_{i 0} \in C_2$: 
\begin{equation} 
(i V_4)_1 = - (i V_4)_2 , \;\; (i V_2)_1 = - (i V_2)_2 , \;\; (i 
V_4')_1 = - (i V_4')_2 . 
\label{sei-ch} 
\end{equation} 
On the other hand, as can be easily be seen from (\ref{shu-pa}), all 
the propagators in $\Sigma_l$ are the same for $z_{i 0} \in C_1$ and 
$z_{i 0} \in C_2$. 

Let $(\Sigma_l)_{i j}$ be $\Sigma_l (x, y; z_1, ..., z_n)$ with $x_0 
\in C_i$ and $y_0 \in C_j$. Similar observation as above shows that 
\begin{eqnarray*} 
& & (\Sigma_l)_{1 i} + (\Sigma_l)_{2 i} = 0 \;\;\;\; (i = 1, 2) 
\;\;\; \mbox{for} \; x_0 > y_0, z_1, ..., z_n , \\ 
& & (\Sigma_l)_{i 1} + (\Sigma_l)_{i 2} = 0 \;\;\;\; (i = 1, 2) 
\;\;\; \mbox{for} \; y_0 > x_0, z_1, ..., z_n . 
\end{eqnarray*} 
Then, we obtain 
\begin{equation} 
\sum_{i, j = 1}^2 (\Sigma_l (x, y; z_1, ..., z_n) )_{i j} = 0 . 
\label{zero} 
\end{equation} 
There are contributions to the self-energy part, $\Sigma_l (x; z_1, 
..., z_n)$ which include only one external vertex. In such cases, 
$(\Sigma_l)_{1 2} = (\Sigma_l)_{2 1}= 0$. Similar argument as above 
shows that 
\begin{equation} 
\sum_{i = 1}^2 (\Sigma_l (x; z_1, ..., z_n))_{i i} = 0 . 
\label{zero-1} 
\end{equation} 
There is an exception to this argument, i.e., a contribution from 
the lowest-order tadpole diagram, $(\Sigma_l (x) )_{i j}$. Explicit 
computation shows that $\sum_{i = 1}^2 (\Sigma_l (x) )_{i i} = 0$ 
for the complex-scalar field theory analyzed in Sections III-V and 
for the $O (N)$ linear $\sigma$ model in the the symmetric phase 
(cf. Section VIIA). This is not the case for the broken-symmetric 
phase (cf. Section VIIB). As a matter of fact, in the region, where 
the curvature of the \lq\lq potential'' is negative, imaginary part 
develops in $\Sigma_{1 1}$ and $\Sigma_{2 2}$: $Im \Sigma_{1 1} = Im 
\Sigma_{2 2}$. Then $\sum_{i = 1}^2 (\Sigma_l (x) )_{i i} \neq 0$ 
but $\sum_{i = 1}^2 Re (\Sigma_l (x) )_{i i} = 0$. Thus, in studying 
this region, one should take this fact into account. 

Inclusion of the vertices $i \hat{\cal V}_c (x, y)$ (Eq.~(\ref{V})) 
that comes from the counter Lagrangian does not invalidate the above 
argument. The vertex $i \hat{\cal V}_c (x, y)$ yields a contribution 
to the self-energy part $\hat{\Sigma}_c (x, y) = - \hat{\cal V}_c 
(x, y)$ (see (\ref{self0})). From (\ref{V}) with (\ref{Apm}), it is 
obvious that 
\[ 
\sum_{i, j = 1}^2 (\Sigma_c (x, y) )_{i j} = 0 . 
\] 
$i \hat{\cal V}_c$ also appears as an internal vertex of 
$\Sigma_l$'s. Using the form (\ref{V}), we can readily see that the 
similar argument as above after (\ref{large-yo}) holds, so that 
(\ref{zero}) and (\ref{zero-1}) hold. 

This completes the proof. 
\subsection*{B.2. Proof of $\Sigma_A (X; P) = [\Sigma_R (X; P)]^*$} 
From the definition of $G$'s, Eq.~(\ref{shu-pa}), it can easily be 
shown that 
\begin{eqnarray} 
i G_{1 1} (x, y) & = & \left[ i G_{2 2} (y, x) \right]^* \nonumber , 
\\ 
i G_{1 2 (2 1)} (x, y) & = & \left[ i G_{1 2 (2 1)} (y, x) \right]^* 
. 
\label{A-11} 
\end{eqnarray} 
This shows that, by taking a complex conjugate of $i G_{j l} (x, 
y)$, the index $j$ [$l$] at the end point $x$ [$y$] changes; $j = 1$ 
(respect. $2$) $\to$ $2$ (respect. $1$) [$l = 1$ (respect. $2$) 
$\to$ $2$ (respect. $1$)]. Let us define 
\[ 
\Sigma_l (x, y) \equiv \int_C \left( \prod_{j = 1}^2 d^{\, 4} z_j 
\right) \Sigma_l (x, y; z_1, ..., z_n) , 
\] 
where $\Sigma_l (x, y; z_1, ..., z_n)$ is as in the previous 
subsection. From (\ref{bun}) and (\ref{bun-1}), we obtain 
\begin{equation} 
i (V_4)_2 = [i (V_4)_1]^* , \;\; i (V_2)_2 = [i (V_2)_1]^* , \;\; 
i (V_4')_2 = [i (V_4')_1]^* . 
\label{B-11} 
\end{equation} 
Thus, again by taking a complex conjugate of $i (V_4)_j$, etc., the 
index $j$ changes; $j = 1$ (respect $2$) $\to$ $2$ (respect $1$). 
For the $O (N)$ linear $\sigma$ model, similar relations hold. As to 
the vertex $i \hat{\cal V}_c (x, y)$, Eq.~(\ref{V}), we have $i 
\hat{\cal V}_c (x, y) = \{ i \hat{\cal V}_c (y, x) \}^*$, so that 
\begin{eqnarray} 
i ({\cal V}_c (x, y) )_{1 1} & = & \{ i ({\cal V}_c (y, x) 
)_{2 2} \}^* , \nonumber \\ 
i ({\cal V}_c (x, y) )_{1 2 (2 1)} & = & \{ i ({\cal V}_c (y, x) 
)_{1 2 (2 1)} \}^* . 
\label{B-22} 
\end{eqnarray} 
Incidentally, this shows that $(\Sigma_c (x, y) )_{i j}$ ($i, \, j = 
1, 2$) in (\ref{self0}) satisfies (\ref{B-22}) with $({\cal V}_c)_{i 
j} \to (\Sigma_c)_{i j}$. 

Using (\ref{A-11})~-~(\ref{B-22}), we can show that 
\begin{eqnarray} 
- i (\Sigma_l (x, y) )_{1 1} & = & \left[ - i (\Sigma_l  (y, x) 
)_{2 2} \right]^* , \nonumber \\ 
- i (\Sigma_l (x, y) )_{1 2 (2 1)} & = & \left[ - i (\Sigma_l 
(y, x) )_{1 2 (2 1)} \right]^* . 
\label{rel} 
\end{eqnarray} 
Fourier transformation on $x - y$ yields 
\begin{eqnarray} 
& (\Sigma_l (X; P) )_{1 1} = - (\Sigma_l (X; P) )_{2 2}^* \, , & 
\; \Sigma_{1 1} (X; P) = - \Sigma_{2 2}^* (X; P) , \nonumber \\ 
& (\Sigma_l (X; P) )_{1 2} \, ,  & \; 
(\Sigma_l (X; P) )_{2 1} \; : \mbox{pure imaginary.} 
\label{rela} 
\end{eqnarray} 
Using this in (\ref{self2}), we obtain $\Sigma_A (X; P) = [\Sigma_R 
(X; P)]^*$. 
\setcounter{equation}{0}
\setcounter{section}{3}
\section*{Appendix C: Useful formula} 
\def\theequation{\mbox{\Alph{section}\arabic{equation}}}
We deal with 
\begin{equation} 
\int d^{\, 4} (x - y) e^{ i P \cdot (x - y)} \int d^{\, 4} u \, 
d^{\, 4} v \, A (x, u) B(u, v) C (v, y) . 
\label{hinagata} 
\end{equation} 
We wish to express this quantity using the macroscopic coordinate 
$X = (x + y) / 2$ up to the first order with respect to the $X$ 
dependence. To this end, we expand $A$ as 
\begin{eqnarray*} 
A (x, u) & = & A ((x + u) / 2; x - u) \\ 
& \simeq & A (X; x - u) + \frac{u - y}{2} \cdot \partial_X A 
(X; x - u) . 
\end{eqnarray*} 
$B$ and $C$ may be expanded similarly. Substituting these into 
(\ref{hinagata}) and Fourier transforming with respect to the 
relative coordinates, we obtain 
\begin{equation} 
\mbox{Eq.~(\ref{hinagata})} \simeq A B C + \frac{i}{2} \left[ 
\frac{\partial A}{\partial P^\mu} \frac{\partial B}{\partial X_\mu} 
C - A \frac{\partial B}{\partial X_\mu} \frac{\partial C}{\partial 
P^\mu} - \frac{\partial A}{\partial X_\mu} \frac{\partial B 
C}{\partial P^\mu} + \frac{\partial A B}{\partial P^\mu} 
\frac{\partial C}{\partial X_\mu} \right] , 
\label{al} 
\end{equation} 
where $A = A (X; P)$, etc. 

The second form we deal with is 
\begin{equation} 
\int d^{\, 4} (x - y) e^{ i P \cdot (x - y)} \int d^{\, 4} u \, 
d^{\, 4} v \, f (x) A (x, u) B(u, v) C (v, y) g (y) , 
\label{hina1} 
\end{equation} 
where $f (x)$ and $g (y)$ depend weakly on $x$. In a similar manner 
as above, 
we obtain 
\begin{eqnarray} 
\mbox{Eq.~(\ref{hina1})} & \simeq & f g A B C + \frac{i}{2} \left[ f 
g \left\{ \frac{\partial A}{\partial P^\mu} \frac{\partial 
B}{\partial X_\mu} C - A \frac{\partial B}{\partial X_\mu} 
\frac{\partial C}{\partial P^\mu} - \frac{\partial A}{\partial 
X_\mu} \frac{\partial B C}{\partial 
P^\mu} + \frac{\partial A B}{\partial P^\mu} \frac{\partial 
C}{\partial X_\mu} \right\} \right. \nonumber \\ 
& & \hspace*{6.5ex} \left. + \left( f 
\stackrel{\leftrightarrow}{\partial_{X_\mu}} g \right) 
\frac{\partial A B C}{\partial P^\mu} \right] , 
\label{al1} 
\end{eqnarray} 
where $f = f (X)$ and $g = g (X)$. 
\setcounter{equation}{0}
\setcounter{section}{4}
\section*{Appendix D: Renormalizability of of the theory based on 
(5.46)} 
\def\theequation{\mbox{\Alph{section}\arabic{equation}}} 
Let us call the perturbation theory based on (\ref{Lag-1}) 
[(\ref{Lag-2})] the theory I [II]. We first observe that, in theory 
II, if we include the term $\chi (x) \phi^\dagger \phi \equiv (M^2 
(x) - m^2) \phi^\dagger \phi$ in ${\cal L}_{r c}$ into ${\cal L}_0$, 
theory II reduces to theory I, which is UV-renormalizable. As stated 
in Section VE, the difference between theory I and theory II lies in 
the fact that, in theory II, $\chi (x) \phi^\dagger \phi$ is 
regarded as a perturbative term. 

Let 
\begin{equation} 
\Gamma (m^2, X) = \sum_{l = l_0}^n \Gamma^{(l)} (m^2, X) 
\label{exp} 
\end{equation} 
be a renormalized and then finite Green function evaluated in theory 
I. In (\ref{exp}), $X$ is the center-of-mass coordinates of the 
\lq\lq reaction'' described by $\Gamma$ and irrelevant arguments 
have been dropped. If we are making a traditional perturbative 
expansion, $\Gamma^{(l)}$ stands for the $O (\lambda^l)$ 
contribution and, if we are making a loop or $\delta$ expansion, 
$\Gamma^{(l)}$ stands for the $l$-loop contribution. The $X$ 
dependence of $\Gamma^{(l)} (m^2, X)$ has come from various sources. 
Among those, the relevant one to the following argument is the 
distribution function (in theory I), which we write 
$f_{\mbox{\scriptsize{I}}} (X; P)$. In fact, $\Gamma^{(l)} (m^2, X)$ 
is written as a sum of integrals over momenta $P$'s, whose 
integrands contain $f_{\mbox{\scriptsize{I}}} (X; P)$, $\partial 
f_{\mbox{\scriptsize{I}}} (X; P) / \partial X_\mu$, etc. In order to 
visualize this dependence, we write (\ref{exp}) as 
\begin{equation} 
\Gamma (m^2, \{ f_I (X) \}) = \sum_{l = l_0}^n \Gamma^{(l)} 
(m^2, \, \{ f_{\mbox{\scriptsize{I}}} (X) \} ) 
\label{exp-1} 
\end{equation} 

Substituting 
\[ 
m^2 = M^2 (X) - \chi (X) , 
\] 
into $\Gamma^{(l)} (m, \{ f_{\mbox{\scriptsize{I}}} (X) \} )$, we 
obtain 
\begin{equation} 
\Gamma^{(l)} (M^2 (X) - \chi (X), \, \{ f_{\mbox{\scriptsize{I}}} 
(X) \} ) . 
\label{exp-2} 
\end{equation} 
As stated in Section VE, in theory II, $\chi (x)$ is one-order 
higher than $M^2 (x)$. More precisely, in traditional perturbative 
expansion, $\chi (x) = O (\lambda)$ and, in loop expansion, $\chi 
(x) = O (\delta)$. Then, we write (\ref{exp-2}) as 
\begin{equation} 
\mbox{Eq.~(\ref{exp-2})} = \Gamma^{(l)} (M^2 (X), \{ 
f_{\mbox{\scriptsize{I}}} (X) \} ) + \sum_{l' = 1}^\infty (- \chi 
(x))^{l'} \frac{\partial^{l'} \Gamma^{(l)} (M^2 (X), \{ 
f_{\mbox{\scriptsize{I}}} (X) \} )}{\partial (M^2 (X))^{l'}} . 
\label{exp-3} 
\end{equation} 
Note that $(- \chi)^{l'} \partial^{l'} / \partial (M^2)^{l'}$ is of 
$(l + l')$th order. 

The distribution functions in theory I and in theory II are 
different. The UV-divergence issue is foreign to them. We can 
rewrite $\Gamma^{(l)} (M^2 (X), \{ f_{\mbox{\scriptsize{I}}} (X) 
\})$ in (\ref{exp-3}), with obvious notation, as 
\begin{eqnarray} 
\Gamma^{(l)} (M^2 (X), \{ f_{\mbox{\scriptsize{I}}} (X) \} ) & = & 
\Gamma^{(l)} (M^2 (X), \{ f_{\mbox{\scriptsize{II}}} (X) \} ) 
\nonumber \\ 
& & + \sum_{l' = 1}^\infty (\{ f_{\mbox{\scriptsize{I}}} (X) \} - \{ 
f_{\mbox{\scriptsize{II}}} (X) \} )^{l'} \frac{\delta^{l'} 
\Gamma^{(l)} (M^2 (X), \{ f_{\mbox{\scriptsize{II}}} (X) \})}{\delta 
(\{ f_{\mbox{\scriptsize{II}}} (X) \})^{l'}} .  
\label{exp-4} 
\end{eqnarray} 
Since, the difference $\chi (x) = M^2 (x) - m^2$ is $O (\lambda)$ 
[$O (\delta)$] for a perturbative [loop] expansion, it is obvious 
that $\{ f_{\mbox{\scriptsize{I}}} (X) \} - \{ 
f_{\mbox{\scriptsize{II}}} (X) \} = O (\lambda)$ [$O (\delta)$]. 
Substituting (\ref{exp-2})~-~(\ref{exp-4}) into (\ref{exp-1}), we 
obtain a new series. Truncating this series at the $n$th order, we 
obtain a perturbation series in theory II. It is obvious from the 
above argument that this series is free from UV divergences. 
\setcounter{equation}{0}
\setcounter{section}{5}
\section*{Appendix E: derivations of (8.53)~-~(8.55)} 
\def\theequation{\mbox{\Alph{section}\arabic{equation}}} 
Here we compute 
\begin{eqnarray} 
{\cal F} (X_0, Y_0; r) & \equiv & - \int_0^\infty d p 
\frac{p \, \sin (p r)}{p^2 - \omega^2} \left[ \frac{\sin \left( 
\sqrt{p^2 - \omega^2} (X_0 + Y_0) \right) - \sin \left( \sqrt{p^2 - 
\omega^2} (X_0 - Y_0) \right)}{\sqrt{p^2 - \omega^2}} \right. 
\nonumber \\ 
& & \mbox{\hspace*{11ex}} \left. - 2 Y_0 \cos \left( \sqrt{p^2 - 
\omega^2} (X_0 - Y_0) \right) \right] 
\label{ryouri} 
\\ 
& = & \frac{1}{4} \int_0^\infty d p \left[ \frac{p}{(p^2 - 
\omega^2)^{3 / 2}} \sum_{\rho, \, \sigma \, \tau = \pm} \rho \sigma 
\tau \, \mbox{exp} \left( i \left[ \rho p r + \sigma \sqrt{p^2 - 
\omega^2} (X_0 + \tau Y_0) \right] \right) \right. \nonumber \\ 
& & \left. - 2 i Y_0 \frac{p}{p^2 - \omega^2} \sum_{\rho, \, \sigma 
= \pm} \rho \, \mbox{exp} \left( i \left[ \rho p r + \sigma 
\sqrt{p^2 - \omega^2} (X_0 - Y_0) \right] \right) \right] . 
\label{ryouri-1} 
\end{eqnarray} 
In the above equations, $X_0 \geq Y_0$ and $\omega$ should be 
understood to be $\omega + i \eta$ $(\eta = 0^+)$. 

Now we pick out from (\ref{ryouri-1}), 
\begin{equation} 
f (r, Z_0) \equiv \int_0^\infty d p \frac{p}{( p^2 - \omega^2)^{3 
/ 2}} \mbox{exp} \left( - i \left[ p r - \sqrt{p^2 - \omega^2} \, 
Z_0 \right] \right) . 
\label{ame-1} 
\end{equation} 
We note that, for $r > Z_0$, 
\begin{equation} 
\oint_{C_-} d p \frac{p}{( p^2 - \omega^2 )^{3 / 2}} \mbox{exp} 
\left( - i \left[ p r - \sqrt{p^2 - \omega^2} \, Z_0 \right] \right) 
= 0 , 
\label{ame} 
\end{equation} 
where $C_-$ is the closed contour in a complex-$p$ plane: 
\[ 
C_- = 0 \to R \to - i R \to 0 \;\;\;\;\;\; (R \to \infty) . 
\] 
Here the contour segment $(0 \to R)$ is on the real axis, $(R \to - 
i R)$ is on $R e^{- i \theta}$ $(\theta = 0 \to \pi / 2)$, and $(- i 
R \to 0)$ is on the imaginary axis. The contribution from the 
contour segment $(R \to - i R)$ to (\ref{ame}) vanishes in the limit 
$R \to \infty$. The contribution from $(- i R \to 0)$ is pure 
imaginary. Then, $f (r, Z_0)$ in (\ref{ame-1}) is pure imaginary. 
Since (\ref{ryouri-1}) is pure real, when (\ref{ame-1}) is used in 
(\ref{ryouri-1}), (\ref{ame-1}) should be canceled by other 
contribution(s), so that we can regard (\ref{ame-1}) as vanishing. 

In the complex-$p$ plane, the integrand of (\ref{ame-1}) has a 
branch point at $p = \omega + i \eta$. We take the branch cut to be 
a straight line $[\omega + i \eta, - \infty + i \eta)$. Then, for $r 
< Z_0$, we have 
\begin{equation} 
\oint_{C_+} d p \frac{p}{( p^2 - \omega^2 )^{3 / 2}} \mbox{exp} 
\left( - i \left[ p r - \sqrt{p^2 - \omega^2} \, Z_0 \right] \right) 
= 0 , 
\label{ame-2} 
\end{equation} 
where the closed contour $C_+$ is defined as 
\begin{eqnarray*} 
C_+ & = & 0 \to R \to i R \to i (\eta + 0^+) \to i (\eta + 0^+) + 
\omega - \alpha \nonumber \\  
& & \to i (\eta - 0^+) + \omega - \alpha \to i (\eta - 0^+) \to 0 
\;\;\;\;\;\;\; (R \to \infty) . 
\end{eqnarray*} 
Here, $(R \to i R)$ is on $R e^{i \theta}$ $(\theta = 0 \to \pi / 
2)$ and $(i (\eta + 0^+) + \omega - \alpha$ $\to$ $i (\eta - 0^+) + 
\omega - \alpha)$ is on a circle of radius $\alpha$, whose center is 
at $p = \omega + i \eta$. From (\ref{ame-2}), we obtain, in the 
limit $\eta \to 0^+$, 
\[ 
f (r, Z_0) = - 2 i \int_0^{\omega - \alpha} d p \, \frac{p e^{- 
i p r}}{(\omega^2 - p^2)^{3 / 2}} \cosh (\sqrt{\omega^2 - p^2} \, 
Z_0) + i \sqrt{\frac{2}{\omega \alpha}} \, e^{- i \omega r} - \pi 
Z_0 e^{- i \omega r} . 
\] 
Here, as above, a pure-imaginary contribution has been dropped. 
Taking the limit $\alpha \to 0^+$, we obtain 
\begin{eqnarray*} 
f (r, Z_0) & = & - i \frac{2}{\omega} \int_0^1 d \xi \, 
\frac{\xi}{(1 - \xi^2)^{3 / 2}} \left[ e^{- i \omega r \xi} \cosh 
(\sqrt{1 - \xi^2} \, \omega Z_0) - e^{- i \omega r} \right] 
\nonumber \\ 
& & + i \frac{2}{\omega} \, e^{- i \omega r} - \pi Z_0 
e^{- i \omega r} . 
\end{eqnarray*} 

Another type of integral in (\ref{ryouri-1}) may be computed 
similarly. Up to a pure-imaginary contribution, we obtain 
\begin{eqnarray*} 
g (r, Z_0) & \equiv & \int_0^\infty d p \, \frac{p}{p^2 - \omega^2} 
\, e^{ - i [p r - \sqrt{p^2 - \omega^2} \, Z_0]} \nonumber \\ 
& = & \theta (Z_0 - r) \left[ i \pi e^{- i \omega r} - 2 \int_0^1 d 
\xi \, \frac{\xi e^{- i \omega r \xi}}{1 - \xi^2} \, \sinh \left( 
\sqrt{1 - \xi^2} \, \omega Z_0\right) \right] . 
\end{eqnarray*} 

With these preliminaries, we can now compute (\ref{ryouri-1}): 
\begin{eqnarray} 
{\cal F} (X_0, Y_0; r) & = & \theta (X_0 + Y_0 - r) \, \theta (r - 
(X_0 - Y_0)) \, \frac{1}{\omega} \left\{ \frac{\pi \omega}{2} (X_0 + 
Y_0) \cos (\omega r) - \sin (\omega r) \nonumber \right. \\ 
& & \left. + \int_0^1 d \xi \frac{\xi}{(1 - \xi^2)^{3 / 2}} \left[ 
\sin (\omega r \xi) \cosh \left( \sqrt{1 - \xi^2} \omega (X_0 + Y_0) 
\right) - \sin (\omega r) \right] \right\} \nonumber \\ 
& & + \theta (X_0 - Y_0 - r) \, \frac{1}{\omega} \left\{ \int_0^1 d 
\xi \frac{\xi \sin (\omega r \xi)}{(1 - \xi^2)^{3 / 2}} \sum_{\tau = 
\pm} \tau \cosh \left( \sqrt{1 - \xi^2} \omega (X_0 + \tau Y_0) 
\right) \right. \nonumber \\ 
& & \left. \left. - 2 \omega Y_0 \int_0^1 d \xi \frac{\xi \sin 
(\omega r \xi)}{1 - \xi^2} \sinh \left( \sqrt{1 - \xi^2} 
\omega (X_0 - Y_0) \right) \right\} \right] . 
\label{iyasa} 
\end{eqnarray} 

It is straightforward to obtain the approximate form for ${\cal F} 
(X_0, Y_0; r)$ in several limits for the arguments: 

${\it 1}$. $X_0 - Y_0 < r < X_0 + Y_0$; $\omega r < \omega (X_0 + 
Y_0) < < 1$:  
\[ 
{\cal F} (X_0, Y_0; r) \simeq \frac{\pi}{2} \left[ X_0 + Y_0 - r 
\left\{ 1 - \frac{\omega^2 (X_0 + Y_0)^2}{4} \right\} \right] . 
\] 

${\it 2}$. $X_0 - Y_0 < r < X_0 + Y_0$; $1 < < \omega (X_0 + Y_0 )$ 
and $\omega r < < \sqrt{\omega (X_0 + Y_0)}$: 
\begin{eqnarray*} 
{\cal F} (X_0, Y_0; r) & \simeq & \frac{\sqrt{2 \pi}}{4} 
\frac{r}{(\omega (X_0 + Y_0))^{3 / 2}} e^{\omega (X_0 + Y_0)} \left[ 
h (\omega (X_0 + Y_0)) - \frac{1}{2} \frac{\omega r^2}{X_0 + Y_0} 
\right] \nonumber \\ 
& & + \frac{\pi}{2} (X_0 + Y_0) \cos \omega r , 
\end{eqnarray*} 
where 
\[ 
h (x) = 1 + \frac{2 1}{8} \frac{1}{x} + ... . 
\] 

${\it 3}$. $r < X_0 - Y_0$; $\omega (X_0 + Y_0 ) < < 1$: 
\[ 
{\cal F} (X_0, Y_0; r) \simeq \frac{\pi}{2} (\omega Y_0)^2 \, r . 
\] 

${\it 4}$. $r < X_0 - Y_0$; $\omega (X_0 - Y_0 ) < < 1 < < \omega 
(X_0 + Y_0)$: 
\[ 
{\cal F} (X_0, Y_0; r) \simeq \frac{\sqrt{2 \pi}}{4} 
\frac{r}{(\omega (X_0 + Y_0))^{3 / 2}} e^{\omega (X_0 + Y_0)} \left[ 
h (\omega (X_0 + Y_0)) - \frac{1}{2} \frac{\omega r^2}{X_0 + Y_0} 
\right] . 
\] 
 

\begin{references} 
\bibitem{sch} 
J. Schwinger, {\it J. Math. Phys.} {\bf 2} (1961), 407; \\ 
L. V. Keldysh, {\it Zh. Eksp. Teor. Fiz.} {\bf 47} (1964), 1515 
({\it Sov. Phys. JETP} {\bf 20} (1965), 1018); \\ 
R. A. Craig, {\it J. Math. Phys.} {\bf 9} (1968), 605; \\ 
J. Rammer and H. Smith, {\it Rev. Mod. Phys.} {\bf 58} (1986), 323; 
\bibitem{chou} K.-C. Chou, Z.-B. Su, B.-L. Hao, and L. Yu, 
{\it Phys. Rep.} {\bf 118} (1985), 1. 
\bibitem{lan} N. P. Landsman and Ch. G. van Weert, {\it Phys. Rep.} 
{\bf 145} (1987), 141. 
\bibitem{ume} H. Umezawa, \lq\lq Advanced Field Theory --- Micro, 
Macro, and Thermal Physics,'' AIP, New York, 1993. 
\bibitem{TFT} Proceedings of Quark Matter 1997, Tsukuba, Japan, to 
appear. 
\bibitem{TFT1} Proceedings of the fifth Workshop on Thermal Field 
Theories and Their Applications, Regensburg, Germany, 1998, to 
appear. 
\bibitem{Gell} M. Gell-Mann and M. L\'evy, {\it Nuovo Cim.} {\bf 16} 
(1960), 705. 
\bibitem{uka} A. Aoki et al., {\it Nucl. Phys. (Proc. Suppl.)} 
{\bf 63} (1998), 397. 
\bibitem{pis} R. D. Pisarski, {\it Phys. Rev. D} {\it 52} (1995), 
R3773; \\ 
S. Chiku and T. Hatsuda, {\it Phys. Rev. D} {\bf 
57} (1998), 6. 
\bibitem{chiku} S. Chiku and T. Hatsuda, {\it Phys. Rev. D} {\bf 58} 
(1998), 076001. 
\bibitem{tsunami} 
K. Rajagopal and F. Wilczek, {\it Nucl. Phys. B} {\bf 404} 
(1993), 577; \\ 
S. Gavin, A. Gocksch, and R. D. Pisarski, {\it Phys. Rev. Lett.} 
{\bf 72} (1994), 2143; \\ 
S. Gavin and B. M\"uller, {\it Phys. Lett. B} {\bf 329} (1994), 486; 
\\ 
Z. Huang and X.-N. Wang, {\it Phys. Rev. D} {\bf 49} (1994), R4335; 
\\ 
J.-P. Blaizot and a. Krzywicki, {\it Phys. Rev. D} {\bf 50} (1994), 
442; \\ 
D. Boyanovsky, H. J. de Vega, and R. Holman, {\it Phys. Rev. D} {\bf 
51} (1995), 734; \\ 
F. Cooper, Y. Kluger, E. Mottola, and J. P. Paz, {\it Phys. Rev. D} 
{\bf 51} (1995), 2377; \\ 
A. Bialas, W. Czyz, and M. Gmyrek, {\it Phys. Rev. D} {\bf 51} 
(1995), 3739; \\ 
M. Asakawa, Z. Huang, and X.-N. Wang, {\it Phys. Rev. Lett.} {\bf 
74} (1995), 3126; \\ 
D. Boyanovsky, M. D'Attanasio, H. J. de Vega, and P. Holman, {\it 
Phys. Rev. D} {\bf 54} (1996), 1748; \\ 
G. Amelino-Camelia, J. D. Bjorken, and S. E. Larsson, {\it Phys. 
Rev. D} {\bf 56} (1997), 6942; \\   
H. Yabu, K. Nozawa, and T. Suzuki, {\it Phys. Rev. D} {\bf 57} 
(1998), 1687; \\ 
D. Boyanovsky, H. J. de Vega, R. Holman, S. P. Kumar, and R. D. 
Pisarski, {\it Phys. Rev. D} {\bf 57} (1998), 3653; \\ 
D. Boyanovsky, F. Cooper, H. J. de Vega, and P. Sodano, {\it Phys. 
Rev. D} {\bf 58} (1998), 025007; \\ 
See also, L. P. Csernai and I. N. Mishustin, {\it Phys. Rev. 
Lett.} {\bf 74} (1995), 5005; \\  
J. Randrup, {\it Phys. Rev. D} {\bf 55} (1997), 1188; \\ 
M. Ishihara, M. Maruyama, and F. Takagi, {\it Phys. Rev. C} {\bf 57} 
(1998), 1440; \\ 
D. Boyanovsky, H. J. de Vega, R. Holman, S. P. Kumar, and R. D. 
Pisarski, hep-ph/9802370, February 1998; \\ 
A. Berera, M. Gleiser, and R. O. Ramos, hep-ph/9803394, March 1998; 
\\ 
E. R. Takano Natti, C.-Y. Lin, A. F. R. de Toledo Piza, and P. L. 
Natti, hep-th/9805060, May 1998. 
\bibitem{bjo} J. D. Bjorken, {\it Acta Phys. Polon. B} {\bf 28} 
(1997), 2773. 
\bibitem{hu} 
E. Calzetta and B. L. Hu, {\it Phys. Rev. D} {\bf 37} (1988), 2878. 
\bibitem{le-b} 
M. Le Bellac, \lq\lq Thermal Field Theory,'' Cambridge University 
Press, Cambridge, 1996. 
\bibitem{ume1} 
Y. Yamanaka, H. Umezawa, K. Nakamura and T. Arimitsu, {\it Int. J. 
Mod. Phys. A} {\bf 9} (1994), 1153; \\ 
Y. Yamanaka and K. Nakamura, {\it Mod. Phys. Lett. A} {\bf 9} 
(1994), 2879; \\ 
H. Chu and H. Umezawa, {\it Int. J. Mod. Phys. A} {\bf 9} (1994), 
1703; 2363. Earlier work is quoted in these papers. 
\bibitem{IZ} C. Itzykson and J.-B. Zuber, \lq\lq Quantum Field 
Theory,'' McGraw-Hill, New York, 1980. 
\bibitem{muta} T. Muta, \lq\lq Foundations of Quantum 
Chromodynamics,'' World Scientific, Singapore, 1987. 
\bibitem{fau} R. Fauser and H. H. Walter, {\it Nucl. Phys. A} 
{\bf 584} (1995), 604. 
\bibitem{pis-f} R. D. Pisarski, {\it Phys. Rev. Lett.} {\bf 63} 
(1989), 1129; \\ 
E. Braaten and R. D. Pisarski, {\it Nucl. Phys. B} {\bf 337} 
(1990), 569; \\ 
J. Frenkel and J. C. Taylor, {\it Nucl. Phys. B} {\bf 334} (1990), 
199. 
\bibitem{thoma} 
M. E. Carrington, H. Defu, and M. H. Thoma, to be published in {\it 
Eur. phys. J. C}, (hep-ph/9708363, August 1997); in \cite{TFT}, 
(hep-ph/9808359, August, 1998). 
\bibitem{nie10} 
A. Ni\'egawa, {\it Phys. Lett. B} {\bf 247} (1990), 351; \\ 
N. Ashida, H. Nakkagawa, A. Ni\'egawa, and H. Yokota, {\it Ann. 
Phys. (N.Y.)} {\bf 215} (1992), 315 [E: {\bf 230} (1994), 161]; 
{\it Phys. Rev. D} {\bf 45} (1992), 2066; \\ 
Asida, N., {\it Int. J. Mod. Phys. A} {\bf 8} (1993), 1729; \\ 
P. V. Landshoff, {\it Phys. Lett. B} {\bf 386} (1996), 291; \\ 
A. Ni\'egawa, {\it Phys. Rev. D} {\bf 57} (1998), 1379; \\ 
See also, H. A. Weldon, {\it Phys. Rev. D} {\bf 28} (1983), 2007. 
\bibitem{mou} H. Matsumoto, I. Ojima, and H. Umezawa, 
{\it Ann. Phys. (N.Y.)} {\bf 152} (1984), 348; \\ 
A. J. Niemi and G. W. semenoff, 
{\it Nucl. Phys. B} {\bf 230} [FS10] (1984), 181. 
\bibitem{aure} M. A. van Eijck and Ch. van Weert, {\it Phys. Lett. 
B} {\bf 278} (1992), 305; \\ 
P. Aurenche and T. Becherrawy, {\it Nucl. Phys. B} {\bf 379} (1992), 
259. 
\bibitem{nie} A. Ni\'egawa, {\it Phys. Rev. D} {\bf 40} (1989), 
1199; \\ 
R. J. Furnstahl and B. D. Serot, {\it Phys. Rev. C} {\bf 44} (1991), 
2141; \\ 
F. Gelis, {\it Z. Phys. C} {\bf 70} (1996), 321; \\ 
H. Mabilat, {\it Z. Phys. C} {\bf 75} (1997), 155. 
\bibitem{nie-pl} A. Ni\'egawa, {\it Phys. Lett. B} {\bf 416} (1998), 
137. 
\bibitem{Hei} P. Zhuang and U. Heinz, {\it Phys. Rev. D} {\bf 57} 
(1998), 6525; \\ 
S. Ochs and U. Heinz, {\it Ann. Phys. (N.Y.)} {\bf 266} (1998), 351. 
\bibitem{lee} B. W. Lee, {\it Nucl. Phys. B} {\bf 9} (1969), 649; \\ 
T. Kugo, {\it Prog. Theor. Phys.} {\bf 57} (1977), 593. 
\bibitem{wein} S. Weinberg, {\it Phys. Rev. D} {\bf 9} (1974), 
3357; \\ 
L. Dolan and R. Jackiw, {\it Phys. Rev. D} {\bf 9} (1974), 3320; \\ 
D. A. Kirzhnits and A. D. Linde, {\it Ann. Phys. (N.Y.)} {\bf 101} 
(1976), 195. 
\bibitem{fen} P. Fendley, {\it Phys. Lett. B} {\bf 196} (1987), 175. 
\bibitem{law} I. D. Lawrie, in \cite{TFT1}, (hep-ph/9809469 
September, 1998). Earlier work is quoted therein. 
\bibitem{jak} R. Jackiw, {\it Phys. Rev. D} {\bf 9} (1974), 1686. 
\bibitem{fuji} Y. Fujimoto, R. Grigjanis, and R. Kobes, {\it Prog. 
Theor. Phys.} {\bf 73} (1985), 434. Earlier work is quoted 
therein. 
\bibitem{eva} T. S. Evans, {\it Z. Phys. C} {\bf 36} (1987), 153. 
\bibitem{hon} J. W. Negele and H. Orland, \lq\lq Quantum 
Many-Particle Systems,'' Addison-Wesley, New York, 1988. 
\bibitem{ste} P. M. Stevenson, Phys. Rev. D23 (1981) 2916. 
\bibitem{grunberg} G. Grunberg, {\it Phys. Lett.} {\bf 95B} (1980), 
70. 
\bibitem{weinberg} E. J. Weinberg and A. Wu, {\it Phys. Rev. D} 
{\bf 36} (1987), 2474. 
\bibitem{morikawa} M. Morikawa, {\it Prog. Theor. Phys.} {\bf 93} 
(1995), 685; \\ 
C. Greiner and M. M\"uller, {\it Phys. Rev. D} {\bf 
55} (1997), 1026; \\ 
A. Berera, M. Gleiser, and R. O. Ramos, in \cite{tsunami}. 
\bibitem{ram} J. Rammer and H. Smith, {\it Rev. Mod. Phys.} {\bf 58} 
(1986), 323. 
\bibitem{alt1} T. Altherr, {\it Phys. Lett. B} {\bf 341} (1995), 
325. 
\bibitem{kobes} R. Kobes, {\it Phys. Rev. D} {\bf 43} (1991), 1269. 
\end{references}
\end{document}